\documentclass[12pt,preprint]{emulateapj}

\shorttitle{ALMA multiple-transition molecular line observations of 
IRAS 20551-4250} 
\shortauthors{Imanishi et al.}

\begin{document}


\title{
ALMA Multiple-Transition Molecular Line Observations of the
Ultraluminous Infrared Galaxy IRAS 20551$-$4250: 
Different HCN, HCO$^{+}$, HNC Excitation and Implications for 
Infrared Radiative Pumping}


\author{Masatoshi Imanishi \altaffilmark{1}}
\affil{National Astronomical Observatory of Japan, National Institutes 
of Natural Sciences (NINS), 2-21-1 Osawa, Mitaka, Tokyo 181-8588, Japan}  
\email{masa.imanishi@nao.ac.jp}

\author{Kouichiro Nakanishi \altaffilmark{1}}
\affil{National Astronomical Observatory of Japan, National Institutes 
of Natural Sciences (NINS), 2-21-1 Osawa, Mitaka, Tokyo 181-8588, Japan}

\and

\author{Takuma Izumi \altaffilmark{2}}
\affil{National Astronomical Observatory of Japan, National Institutes 
of Natural Sciences (NINS), 2-21-1 Osawa, Mitaka, Tokyo 181-8588, Japan}

\altaffiltext{1}{Department of Astronomical Science, The Graduate
University for Advanced Studies (SOKENDAI), Mitaka, Tokyo 181-8588,
Japan} 

\altaffiltext{2}{Institute of Astronomy, School of Science, The
University of Tokyo, 2-21-1 Osawa, Mitaka, Tokyo 181-0015, Japan}


\begin{abstract}
We present our ALMA multi-transition molecular line observational
results for the ultraluminous infrared galaxy, IRAS 20551$-$4250,
which is known to contain a luminous buried AGN and shows detectable
vibrationally excited (v$_{2}$=1f) HCN and HNC emission lines. 
The rotational J=1--0, 4--3, and 8--7 of HCN, HCO$^{+}$, and HNC
emission lines were clearly detected at a vibrational ground level (v=0). 
Vibrationally excited (v$_{2}$=1f) J=4--3 emission lines were detected
for HCN and HNC, but not for HCO$^{+}$.
Their observed flux ratios further support our previously obtained
suggestion, based on J=3--2 data, that (1) infrared radiative pumping
plays a role in rotational excitation at v=0, at least for HCN and HNC,
and (2) HCN abundance is higher than HCO$^{+}$ and HNC.
The flux measurements of the isotopologue H$^{13}$CN, H$^{13}$CO$^{+}$,
and HN$^{13}$C J=3--2 emission lines support the higher HCN abundance
scenario.
Based on modeling with collisional excitation, we constrain the
physical properties of these line-emitting molecular gas, but find that
higher HNC rotational excitation than HCN and HCO$^{+}$ is
difficult to explain, due to the higher effective critical density of HNC. 
We consider the effects of infrared radiative pumping using the
available 5--30 $\mu$m infrared spectrum and find that our observational
results are well explained if the radiation source is located at
30--100 pc from the molecular gas.  
The simultaneously covered very bright CO J=3--2 emission line displays
a broad emission wing, which we interpret as being due to molecular outflow
activity with the estimated rate of $\sim$150 M$_{\odot}$ yr$^{-1}$.  
\end{abstract}


\section{Introduction}

The ubiquity of supermassive black holes (SMBH) at the center of galaxy
stellar spheroidal components and the mass correlation between 
SMBHs and spheroidal stars suggest that SMBHs are an important
ingredient of galaxies \citep{mag98,fer00,gul09,mcc13}.  
In the currently widely accepted cold dark matter-based galaxy formation
scenario, small gas-rich galaxies collide and merge, and then grow into
more massive galaxies \citep{whi78}.
Numerical simulations of such merging processes of gas-rich galaxies
containing SMBHs at their centers have been extensively performed, and
it has been argued that active star formation and mass accretion onto
central SMBHs occur in highly obscured regions during an infrared
luminous phase \citep{hop05,hop06,hop08,deb11}.   

Active mass accretion onto SMBHs emits strong radiation and is observed 
as active galactic nucleus (AGN) activity. 
Luminous AGNs deeply buried in gas/dust-rich infrared luminous merging 
galaxies are now thought to play an essential role in galaxy
formation, through feedback to galaxies
\citep{gra04,spr05,dim05,rob06,sij07,hop08,cio10}; however, observational
understanding of such buried AGNs is not easy, due to dust extinction. 
We must establish a method to detect and investigate the
properties of buried AGNs, by separating these from the surrounding starburst
emission.
Observing at wavelengths where the effects of dust extinction are
small is clearly one of the best ways to study dust-obscured energy sources. 

Molecular rotational J-transition emission line flux ratios at the
(sub)millimeter wavelength can be a powerful tool to study buried
energy sources, because (1) dust extinction is typically negligible,
unless the column density of obscuring material is very high (e.g.,
N$_{\rm H}$ $>>$ 10$^{25}$ cm$^{-2}$) and (2) some molecular lines are
argued to become good signatures of AGN activity. 
In particular, high dipole moment molecules, such as HCN, HCO$^{+}$,
and HNC, are better suited than the widely used low-J transition CO emission
lines to investigate physical properties around hidden energy sources
because nuclear molecular gas in the vicinity of active star formation
and AGN activity is usually in a dense form with $>$10$^{4}$ cm$^{-3}$. 
For example, it was proposed that optically selected AGNs and starbursts
show different molecular line flux ratios, in such a way that in AGNs, 
HCN rotational J-transition emission lines are enhanced, relative to  
HCO$^{+}$ \citep{koh05,kri08}.
Based on pre-ALMA and ALMA observations of the nuclei of
gas/dust-rich luminous infrared galaxies (LIRGs; infrared luminosity
L$_{\rm IR}$ $>$ 10$^{11}$ L$_{\odot}$), which are diagnosed to contain
optically detectable AGNs or optically elusive, but infrared/X-ray
detectable AGNs or no detectable AGNs, it was demonstrated that
(sub)millimeter molecular emission line flux ratios indeed work to
detect the signs of deeply buried AGNs in these LIRGs 
\citep{ima04,ima06,in06,ima07a,ima09a,cos11,ima13a,ima13b,ion13,ima14,izu15,pri15,izu16,ima16a,ima16b,ima16c}.  
Thus, these (sub)millimeter molecular line observations have potential
in the systematic investigation of buried AGNs in gas/dust-rich LIRGs not
only in the local universe, but also in the distant universe, thanks to
the advent of the highly sensitive ALMA observing facility in this
wavelength range.

However, the physical origin of the strong HCN J-transition line
emission in AGNs remains to be fully understood.
An HCN abundance enhancement in molecular gas in the close vicinity of
a buried AGN is a natural explanation for the strong HCN emission 
\citep{yam07,izu16}.
While this high HCN abundance scenario in molecular gas, largely
affected by AGN radiation, is predicted in some parameter range by
chemical calculations \citep{mei05,lin06,har10}, it is not necessarily
true that the HCN abundance is {\it always} higher than HCO$^{+}$
around an AGN \citep{mei05,har13}.
Higher HCN rotational J-excitation in an AGN than a normal
starburst is an alternative explanation because the AGN's higher radiative
energy generation efficiency can increase the temperature of the surrounding
molecular gas and can excite HCN (higher critical density than
HCO$^{+}$ under the same line opacity) more than in a normal starburst
\citep{ima16c}. 
Multiple rotational J-transition line observations are required to
disentangle the abundance and excitation effects \citep{ima16c}.
Flux attenuation by line opacity (not dust extinction) is another
uncertain factor to discuss the hidden energy sources based on 
{\it observed} molecular line flux ratios \citep{cos15}.
Optically thin isotopologue molecular line observations will help us
to estimate these line opacity effects for the main bright molecular
emission lines.

Another possible good AGN indicator is vibrationally excited molecular
emission lines whose energy levels are much higher than the widely
investigated rotationally excited emission lines at a vibrational ground
level (v=0).  
The vibrationally excited (v$_{2}$=1, l=1f; hereafter v$_{2}$=1f)
emission lines of HCN and HNC have recently been detected in several
LIRGs \citep{sak10,ima13b,aal15a,aal15b,cos15,ima16b,ima16c,mar16}.
Since the energy levels of these vibrationally excited states are 
$\sim$1030 K (670 K) for HCN (HNC), it is very difficult to
excite by collision; however, an infrared radiative pumping process can
achieve this, by absorbing $\sim$14 $\mu$m ($\sim$22 $\mu$m) infrared photons
\citep{aal95,sak10}.
Because an AGN can emit mid-infrared (3--30 $\mu$m) continuum emission
more efficiently than a starburst with the same bolometric luminosity
due to AGN-heated hot ($>$100 K) dust emission, if the vibrationally
excited emission lines are strongly detected in LIRGs, then an obscured 
AGN is a plausible origin as the strong infrared continuum emitting
source to vibrationally excite HCN and HNC \citep{aal15b}. 
However, an extreme starburst with a very compact
size remains another possibility \citep{aal15b}.

The ultraluminous infrared galaxy (ULIRG), IRAS 20551$-$4250, 
with infrared luminosity L$_{\rm IR}$ $\sim$ 10$^{12}$L$_{\odot}$ at 
z=0.043 (Table 1) is one such galaxy, where the vibrationally excited
(v$_{2}$=1f) HCN and HNC emission lines have been clearly detected, due to
small observed molecular line widths \citep{ima13b,ima16b}.
The observed HCN-to-HCO$^{+}$ flux ratios at J=3--2 and J=4--3 have been
found to be substantially larger than in starburst-dominated regions
\citep{ima13b,ima16b}. 
This galaxy displays a long merging tail in the southern direction
from the main single nucleus \citep{duc97,rot04}.
Based on optical emission line flux ratios, it is classified as a
LINER/HII-region \citep{duc97}, whereas \citet{yua10} classified it as a 
starburst-AGN composite or HII-region type, depending on the emission
lines used.   
The presence of a buried AGN, which could explain 20--60\% of
the bolometric luminosity, has been suggested, in addition to starburst
activity, based on infrared and X-ray observations 
\citep{fra03,ris06,san08,nar08,nar09,nar10,ima10a,ima11,vei13}.  
Since IRAS 20551$-$4250 displays bright molecular rotational
J-transition emission lines \citep{ima13b,ima16b}, this is an
interesting and valuable object to improve our understanding of the
physical origin of observed molecular emission line properties, based on
multiple rotational J-transition lines for multiple molecules. 

In this study, we present our new ALMA observational results in bands 3
(84--116 GHz), 7 (275--373 GHz), and 9 (602--720 GHz) of the ULIRG, IRAS
20551$-$4250.
The J=1--0, J=4--3, and J=8--7 emission lines of HCN, HCO$^{+}$, and
HNC are covered in bands 3, 7, and 9, respectively.
For J=4--3, vibrationally excited (v$_{2}$=1f) emission lines
were also observed for HCN, HCO$^{+}$, and HNC.   
For J=8--7, vibrationally excited (v$_{2}$=1f) HCN and HNC emission
lines were included in our band 9 data
\footnote{A vibrationally excited J=1--0 line is not present.}.
ALMA band 6 (211--275 GHz) observations of isotopologue lines, 
H$^{13}$CN, H$^{13}$CO$^{+}$, and HN$^{13}$C J=3--2, were also conducted
and their results are included. 
We adopt H$_{0}$ $=$ 71 km s$^{-1}$ Mpc$^{-1}$, $\Omega_{\rm M}$ = 0.27,
and $\Omega_{\rm \Lambda}$ = 0.73 \citep{kom09}, to be consistent with
our previously published ALMA papers for this galaxy.
The physical scale at $z =$ 0.043 is 0.84 kpc arcsec$^{-1}$.
In the absence of a statement about vibrational level, we mean the
vibrational ground level (v=0).

\section{Observations and Data Analysis}

Band 7 (275--373 GHz), 3 (84--116 GHz), 9 (602--720 GHz), and 
6 (211--275 GHz) observations were conducted through our 
ALMA Cycle 2 program 2013.1.00033.S (PI = M. Imanishi), 
Cycle 3 program 2015.1.00028.S (PI = M. Imanishi), 
Cycle 3 program 2015.1.00028.S (PI = M. Imanishi), and 
Cycle 4 program 2016.1.00051.S (PI = M. Imanishi), respectively.
The widest 1.875 GHz band mode and 3840 total channel number were
employed for all observations.
To reduce the data rate, online spectral averaging with a factor of 2
or 4 was applied for some observations. 
Table 2 summarizes these ALMA observations.

For our ALMA band 7 observations, we covered HCO$^{+}$ J=4--3 
(rest-frame frequency is $\nu_{\rm rest}$ = 356.734 GHz), HCN
v$_{2}$=1f J=4--3 ($\nu_{\rm rest}$ = 356.256 GHz), and HCO$^{+}$
v$_{2}$=1f J=4--3 ($\nu_{\rm rest}$ = 358.242 GHz), but had to exclude 
the HCN J=4--3 line ($\nu_{\rm rest}$ = 354.505 GHz), due to the limited
frequency coverage of the ALMA system.  
HCN J=4--3 and HCO$^{+}$ J=4--3 (v=0) lines were observed in our ALMA
Cycle 0 observations \citep{ima13b}.
The bright HCO$^{+}$ J=4--3 (v=0) emission line can be used for
inter-calibration between Cycle 0 and 2 data, by correcting for possible 
absolute flux calibration uncertainty during individual ALMA
observations. 
The very bright CO J=3--2 ($\nu_{\rm rest}$ = 345.796 GHz) emission line
was also included in our Cycle 2 observations.
HNC J=4--3 emission lines at v=0 ($\nu_{\rm rest}$ = 362.630 GHz) and
v$_{2}$=1f ($\nu_{\rm rest}$ = 365.147 GHz) were obtained
independently from the HCO$^{+}$ J=4--3 observations.

For band 9, HCN J=8--7 ($\nu_{\rm rest}$ = 708.877 GHz), 
HCO$^{+}$ J=8--7 ($\nu_{\rm rest}$ = 713.341 GHz), and 
HNC J=8--7 ($\nu_{\rm rest}$ = 725.107 GHz) lines were observed.
The vibrationally-excited 
HCN v$_{2}$=1f J=8--7 ($\nu_{\rm rest}$ = 712.372 GHz) and 
HNC v$_{2}$=1f J=8--7 ($\nu_{\rm rest}$ = 730.131 GHz) lines were also
covered, but HCO$^{+}$ v$_{2}$=1f J=8--7 ($\nu_{\rm rest}$ = 716.354
GHz) line was not.
Our scientific aim was to measure the strengths of the vibrational
ground (v=0) J=8--7 emission lines of HCN, HCO$^{+}$, HNC;
vibrationally excited (v$_{2}$=1f) emission lines were our second
objective. 

In bands 7 and 9, vibrationally excited (v$_{2}$=1, l=1e; hereafter 
v$_{2}$=1e) J=4--3 and J=8--7 lines, respectively, were covered for
HCN, HCO$^{+}$, and HNC. 
However, these frequencies are so close to the bright vibrational ground
(v=0) emission lines for HCN, HCO$^{+}$, and HNC that we could not
extract the faint v$_{2}$=1e emission line components in a reliable
manner. 
These v$_{2}$=1e emission line fluxes will not be discussed in this
paper. 

For band 3, HCN J=1--0 ($\nu_{\rm rest}$ = 88.632 GHz), 
HCO$^{+}$  J=1--0 ($\nu_{\rm rest}$ = 89.189 GHz), and 
HNC J=1--0 ($\nu_{\rm rest}$ = 90.664 GHz) line data were obtained.

For band 6, we targeted H$^{13}$CN ($\nu_{\rm rest}$ = 259.012 GHz), 
H$^{13}$CO$^{+}$ ($\nu_{\rm rest}$ = 260.255 GHz), and HN$^{13}$C J=3--2
($\nu_{\rm rest}$ = 261.263 GHz), because these isotopologue emission
lines are thought to be optically thin and thus can be used to estimate
possible flux attenuation by line opacity (not dust extinction) for the
previously-obtained HCN, HCO$^{+}$, and HNC J=3--2 emission lines 
\citep{ima16b}. 
The bright CS J=5--4 line ($\nu_{\rm rest}$ = 244.936 GHz) was included
in this band 6 observation.

We performed data analysis in the same way as for our
previously obtained ALMA data of IRAS 20551$-$4250
\citep{ima13b,ima16b}. 
We retrieved data calibrated by ALMA and used CASA
(https://casa.nrao.edu) for further data reduction. 
For the spectral window that includes the very bright CO J=3--2
emission line, we employed self-calibration, using the CO J=3--2 emission
line itself for phase calibration. 
Except for this spectral window, we adopted results produced with a
standard phase calibration using phase calibrators, which were provided
by ALMA.   
We first checked the visibility plots to view the signatures of bright
emission lines.  
To estimate the continuum flux levels, we removed channels that 
contained discernible emission lines.
We then subtracted the derived continuum levels, to extract only molecular
line data.
The task ``clean'' was then applied for the molecular line data, by
binning spectral channels to make the velocity resolution 
20--40 km s$^{-1}$.  
Pixel scale was set as 0$\farcs$1 pixel$^{-1}$ for band 3 and 7 data,
but was 0$\farcs$03 pixel$^{-1}$ for band 9 and 6 data, because their
beam sizes were much smaller than those of bands 3 and 7.
The ``clean'' task was also applied for continuum data. 
We then obtained spectra at the nuclear position defined from the
continuum peaks in individual observations. 
When the flux density levels in the spectra were significantly below
zero at the frequency where no emission and/or absorption lines were
expected to be present, continuum was
over-subtracted, possibly due to the inclusion of weak emission lines for
continuum determination. 
In such cases, we redefined line-free channels and created clean maps
of molecular emission lines and continuum.
After confirming that the extracted spectra at line-free channels at
the continuum peak position show flux density fluctuating around zero
level with noise, we adopted these re-analyzed results as final
products. 

\section{Result}

Continuum-J43a (taken with HCO$^{+}$ J=4--3), continuum-J43b 
(taken with HNC J=4--3), continuum-J87 (taken with HCN, HCO$^{+}$, HNC
J=8--7),  
continuum-J10 (taken with HCN, HCO$^{+}$, HNC J=1--0), and 
continuum-J32 (taken with H$^{13}$CN, H$^{13}$CO$^{+}$, HN$^{13}$C J=3--2)
maps are displayed in Figure 1.
In all maps, continuum emission is dominated by a spatially compact
component at the nucleus of IRAS 20551$-$4250.
Table 3 lists continuum fluxes at the peak position.
We had two continuum measurements in band 7 with
slightly different central frequencies ($\nu_{\rm cent}$ = 336.7 GHz
and 343.2 GHz; Table 3).
Both of these provided comparable flux levels with 10.5 and 10.3 (mJy
beam$^{-1}$) (Table 3). 
Band 7 continuum measurements of IRAS 20551$-$4250 were made in
ALMA Cycle 0, and the estimated fluxes were 10.1 (mJy beam$^{-1}$) 
($\nu_{\rm cent}$ = 341.6 GHz) and 9.4 (mJy beam$^{-1}$)
($\nu_{\rm cent}$ = 346.7 GHz) \citep{ima13b}.
These band 7 continuum measurements at similar central frequencies, taken
in ALMA Cycles 0 and 2, agree within $\sim$10\%, supporting the quoted
$<$10\% absolute flux calibration uncertainty in individual ALMA
observations (ALMA Proposer's Guide for Cycles 0 and 2). 

Spectra in bands 7, 9, 3, and 6 at the continuum peak positions within
the beam size of individual data are presented in Figure 2.
The targeted bright emission lines, such as CO J=3--2 line, J=8--7,
J=4--3, and J=1--0 lines of HCN, HCO$^{+}$, HNC, and J=3--2 lines of
H$^{13}$CN, H$^{13}$CO$^{+}$, and HN$^{13}$C were clearly detected.
Additionally, signatures of several fainter emission lines,
including v$_{2}$=1f J=4--3 emission lines of HCN, HCO$^{+}$, HNC and
other serendipitously detected emission lines, were observed.
Because gas-rich (U)LIRGs exhibit many faint
emission lines from molecules and the bulk of the observed frequency
range could be occupied by such lines, particularly in band 7
\citep{cos15}, identifying the detected faint emission lines is
not an easy task. 
Our proposed identifications of faint emission lines are indicated with
arrows in Figure 2. 
In Figure 2(m), the isotopologue HC$^{15}$N J=3--2 line 
($\nu_{\rm rest}$ = 258.157 GHz) is covered and is expected to be
redshifted to the observed frequency of $\nu_{\rm obs}$ $\sim$ 247.5
GHz, just lower than the frequency of the SO emission.
However, its signature is not as clear as that of the other isotopologue
line, H$^{13}$CN J=3--2.
This is reasonable because, in (U)LIRGs, the
$^{14}$N-to-$^{15}$N  abundance ratio ($\sim$440) \citep{wan16} is much
higher than the $^{12}$C-to-$^{13}$C abundance ratio (50--100)
\citep{hen93a,hen93b,mar10,hen14}, so that the HC$^{15}$N J=3--2
emission line is expected to be weaker than H$^{13}$CN J=3--2 by a large
factor. 

For molecular emission lines that are recognizable in the spectra, we 
created integrated intensity (moment 0) maps, by summing spectral
elements displaying discernible signals. 
These maps are shown in Figure 3 for the primarily targeted main emission
lines, and in Appendix A (Figure 17) for serendipitously detected
emission lines. 
All detected molecular lines showed peak positions that agree
with those of the simultaneously taken continuum emission  
within 1 pixel (0$\farcs$1 for bands 7 and 3, or 0$\farcs$03 for bands 9
and 6) in both RA and DEC directions.
These agreements suggest that the serendipitously detected faint
emission lines are likely to be real features, rather than artifacts,
with the exception of HCO$^{+}$ v$_{2}$=1f J=4--3 (band 7), HCN
v$_{2}$=1f J=8--7 (band 9), and HNC v$_{2}$=1f J=8--7 (band 9), which
will be discussed later.  

Figure 4 shows magnified spectra in the vicinity of the primarily targeted
individual molecular emission lines at the continuum peak position
within the beam size, together with the best Gaussian fits. 
The same figures for selected serendipitously detected emission lines
are shown in Appendix A (Figure 18).
Peak flux values in the moment 0 maps and emission line fluxes estimated
from the best Gaussian fits are summarized in Table 4.
Table 5 shows the deconvolved, intrinsic emission sizes of the main
bright molecular lines and continuum, estimated using
the CASA task ``imfit''.

HCO$^{+}$ J=4--3 and HNC J=4--3 (v=0) fluxes were obtained in both our
ALMA Cycle 0 and 2 observations. 
Our ALMA Cycle 0 data provided HCO$^{+}$ J=4--3 and HNC J=4--3 (v=0) 
fluxes of 14$\pm$1 and 5.8$\pm$0.2 (Jy km s$^{-1}$), respectively,
based on Gaussian fit, within 0.6$''$ $\times$ 0.4$''$ beam size 
\citep{ima13b}.
The estimated fluxes based on Gaussian fit in our ALMA Cycle 2 data were
17$\pm$1 and 6.6$\pm$0.2 (Jy km s$^{-1}$) for HCO$^{+}$ J=4--3 (0.9$''$
$\times$ 0.7$''$ beam) and HNC J=4--3 (0.8$''$ $\times$ 0.6$''$ beam),
respectively. 
The fluxes in ALMA Cycle 2 data were 10--20\% higher than in our ALMA
Cycle 0 data.
This could be partly due to larger beam sizes in the ALMA Cycle 2 data, 
if a significant fraction of these emission lines came from a 
spatially extended region with $>$0$\farcs$5 ($>$400 pc).
However, the nuclear HCO$^{+}$ J=4--3 and HNC J=4--3 emission components
were estimated to be spatially compact (Table 5).
The 10--20\% flux discrepancy could be largely accounted for by the
possible absolute flux calibration uncertainty in individual ALMA Cycle
0 and 2 data (maximum $\sim$10\% for each).  

The HCN J=4--3 (v=0) line was not covered in our ALMA Cycle 2 observations.
In ALMA Cycle 0, the HCN J=4--3 and HCO$^{+}$ J=4--3 lines were
simultaneously observed.
For the HCO$^{+}$ J=4--3 (v=0) line, our Cycle 2 band 7 data
(17$\pm$1 Jy km s$^{-1}$) provided $\sim$22\% higher absolute
flux than Cycle 0 (14$\pm$1 Jy km s$^{-1}$).
We thus multiplied the HCN J=4--3 (v=0) flux derived in our Cycle 0
observations (9.5$\pm$0.2 Jy km s$^{-1}$) \citep{ima13b} by
a factor of 1.22, and adopted the re-calibrated HCN J=4--3 flux with
11.6$\pm$0.2 (Jy km s$^{-1}$) for all subsequent discussion in 
this paper.  

The HCN v$_{2}$=1f J=4--3 emission line was also observed in both ALMA Cycles 
0 and 2. 
Its detection significance in the moment 0 map (Figure 3) of our Cycle 2
data was $>$9$\sigma$, which is improved from the $\sim$5$\sigma$
detection in our Cycle 0 moment 0 map \citep{ima13b}, and further
confirms the presence of the detectable HCN v$_{2}$=1f J=4--3 emission line
in IRAS 20551$-$4250.  
The estimated HCN v$_{2}$=1f J=4--3 emission line flux, based on Gaussian
fit, in the Cycle 0 data was 0.39$\pm$0.07 (Jy km s$^{-1}$), which is
converted to 0.47$\pm$0.08 (Jy km s$^{-1}$), after the factor of 1.22
multiplication, because HCN v$_{2}$=1f J=4--3 and HCO$^{+}$ J=4--3 (v=0) 
line data were taken simultaneously in our ALMA observations. 
That in the Cycle 2 data is 0.60$\pm$0.13 (Jy km s$^{-1}$).
The flux measurements for the Cycle 0 and 2 data are consistent. 
Given deeper Cycle 2 data than Cycle 0 data, we adopted the 
HCN v$_{2}$=1f J=4--3 flux estimated with Gaussian fit in the Cycle 2
data (0.60$\pm$0.13 Jy km s$^{-1}$). 

The HNC v$_{2}$=1f J=4--3 emission line for IRAS 20551$-$4250 was first
covered in our ALMA Cycle 2 observations, and its signature is shown in
the band 7 spectrum in Figures 2f and 4. 
Based on the moment 0 map (Figure 3) and Gaussian fit (Table 4), we
regard that the HNC v$_{2}$=1f J=4--3 emission line was detected.
However, the line width of the HNC v$_{2}$=1f J=4--3 emission line (FWHM
$\sim$ 440 km s$^{-1}$) is substantially larger than other bright
molecular emission lines with FWHM $\sim$ 200 km s$^{-1}$ (Table
4). Contamination from other faint molecular emission lines is
possible. Thus, the measured HNC v$_{2}$=1f J=4--3 
emission line flux should be taken as an upper limit.
If we assume the intrinsic line width of the HNC v$_{2}$=1f J=3--2
emission line to be FWHM $\sim$ 200 km s$^{-1}$, then the actual HNC
v$_{2}$=1f J=3--2 emission line flux will be about half of that shown in
Table 4. 

For the HCO$^{+}$ v$_{2}$=1f J=4--3 emission line, we barely see the
3.2$\sigma$ peak at the nuclear position in the moment 0 map (Figure 3).
However, the contour size at the peak was significantly smaller than the
synthesized beam size and emission with similar 3$\sigma$ level contours 
was seen at some other off-nuclear regions (Figure 3), making 
the existence of this 3.2$\sigma$ peak uncertain. 
In Figure 2d, an emission-like feature may be present close to the
expected frequency of the HCO$^{+}$ v$_{2}$=1f J=4--3 line; however, 
its observed peak seems to be slightly offset from the expected frequency at 
z= 0.043 (shown as a downward arrow). 
Based on Gaussian fit, the peak velocity of this emission-like feature
(v$_{\rm opt}$ =  12947$\pm$34 km s$^{-1}$) was significantly higher
than other bright molecular emission lines detected in band 7 
(v$_{\rm opt}$ $\sim$ 12900 km s$^{-1}$), and 
the detection significance was $<$2$\sigma$ (Table 4).
Detection of the HCO$^{+}$ v$_{2}$=1f J=4--3 emission line has never
been reported in any external galaxy. 
We require higher S/N data to improve the constraining of the strength of this
yet-to-be-detected HCO$^{+}$ v$_{2}$=1f J=4--3 emission line.

The HCN v$_{2}$=1f J=8--7 ($\nu_{\rm rest}$ = 712.372 GHz) and HNC
v$_{2}$=1f J=8--7 ($\nu_{\rm rest}$ = 730.131 GHz) lines were
covered in our ALMA band 9 spectrum, while HCO$^{+}$ v$_{2}$=1f J=8--7 
($\nu_{\rm rest}$ = 716.354 GHz) line was not.
In Figure 2(h) and (i), subtle signatures of emission lines are observed
close to the expected frequencies of redshifted HCN v$_{2}$=1f J=8--7
and HNC v$_{2}$=1f J=8--7 lines, respectively. 
However, their detection significance, based on Gaussian fit, is
$<$3$\sigma$.
In the moment 0 map of HCN v$_{2}$=1f J=8--7, we see a 3.1$\sigma$
emission peak at the nuclear position, but the 3$\sigma$ contour was much
smaller than the synthesized beam size (Figure 3). 
Detection of v$_{2}$=1f J=8--7 lines of HCN and HNC was unclear, and we
will need data with higher S/N ratios to discuss their fluxes quantitatively. 

The H$^{13}$CN J=3--2 emission lines were observed in ALMA Cycle 2 (36
min integration) and Cycle 4 (129 min integration).
The flux measurements based on Gaussian fit were 0.37$\pm$0.05 
(Jy km s$^{-1}$) with 0$\farcs$53 $\times $0$\farcs$47
\citep{ima16b} and 0.26$\pm$0.03 (Jy km s$^{-1}$) with
0$\farcs$16 $\times$ 0$\farcs$15 in Cycles 2 and 4, respectively.
The smaller flux measurement in Cycle 4 could be largely explained 
by the smaller beam size and maximum 10\% absolute flux calibration
uncertainty in individual ALMA observations. 
We adopted our Cycle 4 measurement for the following two reasons:
(1) The primary aim of our isotopologue observations is to estimate the
flux attenuation by line opacity for HCN, HCO$^{+}$, HNC J=3--2 emission.
(2) Our Cycle 4 data contain H$^{13}$CN, H$^{13}$CO$^{+}$, HN$^{13}$C
J=3--2 emission line flux measurements with similar beam sizes.
Adopting our Cycle 4 measurement of the H$^{13}$CN J=3--2 flux is a more
straightforward way to compare and correct flux attenuation for the HCN,
HCO$^{+}$, HNC J=3--2 emission lines.

Figure 5 presents intensity-weighted mean velocity (moment 1)
and intensity-weighted velocity dispersion (moment 2) maps of
selected bright emission lines in band 7, CO J=3--2, HCO$^{+}$ J=4--3,
and HNC J=4--3.
The CO J=3--2 emission line was detected not only in the nuclear region,
but also in spatially extended outer regions.
In addition to the much higher flux of CO J=3--2 than other molecular
lines, its lower critical density is likely to contribute 
to the detection of the spatially extended structure because outer
low-density molecular gas can collisionally excite CO J=3--2 more than
HCO$^{+}$ J=4--3 and HNC J=4--3.  
Figure 6 shows moment 1 and 2 maps of HCN, HCO$^{+}$, HNC J=8--7
emission lines in band 9 for the central zoomed regions.
Velocity information is available only for the nuclear compact regions.
The moment 1 and 2 maps for HCN, HCO$^{+}$, HNC J=1--0 emission lines in 
band 3 are presented in Figure 7, where we can obtain velocity
information for spatially extended regions outside the synthesized beam
sizes at least for HCN and HCO$^{+}$. 
In Figure 8, the moment 1 and 2 maps of the CS J=5--4 emission line in band
6, the additional dense molecular gas tracer \citep{gre14}, are presented.
Overall, most of the bright emission lines of dense molecular gas
tracers display similar rotational patterns, showing that the
north-eastern region of the nucleus is more redshifted than the
south-western region. 

Table 6 provides the luminosities of the primarily targeted molecular
emission lines. 
In addition to the HCN, HCO$^{+}$, HNC emission lines at various
rotational J-transitions at v=0 and v$_{2}$=1f, the isotopologue
H$^{13}$CN, H$^{13}$CO$^{+}$, HN$^{13}$C J=3--2, CO J=3--2, and CS
J=5--4 emission lines at v=0 are tabulated. 

\section{Discussion}

\subsection{Molecular gas morphology and dynamics}

Continuum and molecular line emission are dominated by a nuclear compact
component; however, in the brightest CO J=3--2 integrated-intensity (moment
0) map in Figure 3, a spatially extended structure is seen in the
south-eastern direction from the nucleus. 
A similar extended structure is seen in the stellar emission probed in
the near-infrared K-band (2.2 $\mu$m) image \citep{duc97}, suggesting
that this extended CO J=3--2 emission originates in the host galaxy. 

In the intensity-weighted mean velocity (moment 1) maps of CO
J=3--2, J=1--0, 3--2, 4--3, 8--7 of HCN, HCO$^{+}$, HNC, and CS J=5--4
in Figures 5--8 and \citet{ima16b}, as well as CO J=1--0 in
\citet{ued14}, the overall dynamics is dominated by rotational motion in
such a way that the north-eastern part is redshifted and the
south-western part is blueshifted, relative to the nucleus.
However, in the moment 1 map of the brightest CO J=3--2 emission line, a
dynamically decoupled component from the overall rotation is seen at the
south-westernmost region (the clump at the lower-right edge in Figure 5).
A plausible explanation is that some type of merger event happened
previously, which is quite reasonable given that IRAS 20551$-$4250 is a
ULIRG, and ULIRGs are usually driven by gas-rich galaxy mergers
\citep{sam96}.  

In Figure 4, although we fit emission lines with single Gaussian
component, when we examine the line profiles of the very bright CO J=3--2
and HCO$^{+}$ J=4--3 emission lines in more detail, skew patterns are
recognizable. 
To investigate the skewed asymmetric line profiles in more detail,
we fit these bright emission lines with a Gaussian
component, only using data points at the red part of the emission peak
(Figures 9a and 9b) and at the blue part of the emission peak (Figure
9c and 9d). 
When we fit the red component, the data in the blue part indicates an 
excess, compared to the best Gaussian fit.
Conversely, when we fit with a Gaussian, using data at the blue
part of the emission peak only, the extrapolation of the best fit
Gaussian to the redder part is higher than the actual data, particularly
for CO J=3--2.
The full-width at half maximum (FWHM) values of the best fit Gaussian 
are 260 km s$^{-1}$ and 165 km s$^{-1}$ for the blue and red components 
of the CO J=3--2 emission line, respectively. 
These are 205 km s$^{-1}$ and 160 km s$^{-1}$ for the blue and red
components of the HCO$^{+}$ J=4--3 emission line, respectively. 
We thus quantitatively confirm that the line widths are larger for 
the blue components than the red components both for CO J=3--2 and
HCO$^{+}$ J=4--3.
Figure 10 displays the contours of the blue and red components
of the CO J=3--2 and HCO$^{+}$ J=4--3 emission lines. 
The red component is slightly offset to the north-eastern direction from
the blue component; this could be explained by the overall rotational
motion of IRAS 20551$-$4250.   
A natural interpretation for this skewed profile is that turbulence is
stronger at the blueshifted molecular gas-emitting region, which
broadens the line width at the blue side.
In the CO J=3--2 moment 1 map in Figure 5, the signature of a
merger-induced distinct emission component is seen at the blueshifted
south-western part of the nucleus.
This possible merger-induced turbulence component may contribute to the
observed broader emission line profile for the blue component.  

In Figure 11(a), we show a zoom-in of the bottom part of the very bright
CO J=3--2 emission line, with the best Gaussian fit of this line (Figure
4 and Table 4) overplotted.
There is a clear excess at both the blue
and red sides of the CO J=3--2 emission line, although at the red side, 
possible contamination from the H$^{13}$CN J=4--3 ($\nu_{\rm rest}$ =
345.340 GHz) emission line makes quantitative discussion of the broad CO
J=3--2 emission component difficult.
This kind of profile is typically interpreted to be due to the broad emission
component by outflow activity \citep{fer10,ala11,mai12,aal12,cic14,gar15}. 
Figure 11(b) displays the contours of the emission at the blue and red
parts of the {\it broad} emission component, defined in Figure 11(a).
The peak position of the red broad component is slightly shifted to the
south-eastern part compared to that of the blue broad component (1 pix
left and 1 pix lower with the pixel scale of 0$\farcs$1 pix$^{-1}$).  
Considering the peak positional accuracy of the blue and red {\it broad}
components with (beam-size)/(signal-to-noise ratio), the significance of 
this positional displacement is marginal. 
However, this pattern is different from the global rotation of IRAS
20551$-$4250. 
We may be witnessing CO J=3--2 outflow toward (away from) us being
ejected in the north-western (south-eastern) direction from the nucleus.

In Figure 11(a), we added a Gaussian with FWHM = 540 km s$^{-1}$ and
velocity peak at 12,886 km s$^{-1}$ for the broad outflow component, and
another Gaussian with FWHM = 190 km s$^{-1}$ to incorporate the
H$^{13}$CN J=4--3 emission line at the red part of the bright CO
J=3--2 emission line. 
The Gaussian fit fluxes of the broad CO J=3--2 emission line
component and the H$^{13}$CN J=4--3 emission line at the tail of the
very bright CO J=3--2 emission line are estimated to be 
$\sim$7.7 (Jy km s$^{-1}$) and $\sim$0.9 (Jy km s$^{-1}$), 
respectively.  

We here estimate a molecular outflow rate from our CO J=3--2 emission
line data. 
The peak flux value of the blue broad emission component in Figure 11(b)
is $\sim$0.4 (Jy km s$^{-1}$). 
We adopt the value of the blue broad component, because the red broad
component is likely to be contaminated by the H$^{13}$CN J=4--3
emission line. 
Assuming that the outflow-origin red broad component is comparable to
the blue broad component, we obtain $\sim$0.8 (Jy km s$^{-1}$) for
outflow-origin CO J=3--2 emission. 
This is a factor of $\sim$10 smaller than the Gaussian fit flux of the
broad CO J=3--2 emission line component, but we adopt the value of
$\sim$0.8 (Jy km s$^{-1}$) to obtain the conservative estimate of a
molecular outflow rate.  
We obtain the CO J=3--2 luminosity with L$'$$_{\rm CO\ J=3-2}$ $\sim$ 7.3
$\times$ 10$^{6}$ (K km s$^{-1}$ pc$^{2}$) \citep{sol05}.
Assuming that CO J=3--2 emission is optically thick and thermalized, and
adopting the ULIRG-like CO luminosity to molecular mass (M$_{\rm H2}$)
conversion factor with $\sim$0.8 M$_{\odot}$ (K km s$^{-1}$
pc$^{2}$)$^{-1}$ \citep{cic14}, we obtain a molecular outflow mass of 
M$_{\rm outf}$ $\sim$ 5.8 $\times$ 10$^{6}$ M$_{\odot}$.   
In Figure 11(b), the peak position difference between the blue broad and
red broad emission component is $\sim$0.14$''$ or $\sim$120 (pc).
Adopting the outflow peak position offset with R $\sim$ 60 (pc) from the 
nucleus and outflow velocity with V = 500 km s$^{-1}$ (in Figure 
11(a), the broad wing component extends to approximately $\pm$500 km
s$^{-1}$ with respect to the systemic velocity of v$_{\rm opt}$ = 12900
km s$^{-1}$), we obtain a molecular outflow rate with 
\.{M}$_{\rm outf}$ $\sim$ 150 (M$_{\odot}$ yr$^{-1}$), where we adopt
the relation of \.{M}$_{\rm outf}$ = 3 $\times$ M$_{\rm outf}$ $\times$
V/R \citep{mai12,cic14}. 
Assuming $\sim$30\% AGN contribution to the infrared luminosity of 
IRAS 20551$-$4250 (i.e., L$_{\rm AGN}$ $\sim$ 1.1 $\times$ 10$^{45}$ 
erg s$^{-1}$), the derived molecular outflow rate with \.{M}$_{\rm outf}$
$\sim$ 150 (M$_{\odot}$ yr$^{-1}$) agrees within a factor of $\sim$2
with the relation seen in other ULIRGs \citep{cic14}.
The molecular outflow kinetic power is estimated to be P$_{\rm outf}$
$\equiv$ 0.5 $\times$ \.{M}$_{\rm outf}$ $\times$ V$^{2}$ $\sim$ 1.2
$\times$ 10$^{43}$ (erg s$^{-1}$), which is $\sim$1\% of the AGN luminosity.
The molecular outflow momentum rate is \.{P}$_{\rm outf}$ 
$\equiv$ \.{M}$_{\rm outf}$ $\times$ V $\sim$ 4.6 $\times$ 10$^{30}$ 
(kg m s$^{-2}$), which is $\sim$12 $\times$ L$_{\rm AGN}$/c. 
These values are comparable to those observed in other ULIRGs with
detectable molecular outflow activity \citep{cic14}. 
Note that all of molecular outflow mass (M$_{\rm outf}$), molecular
outflow rate (\.{M}$_{\rm outf}$), molecular outflow kinetic power 
(P$_{\rm outf}$), and molecular outflow momentum rate (\.{P}$_{\rm outf}$)
could increase by an order of magnitude, if we use the Gaussian fit flux
of the broad CO J=3--2 emission line component.    

\subsection{Isotopologue molecular lines and opacity estimate}

From Table 4 and \citet{ima16b}, we obtained the ratios of
HCN-to-H$^{13}$CN J=3--2 flux in (Jy km s$^{-1}$) to be $\sim$22$\pm$3, 
HCO$^{+}$-to-H$^{13}$CO$^{+}$ J=3--2 flux in (Jy km s$^{-1}$) to be
$\sim$60$\pm$9, and HNC-to-HN$^{13}$C J=3--2 flux in (Jy km s$^{-1}$) to
be $\sim$45$\pm$15.
We must note that the detection significance of HN$^{13}$C
J=3--2 is only $\sim$3$\sigma$ in both Gaussian fit in the spectrum and
moment 0 map. 
Thus, discussion of HNC could be more uncertain than HCN and HCO$^{+}$. 
Adopting the $^{12}$C-to-$^{13}$C abundance ratios in ULIRGs with 50--100
\citep{hen93a,hen93b,mar10,hen14}, we find that 
the flux attenuation by line opacity for HCN J=3--2 is estimated to be a
factor of 2--5, while those of HCO$^{+}$ J=3--2 and HNC J=3--2 are 
a factor of $\sim$1--1.5 and $\sim$1--2, respectively, where we assume that 
isotopologue emission lines are optically thin \citep{jim16}.

In a geometry where radiation sources and molecular gas are spatially
well mixed, the flux attenuation and optical depth ($\tau$) are related
as $\frac{\tau}{1 - exp(-\tau)}$. 
From this relationship, we obtain $\tau$ = 2--5 for HCN J=3--2. 
Using the estimated H$^{13}$CN J=4--3 flux with $\sim$0.9 (Jy km
s$^{-1}$) ($\S$4.1), we obtain the ratio of HCN-to-H$^{13}$CN J=4--3
flux in (Jy km s$^{-1}$) to be $\sim$13.
The flux attenuation by line opacity for HCN J=4--3 is estimated to be
4--7 or $\tau$=4--7, which is comparable to that for HCN J=3--2
($\tau$=2--5). 
In summary, our isotopologue observations suggest that the HCN J=3--2 and
J=4--3 emission lines are considerably flux-attenuated by line opacity.

In contrast, the relatively small flux attenuation with $<$2
suggests that the line opacity for HCO$^{+}$ J=3--2 and HNC J=3--2
is $\tau$ $<$ 1.5.
Based on the observed HCN, HCO$^{+}$, HNC J=3--2 fluxes at
v$_{2}$=1f and v=0, and calculation of vibrational excitation by 
infrared radiative pumping, using the available 5--30 $\mu$m spectrum of
IRAS 20551$-$4250, \citet{ima16b} argued that HCN abundance is
higher than those of HCO$^{+}$ and HNC. 
This higher HCN abundance scenario is a plausible explanation 
for the higher HCN line opacity than HCO$^{+}$ and HNC at J=3--2.

\subsection{Infrared radiative pumping and independent molecular 
abundance estimate} 

Based on the fluxes of vibrationally excited (v$_{2}$=1f) and
vibrational ground (v=0) HCN, HCO$^{+}$, HNC emission lines at J=3--2,
it has been estimated that the infrared radiative pumping mechanism
plays a role in rotational excitation at v=0 in IRAS 20551$-$4250 at
least for HCN and HNC \citep{ima16b}. 
Namely, once HCN and HNC are vibrationally excited to v$_{2}$=1 by
absorbing infrared $\sim$14 $\mu$m and $\sim$22 $\mu$m photons,
respectively, they are decayed back to v=0 and their rotational
J-transition fluxes at v=0 can be higher than those for collisional
excitation alone \citep{ran11}. 
We here attempt to confirm this result, based on our newly collected ALMA
data.
The J=1--0 line is not present at the vibrationally excited level
(v$_{2}$=1).  
For J=8--7, v$_{2}$=1f emission lines are not clearly detected, mainly due to
large background noise at this high ALMA frequency range. 
We thus focus on J=4--3 data, and follow the same logical steps as those
made by \citet{ima16b} for J=3--2 data. 

Infrared radiative pumping is estimated to play a significant role
if the following condition is met:  
\begin{equation}
T_{ex-vib} > \frac{T_{0}}{ln\frac{A_{v2=1-0,vib}}{A_{J=4-3,rot}}}.
\end{equation} 
where T$_{ex-vib}$ is the v$_{2}$=1f vibrational excitation temperature,
T$_{0}$ is the energy level at v$_{2}$=1f J=4, A$_{v2=1-0,vib}$ is the
Einstein A coefficient from v$_{2}$=1f to v=0, and A$_{J=4-3,rot}$ is
the Einstein A coefficient from J=4 to J=3 at v=0 \citep{car81,sak10,mil13}. 

Adopting the values in Table 7, and A$_{v2=1-0,vib}$ = $\sim$1.7 s$^{-1}$,  
$\sim$3.0 s$^{-1}$, and $\sim$5.2 s$^{-1}$ for HCN, HCO$^{+}$, and HNC,
respectively \citep{deg86,aal07a,mau95}, we obtain $\sim$160 (K),
$\sim$185 (K), and $\sim$210 (K) in the right-hand column of
Eqn. (1) for HCN, HCO$^{+}$, and HNC, respectively. 
The observed T$_{\rm ex-vib}$ values in the left-hand column of
Eqn. (1), derived based on e.g., Eqn. (2) of \citet{ima16b}, 
are $\sim$360 (K), $<$280 (K), and $\sim$360 (K) for HCN, HCO$^{+}$, and HNC,
respectively (Table 8).
For HCN and HNC, the condition is fulfilled. 
If the HNC v$_{2}$=1f J=4--3 flux is approximately half that shown in Table 4
($\S$3), the left-hand column of Eqn. (1) for HNC is $\sim$260 (K),
which is still higher than the right-hand column.
For HCO$^{+}$, the upper limit of T$_{\rm ex-vib}$ still allows the
condition to be met.

The left-hand column of Eqn. (1) will decrease if flux attenuation by
line opacity is significant for some v=0 lines and the intrinsic
line opacity-corrected flux at v=0 and v$_{2}$=1f is used for the
calculation.   
From our isotopologue observations, we see that HCN J=4--3 flux is
estimated to be attenuated by line opacity with a factor of 4--7.
Adopting a factor of 5 flux attenuation for HCN J=4--3, the T$_{\rm ex}$
value of HCN decreases only to $\sim$230 (K). 
Eqn. (1) is still fulfilled for HCN, not changing the above conclusion.
For HCO$^{+}$ and HNC, flux attenuation by line opacity is not
significant at J=3--2 \citep{ima16b}.
Assuming that this is also the case for J=4--3, the condition in
Eqn. (1) is still valid for HNC J=4--3.
Thus, it is confirmed from our new ALMA J=4--3 data that the role of 
infrared radiative pumping is significant in IRAS 20551$-$4250 at least
for HCN and HNC. 

As discussed in \citet{ima16b}, if the column density at v=0 is the
same for HCN, HCO$^{+}$, and HNC, and if all molecular lines are emitted
from the same regions, the ratio of the infrared radiative pumping rate
among HCN, HCO$^{+}$, and HNC in IRAS 20551$-$4250 is calculated to be  
\begin{eqnarray}
HCN:HCO^{+}:HNC (P_{IR}) & = & 1:0.9:27, 
\end{eqnarray}
and the v$_{2}$=1 to v=0 column density ratio among HCN, HCO$^{+}$,
and HNC is calculated to be  
\begin{eqnarray}
HCN:HCO^{+}:HNC (N_{v_{2}=1}}/{N_{v=0}) & = & 1:0.5:9, 
\end{eqnarray}
based on the Einstein A coefficients of
HCN, HCO$^{+}$, HNC in Table 7 and the infrared 5--35 $\mu$m spectrum of
IRAS 20551$-$4250 obtained with Spitzer IRS \citep{ima16b}.
In essence, the number of infrared photons at $\sim$22 $\mu$m 
(corresponding to the vibrational-rotational transitions for HNC) is
higher than at 12--14 $\mu$m (corresponding to the
vibrational-rotational transitions of HCN and HCO$^{+}$), such that it is
naively expected that HNC is more vibrationally excited by infrared
radiative pumping than HCN and HCO$^{+}$.

The ratio of the column density at the 
v$_{2}$=1f J=4 level (N$_{v_{2}=1f, J=4}$) among HCN, HCO$^{+}$, and HNC
is derived with $\propto$ $\frac{flux}{A_{ul}}$, where the line flux
from the upper (u) to lower (l) transition level is in units of 
(Jy km s$^{-1}$) and  $A_{ul}$ is the Einstein A coefficient for
spontaneous emission from the upper (u) to lower (l) level
\citep{gol99,izu13,ima16b}. 
From observational fluxes in Table 4 and Einstein A coefficients in
Table 7, we obtain the column density ratio at J=4 at v$_{2}$=1 
\begin{eqnarray}
HCN:HCO^{+}:HNC (N_{v_{2}=1f, J=4}) & = & 1:<0.2:0.7,
\end{eqnarray}
where we adopt that the actual flux of HNC v$_{2}$=1f J=3--2 emission
line is half of the Gaussian fitted result from Table 4 (see $\S$3).

Assuming that the fraction of J=4 level, relative to all J-levels at
v=0, does not differ significantly among HCN, HCO$^{+}$, and HNC 
\footnote{The J=4--3 to J=3--2 flux ratios for HCN, HCO$^{+}$, and HNC 
are all comparable to each other \citep{ima16b}.
Given that the molecular population is typically dominated by low J levels at
J $<$ 5, this assumption is reasonable for IRAS 20551$-$4250. 
},
and that v$_{2}$=1f J-transition emission is optically thin
\citep{ima16b}, Equations (3) (prediction) and (4) (observation) can be
reconciled if the column density at v=0 (N$_{v}$=0) (and thereby
abundance) for HCN is higher than HCO$^{+}$ and HNC by factors of
$>$2.5 and $\sim$10, respectively.  
Thus, we obtain an abundance ratio of 
\begin{eqnarray}
HCN:HCO^{+}:HNC (abundance) & = & 1:<0.4:0.1.
\end{eqnarray}
The HCN-to-HCO$^{+}$ abundance ratio with $>$2.5 and 
HCN-to-HNC abundance ratio with $\sim$10 are comparable to those derived
from J=3--2 data \citep{ima16b}. 
The HCN-to-HNC abundance ratios derived in this manner at J=4--3 and J=3--2
($\sim$10) appear to be larger than that from isotopologue observations 
($\sim$2).
This discrepancy could be partly explained by the possibly large
ambiguity of the HN$^{13}$C J=3--2 flux, due to its marginal
($\sim$3$\sigma$) detection ($\S$4.2).
Although a quantitatively accurate estimate of the actual HCN-to-HNC
abundance ratio is subject to further investigation, two independent
methods consistently suggest that HCN abundance is higher than 
HCO$^{+}$ and HNC, by at least a few times.

\subsection{Molecular gas properties}

To obtain a better understanding of the physical properties of these
line-emitting molecular gas, we use the RADEX software \citep{van07},
which treats the non-LTE analysis and predicts molecular line flux
ratios at v=0, by solving the excitation of rotational transitions at
v=0 by collision, based on the large velocity gradient (LVG) approximation.
Free parameters are the number density of hydrogen molecule
(n$_{\rm H2}$), kinetic temperature of H$_{2}$ (T$_{\rm kin}$), column
density of a particular molecule divided by its line width 
(N$_{\rm mol}$/$\Delta$v), and background radiation temperature 
(T$_{\rm bak}$), which is set as $\sim$3 K.  

Infrared radiative pumping to v$_{2}$=1 should play a
role for molecular rotational excitation at v=0, in addition to
collision, in IRAS 20551$-$4250 ($\S$4.3). 
However, with the limited amount of available molecular rotational
J-transition line data in our study, separating the contributions from
collisional excitation and infrared radiative pumping with sufficient
reliability is not easy. 
Even for collisional excitation alone, it is sometimes argued that 
(1) HCN and HCO$^{+}$ probe substantially different phases of molecular
gas, and 
(2) there exist multiple temperature components for
HCN-, HCO$^{+}$-, HNC-emitting molecular gas in LIRGs \citep{gre09}. 
These collisional excitation models with increased numbers of parameters
and infrared radiative pumping effects are virtually
indistinguishable.  
\citet{vol16} estimated that, in LIRGs, the infrared radiative pumping
has a measurable, but relatively small (20--30\%), effect of enhancing the
HCN J=1--0 flux, when compared to collisional excitation alone, 
although the effect for higher-J was not estimated and could be higher.  
This 20--30\% level of flux alternation is comparable to the uncertainty
of different J-transition line flux ratios, coming from ALMA's absolute 
flux calibration uncertainty in individual observations in each
band (maximum $\sim$10$\%$). 
Thus, we first derive rough values of the molecular gas parameters based
on collisional excitation, and will later investigate how they change by
including the infrared radiative pumping.

We vary the H$_{2}$ number density (n$_{\rm H2}$) between 10$^{3}$
and 10$^{8}$ cm$^{-3}$, because HCN, HCO$^{+}$, and HNC are well
known to be dense ($>$10$^{3}$ cm$^{-3}$) molecular gas tracers, due to
their high dipole moments and resulting high critical densities. 
The kinetic temperature range (T$_{\rm kin}$) is set from 10 K to 500 K,
where the lowest kinetic temperature of 10 K corresponds to cool
molecular gas far from energy sources, whereas the highest kinetic
temperature of 500 K reflects compact ($<$10 pc), warm molecular gas in
the close vicinity of the putative AGN. 
Since the molecular line widths of HCN, HCO$^{+}$, and HNC at J=1--0,
3--2, 4--3, and 8--7 are $\sim$300 km s$^{-1}$ at 10\% of the peak
intensity \citep{ima13b,ima16b}, we adopt $\Delta$v = 300 km s$^{-1}$
for the line width in our RADEX calculation. 
We include T$_{\rm bak}$ $\sim$ 3 K cosmic microwave background radiation
in the calculations. 
The column density of individual molecular gas is not directly derived
from observations. 
The XMM-Newton 2--10 keV X-ray observation of IRAS 20551$-$4250 shows
that the hydrogen column density toward the putative X-ray-emitting
obscured AGN is estimated to be N$_{\rm H}$ = 7.9$^{+6.9}_{-1.9}$ 
$\times$ 10$^{23}$ cm$^{-2}$ \citep{fra03}. 
Since the abundance ratios of HCN, HCO$^{+}$ and HNC, relative to H, 
in warm molecular gas in LIRGs are estimated to be $\sim$10$^{-8}$
\citep{gre09,sai17}, we first set the column density (N$_{\rm mol}$) of
HCN, HCO$^{+}$, and HNC to be 1 $\times$ 10$^{16}$ cm$^{-2}$. 

We summarize our RADEX calculation results in Figure 12, where the
ratios are derived from the comparison of observed fluxes in (Jy km
s$^{-1}$) (Table 9).
The parameter ranges that can reproduce the observed flux ratios
of individual J-transitions are shown as thick lines.
For n$_{\rm H2}$ = a few $\times $10$^{5}$ cm$^{-3}$, 
T$_{\rm kin}$ = 70--100 K, $\Delta$v = 300 km s$^{-1}$, and 
HCN column density N$_{\rm HCN}$ = 1 $\times$ 10$^{16}$ cm$^{-2}$, 
which can reproduce the observed HCN J=3--2 to J=1--0 flux ratio in
Figure 12, our RADEX calculations show that the HCN J=3--2 and J=4--3
line optical depths are $\tau$ $\sim$ 3, roughly comparable to
the observationally derived values from isotopologue line data ($\S$4.2). 
Thus, the choice of N$_{\rm HCN}$ = 1 $\times$ 10$^{16}$ cm$^{-2}$ is
regarded as appropriate. 

Since HCO$^{+}$ and HNC abundance are estimated to be at least a few
times lower than HCN in IRAS 20551$-$4250 ($\S$4.3), we show the same 
RADEX calculation results in Appendix A (Figure 19) for the HCO$^{+}$
and HNC column density of 3 $\times$ 10$^{15}$ cm$^{-2}$.
The overall trends change little.  
In interpreting these results, we must consider two possible
caveats. 

First, the line-emitting volume can differ between the high-J and
low-J transition lines.
AGNs are typically located at galactic center regions where typical
molecular gas density is expected to be higher, due to nuclear molecular
gas concentration, than that in outer off-nuclear regions.
Additionally, in an AGN, the surrounding molecular gas and dust has a
strong temperature gradient such that the temperature at the inner
part is higher than that at the outer part.
Consequently, dense and warm molecular gas is concentrated in the inner
part of the nuclear region from where high-J transition lines can be
efficiently emitted. 
More diffuse, low-density cool gas is distributed in more spatially
extended regions, which can strongly emit low-J transition lines, but 
high-J transition line emission is relatively limited.
Thus, the probed effective molecular gas volume is very likely to be
smaller at higher-J than lower-J transition lines, whereas we assume that
both lines come from the same volume in the RADEX calculations.
If we focus on the volume of high-J line-emitting molecular gas and 
compare flux ratios {\it within that volume}, then the actual high-J to
low-J line flux ratios will be higher than the observed ratios.
Thus, the derived temperature and density from the observed flux ratios
and RADEX calculations are taken as {\it lower limits} for the actual
values for the high-J line-emitting molecular gas.

Second, since vibrational transition is not included in our RADEX
calculations, infrared radiative pumping is not properly taken into
account. 
For IRAS 20551$-$4250, it is quantitatively demonstrated by
\citet{ima16b} and this study ($\S$4.3) that infrared radiative
pumping plays a role in rotational excitation at v=0, through the decay
from v$_{2}$=1 to v=0, at least for HCN and HNC, and 
could realize higher rotational excitation at v=0 than collisional
excitation alone, for a given H$_{2}$ number density and kinetic
temperature. 
Owing to infrared radiative pumping, lower values of H$_{2}$ number
density and kinetic temperature may suffice to reproduce the observed
flux ratios.
Thus, the H$_{2}$ number density and kinetic temperature should be taken
as {\it upper limits}. 
This second ambiguity works in opposition to the first, 
and it is not easy to disentangle the two in a quantitatively reliable
manner.

Regarding the first issue, the derived number density and kinetic
temperature using low-J lines largely reflect a large volume of
spatially extended low-density, low-temperature molecular gas,
resulting in low values.    
If we compare the flux ratios at high-J transitions only, a small volume 
of dense and high-temperature molecular gas is preferentially probed,
and the derived density and temperature can be high.
In Figure 12, we find that the density and kinetic temperature that
reproduce the observed flux ratios increase from ``J=3--2 to J=1--0'' 
to ``J=4--3 to J=3--2'' or ``J=8--7 to J=4--3'' or ``J=8--7 to J=3--2'' 
for HCN, HCO$^{+}$, HNC, namely, the thick curved line moves from the
lower-left to upper-right regions.
This is a natural consequence, that higher-J lines selectively
probe dense and warm molecular gas with a small emitting volume. 

In Figure 12, the J=3--2 to J=1--0 flux comparison is expected to
reflect the largest molecular gas volume, including outer low-density and
low-temperature gas.
For the kinetic temperature of T$_{\rm kin}$ $>$ 25 K, H$_{2}$ number
density of n$_{\rm H2}$ = 10$^{5}$--10$^{6}$ cm$^{-3}$ 
(10$^{4}$--10$^{5}$ cm$^{-3}$) seems required for HCN and HNC
(HCO$^{+}$).
The lower required density for HCO$^{+}$ is reasonable, given the
lower critical density of HCO$^{+}$ than HCN and HNC, under similar
line opacity \citep{mei07,gre09}. 

This possible difference of the probed molecular line-emitting volumes
can be small, if we compare the flux ratios between adjacent
J-transitions, such as J=4--3 and J=3--2.  
The typical density and temperature of molecular gas that emits J=4--3
and J=3--2 lines are $>$10$^{6}$ cm$^{-3}$ and $>$50 K for HCN,
HCO$^{+}$, and HNC (Figure 12).
We thus quantitatively confirm that these J=4--3 and J=3--2 emission
lines selectively probe dense and warm molecular gas in the close
vicinity of energy sources. 

The derived density and kinetic temperature that reproduce the observed  
``J=8--7 to J=3--2'' and ``J=8--7 to J=4--3'' flux ratios are lower
than those for ``J=4--3 to J=3--2'' flux ratios for HCN, HCO$^{+}$, and
HNC. 
This is opposite to the expected trend that the derived density becomes
higher when we compare flux ratios at higher-J transition. 
A plausible explanation for this is that the J=8--7 emitting volume is
substantially smaller than those of J=4--3 and J=3--2. 
This can result in much smaller ``J=8--7 to J=3--2'' and ``J=8--7 to
J=4--3'' flux ratios than expected from the same emitting volume
assumption, and thereby smaller derived density and kinetic temperature. 

The dense molecular mass, derived from the observed HCN J=1--0
luminosity (Table 6), is estimated to be $\sim$2 $\times$ 10$^{9}$ M$_{\odot}$,
where we adopt the relationship between dense molecular mass and HCN J=1--0
luminosity, M$_{\rm dense}$ = 10 $\times$ HCN J=1--0 luminosity 
[M$_{\odot}$ (K km s$^{-1}$ pc$^{2}$)$^{-1}$] \citep{gao04}.
Conversely, the hydrogen column density toward the X-ray emitting
AGN in IRAS 20551$-$4250 is N$_{\rm H}$ $\sim$ 8 $\times$ 10$^{23}$
cm$^{-2}$ \citep{fra03}. 
We here assume that molecular gas distributes spherically and the AGN is
located at the center.
If all X-ray-obscuring material is in a dense molecular form that is
probed with HCN J=1--0, we obtain the following two relations:  
\begin{eqnarray}
r (pc) \times n (cm^{-3}) \sim 8 \times 10^{23} (cm^{-2}), \\
4/3 \times \pi \times r^{3} (pc^{3}) \times n (cm^{-3}) \times
m_{H2} \sim 2 \times 10^{9} M_{\odot},  
\end{eqnarray}
where r (pc) is the radius of the sphere, n (cm$^{-3}$) is volume
averaged gas number density, and m$_{\rm H2}$ is the mass of molecular 
hydrogen (=3.3 $\times$ 10$^{-24}$ g).
From Equations (6) and (7), we obtain r $\sim$ 200 (pc) and n $\sim$
10$^{3}$ (cm$^{-3}$). 
However, if the kinetic temperature is $>$30 (K), 
the typical number density of HCN J=1--0 emitting gas, derived from 
the observed HCN J=3--2 to J=1--0 flux ratio and RADEX calculation, 
is 10$^{5}$--10$^{6}$ cm$^{-3}$ (Figure 12).
These results can be reconciled if nuclear molecular clouds in IRAS
20551$-$4250 consist of dense (10$^{5}$--10$^{6}$ cm$^{-3}$) clumps
with a small volume filling factor of 0.1--1\%, as argued by
\citet{sol87}.
The slightly low estimated density of 10$^{4}$--10$^{5}$ cm$^{-3}$
from HCO$^{+}$ observations (Figure 12) could be reconciled if
HCO$^{+}$ probes more extended lower-density envelopes of each clump
\citep{ima16b}, because HCO$^{+}$ emission is observationally found to
be more diffuse than HCN \citep{ima07a,sai15,sai17}.
This kind of clumpy molecular structure can naturally explain similar
profiles of various molecular emission lines with different line opacity 
\citep{sol87}, as observed in IRAS 20551$-$4250 \citep{ima16b}.  

\subsection{Molecular line flux ratio}

Figure 13 plots the observed HCN-to-HNC and HCN-to-HCO$^{+}$ flux ratios
at J=1--0, J=3--2, J=4--3, and J=8--7.
For J=3--2 and J=4--3, line-opacity corrected intrinsic flux ratios are
also shown. 
In the intrinsic flux ratios, IRAS 20551$-$4250 deviates even farther
away from starburst-dominated regions, suggesting that the line opacity
correction, particularly for high-HCN-abundance AGN-containing
objects, makes our HCN-based molecular line flux ratio method even more
powerful in distinguishing AGNs from starbursts.

On the ordinate, the observed HCN-to-HCO$^{+}$ flux ratios are comparable 
at J=1--0, J=3--2, and J=4--3, but are slightly lower than that at J=8--7.
Since the critical density of HCN is higher than HCO$^{+}$ at individual
J-transitions, the higher HCN-to-HCO$^{+}$ flux ratio at higher
J-transition is difficult to explain with collisional excitation under
the same line opacity.
However, the estimated higher line opacity of HCN reduces its effective 
critical density with 1/$\tau$.
Furthermore, due to higher HCN abundance than HCO$^{+}$,
line-opacity-corrected intrinsic HCN-to-HCO$^{+}$ flux ratios will be
higher than those observed. 
The intrinsic HCN-to-HCO$^{+}$ flux ratios at J=3--2 and J=4--3 are 
at least higher than the observed ratio at J=8--7.
Since the line opacity at J=1--0 is estimated to be higher than J=8--7
with RADEX under the parameter range of 
n$_{\rm H2}$ = 10$^{5}$ cm$^{-3}$, 
T$_{\rm kin}$ = 50--100 K, 
N$_{\rm HCN}$ = 1 $\times$ 10$^{16}$ cm$^{-2}$, 
and $\Delta$v = 300 km s$^{-1}$, the intrinsic HCN-to-HCO$^{+}$ flux
ratio at J=1--0 will move to the upper side more than at J=8--7.
Both of these factors could help to explain the result. 

On the abscissa, the observed HCN-to-HNC flux ratio at J=1--0 is
higher than those at J=3--2, J=4--3, and J=8--7. 
Figure 14 is the plot of the ``high-J to low-J'' flux ratios for HCN,
HCO$^{+}$, and HNC, where the combination of ``J=3--2 to J=1--0'', 
``J=4--3 to J=1--0'', ``J=8--7 to J=1--0'', ``J=4--3 to J=3--2'', 
``J=8--7 to J=3--2'', and ``J=8--7 to J=4--3'' are selected. 
The actual ratios are summarized in Table 9.
We see a trend that the ratios are a factor of $\sim$2 higher for HNC
than HCN and HCO$^{+}$ for the first three combinations (including 
J=1--0 as the low-J transition), but comparable for the latter three 
combinations (not including J=1--0).
HCN and HNC have comparable critical densities at individual J-transition 
under a similar line opacity \citep{gre09}.
The derived higher HCN opacity ($\S$4.2 and 4.3) will make the effective
HCN critical density lower than HNC, in which case the 
``high-J to low-J'' flux ratios for HNC are expected to be lower than
those for HCN, when collisional excitation is considered.
This is opposite to the trends observed in Figure 14.

The excitation temperatures of rotational J-transition at v=0 are
summarized in Table 8, where the differences of beam sizes are not
considered, to be consistent with our previous calculations
\citep{ima16b}. 
This assumes that the intrinsic emission size is the same among 
HCN, HCO$^{+}$, HNC and is smaller than the beam sizes in all
observations.     
The rotational excitation temperatures within v=0 are slightly
higher for HNC than for HCN (Table 8).

We suggest that the following two mechanisms could help reduce the
discrepancy between observations and predictions.
First, we find from our RADEX calculations that at high density 
(n$_{\rm H2}$ $>$ 3 $\times$ 10$^{5}$ cm$^{-3}$) and high temperature
(T$_{\rm kin}$ = 50--100 K),  
a maser phenomenon occurs at J=1--0 first for HCN and HCO$^{+}$, and
then later for HNC. 
For example, for the column density of 1 $\times$ 10$^{16}$ cm$^{-2}$, 
$\Delta$v = 300 km s$^{-1}$, and n$_{\rm H2}$ = 1--3 $\times$ 10$^{6}$
cm$^{-3}$, the maser phenomenon is predicted to occur at T$_{\rm kin}$ 
$>$ 30 K for HCN J=1--0, but only at T$_{\rm kin}$ $>$ 70 K for HNC
J=1--0.  
Also with n$_{\rm H2}$ = 3 $\times$ 10$^{5}$ cm$^{-3}$, HCN J=1--0 shows
the maser phenomenon at T$_{\rm kin}$ = 50--100 K, but HNC J=1--0 does
not. 
The maser phenomenon could boost J=1--0 emission.
If the maser occurs only for HCN J=1--0, but not for HNC J=1--0,
under some parameter range, the ``high-J to J=1--0'' flux ratios for HCN
could be smaller than HNC.   
HCO$^{+}$ J=1--0 also starts to show the maser phenomenon earlier than
HNC from lower density and temperature.
The higher values of ``high-J to J=1--0'' flux ratios for HNC than
HCO$^{+}$ could also be explained by the maser phenomenon in some
parameter range. 

Our second scenario is infrared radiative pumping.
In IRAS 20551$-$4250, it is calculated that the infrared radiative
pumping rate is higher for HNC than for HCN and HCO$^{+}$ 
($\S$4.3 and \citet{ima16b}).
This could boost high-J HNC emission more than HCN and HCO$^{+}$,
through decay back from v$_{2}$=1 to v=0.
Given that this infrared radiative pumping is estimated to play a role
in IRAS 20551$-$4250 at least for HCN and HNC ($\S$4.3), its effect
must be evaluated in an appropriate manner. 

\subsection{Inclusion of infrared radiative pumping}

Our ultimate goal is to quantitatively estimate (1) how much
fraction of molecular excitation at individual J-transition levels is 
due to collision and infrared radiative pumping, and (2) the increase of 
molecular line fluxes at individual J-transitions, brought by infrared
radiative pumping, when compared to collisional excitation alone.
However, this is difficult to achieve with currently available data,
because of a large freedom of parameters for collisional excitation.
First, the possibly different emitting volumes at different
J-transitions ($\S$4.4) are not constrained in a meaningful manner. 
Next, the typical temperature and density probed by different
J-transition lines are different ($\S$4.4), so that different number of
multiple temperature and density components needs to be taken into
account when we compare flux ratios among different J-transition lines.
We thus investigate quantitatively how infrared radiative pumping
reproduces the observed trend of higher values of ``high-J to 
low-J'' flux ratios for HNC than HCN shown in Figure 14, by fixing molecular
gas physical parameters to canonical values. 
We use MOLPOP \citep{eli06}, which handles the exact solution of
radiative transfer problem in multi-level atomic systems. 
The $\sim$3 K cosmic microwave background radiation is always included
in MOLPOP.  
The large velocity gradient (LVG) method is prepared as an option, and
we adopt it to compare the output with our RADEX calculation results.
We have quantitatively confirmed that, under LVG, both MOLPOP and RADEX 
outputs about molecular line flux ratios among J=1--0, J=3--2, J=4--3,
and J=8--7 agree within 20--30\% for various H$_{2}$ number density,
H$_{2}$ kinetic temperature, and molecular column density divided by
line width.  

To calculate the role of infrared radiative pumping, we require Einstein A
coefficient values for vibrational (v)-rotational (J) transitions between 
(v$_{2}$,J) = (1,J) and (0, J$\pm$1 or J).
We find the required information for HCN and HNC, but not for HCO$^{+}$, in
the GEISA database \citep{jac16}.
We thus perform this calculation for HCN and HNC.
Collision rates between J-transition levels at v$_{2}$=1 are not
available, so we tentatively adopt the same values as v=0. 
Since vibrational-rotational transitions are much faster than the
transitions between J-levels within v$_{2}$=1 \citep{ima16b}, possible 
small ambiguities in the collision rates at v$_{2}$=1 do not significantly
affect our final results. 
Collisional transitions between v$_{2}$=1 and v=0 are also not taken
into account, due to their very large energy gaps. 

In MOLPOP, we first set n$_{\rm H2}$ = 10$^{5}$ cm$^{-3}$, 
T$_{\rm kin}$ = 100 (K), $\Delta$v = 300 km s$^{-1}$, and the HCN and
HNC column density of 10$^{16}$ cm$^{-2}$ to calculate molecular line
fluxes by collisional excitation. 
We confirm that the maser phenomenon does not happen at J=1--0.
We choose the small side of the estimated density range for HCN and HNC,
n$_{\rm H2}$ = 10$^{5}$ cm$^{-3}$ for our calculations, because the maser
phenomenon happens at higher n$_{\rm H2}$ values, which complicates
interpretations. 
We then add infrared radiation at 5--30 $\mu$m in the power-law form of 
F$_{\nu}$ (Jy) = 0.0024 $\times$ $\lambda$($\mu$m) $^{2.0}$ (the dashed
straight line in \citet{ima16b}).
In this form, the number of infrared photons at $\sim$22 $\mu$m is
larger than at 12--14 $\mu$m.
The integrated luminosity between 5--30 $\mu$m is estimated to be 
$\sim$2 $\times$ 10$^{11}$L$_{\odot}$ at a distance of 187 Mpc 
(z = 0.043).
Plane-parallel geometry is adopted, and the infrared 5--30 $\mu$m
radiation source is placed at 30 pc and 100 pc from the
molecular gas. 
We then investigate if the high-J to low-J flux ratios of HNC are
higher than HCN, particularly for ``J=3--2 to J=1--0'', 
``J=4--3 to J=1--0'', and ``J=8--7 to J=1--0'', as observed in Figure
14.
Figure 15(a) plots the values of ``high-J to low-J flux ratio of HNC,
divided by the same ratio of HCN'' in the ordinate.
Since the infrared radiative pumping effects are calculated to be higher
for HNC than HCN, the high-J to low-J flux ratios for HNC are expected
to increase more than those for HCN, by the inclusion of infrared
radiation. 
Namely, the values in the ordinate are expected to be higher in the case
of stronger infrared radiation than the case with no infrared radiation
(i.e., collisional excitation only).
We find that the inclusion of infrared radiative pumping 
better reproduces the observed trend of higher values of 
``high-J to low-J flux ratios'' for HNC than HCN, for the combinations  
of ``J=3--2 to J=1--0'', ``J=4--3 to J=1--0'', and ``J=8--7 to
J=1--0'' observed in Figure 14.

In Figure 15(b), we display a similar plot, but with a factor of 
$\sim$3 smaller HNC column density than HCN, as indicated in $\S$4.3.
Since an additional parameter is introduced, the behavior becomes
complicated, but the general trend of enhanced high-J to J=1--0 flux
ratios for HNC than HCN is observed.
We thus argue that the observed trend in IRAS 20551$-$4250 can be
largely accounted for by infrared radiative pumping.

Although collisional excitation is usually adopted to interpret the 
observed molecular J-transition emission line fluxes in galaxies, it has
been widely argued that the infrared radiative pumping mechanism can
play an important role in AGN-containing galaxies, because AGNs can emit
strong mid-infrared 5--30 $\mu$m continuum photons, due to AGN-heated
hot dust grains.  
In this AGN-containing ULIRG, IRAS 20551$-$4250, we have demonstrated 
that the infrared radiative pumping indeed plays a role for molecular
rotational J-level excitation at least for HCN and HNC, which
suggests that this is the case in other AGN-containing ULIRGs as well.  
However, in general ULIRGs, the similar estimates of the infrared
radiative pumping effects, through the v$_{2}$=1f emission line flux
measurements, are prohibitively difficult because HCN J=3--2 and J=4--3
emission lines at the v$_{2}$=1 level are spectrally highly overlapped
with the nearby, much brighter HCO$^{+}$ J=3--2 and J=4--3 emission
lines at the v=0 level, respectively \citep{aal15a,ima16c}. 
IRAS 20551$-$4250 is unique in that, thanks to its small observed 
molecular emission line widths, the flux estimates of the HCN v$_{2}$=1f 
J=3--2 and J=4--3 emission lines are possible, by clearly separating
from the bright HCO$^{+}$ (v=0) emission lines.
Further detailed investigations of IRAS 20551$-$4250 with additional
emission lines at v=0 and v$_{2}$=1 will help better constrain the
quantitative role of infrared radiative pumping in this
ULIRG as well as other AGN-containing ULIRGs.

\subsection{Infrared to radio spectral energy distribution}

Figure 16 shows the spectral energy distribution (SED) of IRAS
20551$-$4250 in the infrared and radio.
The q-value, defined as the far-infrared-to-radio flux ratio, is q
$\sim$ 2.7 (Table 1).
This is not smaller than for starburst-dominated galaxies and many
radio-quiet AGNs (q $\sim$ 2.3--2.4) \citep{con91,bar96,cra96,roy98},
suggesting that the infrared-to-radio SED of IRAS 20551$-$4250 is an
example of a radio-quiet AGN and starburst composite. 

Continuum data points in ALMA bands 7 (275--373 GHz) and 9 (602--720 GHz)
are well fitted to the extrapolation from the infrared dust thermal
radiation at 2000--7500 GHz (40--150 $\mu$m).
ALMA band 3 (84--116 GHz) continuum data points roughly agree with 
the extrapolation from lower-frequency synchrotron emission. 
The thermal free-free emission from starburst regions in IRAS
20551$-$4250 is also present \citep{ima13b,ima16b}, but mainly
contributes to the frequency range at the minimum flux density
($\sim$200 GHz), due to its relatively flat spectral shape.
Overall, continuum flux measurements with our ALMA data in the
(sub)millimeter wavelength range are consistent with those at other
wavelengths taken with other observing facilities, further confirming
that most activity (AGN and/or starburst) in IRAS 20551$-$4250 is 
spatially compact, and their emission is well recovered with our ALMA 
interferometric data.

\section{Summary}

We conducted ALMA band 3, 6, 7, and 9 observations of the
buried-AGN-hosting ULIRG, IRAS 20551$-$4250, in Cycles 2, 3, and 4.
When our new data were combined with our previously collected ALMA data, we
obtained information on flux and its sensitive upper limit for
J=1--0, J=3--2, J=4--3, and J=8--7 emission lines of HCN, HCO$^{+}$, HNC,
CO J=3--2, vibrationally excited (v$_{2}$=1f) J=3--2 and J=4--3 emission
lines of HCN, HCO$^{+}$, HNC, v$_{2}$=1f J=8--7 emission lines of HCN
and HNC, J=3--2 emission lines of isotopologue H$^{13}$CN,
H$^{13}$CO$^{+}$, HN$^{13}$C, and some other serendipitously detected
emission lines.    
From our previous ALMA observations of HCN, HCO$^{+}$, HNC at J=3--2 and
calculations of infrared radiative pumping, it had been argued that HCN
abundance was higher than HCO$^{+}$ and HNC, and that an infrared
radiative pumping played a role in molecular gas rotational (J)
excitation at the vibrational ground level (v=0) in IRAS 20551$-$4250. 
By adding our new ALMA data, we found the following main results. 

\begin{enumerate}

\item We detected v$_{2}$=1f emission lines at J=4--3 for HCN and HNC,
but not for HCO$^{+}$, in the same manner as our previous J=3--2
observations.
HCO$^{+}$ v$_{2}$=1f fluxes are smaller than HCN v$_{2}$=1f and HNC
v$_{2}$=1f fluxes at both J=3--2 and J=4--3. 
The estimated vibrational excitation temperature from J=4--3 supported 
our previous argument based on J=3--2 data that the infrared radiative
pumping plays a role in rotational excitation at v=0, at least for HCN
and HNC.

\item There are possible signatures of v$_{2}$=1f J=8--7 emission lines
of HCN and HNC in our ALMA band 9 spectrum; however, their detection
significance is at most $\sim$3$\sigma$, in both their integrated
intensity (moment 0) maps and Gaussian fits in the spectra.
Confirmation with deeper data will be necessary for further discussion
of these lines. 

\item The J=4--3 fluxes of HCN, HCO$^{+}$, and HNC at v$_{2}$=1f and
v=0 can be explained by a higher HCN abundance than HCO$^{+}$ and
HNC, as was previously argued from J=3--2 data.   

\item The comparison of fluxes between HCN, HCO$^{+}$, HNC and their 
isotopologues (H$^{13}$CN, H$^{13}$CO, HN$^{13}$C) at J=3--2 also suggests
higher flux attenuation by line opacity for HCN than for HCO$^{+}$ and HNC,
supporting the above higher HCN abundance scenario than 
HCO$^{+}$ and HNC.

\item Line-opacity-corrected intrinsic molecular line flux ratios among 
HCN, HCO$^{+}$, and HNC were derived at J=3--2 and J=4--3. 
We confirmed that IRAS 20551$-$4250 has higher intrinsic 
HCN-to-HCO$^{+}$ and HCN-to-HNC flux ratios than the observed 
ratios. The difference from starburst-dominated regions was even larger
if the intrinsic ratios are used for IRAS 20551$-$4250.

\item The bright CO J=3--2 emission line enabled us to probe the
dynamical properties of spatially extended molecular gas in detail, and
showed a dynamically distinct region at the south-western part of the
nucleus, compared to overall rotational motion.
Our molecular observations support the scenario that some merger event
occurred in the past in IRAS 20551$-$4250.  

\item A broad (FWHM $\sim$ 500 km s$^{-1}$) emission line component was
detected in the bright CO J=3--2 emission line, which we interpret to
be caused by molecular outflow activity in this galaxy. 
We estimate a molecular outflow mass M$_{\rm outf}$ $\sim$ 5.8 $\times$
10$^{6}$ M$_{\odot}$, a molecular outflow rate \.{M}$_{\rm outf}$ $\sim$
150 (M$_{\odot}$ yr$^{-1}$), a molecular outflow kinetic power 
P$_{\rm outf}$ $\sim$ 1\% of AGN luminosity (L$_{\rm AGN}$), and 
a molecular outflow momentum rate \.{P}$_{\rm outf}$ $\sim$ 12 
$\times$ L$_{\rm AGN}$/c.

\item We conducted RADEX calculations and derived H$_{2}$ number density 
and kinetic temperature that reproduce the observed flux ratios at
different rotational J-transition lines of HCN, HCO$^{+}$, HNC at v=0,
by collisional excitation.
The derived number density and kinetic temperature tend to be higher
when we use higher-J transition lines, supporting the expectation that
higher-J transition lines preferentially probe higher-density and
higher-temperature molecular gas in more compact areas.
The comparison between J=3--2 and J=1--0 suggests that the typical
molecular gas density that emits the HCN J=1--0 emission line is
$\sim$10$^{5}$--10$^{6}$ (cm$^{-3}$). 
The X-ray-absorbing hydrogen column density and 
estimated mass of dense molecular gas from HCN J=1--0 luminosity suggest
that the volume averaged number density and size of HCN J=1--0 emitting
molecular gas are $\sim$10$^{3}$ (cm$^{-3}$) and $\sim$200 (pc),
respectively. 
We infer that the molecular clouds in the IRAS 20551$-$4250 nucleus are
in a clumpy form in which dense clumps with $\sim$10$^{5}$--10$^{6}$
(cm$^{-3}$) occupy 0.1--1\% of the total volume of the nuclear molecular
gas.  

\item From our J=8--7, J=4--3, J=3--2, and J=1--0 data of
HCN, HCO$^{+}$, and HNC, higher rotatioal excitation was seen for
HNC than for HCN and HCO$^{+}$.
This is difficult to explain with collisional excitation alone, due to
the estimated higher effective critical density of HNC than HCN and 
HCO$^{+}$.
Infrared radiative pumping could explain this result because of
higher infrared pumping efficiency for HNC than for HCN and HCO$^{+}$, due
to a larger amount of infrared photons at $\sim$22 $\mu$m (wavelength for
vibrational-rotational transitions between v=0 and v$_{2}$=1 for HNC) 
than at 12--14 $\mu$m (those for HCN and HCO$^{+}$).
Besides the vibrational excitation temperature, this is another 
indication that infrared radiative pumping is at work in IRAS
20551$-$4250.
Given that Einstein A coefficients for vibrational-rotational
transitions are available for HCN and HNC, we performed quantitative 
calculations of the effects of the infrared radiative pumping process 
for HCN and HNC, using the MOLPOP software.
We found that the observed higher HNC rotational excitation was
largely reproduced if an energy source with observationally constrained
infrared spectral shape was placed at 30--100 pc from the molecular gas.

\end{enumerate}

IRAS 20551$-$4250 is a unique ULIRG in that the small observed 
molecular emission line widths allow us to clearly separate the HCN
v$_{2}$=1f J=3--2 and J=4--3 emission lines from the nearby, much
brighter HCO$^{+}$ (v=0) J=3--2 and J=4--3 emission lines.
Thus, molecular line fluxes at the v$_{2}$=1 level and the effects of 
infrared radiative pumping can be investigated with small quantitative
uncertainty.   
Given that the infrared radiative pumping can have a significant effect
to other AGN-containing ULIRGs, but that its quantitative estimate is
more difficult due to the spectral overlap between HCN v$_{2}$=1f and
HCO$^{+}$ v=0 emission lines at J=3--2 and J=4--3, further detailed
investigations of IRAS 20551$-$4250 will provide an important clue to
understand how the infrared radiative pumping contributes to molecular
rotational J-transition line fluxes at v=0, in addition to the usually
assumed collisional excitation, in AGN-containing ULIRGs.

We thank the anonymous referee for his/her useful comment 
which helped a lot to improve the clarity of this manuscript.
We are grateful to Dr. H. Nagai, A. Kawamura, F. Egusa for their kind help
regarding ALMA data retrieval and reduction.
We are thankful to Dr. M. Elitzur and A. Asensio Ramos for their kind
advice about the use of the MOLPOP code.  
M.I. was supported by JSPS KAKENHI Grant Number 23540273, 15K05030 and
the ALMA Japan Research Grant of the NAOJ Chile Observatory,
NAOJ-ALMA-0001, 0023, 0072. 
T.I. was thankful for the fellowship received from the Japan Society for
the Promotion of Science (JSPS).
This paper makes use of the following ALMA data:
ADS/JAO.ALMA\#2013.1.00033.S, \#2015.1.00028.S, and \#2016.1.00051.S. 
ALMA is a partnership of ESO (representing
its member states), NSF (USA) and NINS (Japan), together with NRC
(Canada), NSC and ASIAA (Taiwan), and KASI (Republic of Korea), in
cooperation with the Republic of Chile. The Joint ALMA Observatory is
operated by ESO, AUI/NRAO, and NAOJ.
Data analysis was in part carried out on the open use data analysis
computer system at the Astronomy Data Center, ADC, of the National
Astronomical Observatory of Japan. 
This research has made use of NASA's Astrophysics Data System and the
NASA/IPAC Extragalactic Database (NED) which is operated by the Jet
Propulsion Laboratory, California Institute of Technology, under
contract with the National Aeronautics and Space Administration.


\begin{deluxetable}{lcrrrrcccl}[!bht]
\tabletypesize{\scriptsize}
\tablecaption{Observed properties of IRAS 20551$-$4250 \label{tbl-1}}
\tablewidth{0pt}
\tablehead{
\colhead{Object} & \colhead{Redshift}   & 
\colhead{f$_{\rm 12}$}   & 
\colhead{f$_{\rm 25}$}   & 
\colhead{f$_{\rm 60}$}   & 
\colhead{f$_{\rm 100}$}  & 
\colhead{log L$_{\rm IR}$} & 
\colhead{log L$_{\rm FIR}$} & 
\colhead{S$_{\rm 1.4 GHz}$}  & \colhead{q} \\
\colhead{} & \colhead{}   & \colhead{(Jy)} & \colhead{(Jy)} 
& \colhead{(Jy)} & \colhead{(Jy)}  & \colhead{(L$_{\odot}$)}  &
\colhead{(L$_{\odot}$)} & \colhead{(mJy)} & \colhead{}  \\
\colhead{(1)} & \colhead{(2)} & \colhead{(3)} & \colhead{(4)} & 
\colhead{(5)} & \colhead{(6)} & \colhead{(7)} & \colhead{(8)} & 
\colhead{(9)} & \colhead{(10)} 
}
\startdata
IRAS 20551$-$4250 & 0.043 & 0.28 & 1.91 & 12.78 & 9.95  & 12.0 & 11.9 & 31.0 & 2.7 \\   
\enddata

\tablecomments{
Col.(1): Object name. 
Col.(2): Redshift. 
Col.(3)--(6): f$_{12}$, f$_{25}$, f$_{60}$, and f$_{100}$ are 
{\it IRAS} fluxes at 12 $\mu$m, 25 $\mu$m, 60 $\mu$m, and 100 $\mu$m,
respectively, taken from the IRAS {\it FSC} catalog. 
Col.(7): Decimal logarithm of infrared (8$-$1000 $\mu$m) luminosity
in units of solar luminosity (L$_{\odot}$), calculated with
$L_{\rm IR} = 2.1 \times 10^{39} \times$ D(Mpc)$^{2}$
$\times$ (13.48 $\times$ $f_{12}$ + 5.16 $\times$ $f_{25}$ +
$2.58 \times f_{60} + f_{100}$) (ergs s$^{-1}$) \citep{sam96}.
Col.(8): Decimal logarithm of far-infrared (40$-$500 $\mu$m) luminosity
in units of solar luminosity (L$_{\odot}$), calculated with
$L_{\rm FIR} = 2.1 \times 10^{39} \times$ D(Mpc)$^{2}$
$\times$ ($2.58 \times f_{60} + f_{100}$) (ergs s$^{-1}$) \citep{sam96}.
Col.(9): Radio 1.4 GHz flux in (mJy) \citep{con96}.
Col.(10): Decimal logarithm of the far-infrared-to-radio flux ratio,
defined as a q-value \citep{con91}.
}

\end{deluxetable}

\begin{deluxetable}{llccc|ccc}[!bht]
\tabletypesize{\scriptsize}
\tablecaption{Log of our ALMA observations \label{tbl-2}} 
\tablewidth{0pt}
\tablehead{
\colhead{Band} & \colhead{Date} & \colhead{Antenna} & 
\colhead{Baseline} & \colhead{Integration} & \multicolumn{3}{c}{Calibrator} \\ 
\colhead{} & \colhead{(UT)} & \colhead{Number} & \colhead{(m)} &
\colhead{(min)} & \colhead{Bandpass} & \colhead{Flux} & \colhead{Phase}  \\
\colhead{(1)} & \colhead{(2)} & \colhead{(3)} & \colhead{(4)} &
\colhead{(5)} & \colhead{(6)} & \colhead{(7)}  & \colhead{(8)} 
}
\startdata 
Band-7a (HCO$^{+}$ J=4--3) & 2014 June 8 & 34 & 28--646 & 15 & J1924$-$2914 & Titan & J2056$-$4714 \\   
& 2015 April 3 & 37 & 15--328 & 15 & J1924$-$2914 & Titan & J2056$-$4714 \\   
Band-7b (HNC J=4--3) & 2014 June 8 & 34 & 28--646 & 39 & J1924$-$2914 & J2056$-$4714 & J2056$-$4714 \\   
& 2015 April 29 & 39 & 15--349 & 39 & J2056$-$4714 & Titan & J2056$-$4714 \\   
Band 9 (HCN, HCO$^{+}$, HNC J=8--7) & 2016 May 18 & 42 & 15--640 & 39 &
J2253$+$1608 & Pallas & J2056$-$4714 \\   
Band-3 (HCN, HCO$^{+}$, HNC J=1--0) & 2016 May 27 & 41 & 15--784 & 39 &
J2056$-$4714 & J2056$-$4714 & J2049$-$4020 \\   
& 2016 August 15 & 38 & 15--1462 & 47 & J2056$-$4714 & J2056$-$4714 & J2049$-$4020 \\   
Band-6 (H$^{13}$CN, H$^{13}$CO$^{+}$, HN$^{13}$C J=3--2) & 2016 October 1
& 41 & 15--3248 & 40 & J1924$-$2914 & J2056$-$4714 & J2056$-$4714 \\    
 & 2016 October 5 & 44 & 19--3248 & 40 &
J2056$-$4714 & J2056$-$4714 & J2056$-$4714 \\   
 & 2016 October 5--6 & 44 & 19--3248 & 40 & J2056$-$4714 & J2056$-$4714 & J2056$-$4714 \\   

\enddata

\tablecomments{ 
Col.(1): Observed band and primarily targeted emission lines.
Col.(2): Observing date in UT. 
Col.(3): Number of antennas used for observations. 
Col.(4): Baseline length in meters. Minimum and maximum baseline lengths are shown.  
Col.(5): Net on-source integration time in min.
Cols.(6), (7), and (8): Bandpass, flux, and phase calibrators for the 
target source, respectively.
}

\end{deluxetable}

\begin{deluxetable}{ccrccl}[!bht]
\tabletypesize{\scriptsize}
\tablecaption{Continuum emission properties \label{tbl-3}}
\tablewidth{0pt}
\tablehead{
\colhead{Continuum} & \colhead{Frequency} & \colhead{Flux} & 
\colhead{Peak Coordinate} & \colhead{rms} & \colhead{Synthesized beam} \\
\colhead{} & \colhead{(GHz)} & \colhead{(mJy beam$^{-1}$)} & 
\colhead{(RA,DEC)J2000} & \colhead{(mJy beam$^{-1}$)} & 
\colhead{(arcsec $\times$ arcsec) ($^{\circ}$)} \\  
\colhead{(1)} & \colhead{(2)} & \colhead{(3)}  & \colhead{(4)}  &
\colhead{(5)} & \colhead{(6)}  
}
\startdata 
J10 & 84.1--87.9, 96.1--99.9 & 2.3 (61$\sigma$) & 
(20 58 26.80, $-$42 39 00.3) & 0.038 & 1.1$\times$0.9 (68$^{\circ}$) \\
J32 & 232.1--235.7, 247.3--250.9 & 1.8 (68$\sigma$) & 
(20 58 26.80, $-$42 39 00.3) & 0.027 & 0.18$\times$0.17 (46$^{\circ}$) \\
J43a & 329--332.5, 340.7--344.4  & 10.5 (59$\sigma$) & (20 58 26.80, $-$42 39 00.3) & 0.18
& 0.84$\times$0.68 (82$^{\circ}$) \\
J43b & 335.3--338.9, 347.2--351.0 &  10.3 (81$\sigma$) & (20 58 26.80,
$-$42 39 00.3) & 0.13 & 0.66$\times$0.53 ($-$81$^{\circ}$) \\   
J87 & 678.7--684.4, 694.2--701.0  & 75.7 (29$\sigma$) & 
(20 58 26.80, $-$42 39 00.3) & 2.6 & 0.23$\times$0.21 (89$^{\circ}$) \\
\enddata

\tablecomments{
Col.(1): Continuum. J10, J32, J43a, J43b, and J87 were taken with
``HCN, HCO$^{+}$, HNC J=1--0'', ``H$^{13}$CN, H$^{13}$CO$^{+}$, HN$^{13}$C
J=3--2'', ``HCO$^{+}$ J=4--3'', ``HNC J=4--3'', and ``HCN, HCO$^{+}$,
HNC J=8--7'', respectively. 
Col.(2): Observed frequency in (GHz).
Col.(3): Flux in (mJy beam$^{-1}$) at the emission peak. 
The detection significance relative to the rms noise is shown in
parentheses. 
Possible systematic ambiguity, coming from ALMA absolute flux
calibration uncertainty and choice of frequency range for continuum
determination, is not included.  
Col.(4): The coordinate of the continuum emission peak in J2000.
Col.(5): The rms noise level (1$\sigma$) in (mJy beam$^{-1}$).
Col.(6): Synthesized beam in (arcsec $\times$ arcsec) and position angle
in ($^{\circ}$). 
The position angle is 0$^{\circ}$ in the north--south direction
and increases in the counterclockwise direction.
}

\end{deluxetable}

\begin{deluxetable}{ll|lll|cccc}
\tabletypesize{\scriptsize}
\tablecaption{Molecular Line Flux \label{tbl-4}} 
\tablewidth{0pt}
\tablehead{
\colhead{Line} & \colhead{$\nu_{\rm rest}$} & 
\multicolumn{3}{c}{Integrated intensity (moment 0) map} & 
\multicolumn{4}{c}{Gaussian line fit} \\  
\colhead{} & \colhead{(GHz)} & \colhead{Peak} & \colhead{rms} & 
\colhead{Beam} & \colhead{Velocity} &
\colhead{Peak} & \colhead{FWHM} & \colhead{Flux} \\ 
\colhead{} & \colhead{} & \multicolumn{2}{c}{(Jy beam$^{-1}$ km s$^{-1}$)} &
\colhead{($''$ $\times$ $''$) ($^{\circ}$)} &
\colhead{(km s$^{-1}$)} & \colhead{(mJy)} & \colhead{(km s$^{-1}$)} &
\colhead{(Jy km s$^{-1}$)} \\  
\colhead{(1)} & \colhead{(2)} & \colhead{(3)} & \colhead{(4)} & 
\colhead{(5)} & \colhead{(6)} & \colhead{(7)} & \colhead{(8)} &
\colhead{(9)} 
}
\startdata 
CO J=3--2 & 345.796& 156.9 (64$\sigma$) & 2.46 & 0.83$\times$0.65
($-$84$^{\circ}$) & 12886$\pm$1 & 717$\pm$7 & 221$\pm$3 & 162$\pm$3 \\
HCO$^{+}$ J=4--3 & 356.734 & 16.1 (70$\sigma$) & 0.23 &
0.85$\times$0.67 (74$^{\circ}$) & 12884$\pm$1 & 85$\pm$1 & 194$\pm$2 &
16.9$\pm$0.2 \\  
HCN J=4--3 v$_{2}$=1f & 356.256 & 0.49 (9.5$\sigma$) & 0.052 &
0.86$\times$0.67 (74$^{\circ}$) & 12915$\pm$12 & 2.9$\pm$0.4 &
200$\pm$32 & 0.60$\pm$0.13 \\     
HCO$^{+}$ J=4--3 v$_{2}$=1f \tablenotemark{a} & 358.242 & 0.20
(3.2$\sigma$) & 0.062 & 0.81$\times$0.66 (84$^{\circ}$) & 12947$\pm$34 &
1.4$\pm$0.6 & 146$\pm$58 & 0.21$\pm$0.13 \\  
SO 8(8)--7(7) \tablenotemark{b} & 344.311 & 0.55 (9.8$\sigma$) & 0.056 &
0.85$\times$0.70 (84$^{\circ}$) & 12908$\pm$10 & 4.3$\pm$0.7 &
148$\pm$24 & 0.64$\pm$0.15 \\     
SO$_{2}$$+$SO \tablenotemark{c} & 346.524--528 \tablenotemark{c} &
0.96 (14$\sigma$) & 0.068 & 0.83$\times$0.65 ($-$84$^{\circ}$) &
--- & --- & --- & --- \\ 
HOC$^{+}$ J=4--3 \tablenotemark{d} & 357.922 & 0.46 (6.9$\sigma$) & 0.067 &
0.86$\times$0.69 (82$^{\circ}$) & 12875$\pm$21 & 3.0$\pm$0.5 &
213$\pm$62 & 0.66$\pm$0.22 \\    
CH$_{3}$CCH & 358.709--818 \tablenotemark{e} & 0.45 (7.0$\sigma$) & 0.065 &
0.81$\times$0.66 (84$^{\circ}$) & 12900$\pm$16 & 2.6$\pm$0.5 &
185$\pm$39 & 0.49$\pm$0.14 \\ \hline
HNC J=4--3 & 362.630 & 6.4 (70$\sigma$) & 0.091 & 0.75$\times$0.58
($-$78$^{\circ}$) & 12890$\pm$2 & 37$\pm$1 & 177$\pm$4 & 6.6$\pm$0.2 \\  
HNC J=4--3 v$_{2}$=1f \tablenotemark{f} & 365.147 & 
0.64 (11.5$\sigma$)\tablenotemark{f} & 0.056\tablenotemark{f} &
0.66$\times$0.53 ($-$83$^{\circ}$) & 12890 (fix) & 2.2$\pm$0.4 & 443
(fix) & 0.98$\pm$0.19 \\  
H$_{2}$CO 5(1,5)--4(1,4) & 351.769 & 1.2 (24$\sigma$) & 0.051 &
0.63$\times$0.52 ($-$81$^{\circ}$) & 12879$\pm$4 & 6.8$\pm$0.2 &
213$\pm$9 & 1.5$\pm$0.1 \\ 
HOCO$^{+}$ 17(1,16)--16(1,15)\tablenotemark{g} & 364.804 & 2.5
(42$\sigma$) & 0.059 & 0.81$\times$0.64 ($-$83$^{\circ}$) & 12898$\pm$2
& 13$\pm$1 & 181$\pm$4 & 2.5$\pm$0.1 \\ \hline
HCN J=8--7 & 708.877 & 12.9 (7.1$\sigma$) & 1.8 & 
0.24$\times$0.21 (88$^{\circ}$) & 12930$\pm$9 & 51$\pm$3 & 262$\pm$19 &
13.7$\pm$1.3 \\   
HCO$^{+}$ J=8--7 & 713.341 & 11.9 (9.9$\sigma$) & 1.2 &
0.23$\times$0.21 ($-$88$^{\circ}$) & 12884$\pm$4 & 77$\pm$4 & 159$\pm$10
& 12.5$\pm$1.0 \\   
HNC J=8--7 & 725.107 & 6.9 (4.8$\sigma$) & 1.4 & 0.23$\times$0.20
(90$^{\circ}$) & 12904$\pm$10 & 43$\pm$5 & 164$\pm$21 & 7.2$\pm$1.3 \\ 
HCN J=8--7 v$_{2}$=1f & 712.372 & 1.0 (3.1$\sigma$) & 0.32
\tablenotemark{h} & 0.23$\times$0.21 ($-$88$^{\circ}$) & 12918$\pm$12 &
15$\pm$5 & 61$\pm$25 & 1.0$\pm$0.5 \\   
HNC J=8--7 v$_{2}$=1f & 730.131 & 0.88 (1.7$\sigma$) & 0.52
\tablenotemark{h} & 0.23$\times$0.20 (88$^{\circ}$) & 12871$\pm$30 & 11
(fix) & 67 (fix) & 0.7 \\  
\hline  
HCN J=1--0 & 88.632 & 1.4 (33$\sigma$) & 0.041 & 1.2$\times$0.9
(69$^{\circ}$) & 12881$\pm$2 & 7.1$\pm$0.2 & 195$\pm$5 & 1.4$\pm$0.1 \\  
HCO$^{+}$ J=1--0 & 89.189 & 2.2 (40$\sigma$) & 0.054 & 1.2$\times$0.9
(72$^{\circ}$) & 12888$\pm$1 & 11$\pm$1 & 202$\pm$3 & 2.2$\pm$0.1 \\   
HNC J=1--0 & 90.664 & 0.36 (11$\sigma$) & 0.034 & 1.0$\times$0.7
(61$^{\circ}$) & 12878$\pm$5 & 2.2$\pm$0.1 & 169$\pm$13 & 0.37$\pm$0.04
\\ \hline
H$^{13}$CN J=3--2 & 259.012 & 0.25 (6.8$\sigma$) & 0.036 &
0.16$\times$0.15 (35$^{\circ}$) & 12903$\pm$6 & 1.4$\pm$0.1 & 181$\pm$16
& 0.26$\pm$0.03 \\
H$^{13}$CO$^{+}$ J=3--2 & 260.255 & 0.13 (5.2$\sigma$) & 0.024 &
0.16$\times$0.16 (59$^{\circ}$) & 12901$\pm$8 & 0.86$\pm$0.09 &
159$\pm$18 & 0.14$\pm$0.02 \\
HN$^{13}$C J=3--2 & 261.263 & 0.066 (2.8$\sigma)$ & 0.024 &
0.16$\times$0.16 (59$^{\circ}$) & 12911$\pm$20 & 0.34$\pm$0.07 &
204$\pm$51 & 0.071$\pm$0.023 \\
CS J=5--4 & 244.936 & 1.2 (40$\sigma$) & 0.030 & 0.19$\times$0.17
(38$^{\circ}$) & 12901$\pm$2 & 7.5$\pm$0.1 & 167$\pm$4 & 1.3$\pm$0.1 \\ 
SO 6(6)--5(5) & 258.256 & 0.34 (10$\sigma$) & 0.034 & 0.16$\times$0.15
(35$^{\circ}$) & 12909$\pm$4 & 1.9$\pm$0.1 & 187$\pm$10 & 
0.37$\pm$0.03 \\
SiO J=6--5 & 260.518 & 0.22 (5.8$\sigma$) & 0.037 & 0.16$\times$0.16
(59$^{\circ}$) & 12902$\pm$7 & 1.2$\pm$0.1 & 188$\pm$15 & 0.24$\pm$0.03 \\ 
HC$_{3}$N J=27--26 & 245.606 & 0.15 (6.4$\sigma)$ & 0.023 &
0.19$\times$0.17 (38$^{\circ}$) & 12900$\pm$5 & 0.99$\pm$0.07 & 154$\pm$13 &
0.16$\pm$0.02 \\ 

\enddata

\tablenotetext{a}{
SO$_{2}$ 20(0,20)--19(1,19) emission line ($\nu_{\rm rest}$ = 358.216 GHz) 
might contaminate.}

\tablenotetext{b}{
HC$^{15}$N J=4--3 emission line ($\nu_{\rm rest}$ = 344.200 GHz) might
contaminate.}

\tablenotetext{c}{
Combination of SO$_{2}$ 16(4,12)--16(3,13) ($\nu_{\rm rest}$ = 346.524
GHz) and SO 9(8)--8(7) ($\nu_{\rm rest}$ = 346.528 GHz).}

\tablenotetext{d}{
SO$_{2}$ 6(4,2)--6(3,3) emission line ($\nu_{\rm rest}$ = 357.926 GHz) 
might contaminate.}

\tablenotetext{e}{
Combination of 
CH$_{3}$CCH 21(4)--20(4) ($\nu_{\rm rest}$ = 358.709 GHz), 
CH$_{3}$CCH 21(3)--20(3) ($\nu_{\rm rest}$ = 358.757 GHz), 
CH$_{3}$CCH 21(2)--20(2) ($\nu_{\rm rest}$ = 358.791 GHz), 
CH$_{3}$CCH 21(1)--20(1) ($\nu_{\rm rest}$ = 358.811 GHz), and 
CH$_{3}$CCH 21(0)--20(0) ($\nu_{\rm rest}$ = 358.818 GHz).}

\tablenotetext{f}{
All emission line components with FWHM $\sim$ 443 km s$^{-1}$ are
integrated.}

\tablenotetext{g}{
H$_{3}$O$^{+}$ 3(2)--2(2) emission line ($\nu_{\rm rest}$ = 364.797 GHz) 
might contaminate.}

\tablenotetext{h}{
The rms noise level was determined from the 40--100 pixels annular region
around the center of the moment 0 map.}

\tablecomments{ 
Col.(1): Observed molecular line. 
Primarily targeted lines are listed first, followed by serendipitously detected
lines.
Col.(2): Rest-frame frequency of each molecular line in (GHz). 
Col.(3): Integrated intensity in (Jy beam$^{-1}$ km s$^{-1}$) at the 
emission peak. 
Detection significance relative to the rms noise (1$\sigma$) in the 
moment 0 map is shown in parentheses. 
Possible systematic uncertainty is not included. 
Col.(4): rms noise (1$\sigma$) level in the moment 0 map in 
(Jy beam$^{-1}$ km s$^{-1}$), derived from the standard deviation 
of sky signals in each moment 0 map. 
Col.(5): Synthesized beam in (arcsec $\times$ arcsec) and position angle
in ($^{\circ}$).
Position angle is 0$^{\circ}$ in the north-south direction, 
and increases in the counterclockwise direction. 
Cols.(6)--(9): Gaussian fits of emission lines in the spectra at the 
continuum peak position, within the beam size. 
Col.(6): Optical LSR velocity (v$_{\rm opt}$) of emission peak in (km
s$^{-1}$).  
Col.(7): Peak flux in (mJy). 
Col.(8): Observed FWHM in (km s$^{-1}$) in Figure 4.  
Col.(9): Flux in (Jy km s$^{-1}$). The observed FWHM in 
(km s$^{-1}$) in column 8 is divided by ($1+z$) to obtain the
intrinsic FWHM in (km s$^{-1}$).
}

\end{deluxetable}

\begin{deluxetable}{lcc}
\tablecaption{Intrinsic emission size after deconvolution for selected 
bright emission lines and continuum \label{tbl-5}}
\tablewidth{0pt}
\tablehead{
\colhead{Line} & \colhead{(mas $\times$ mas) ($^{\circ}$)} &
\colhead{Beam (arcsec $\times$ arcsec) } \\
\colhead{(1)} & \colhead{(2)} & \colhead{(3)} 
}
\startdata
CO J=3--2 & 828$\pm$123, 561$\pm$127 (131$\pm$23) & 0.83 $\times$ 0.65 \\
HCO$^{+}$ J=4--3 & 306$\pm$35, 275$\pm$37 (96$\pm$76) & 0.85 $\times$ 0.67 \\
cont43a & 348$\pm$52, 336$\pm$63 (136$\pm$88) & 0.84 $\times$ 0.68 \\ \hline
HNC J=4--3 & 206$\pm$30, 105$\pm$55 (152$\pm$17) & 0.75 $\times$ 0.58 \\
cont43b & 326$\pm$26, 284$\pm$22 (89$\pm$48) & 0.66 $\times$ 0.53 \\ \hline
HCN J=8--7 & 218$\pm$44, 135$\pm$43 (84$\pm$38) & 0.24 $\times$ 0.21\\
HCO$^{+}$ J=8--7 & 246$\pm$36, 179$\pm$34 (70$\pm$34) & 0.23 $\times$ 0.21 \\
HNC J=8--7 & 141$\pm$57, 65$\pm$50 (72$\pm$74) & 0.23 $\times$ 0.20 \\
cont87 & 261$\pm$16, 225$\pm$15 (56$\pm$23) & 0.23 $\times$ 0.21 \\ \hline
HCN J=1--0 & 711$\pm$138, 655$\pm$170 (108$\pm$82) & 1.19 $\times$ 0.95 \\
HCO$^{+}$ J=1--0 & 763$\pm$136, 609$\pm$221 (142$\pm$83) & 
1.22 $\times$ 0.95 \\
HNC J=1--0 & 320$\pm$222, 45$\pm$186 (71$\pm$22) & 1.02 $\times$ 0.73 \\
cont10 & 510$\pm$81, 468$\pm$87 (84$\pm$87) & 1.09 $\times$ 0.87 \\ \hline
CS J=5--4 & 201$\pm$13, 152$\pm$13 (80$\pm$13) & 0.19 $\times$ 0.17 \\
cont32 & 198$\pm$13, 170$\pm$12 (60$\pm$28) & 0.18 $\times$ 0.17\\ \hline

\enddata

\tablecomments{
Col.(1): Emission line or continuum.
Col.(2): Intrinsic emission size in mas, after deconvolution using
the CASA task ``imfit''. 
The position angle in ($^{\circ}$) is shown in parentheses.
Col.(3): Synthesized beam size in arcsec $\times$ arcsec, shown for
reference.
}

\end{deluxetable}

\begin{deluxetable}{lccc}
\tabletypesize{\scriptsize}
\tablecaption{Luminosity of selected molecular emission lines \label{tbl-6}}
\tablewidth{0pt}
\tablehead{
\colhead{Line} & \colhead{Flux (Jy km s$^{-1}$)} & 
\colhead{10$^{4}$ (L$_{\odot}$)} & 
\colhead{10$^{7}$ (K km s$^{-1}$ pc$^{2}$)} 
}
\startdata
HCN J=1--0 & 1.4$\pm$0.1 & 0.44$\pm$0.03 & 19.6$\pm$1.4 \\
HCN J=3--2 & 5.9$\pm$0.1 \tablenotemark{a} & 5.5$\pm$0.1 \tablenotemark{a} &
9.2$\pm$0.2 \tablenotemark{a} \\
HCN J=4--3 & 11.6$\pm$0.2 \tablenotemark{b} & 14.5$\pm$0.3
\tablenotemark{b} & 10.1$\pm$0.2 \tablenotemark{b} \\ 
HCN J=8--7 & 13.7$\pm$1.3 & 34.1$\pm$3.2 & 3.0$\pm$0.3 \\ 
HCN v$_{2}$=1f J=3--2 & 0.25$\pm$0.07 \tablenotemark{a} & 0.23$\pm$0.07
\tablenotemark{a} & 0.35$\pm$0.10 \tablenotemark{a} \\ 
HCN v$_{2}$=1f J=4--3 & 0.60$\pm$0.13 & 0.75$\pm$0.16 & 0.52$\pm$0.11 \\ \hline
HCO$^{+}$ J=1--0 & 2.2$\pm$0.1 & 0.69$\pm$0.03 & 30.4$\pm$1.4 \\
HCO$^{+}$ J=3--2 & 8.4$\pm$0.1 \tablenotemark{a}  & 7.9$\pm$0.1 \tablenotemark{a} 
& 12.9$\pm$0.2 \tablenotemark{a} \\
HCO$^{+}$ J=4--3 & 16.9$\pm$0.2 & 21.2$\pm$0.3 & 14.6$\pm$0.2 \\
HCO$^{+}$ J=8--7 & 12.5$\pm$1.0 & 31.3$\pm$2.5 & 2.7$\pm$0.2 \\ 
HCO$^{+}$ v$_{2}$=1f J=3--2 &  $<$0.088 \tablenotemark{a} & $<$0.084
\tablenotemark{a} & $<$0.13 \tablenotemark{a} \\ 
HCO$^{+}$ v$_{2}$=1f J=4--3 & $<$0.20 & $<$0.26 & $<$0.18 \\ \hline 
HNC J=1--0 & 0.37$\pm$0.04 & 0.12$\pm$0.01 & 4.9$\pm$0.5 \\
HNC J=3--2 & 3.2$\pm$0.1 \tablenotemark{a} & 3.1$\pm$0.1
\tablenotemark{a} & 4.7$\pm$0.1 \tablenotemark{a} \\ 
HNC J=4--3 & 6.6$\pm$0.2 & 8.4$\pm$0.3 & 5.5$\pm$0.2 \\
HNC J=8--7 & 7.2$\pm$1.3 & 18.3$\pm$3.3 & 1.5$\pm$0.3 \\ 
HNC v$_{2}$=1f J=3--2 & 0.20$\pm$0.07 \tablenotemark{a} & 0.19$\pm$0.07
\tablenotemark{a} & 0.27$\pm$0.09 \tablenotemark{a} \\
HNC v$_{2}$=1f J=4--3 & 0.98$\pm$0.19 & 1.3$\pm$0.2 & 0.81$\pm$0.16 \\ \hline
H$^{13}$CN J=3--2 & 0.26$\pm$0.03 & 0.24$\pm$0.03 & 0.43$\pm$0.05 \\
H$^{13}$CO$^{+}$ J=3--2 & 0.14$\pm$0.02 & 0.13$\pm$0.02 & 0.23$\pm$0.03 \\
HN$^{13}$C J=3--2 & 0.071$\pm$0.023 & 0.065$\pm$0.021 & 0.11$\pm$0.04 \\
\hline
CO J=3--2 & 162$\pm$3 & 197$\pm$4 & 149$\pm$3 \\
CS J=5--4 & 1.3$\pm$0.1 & 1.1$\pm$0.1 & 2.4$\pm$0.2 \\

\enddata

\tablenotetext{a}{Taken from \citet{ima16b}.}

\tablenotetext{b}{Originally taken from ALMA Cycle 0 data by
\citet{ima13b} and multiplied by a factor of 1.22 to correct for the
flux difference between Cycle 0 and 2 data. 
See $\S$3 for more detail.}

\tablecomments{
Col.(1): Emission line.
Col.(2): Adopted values for the observed flux in (Jy km s$^{-1}$) shown
for reference.  
Col.(3): Luminosity in units of (L$_{\odot}$), calculated with
Equation (1) of \citet{sol05}.
Col.(4): Luminosity in units of (K km s$^{-1}$ pc$^{2}$), 
calculated with Equation (3) of \citet{sol05}.
}

\end{deluxetable}

\begin{deluxetable}{lcrcc}
\tabletypesize{\small}
\tablecaption{Parameters for molecular transition lines \label{tbl-7}}
\tablewidth{0pt}
\tablehead{
\colhead{Line} & \colhead{Frequency} & \colhead{E$_{\rm u}$/k$_{\rm B}$} & 
\colhead{A$_{\rm ul}$} & \colhead{Flux} \\
\colhead{} & \colhead{(GHz)} & \colhead{(K)} & 
\colhead{(10$^{-3}$ s$^{-1}$)} & \colhead{(Jy km s$^{-1}$)} \\
\colhead{(1)} & \colhead{(2)} & \colhead{(3)}  & \colhead{(4)}  &
\colhead{(5)}  
}
\startdata 
HCN J=1--0 & 88.632 & 4.3 & 0.024 & 1.4$\pm$0.1 \\
HCN J=3--2 & 265.886 & 25.5 & 0.84 & 5.9$\pm$0.1 \\
HCN J=4--3 & 354.505 & 42.5 & 2.1 & 11.6$\pm$0.2 \\  
HCN J=8--7 & 708.877 & 153.1 & 17.4 & 13.7$\pm$1.3 \\
HCN J=3--2, v$_{2}$=1f & 267.199 & 1050.0 & 0.73 & 0.25$\pm$0.07 \\
HCN J=4--3, v$_{2}$=1f & 356.256 & 1067.1 & 1.9 & 0.60$\pm$0.13 \\ \hline
HCO$^{+}$ J=1--0 & 89.189 & 4.3 & 0.042 & 2.2$\pm$0.1 \\
HCO$^{+}$ J=3--2 & 267.558 & 25.7 & 1.5 & 8.4$\pm$0.1 \\
HCO$^{+}$ J=4--3 & 356.734 & 42.8 & 3.6 & 16.9$\pm$0.2 \\  
HCO$^{+}$ J=8--7 & 713.341 & 154.1 & 30.2 & 12.5$\pm$1.0 \\
HCO$^{+}$ J=3--2, v$_{2}$=1f & 268.689 & 1217.4 & 1.3 & $<$0.088 \\
HCO$^{+}$ J=4--3, v$_{2}$=1f & 358.242 & 1234.6 & 3.4 & $<$0.20 \\ \hline
HNC J=1--0 & 90.664 & 4.4 & 0.027 & 0.37$\pm$0.04 \\
HNC J=3--2 & 271.981 & 26.1 & 0.93 & 3.2$\pm$0.1 \\
HNC J=4--3  & 362.630 & 43.5 & 2.3 & 6.6$\pm$0.2 \\  
HNC J=8--7 & 725.107 & 156.6 & 19.4 & 7.2$\pm$1.3 \\
HNC J=3--2, v$_{2}$=1f & 273.870 & 692.0 & 0.85 & 0.20$\pm$0.07 \\
HNC J=4--3, v$_{2}$=1f & 365.147 & 709.6 & 2.2 & 0.98$\pm$0.19 \\ 
\enddata

\tablecomments{
Col.(1): Molecular transition line.
Col.(2): Rest-frame frequency in (GHz).
Col.(3): Upper energy level in (K).
Col.(4): Einstein A coefficient for spontaneous emission in 
(10$^{-3}$ s$^{-1}$). 
Values in Cols. (3) and (4) are from the Cologne Database of
Molecular Spectroscopy (CDMS) \citep{mul05} via Splatalogue
(http://www.splatalogue.net).
Col.(5): Flux in (Jy km s$^{-1}$) estimated based on Gaussian fit 
(Table 4, column 9 of this paper; Imanishi \& Nakanishi 2013b; Imanishi
et al. 2016b). 
For the undetected HCO$^{+}$ v$_{2}$=1f J=3--2 and J=4--3 emission
lines, upper limits based on the integrated intensity (moment 0) maps
are adopted.
}

\end{deluxetable}

\begin{deluxetable}{lcc}
\tablecaption{Excitation temperature for IRAS 20551$-$4250 \label{tbl-8}}
\tablewidth{0pt}
\tablehead{
\colhead{Molecule} & \colhead{(v$_{2}$,J; v, J)} & 
\colhead{T$_{\rm ex}$ (K)} \\ 
\colhead{(1)} & \colhead{(2)} & \colhead{(3)} 
}
\startdata 
HCN & (1f,4; 0,4) & 360  \\  
HCO$^{+}$ & (1f,4; 0,4) & $<$280 \\ 
HNC & (1f,4; 0,4) & 360  \\ \hline
HCN & (0,3; 0,1) & 7  \\
    & (0,4; 0,3) & 36 \\
    & (0,8; 0,4) & 42 \\ \hline
HCO$^{+}$ & (0,3; 0,1) & 7 \\
    & (0,4; 0,3) & 38 \\
    & (0,8; 0,4) & 36 \\ \hline
HNC & (0,3; 0,1) & 10 \\
    & (0,4; 0,3) & 41 \\
    & (0,8; 0,4) & 42 \\ 
\enddata

\tablecomments{
Col.(1): Molecule.
Col.(2): Transition.
Col.(3): Excitation temperature (T$_{\rm ex}$) in (K), derived from
observed fluxes.
Different beam sizes among different J-transition lines are not
considered, to be consistent with our previous estimates \citep{ima16b}.
}
\end{deluxetable}

\begin{deluxetable}{cccc}
\tablecaption{Flux ratios for HCN, HCO$^{+}$, and HNC \label{tbl-9}}
\tablewidth{0pt}
\tablehead{
\colhead{J-transition} & \colhead{HCN} & \colhead{HCO$^{+}$} & \colhead{HNC} \\
\colhead{(1)} & \colhead{(2)} & \colhead{(3)} & \colhead{(4)} 
}
\startdata
3--2/1--0 & 4.2$\pm$0.6 & 3.8$\pm$0.5 & 8.6$\pm$1.2 \\
4--3/1--0 & 8.3$\pm$1.2 & 7.6$\pm$1.1 & 17.8$\pm$2.5 \\
8--7/1--0 & 9.8$\pm$1.4 & 5.6$\pm$0.8 & 19.3$\pm$4.0 \\
4--3/3--2 & 2.0$\pm$0.3 & 2.0$\pm$0.3 & 2.1$\pm$0.3 \\
8--7/3--2 & 2.3$\pm$0.3 & 1.5$\pm$0.2 & 2.2$\pm$0.5 \\
8--7/4--3 & 1.2$\pm$0.2 & 0.74$\pm$0.10 & 1.1$\pm$0.2 \\
\enddata

\tablecomments{
Ratios of flux in (Jy km s$^{-1}$).
Col.(1): J-transition. 
Col.(2): HCN.
Col.(3): HCO$^{+}$.
Col.(4): HNC.
For all flux ratios, 10$\%$ absolute calibration uncertainty in
individual ALMA observations is included.
}
\end{deluxetable}

\begin{figure}
\begin{center}
\includegraphics[angle=0,scale=.391]{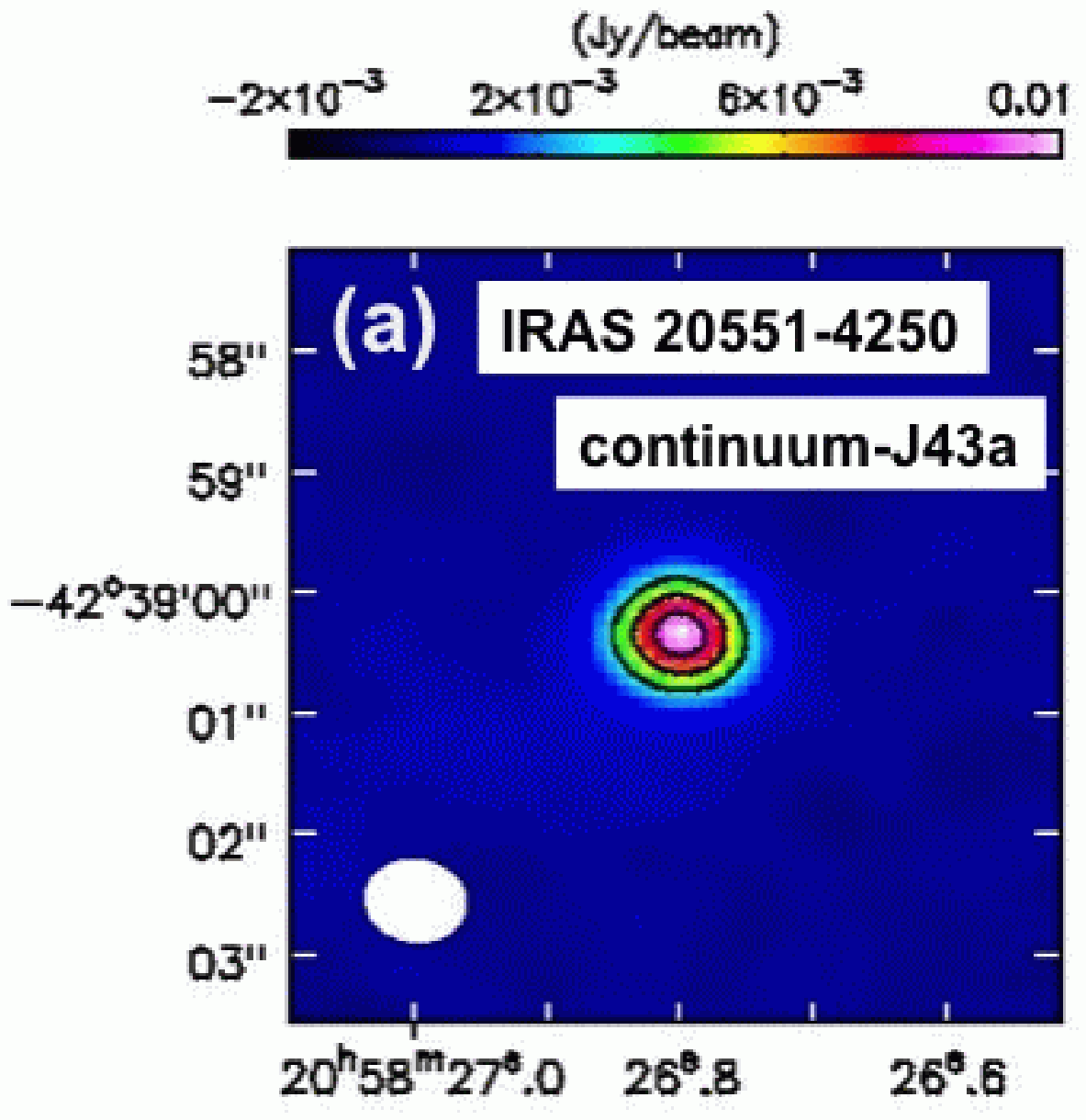} 
\includegraphics[angle=0,scale=.391]{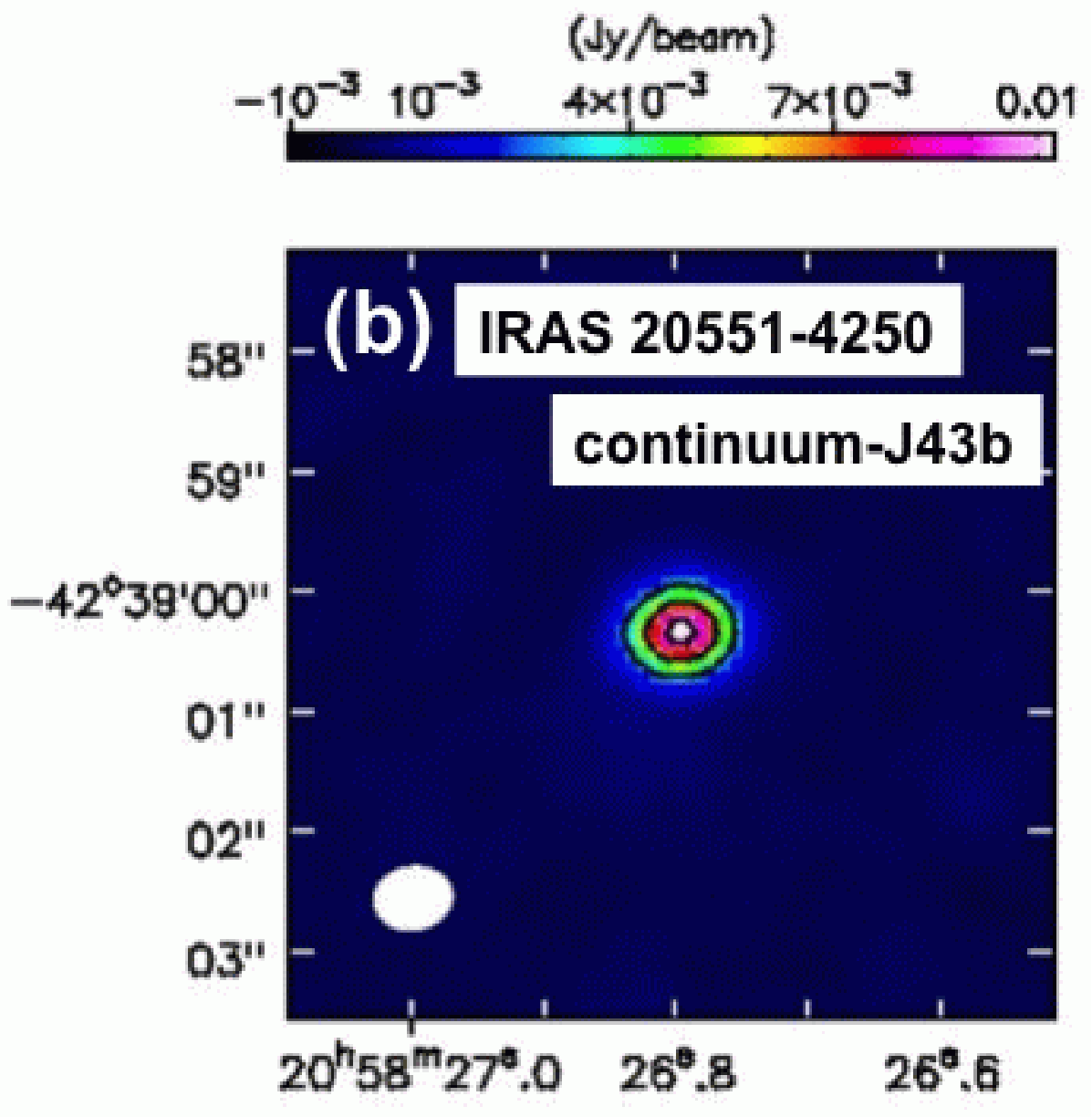} 
\includegraphics[angle=0,scale=.391]{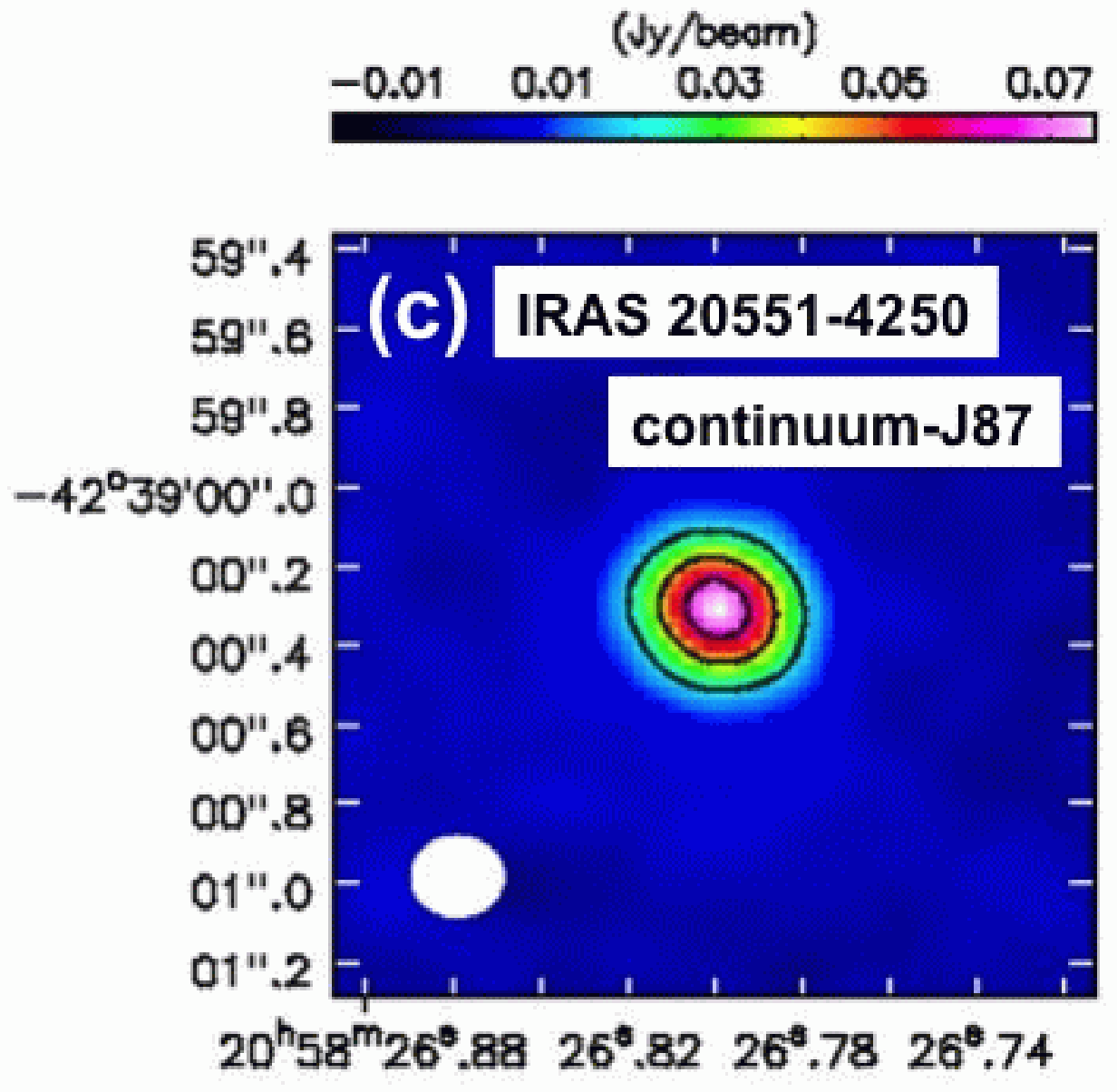} \\
\includegraphics[angle=0,scale=.391]{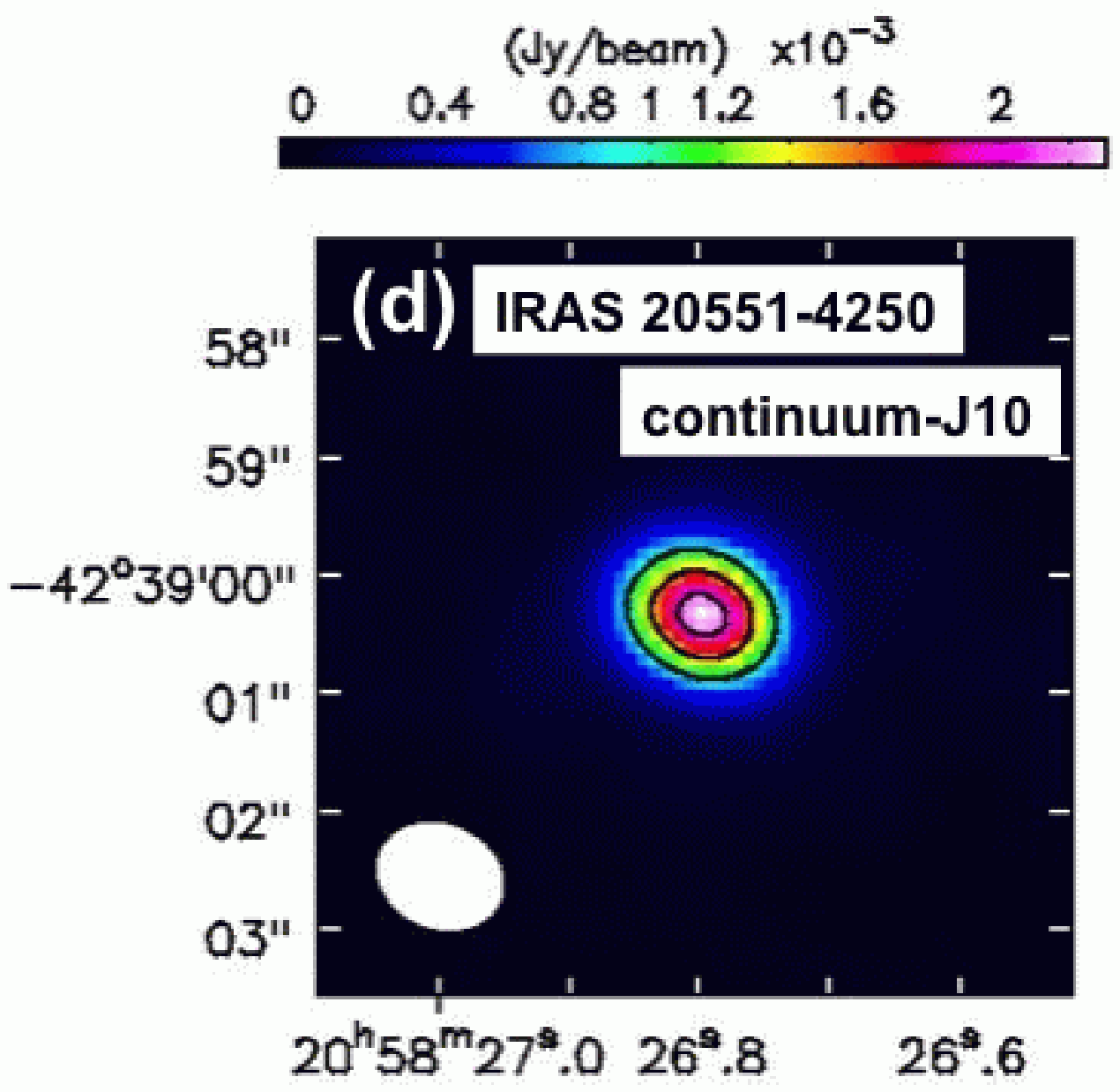} 
\includegraphics[angle=0,scale=.391]{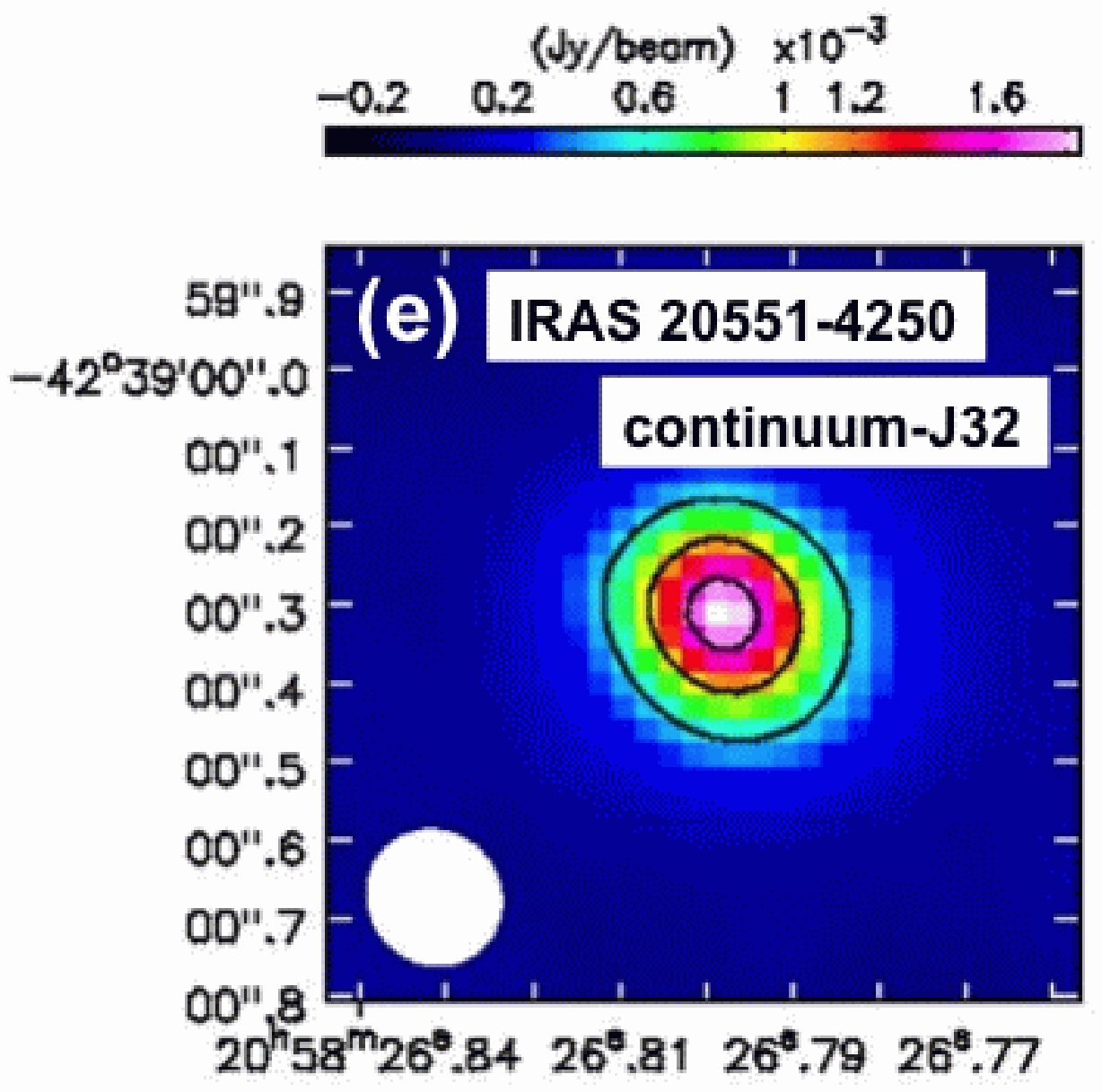}  \\
\end{center}
\vspace{-0.4cm}
\caption{
Continuum emission maps in bands 7, 9, 3, and 6. 
The abscissa and ordinate are right ascension (J2000) and declination 
(J2000), respectively. 
(a): Continuum-J43a (central observed frequency 
$\nu_{\rm center}$ $\sim$ 336.7 GHz) data taken with the HCO$^{+}$
J=4--3 observations in band 7. 
(b): Continuum-J43b ($\nu_{\rm center}$ $\sim$ 343.2 GHz)
data taken with the HNC J=4--3 observations in band 7. 
(c): Continuum-J87 ($\nu_{\rm center}$ $\sim$ 689.9 GHz)
data taken with the HCN, HCO$^{+}$, HNC J=8--7 observations in band 9.
(d): Continuum-J10 ($\nu_{\rm center}$ $\sim$ 92 GHz)
data taken with HCN, HCO$^{+}$, HNC J=1--0 observations in band 3.
(e): Continuum-J32 ($\nu_{\rm center}$ $\sim$ 241.5 GHz)
data taken with H$^{13}$CN, H$^{13}$CO$^{+}$, HN$^{13}$C J=3--2
observations in band 6. 
The contours represent 20$\sigma$, 35$\sigma$, and 50$\sigma$ for
continuum-J43a, 
25$\sigma$, 50$\sigma$, 75$\sigma$ for continuum-J43b,
8$\sigma$, 16$\sigma$, 24$\sigma$ for continuum-J87, 
25$\sigma$, 40$\sigma$, 55$\sigma$ for continuum-J10, and 
20$\sigma$, 40$\sigma$, 60$\sigma$ for continuum-J32.  
Beam sizes are shown as filled circles in the lower-left region.
The displayed sizes in bands 9 and 6 are different from those of bands 7
and 3, because of smaller beam sizes in the former.
}
\end{figure}

\begin{figure}
\begin{center}
\includegraphics[angle=0,scale=.41]{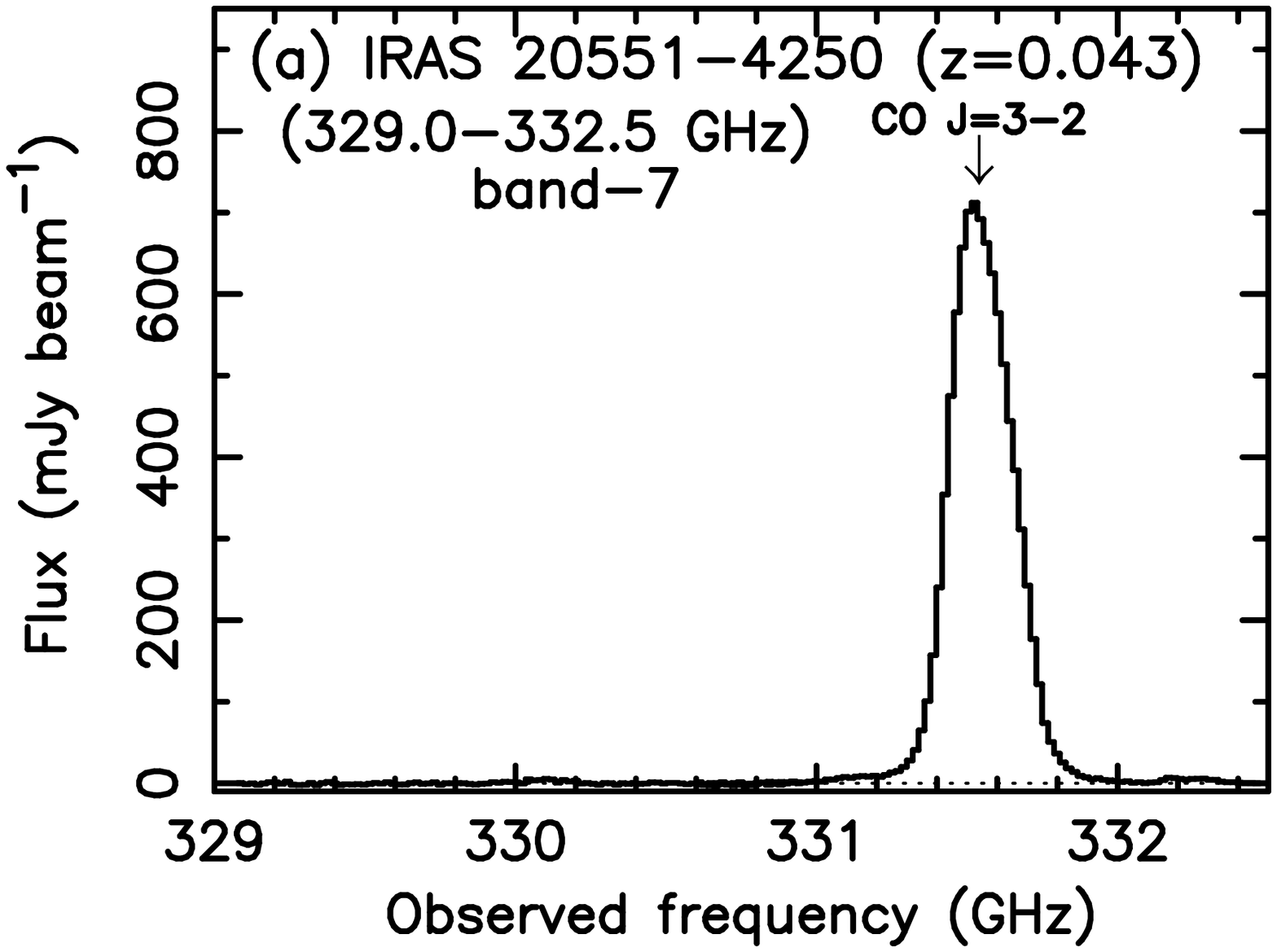} 
\includegraphics[angle=0,scale=.41]{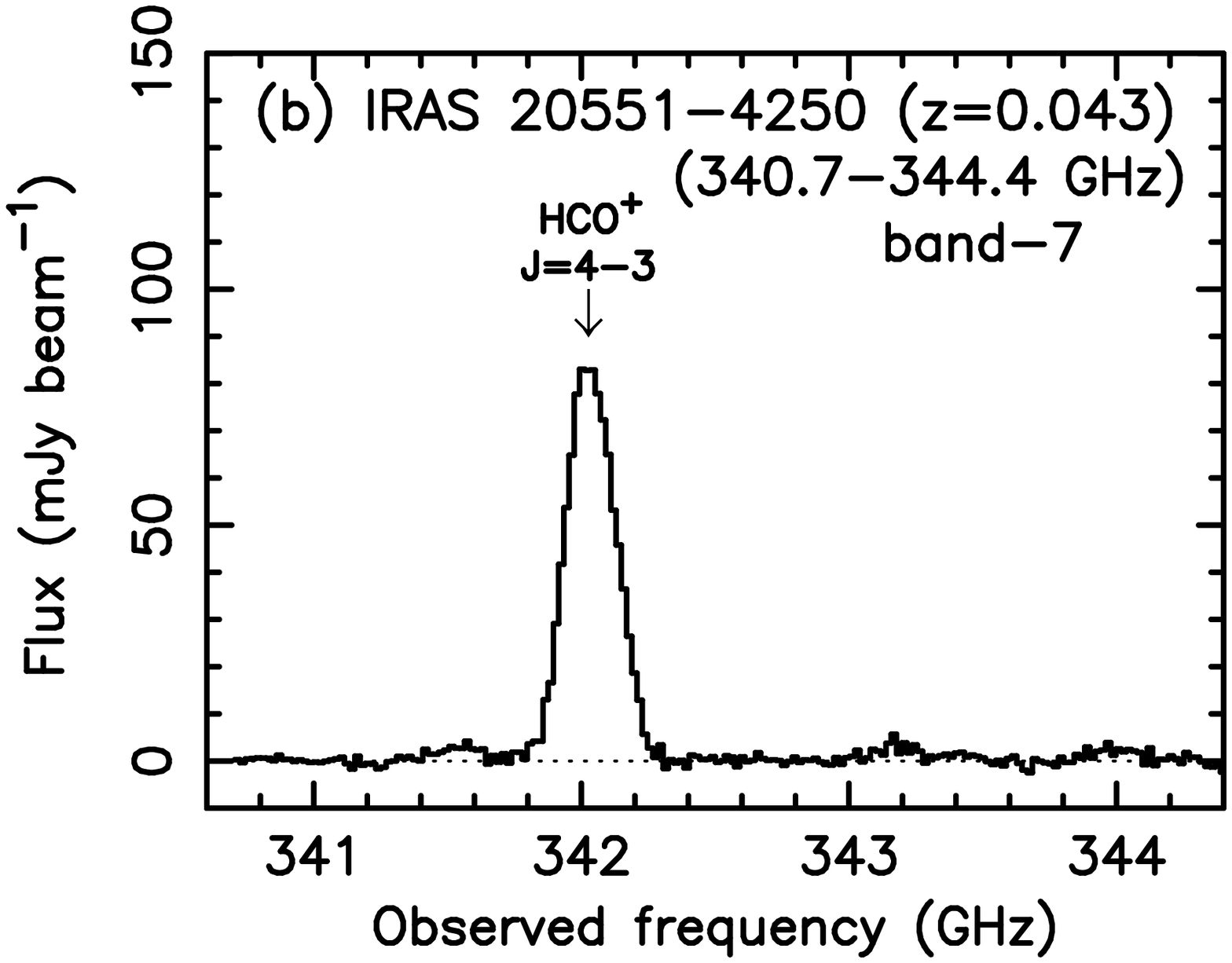} \\ 
\includegraphics[angle=0,scale=.41]{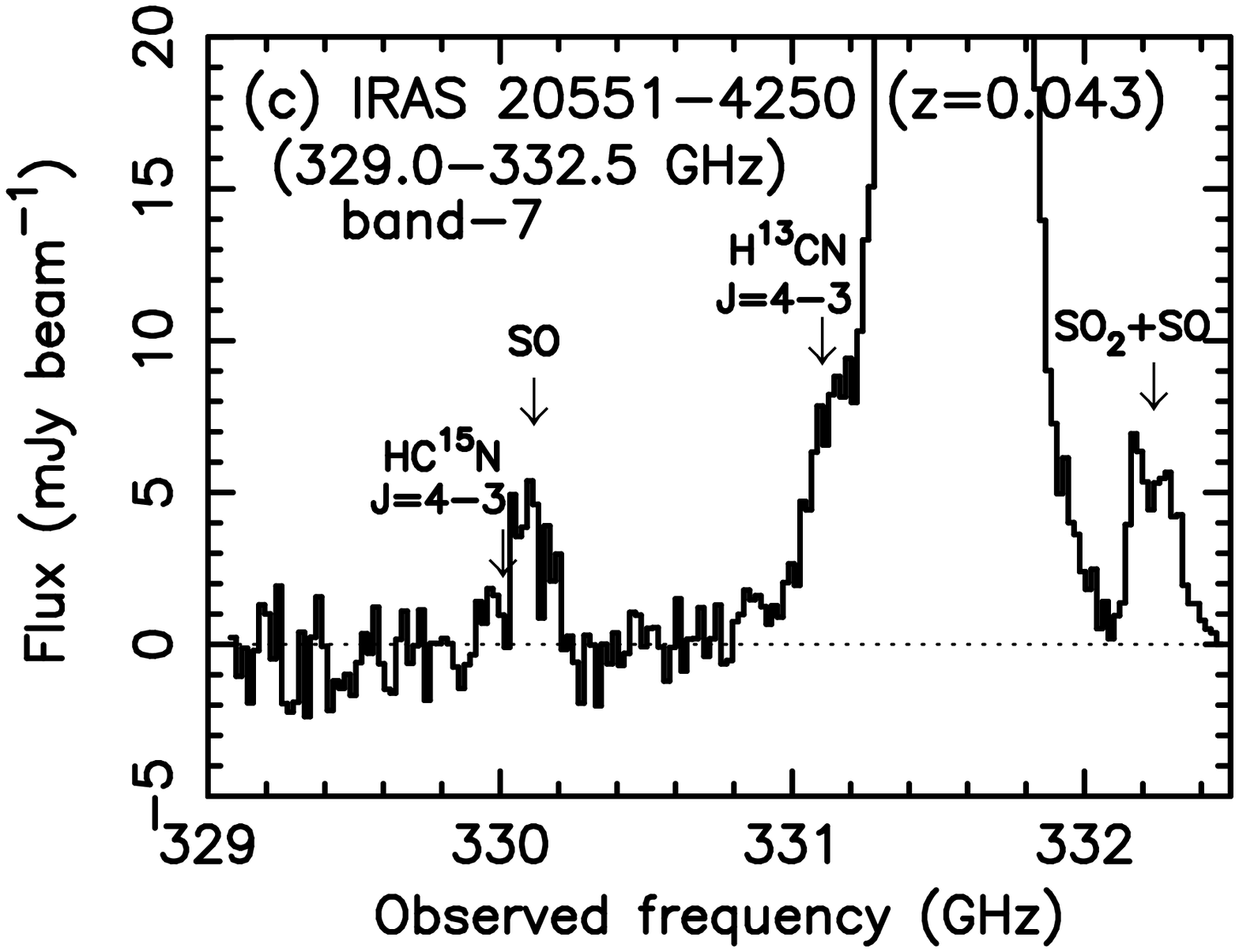} 
\includegraphics[angle=0,scale=.41]{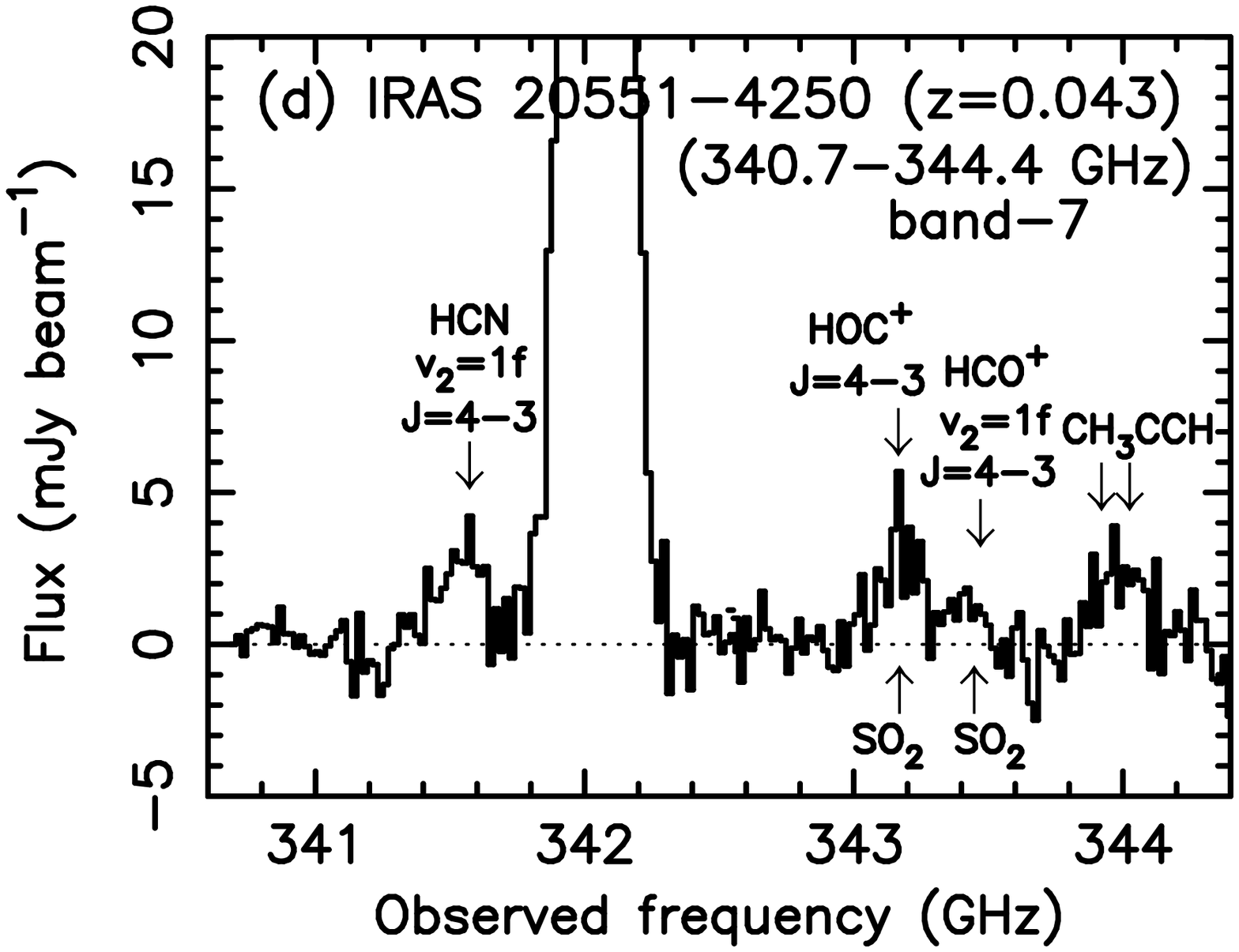} \\ 
\includegraphics[angle=0,scale=.41]{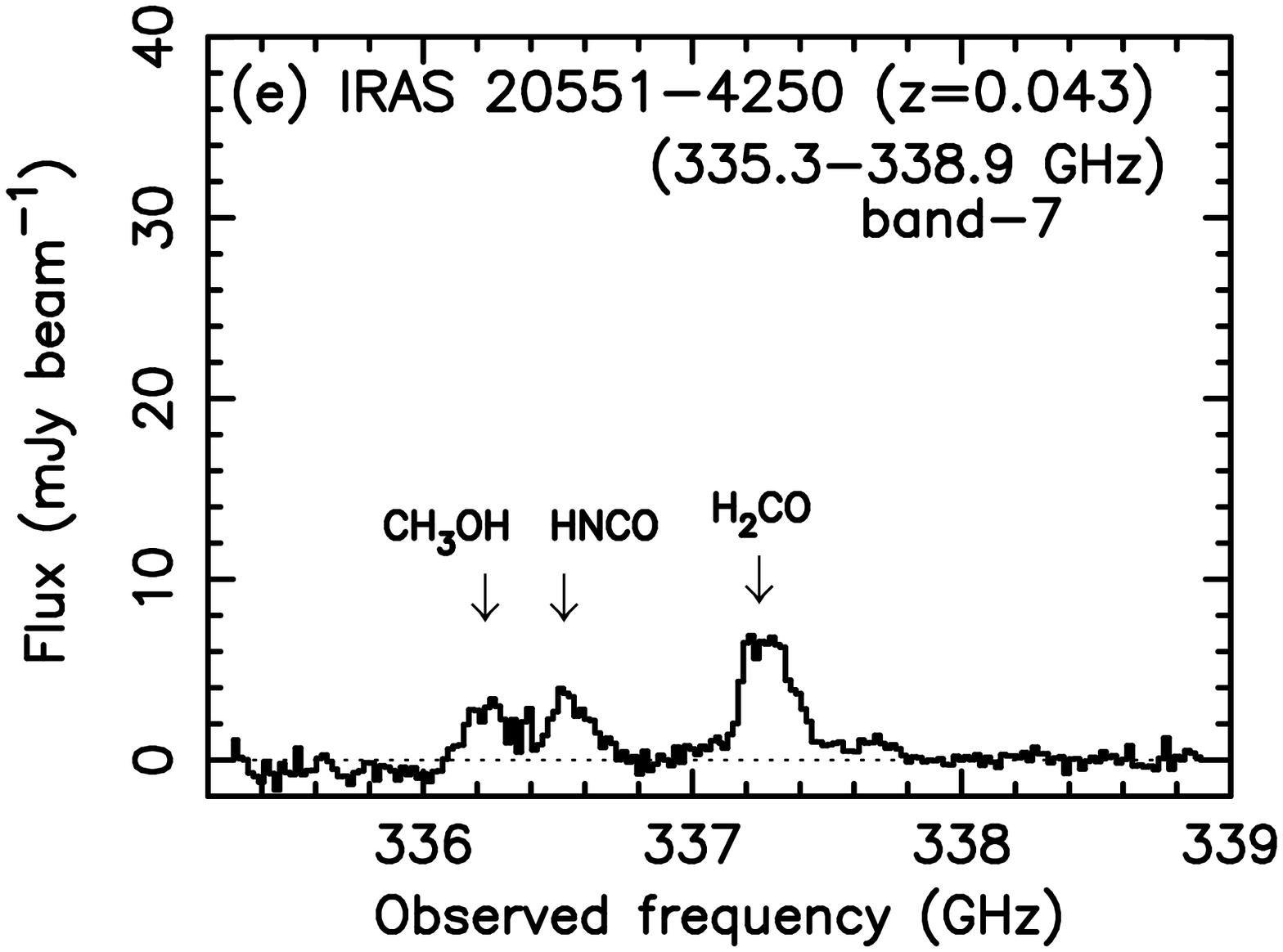} 
\includegraphics[angle=0,scale=.41]{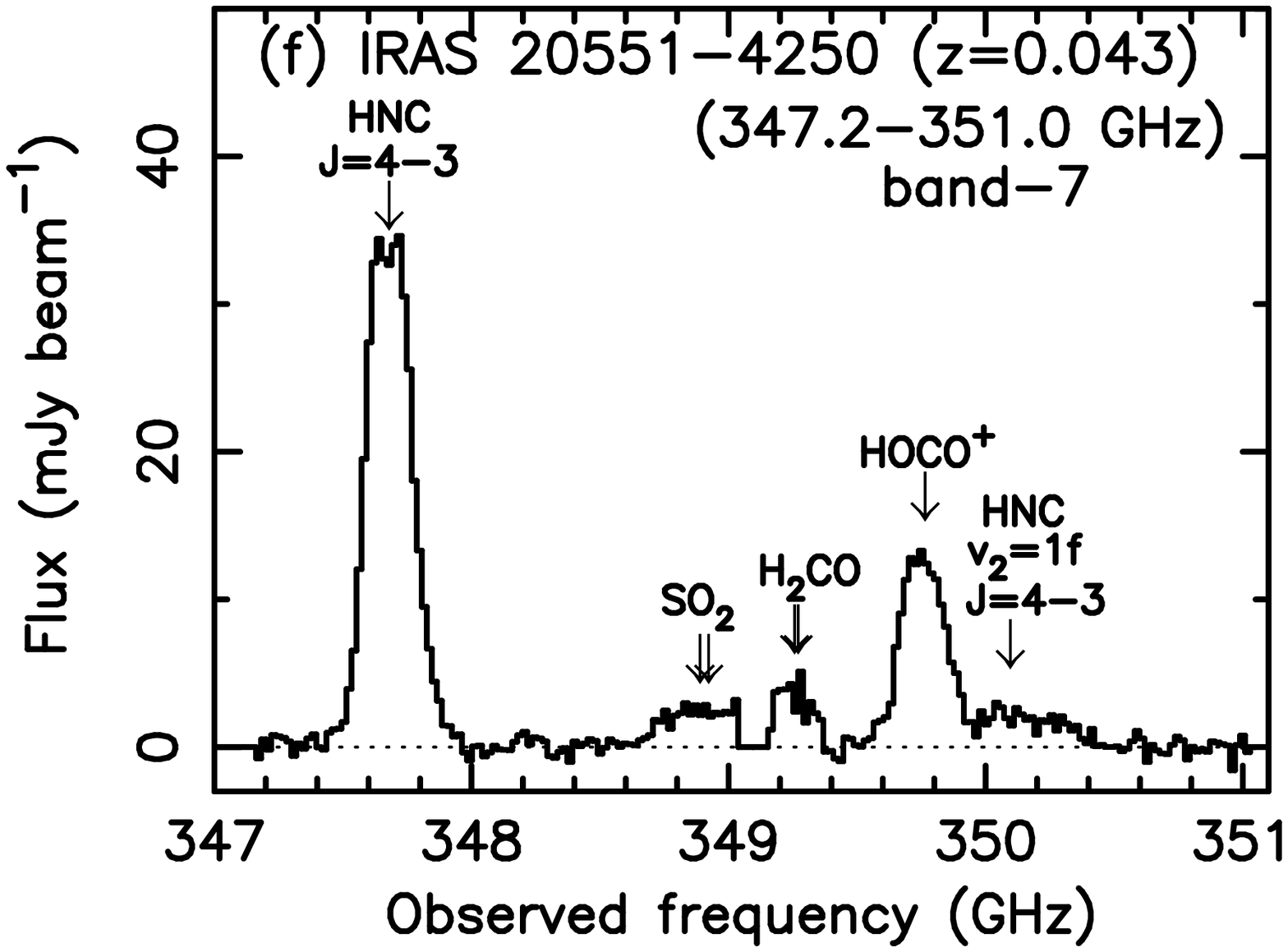} \\ 
\includegraphics[angle=0,scale=.41]{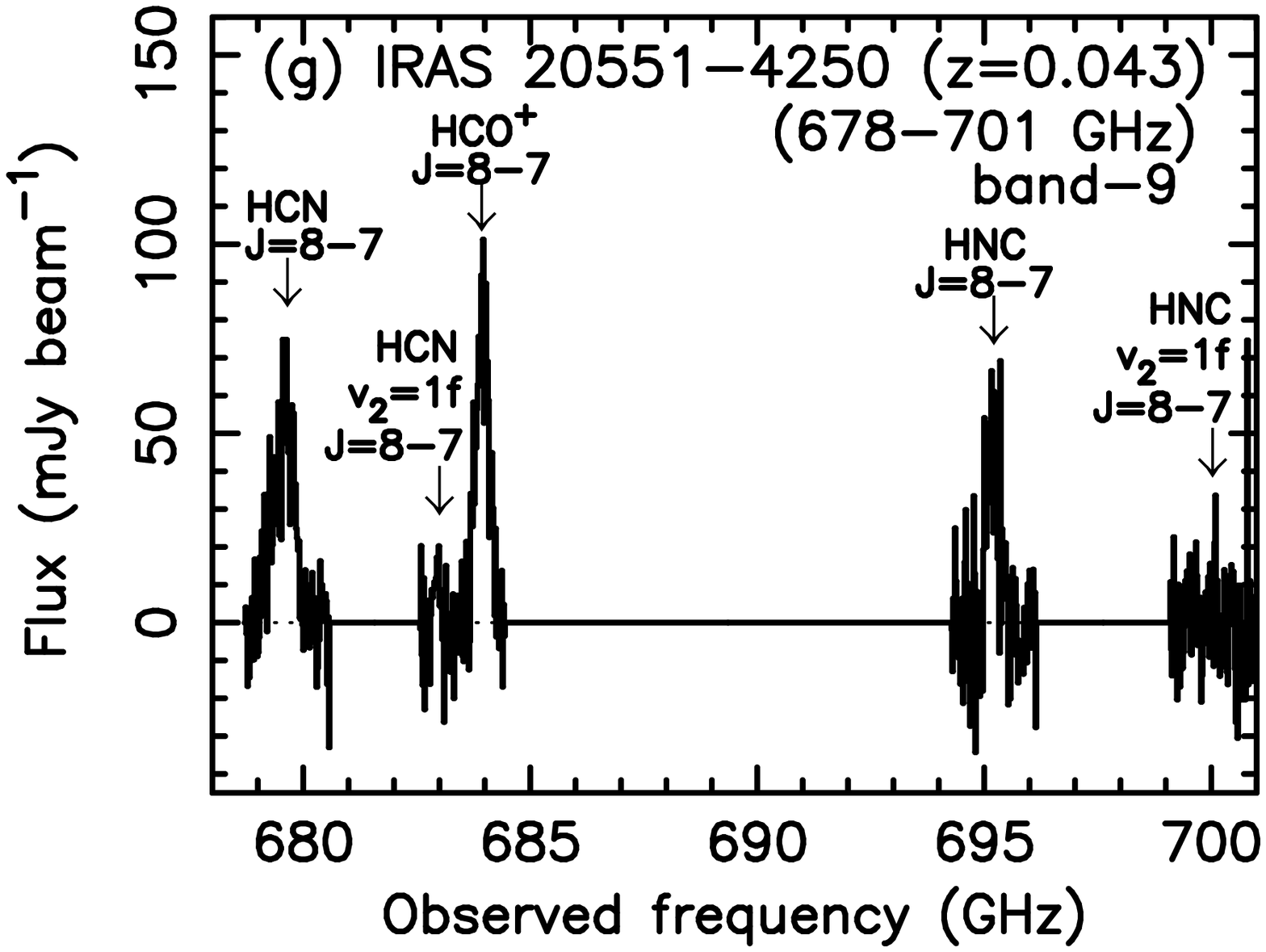} 
\includegraphics[angle=0,scale=.41]{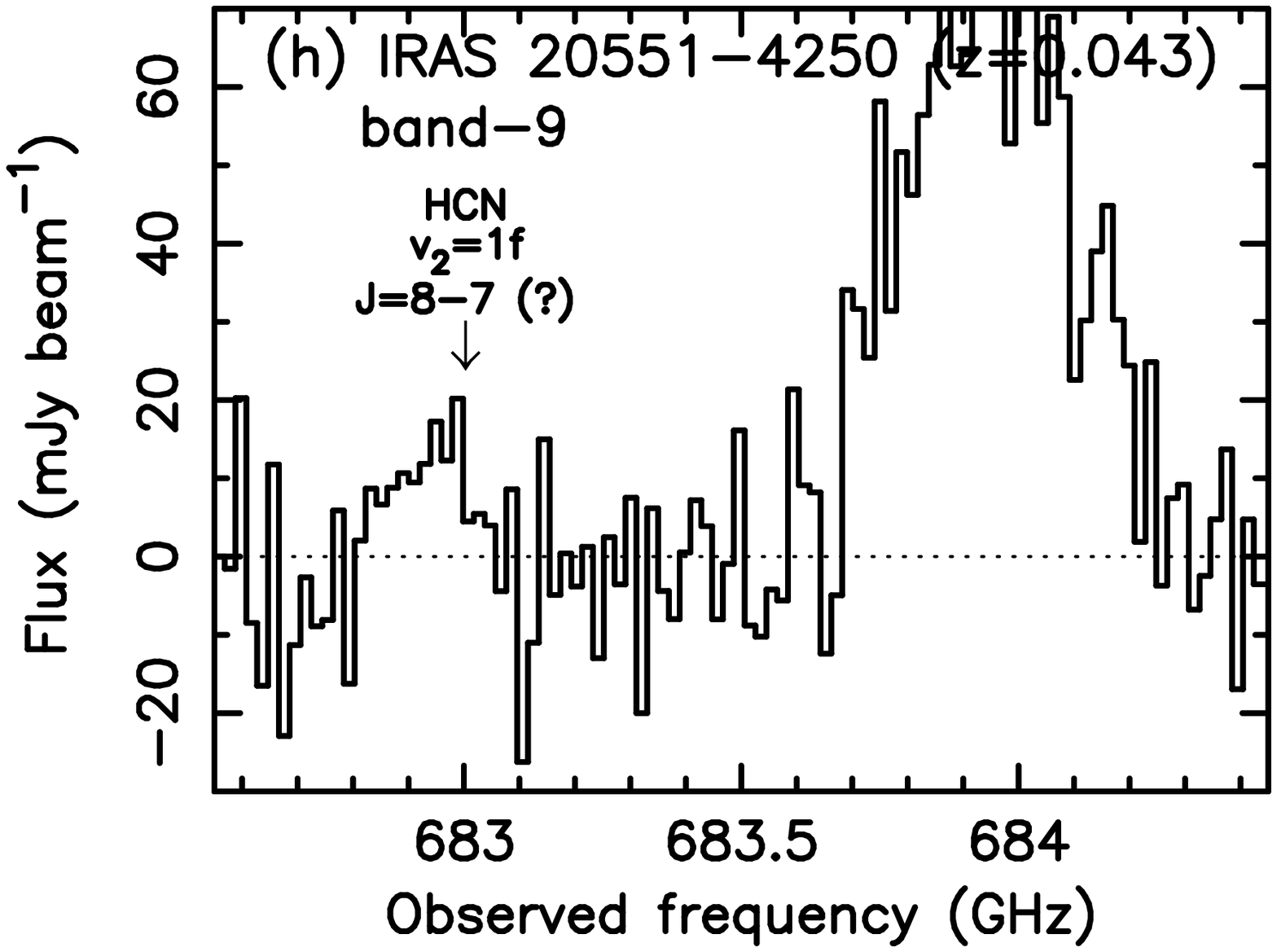} \\
\end{center}
\end{figure}

\clearpage

\begin{figure}
\begin{center}
\includegraphics[angle=0,scale=.41]{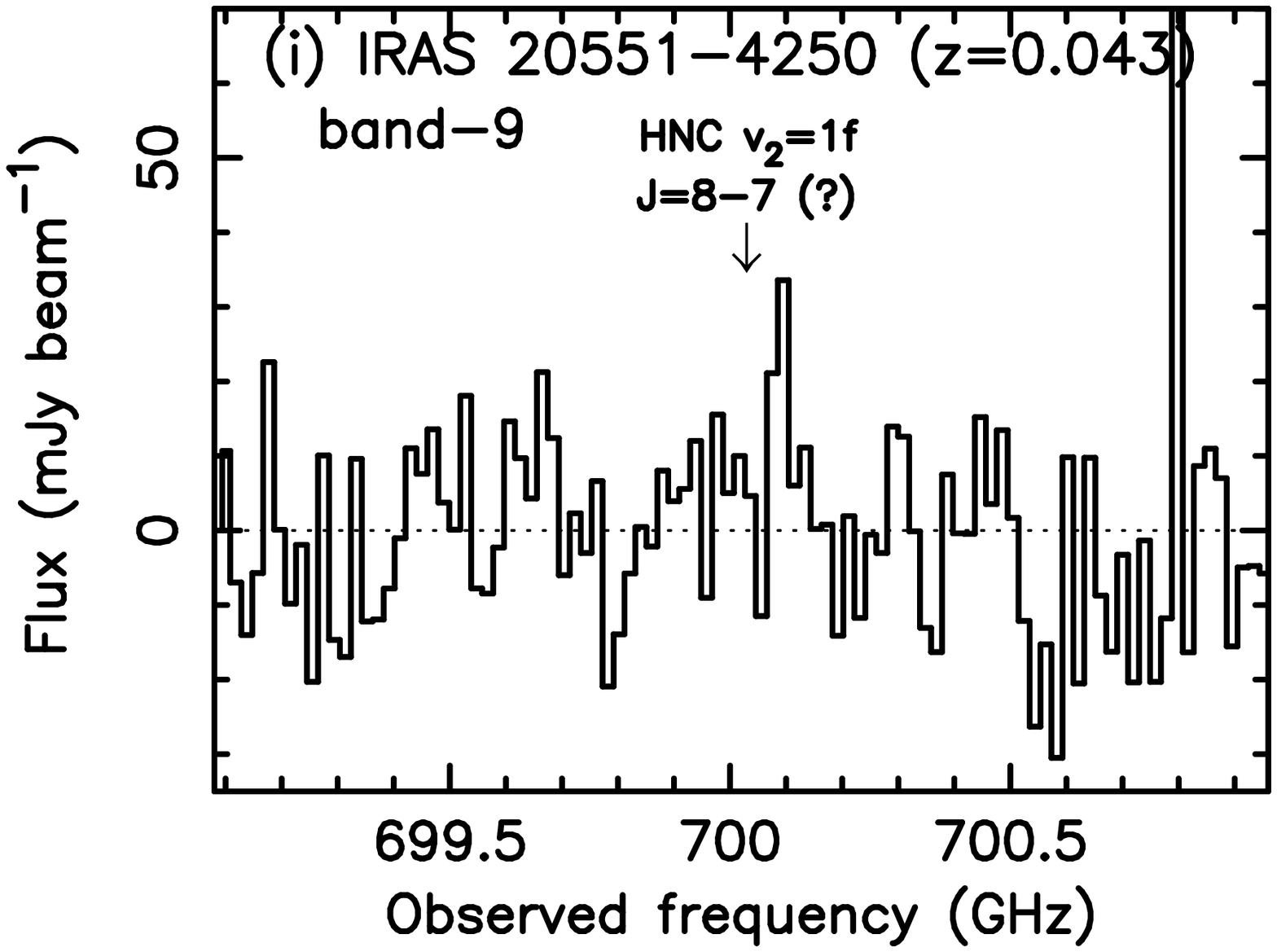} 
\includegraphics[angle=0,scale=.41]{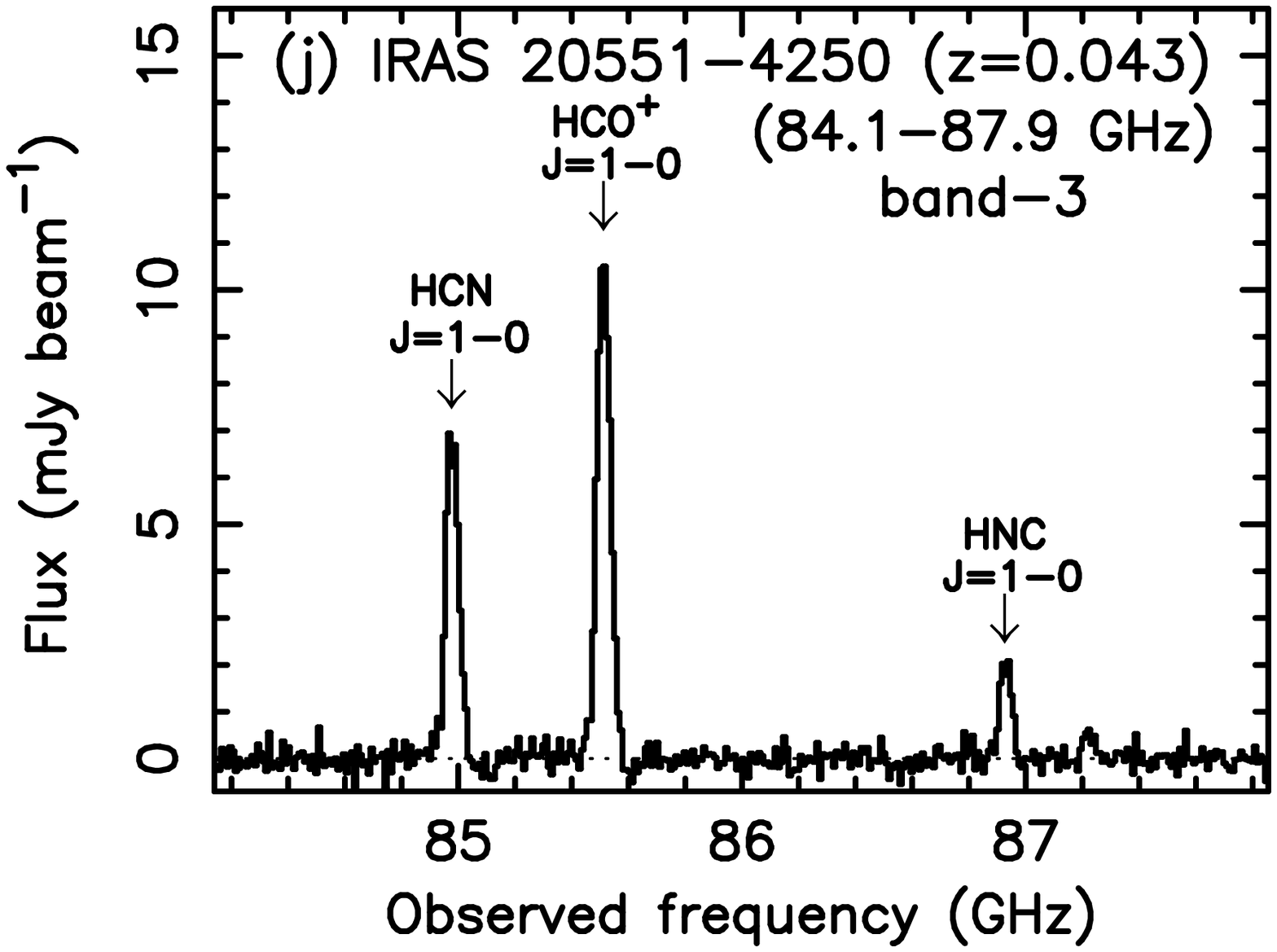} \\
\includegraphics[angle=0,scale=.41]{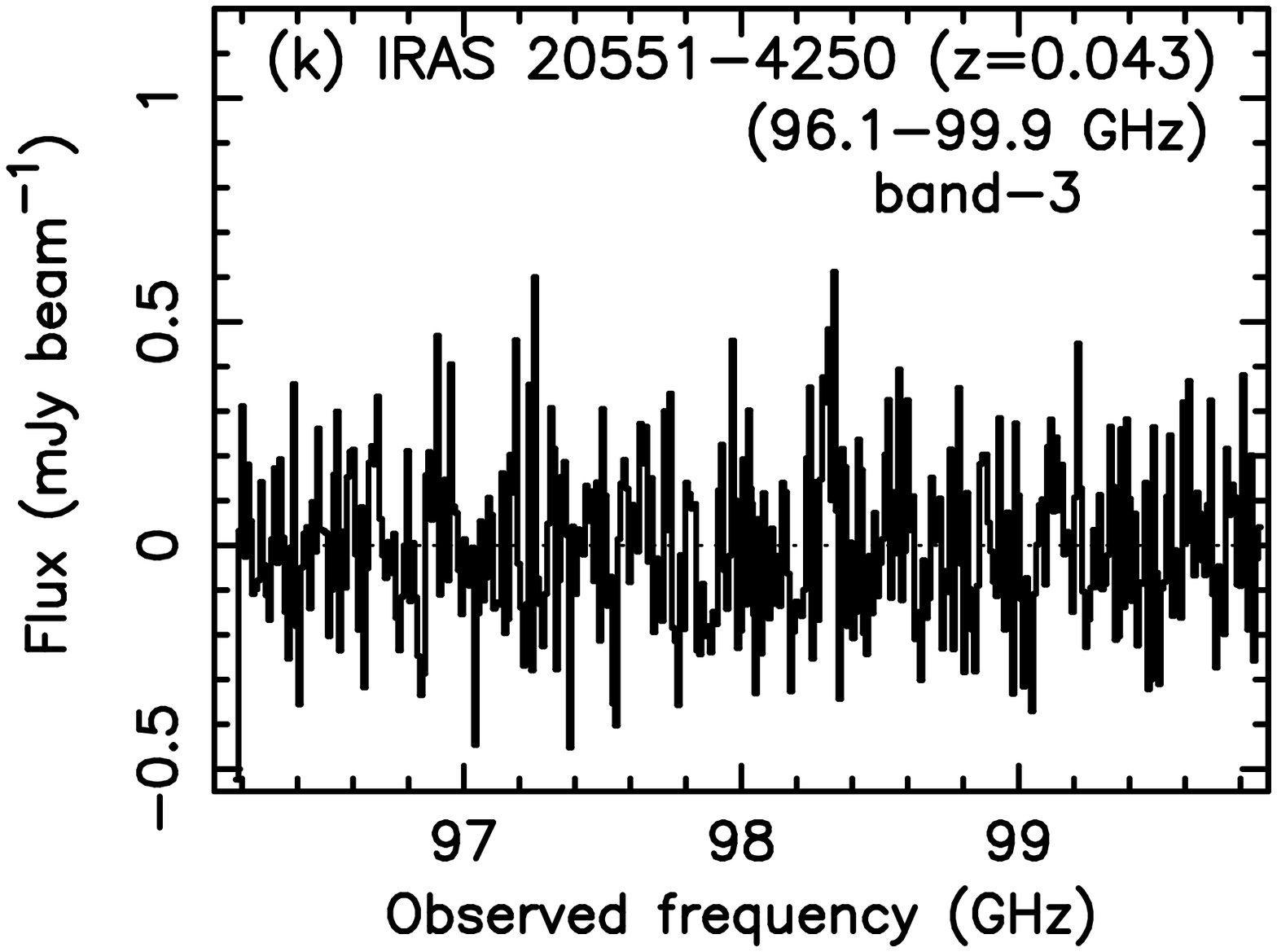}  
\includegraphics[angle=0,scale=.41]{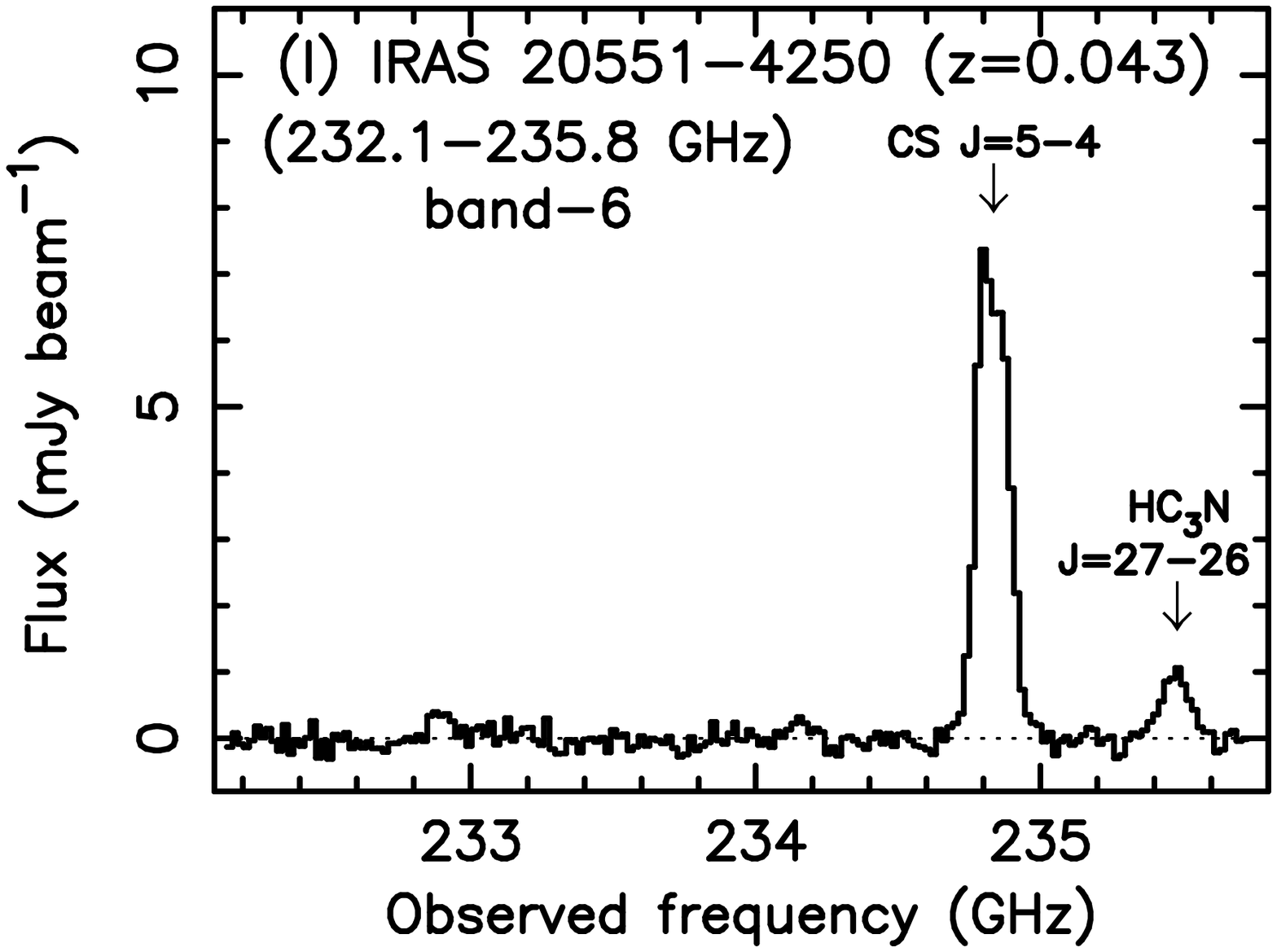}  \\


\includegraphics[angle=0,scale=.41]{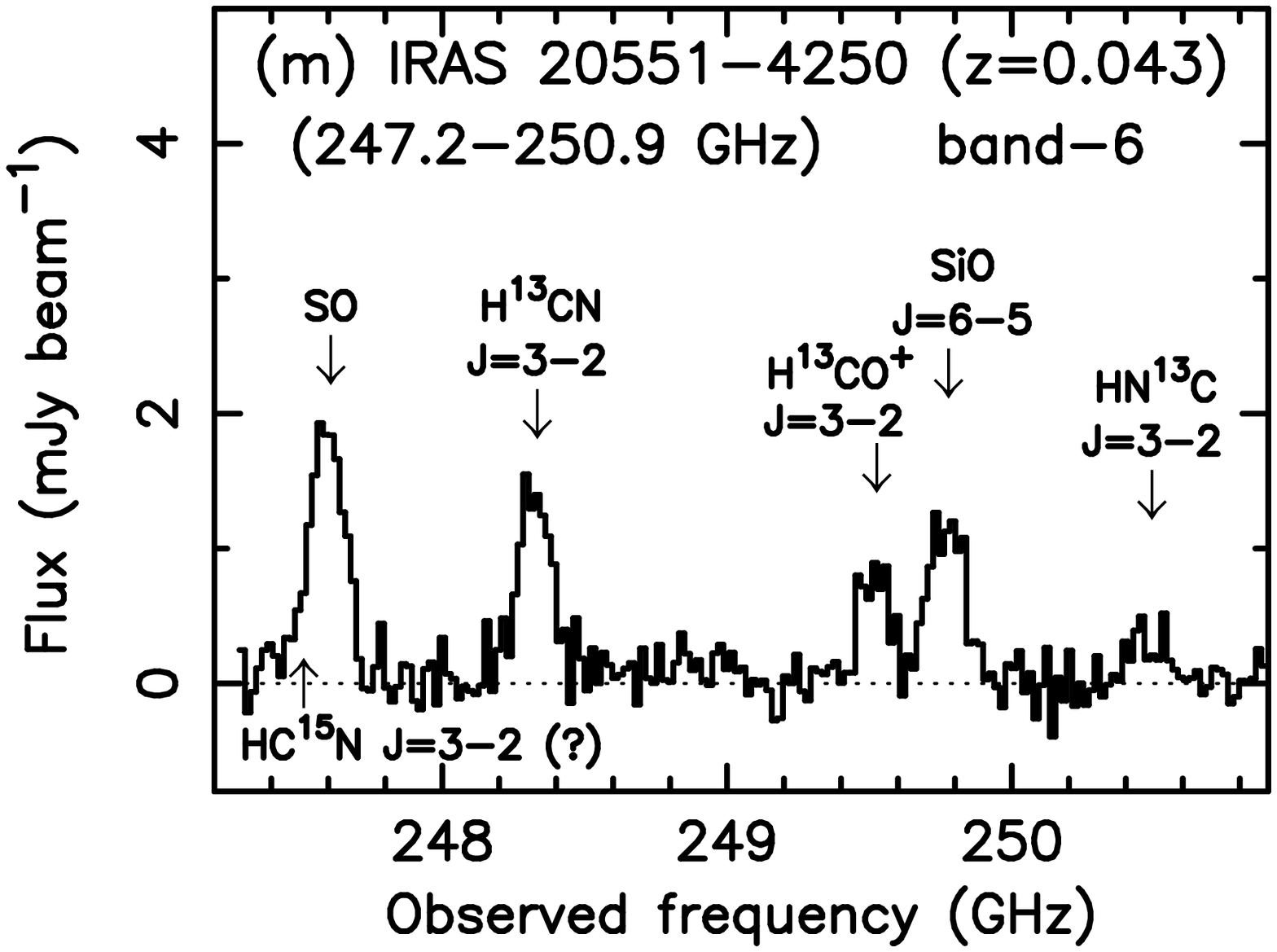}  
\caption{
Spectra at the continuum peak positions, within the beam size.
(a) and (b) are band 7 spectra taken with HCO$^{+}$ J=4--3 observations.   
(c) and (d) are magnified spectra of (a) and (b), respectively, to show
serendipitously detected weak emission lines in more detail.   
In (c), in addition to the primarily targeted major emission lines, 
downward arrows are shown at the expected redshifted observed frequency with
z= 0.043 for some faint molecular lines, 
SO 8(8)--7(7) ($\nu_{\rm rest}$ = 344.311 GHz) and   
SO$_{2}$ 16(4,12)--16(3,13) ($\nu_{\rm rest}$ = 346.524 GHz) + 
SO 9(8)--8(7) ($\nu_{\rm rest}$ = 346.528 GHz).
In (d), down arrows are shown for 
HOC$^{+}$ J=4--3 ($\nu_{\rm rest}$ = 357.922 GHz) and 
CH$_{3}$CCH ($\nu_{\rm rest}$ = 358.709--818 GHz), 
and up arrows are shown for 
SO$_{2}$ 6(4,2)--6(3,3) ($\nu_{\rm rest}$ = 357.926 GHz), and 
SO$_{2}$ 20(0,20)--19(1,19) ($\nu_{\rm rest}$ = 358.216 GHz).
(e) and (f) are band 7 spectra obtained with HNC J=4--3.
In (e), down arrows are shown for 
CH$_{3}$OH 4(0,4)--3($-$1,3) ($\nu_{\rm rest}$ = 350.688 GHz),  
HNCO 16(5,11)--15(5,10) + HNCO 16(5,12)--15(5,11) ($\nu_{\rm rest}$ =
350.993 GHz), and 
H$_{2}$CO 5(1,5)--4(1,4) ($\nu_{\rm rest}$ = 351.769 GHz). 
In (f), down arrows are shown for 
SO$_{2}$ 24(1,23)--24(0,24) ($\nu_{\rm rest}$ = 363.891 GHz) +  
SO$_{2}$ 23(2,22)--23(1,23) ($\nu_{\rm rest}$ = 363.926 GHz), 
H$_{2}$CO 5(3,3)--4(3,2) ($\nu_{\rm rest}$ = 364.275 GHz) + 
H$_{2}$CO 5(3,2)--4(3,1) ($\nu_{\rm rest}$ = 364.289 GHz), and 
HOCO$^{+}$ 17(1,16)--16(1,15) ($\nu_{\rm rest}$ = 364.804 GHz). 
(g) is a band 9 spectrum. 
(h) and (i) are magnified band 9 spectra around HCN v$_{2}$=1f J=8--7 and 
HNC v$_{2}$=1f J=8--7 emission lines, respectively.
(j) and (k) are band 3 spectra. 
(l) and (m) are band 6 spectra.
In (l), a down arrow is shown for HC$_{3}$N J=27--26 
($\nu_{\rm rest}$ = 245.606 GHz).  
In (m), down arrows are shown for SO 6(6)--5(5) 
($\nu_{\rm rest}$ = 258.256 GHz) and 
SiO J=6--5 ($\nu_{\rm rest}$ = 260.518 GHz).
}
\end{center}
\end{figure}

\begin{figure}
\begin{center}                                  
\includegraphics[angle=0,scale=.41]{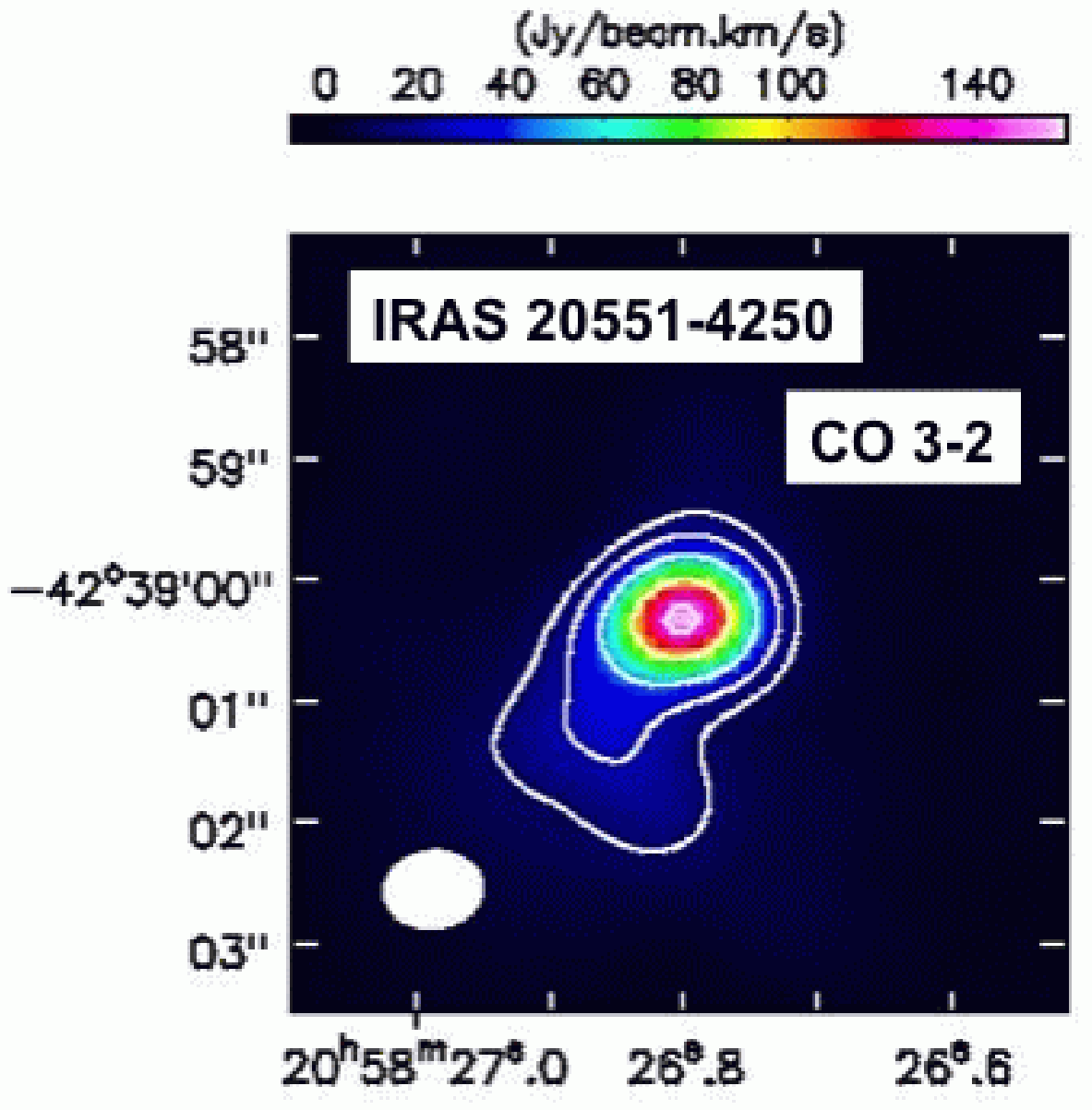} 
\includegraphics[angle=0,scale=.41]{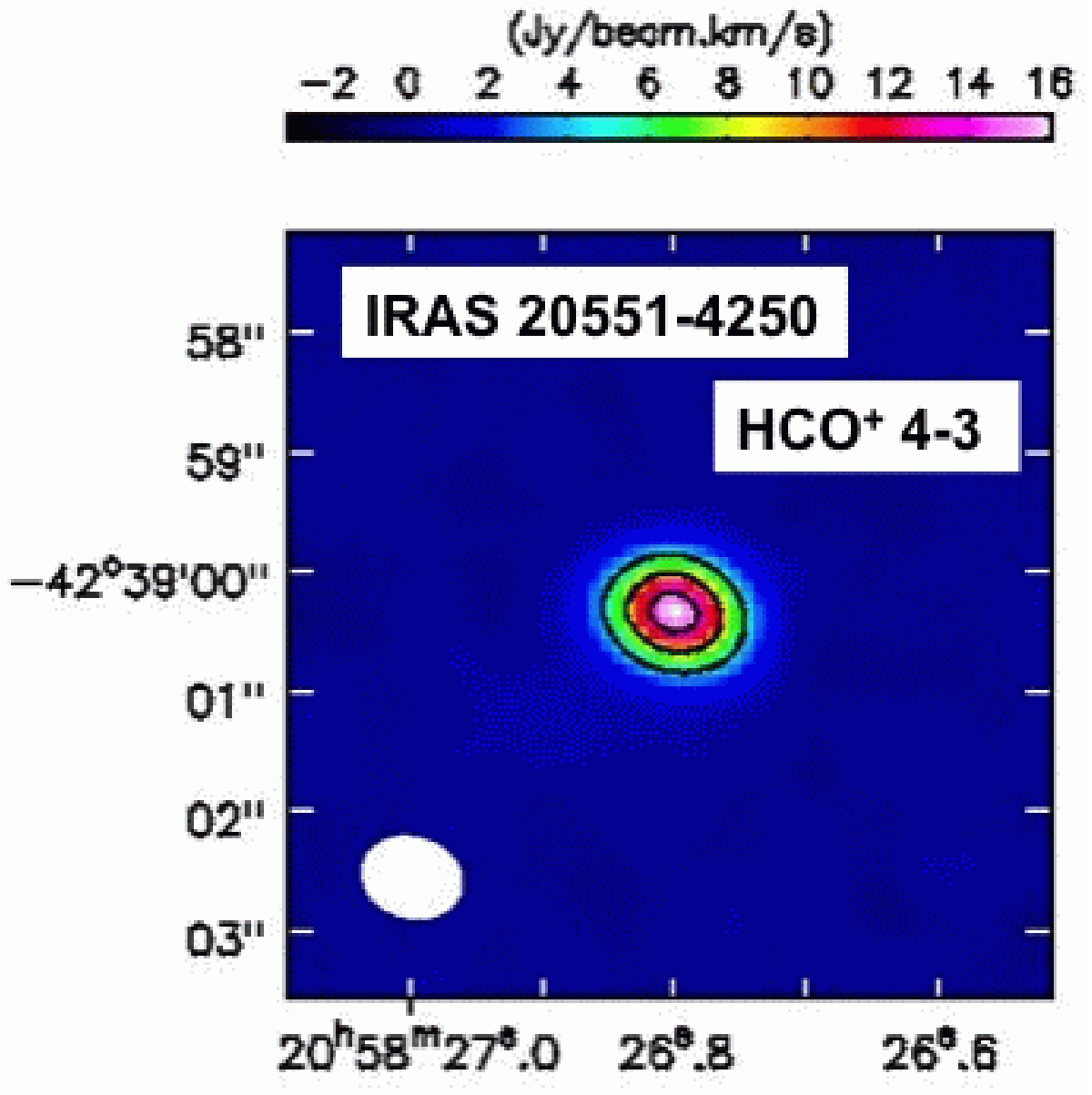} 
\includegraphics[angle=0,scale=.41]{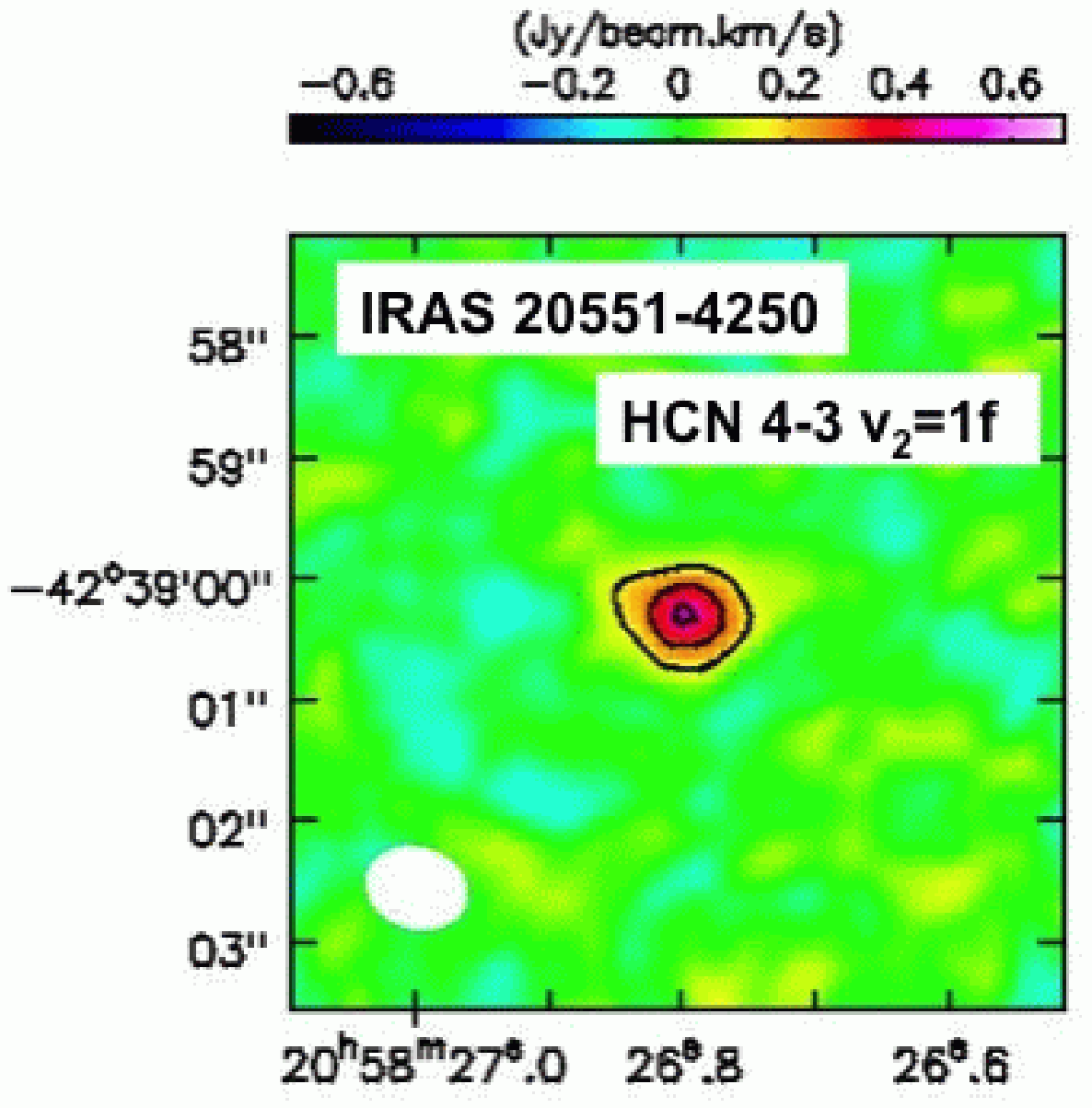} \\ 
\includegraphics[angle=0,scale=.41]{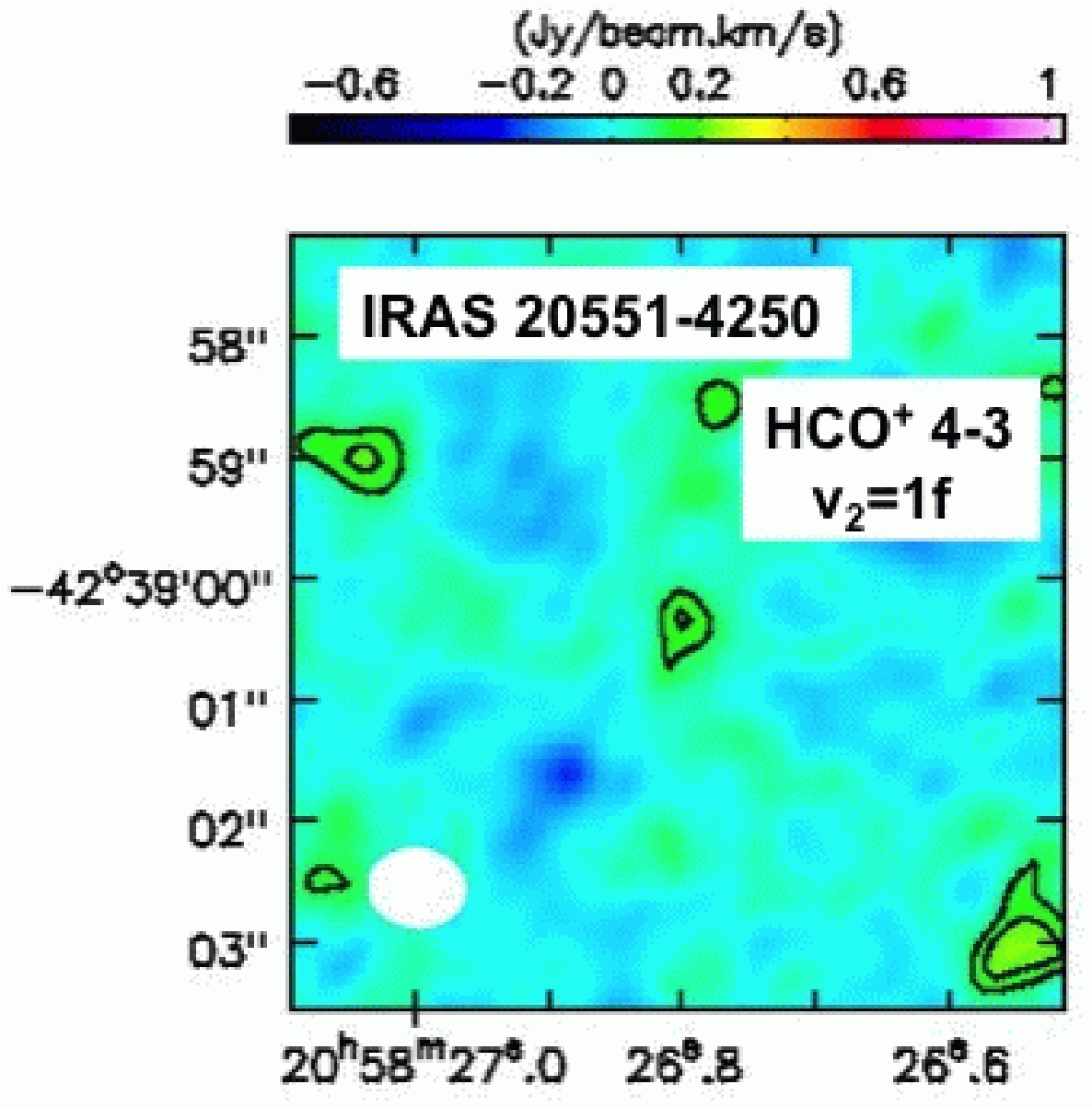}  
\includegraphics[angle=0,scale=.41]{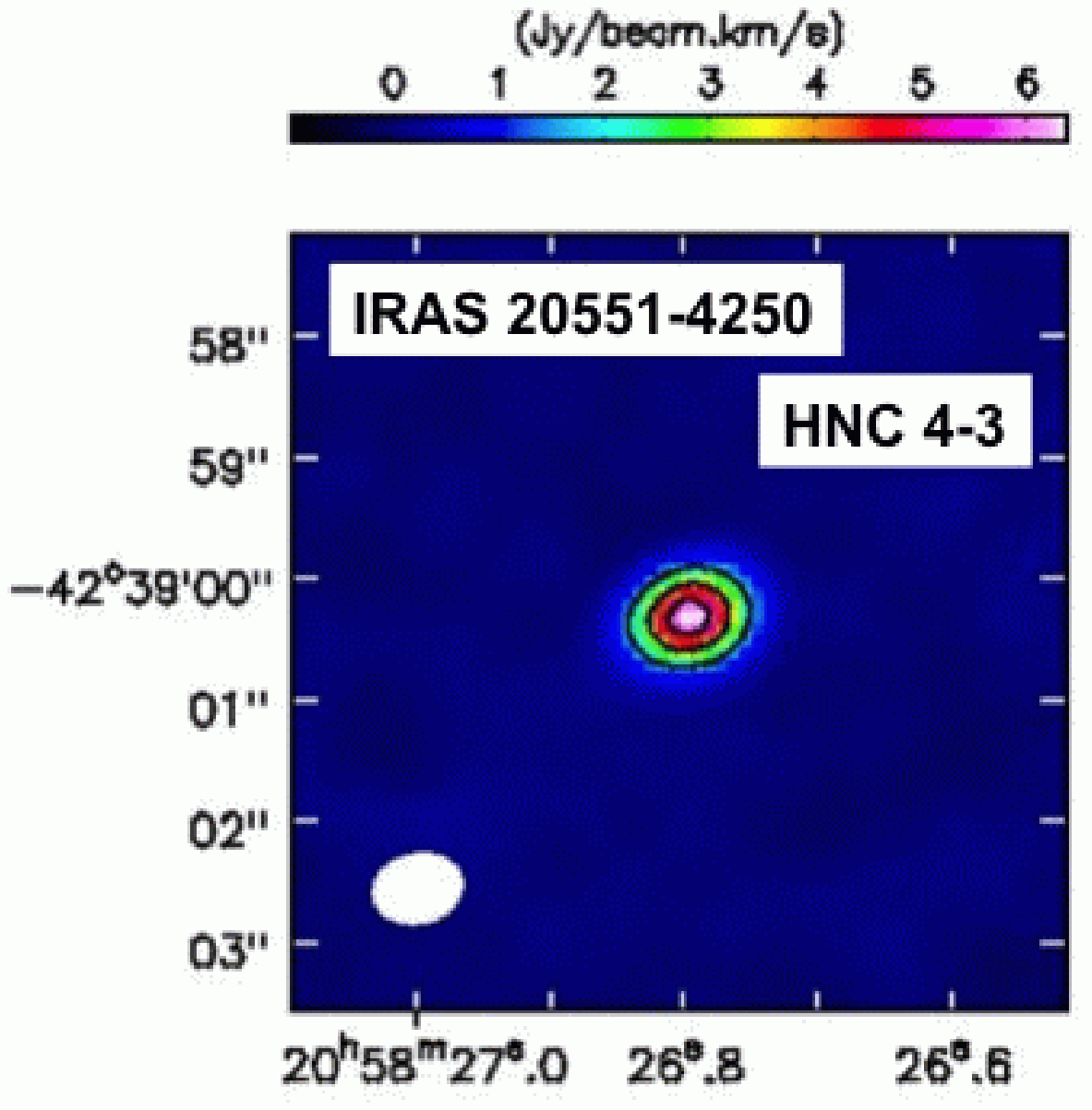} 
\includegraphics[angle=0,scale=.41]{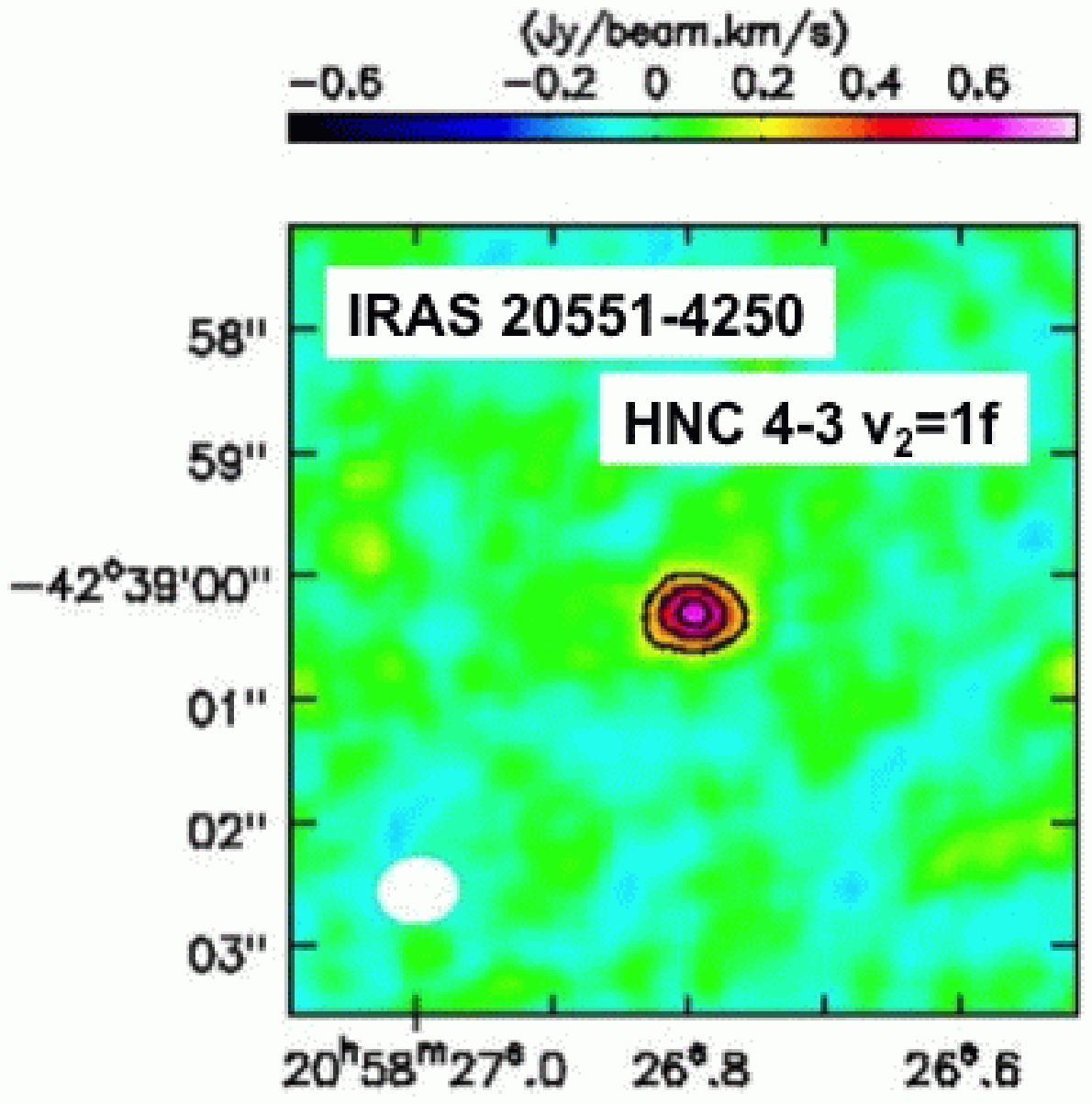} \\
\includegraphics[angle=0,scale=.41]{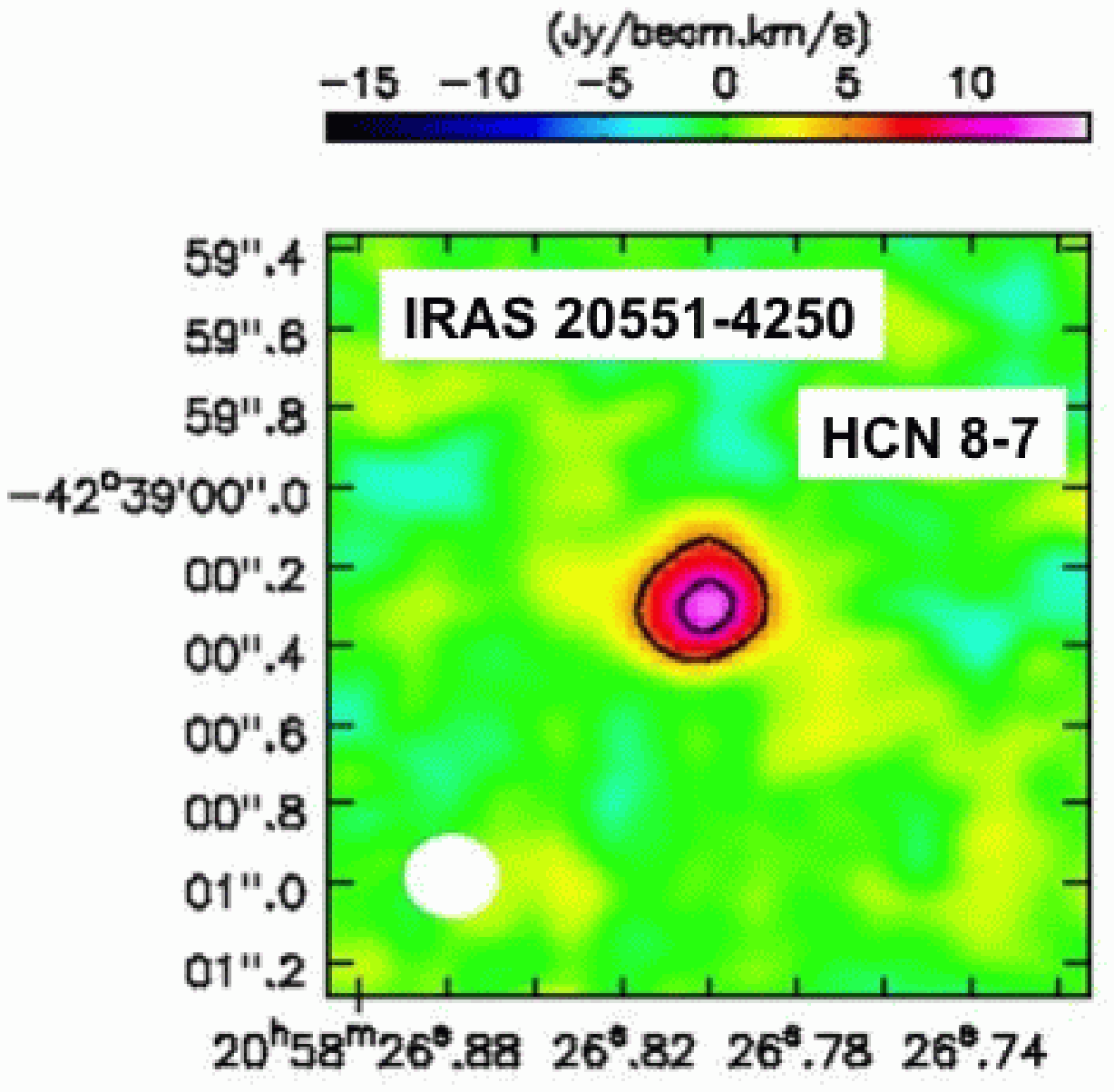} 
\includegraphics[angle=0,scale=.41]{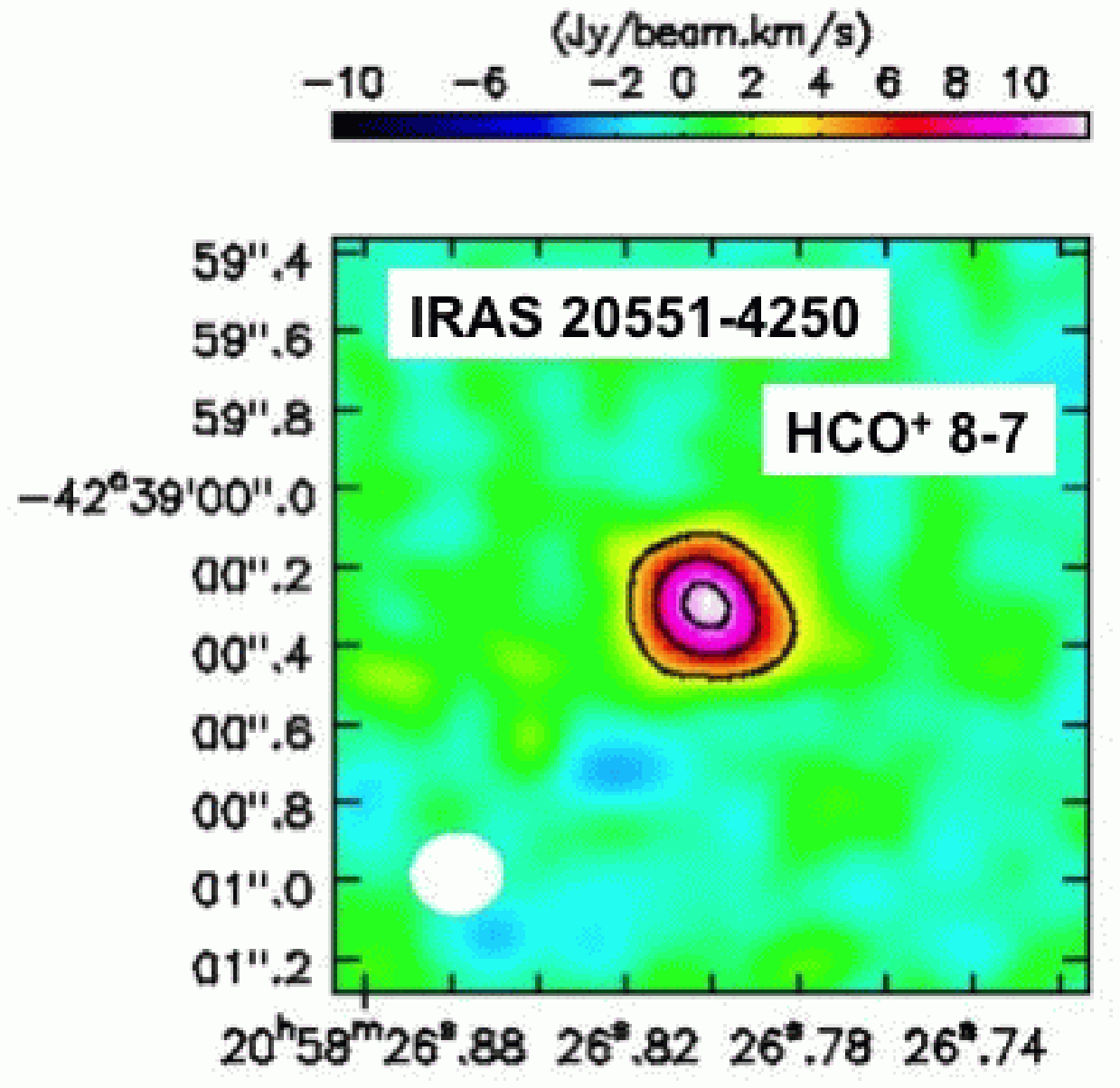} 
\includegraphics[angle=0,scale=.41]{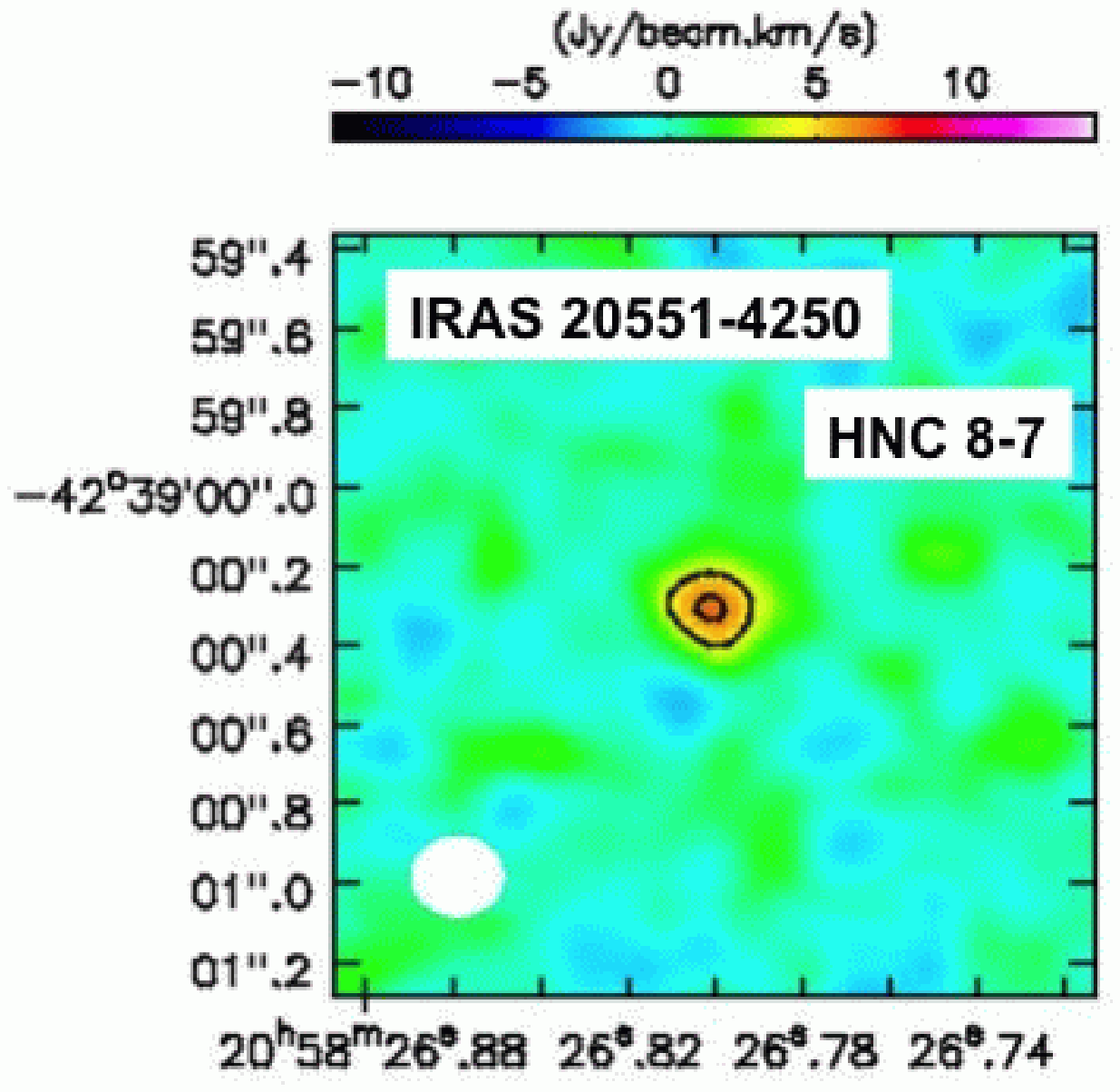} \\
\end{center}
\end{figure}

\clearpage

\begin{figure}
\begin{center}
\includegraphics[angle=0,scale=.41]{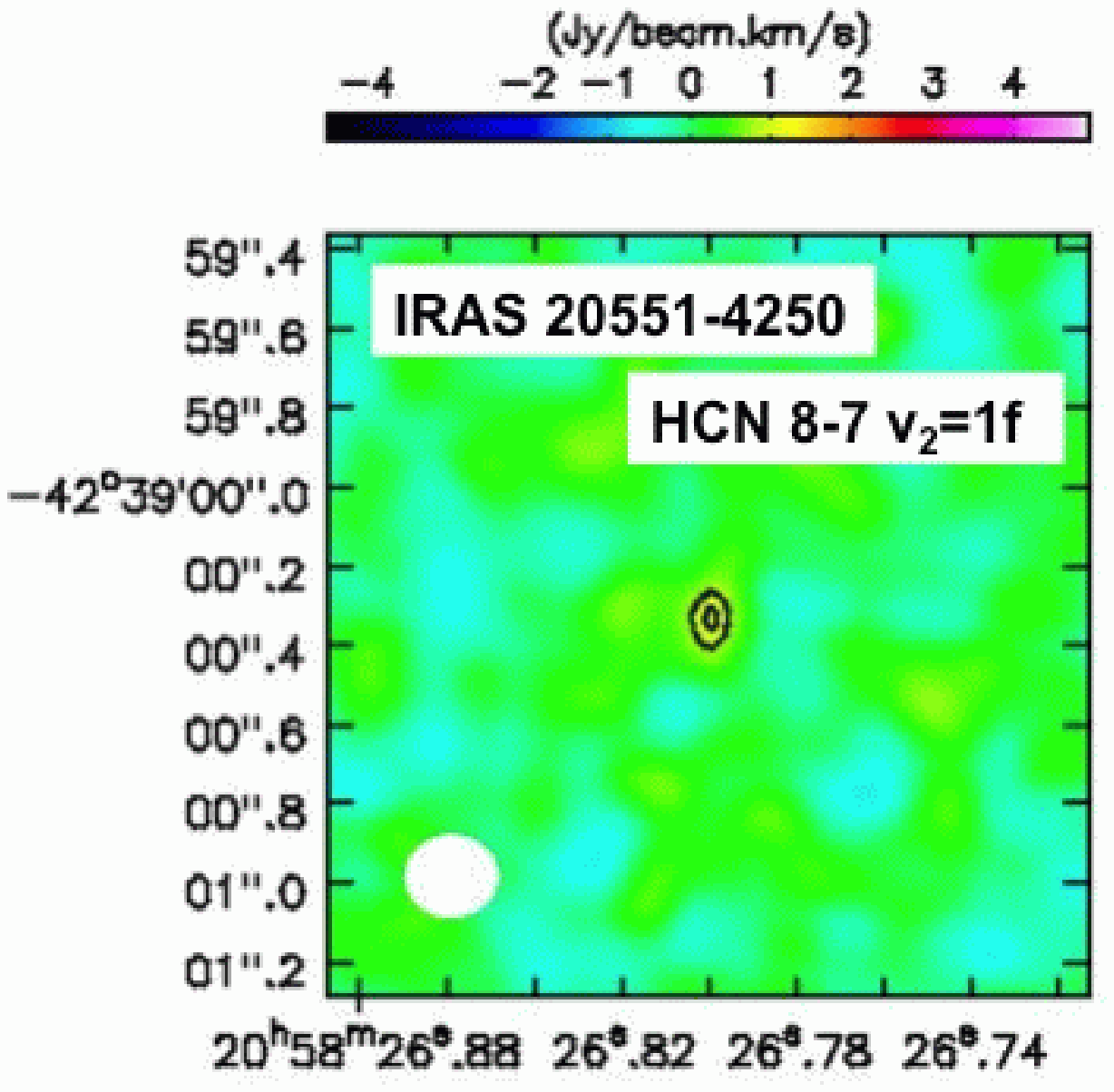}  
\includegraphics[angle=0,scale=.41]{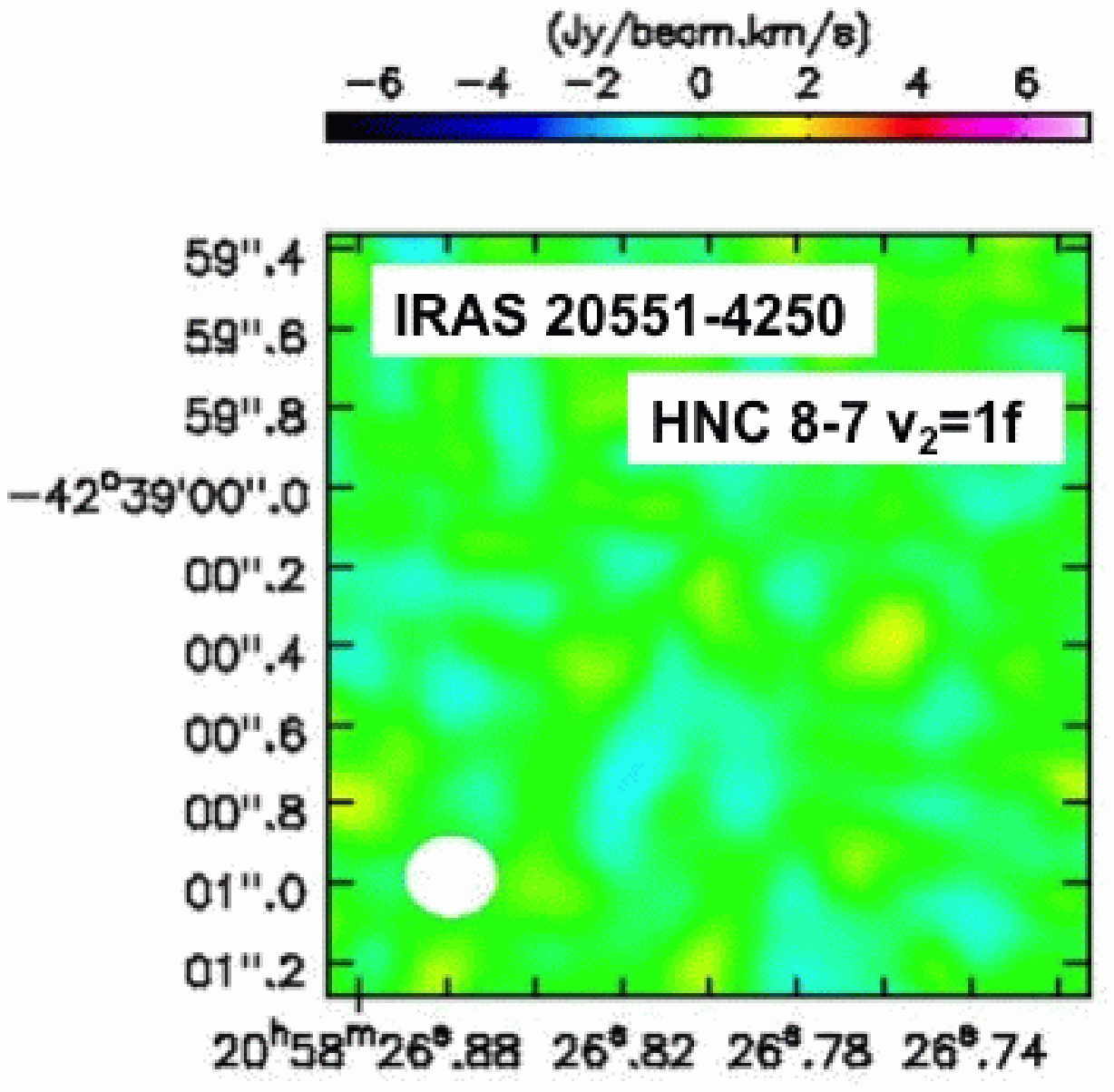} 
\includegraphics[angle=0,scale=.38]{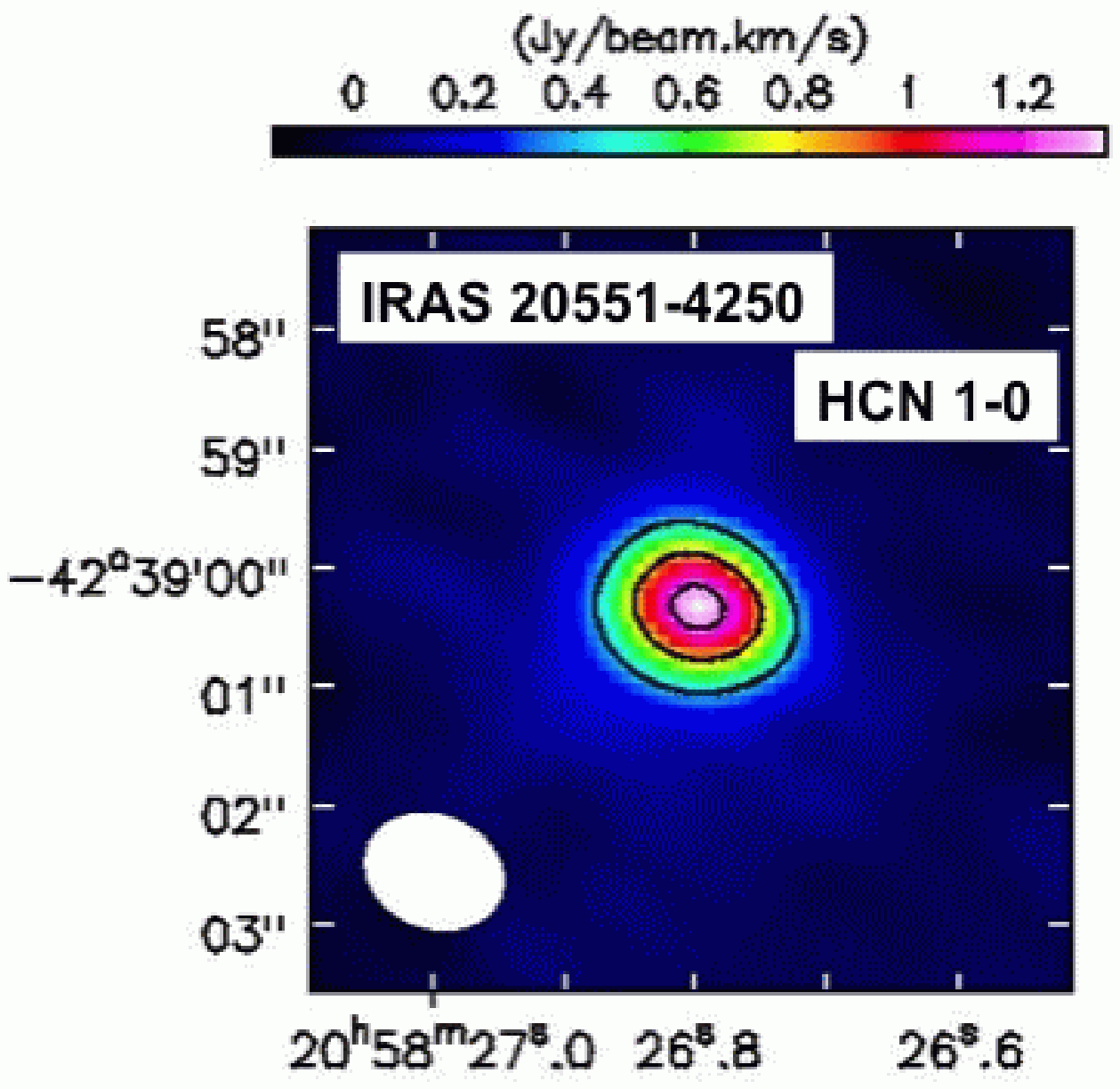} \\ 
\includegraphics[angle=0,scale=.38]{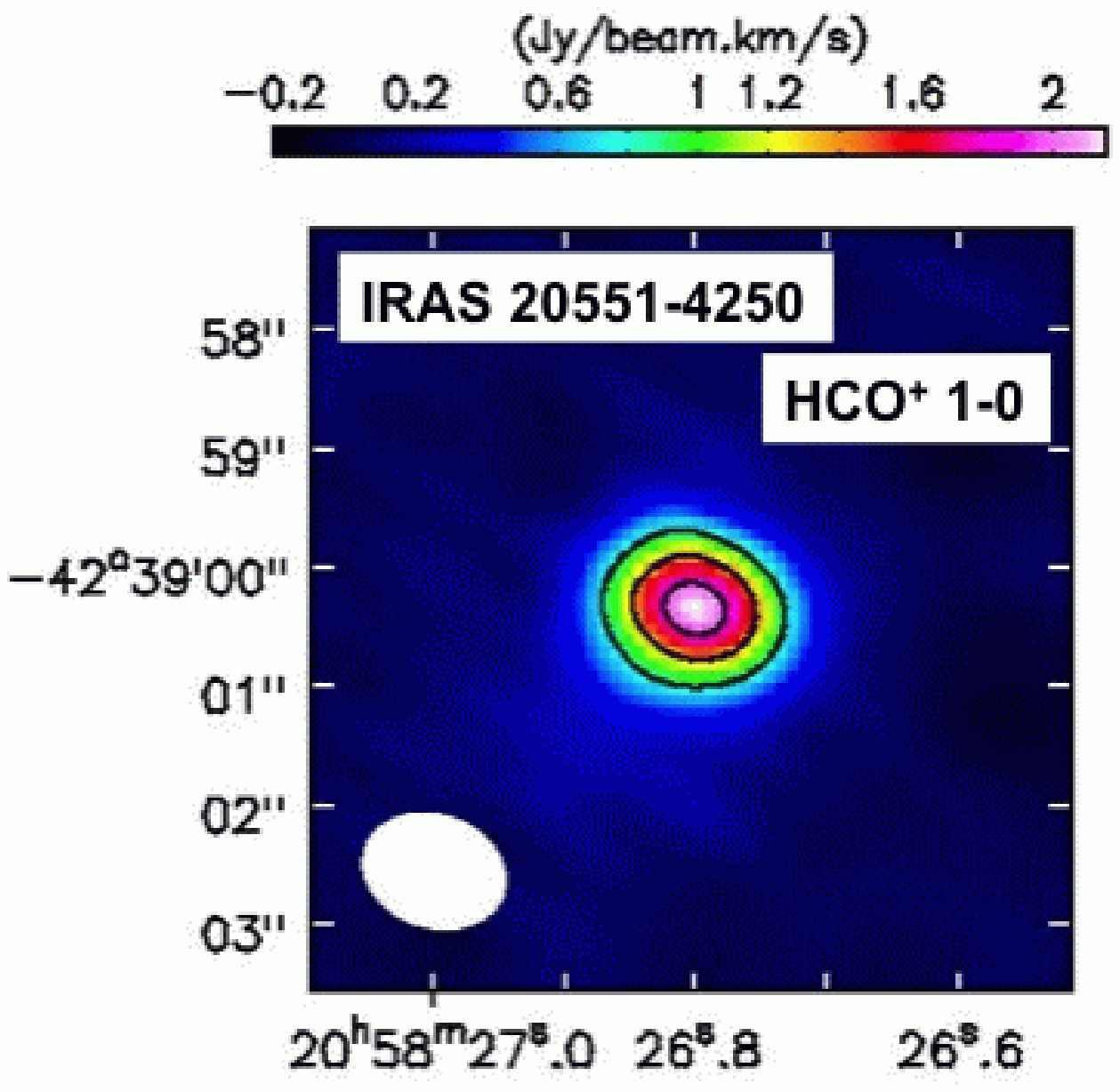} 
\includegraphics[angle=0,scale=.38]{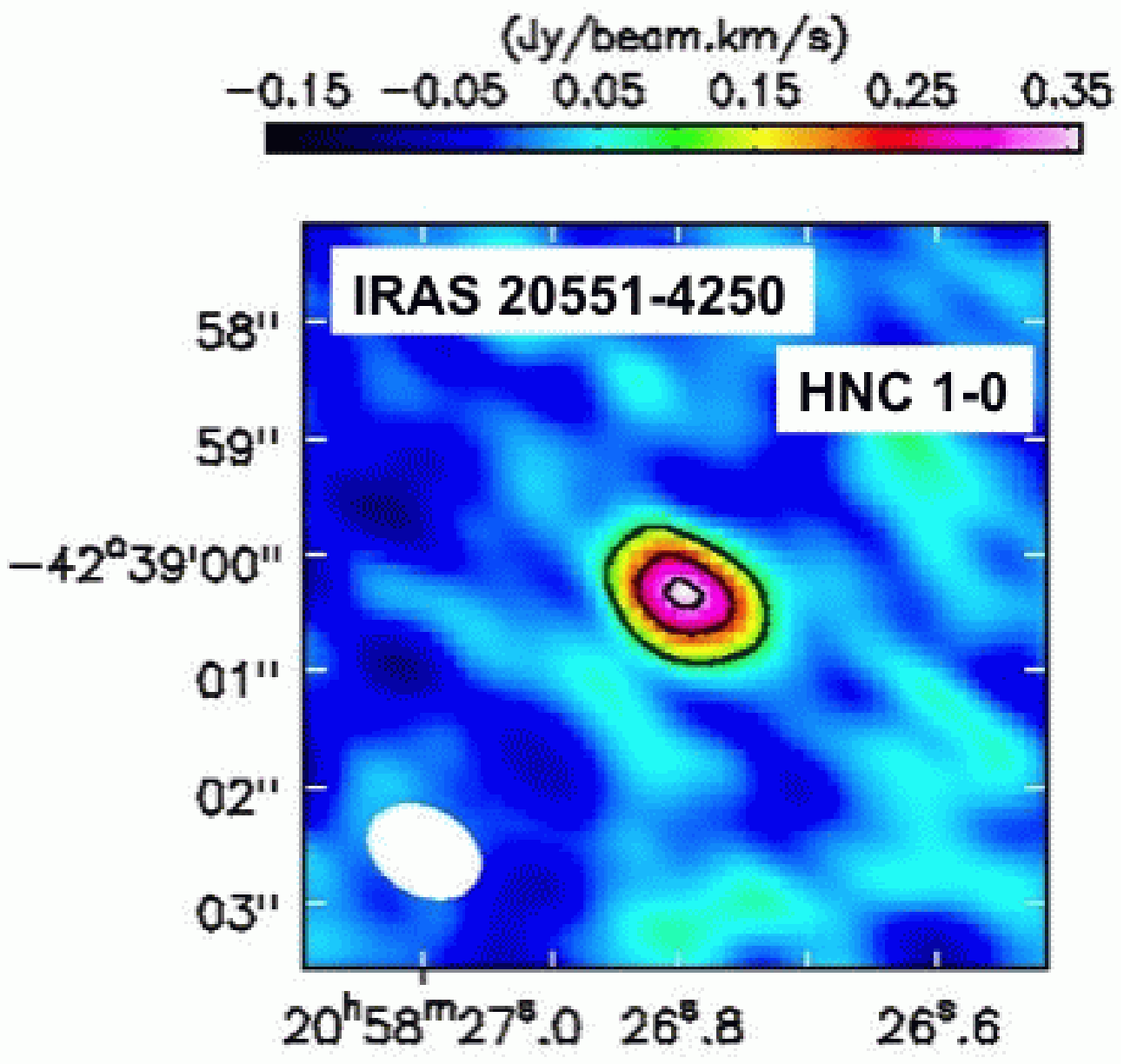}
\includegraphics[angle=0,scale=.41]{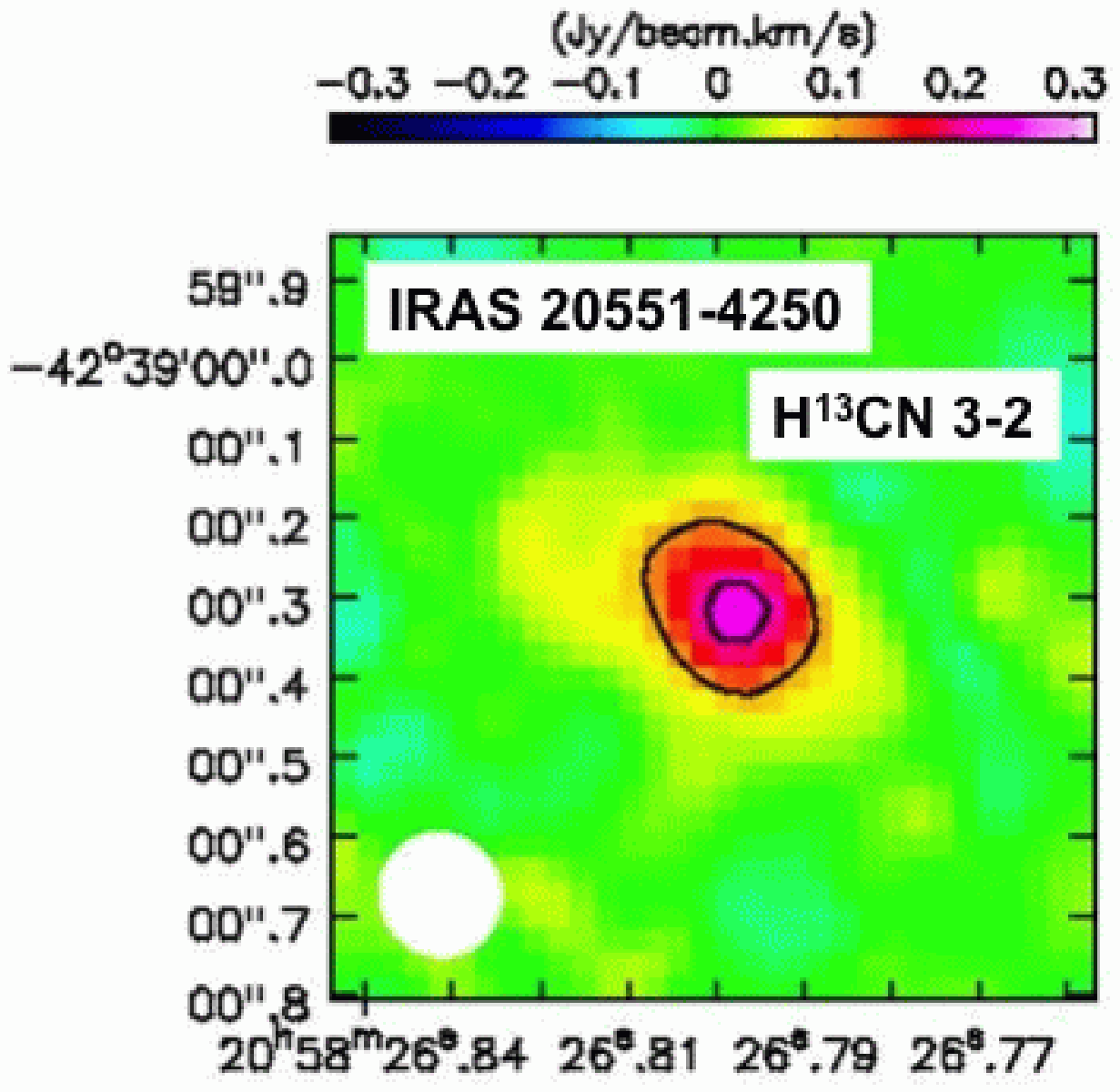} \\
\includegraphics[angle=0,scale=.41]{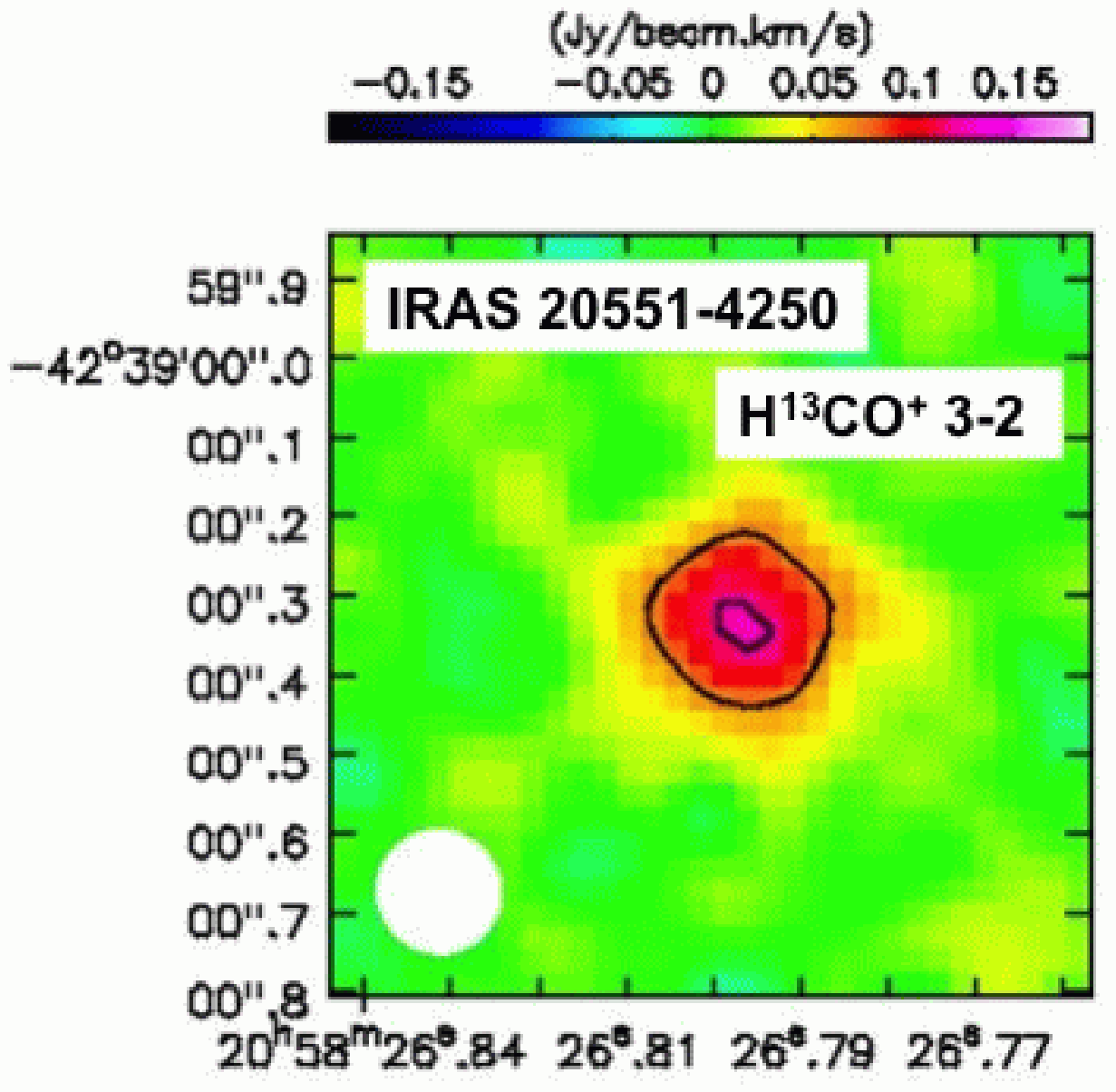} 
\includegraphics[angle=0,scale=.41]{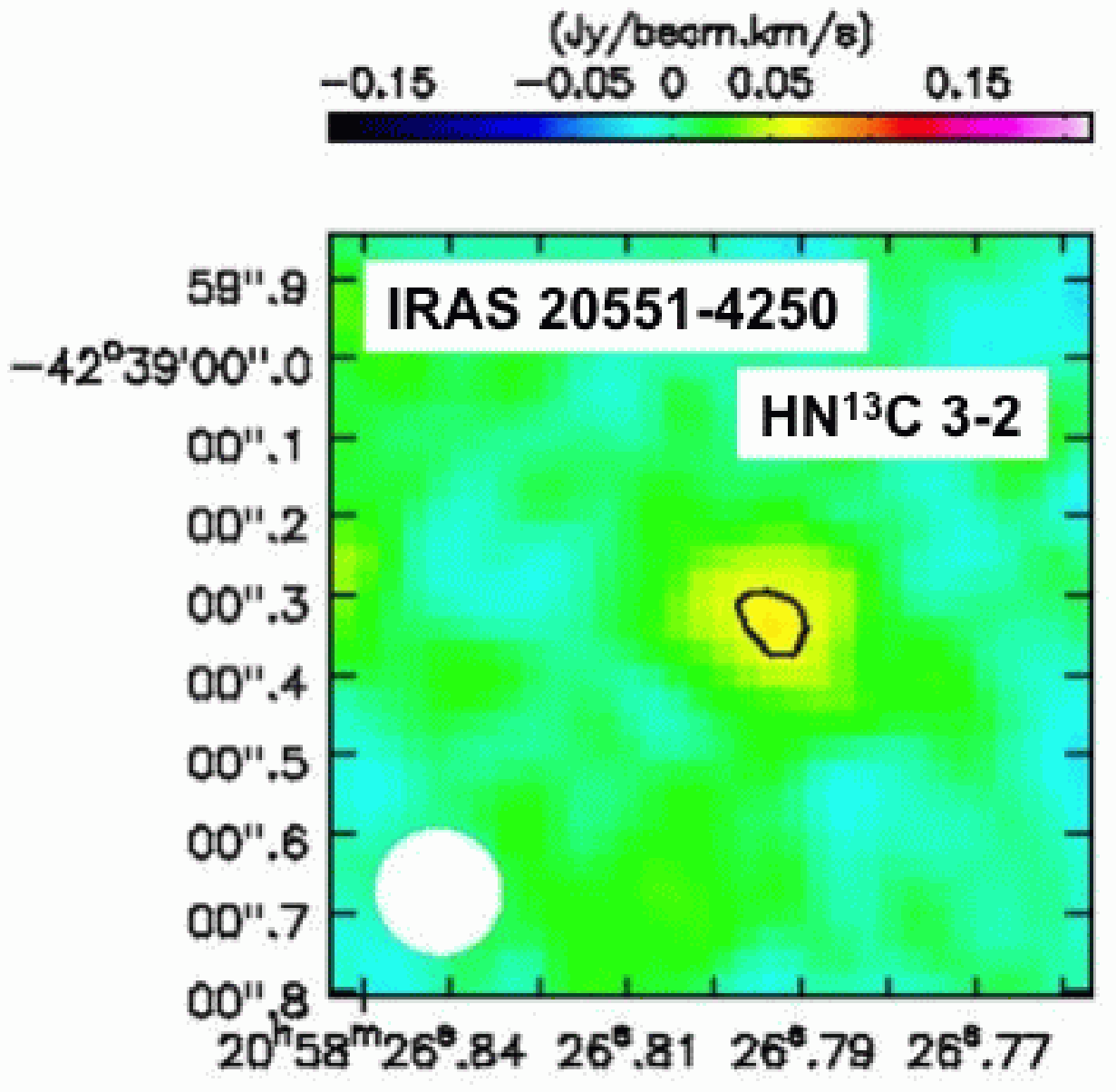} 
\includegraphics[angle=0,scale=.41]{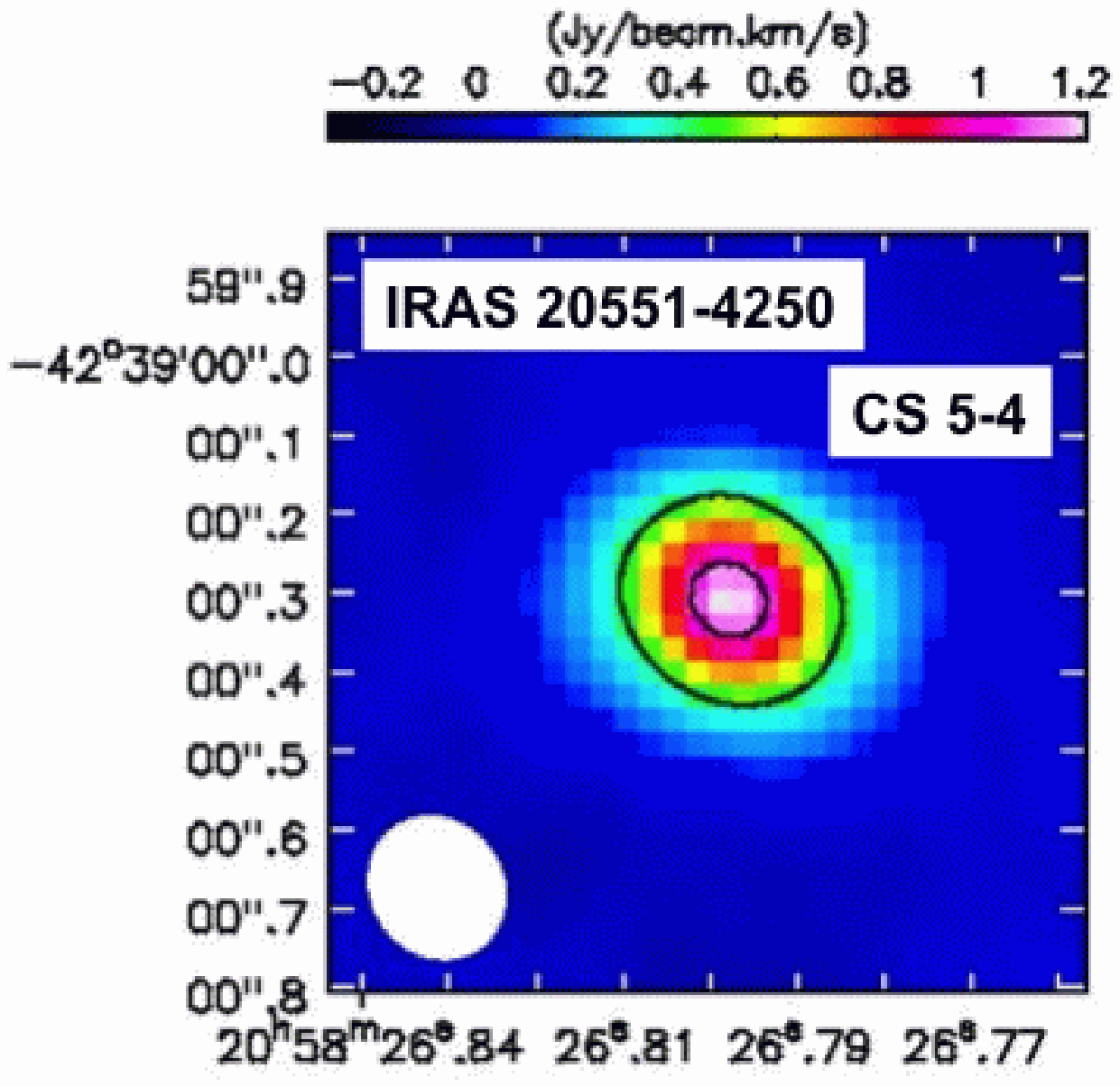} \\


\caption{
Integrated intensity (moment 0) maps of the primarily targeted
molecular emission lines in IRAS 20551$-$4250.  
The abscissa and ordinate are right ascension (J2000) and declination 
(J2000), respectively. 
Molecular lines observed in ALMA band 7 observations are displayed
first (first six images), followed by those in band 9 (five images), band 3 (three images), and band 6
(four images). 
The contours represent 5$\sigma$, 10$\sigma$, 20$\sigma$,
40$\sigma$, 60$\sigma$ for CO J=3--2, 
20$\sigma$, 40$\sigma$, 60$\sigma$ for HCO$^{+}$ J=4--3, 
3$\sigma$, 6$\sigma$, 9$\sigma$ for HCN v$_{2}$=1f J=4--3, 
2$\sigma$, 3$\sigma$ for HCO$^{+}$ v$_{2}$=1f J=4--3, 
20$\sigma$, 40$\sigma$, 60$\sigma$ for HNC J=4--3, 
4$\sigma$, 7$\sigma$, 10$\sigma$ for HNC v$_{2}$=1f J=4--3, 
3$\sigma$, 6$\sigma$ for HCN J=8--7, 
3$\sigma$, 6$\sigma$, 9$\sigma$ for HCO$^{+}$ J=8--7, 
3$\sigma$, 4.5$\sigma$ HNC J=8--7, 
2.5$\sigma$, 3$\sigma$ HCN v$_{2}$=1f J=8--7, 
10$\sigma$, 20$\sigma$, 30$\sigma$ for HCN J=1--0, 
15$\sigma$, 25$\sigma$, 35$\sigma$ for HCO$^{+}$ J=1--0, 
3$\sigma$, 7$\sigma$, 10$\sigma$ for HNC J=1--0, 
3$\sigma$, 6$\sigma$ for H$^{13}$CN J=3--2, 
3$\sigma$, 5$\sigma$ for H$^{13}$CO$^{+}$ J=3--2, 
2.5$\sigma$ for HN$^{13}$C J=3--2, and  
15$\sigma$, 35$\sigma$ for CS J=5--4. 
For HNC v$_{2}$=1f J=8--7, no contours with $>$2$\sigma$ are seen.
The 1$\sigma$ levels are different for different molecular lines, and
are summarized in Table 4. 
Beam sizes are shown as filled circles in the lower-left region.
The displayed areas differ depending on the beam size.
}
\end{center}
\end{figure}

\begin{figure}
\begin{center}
\includegraphics[angle=0,scale=.274]{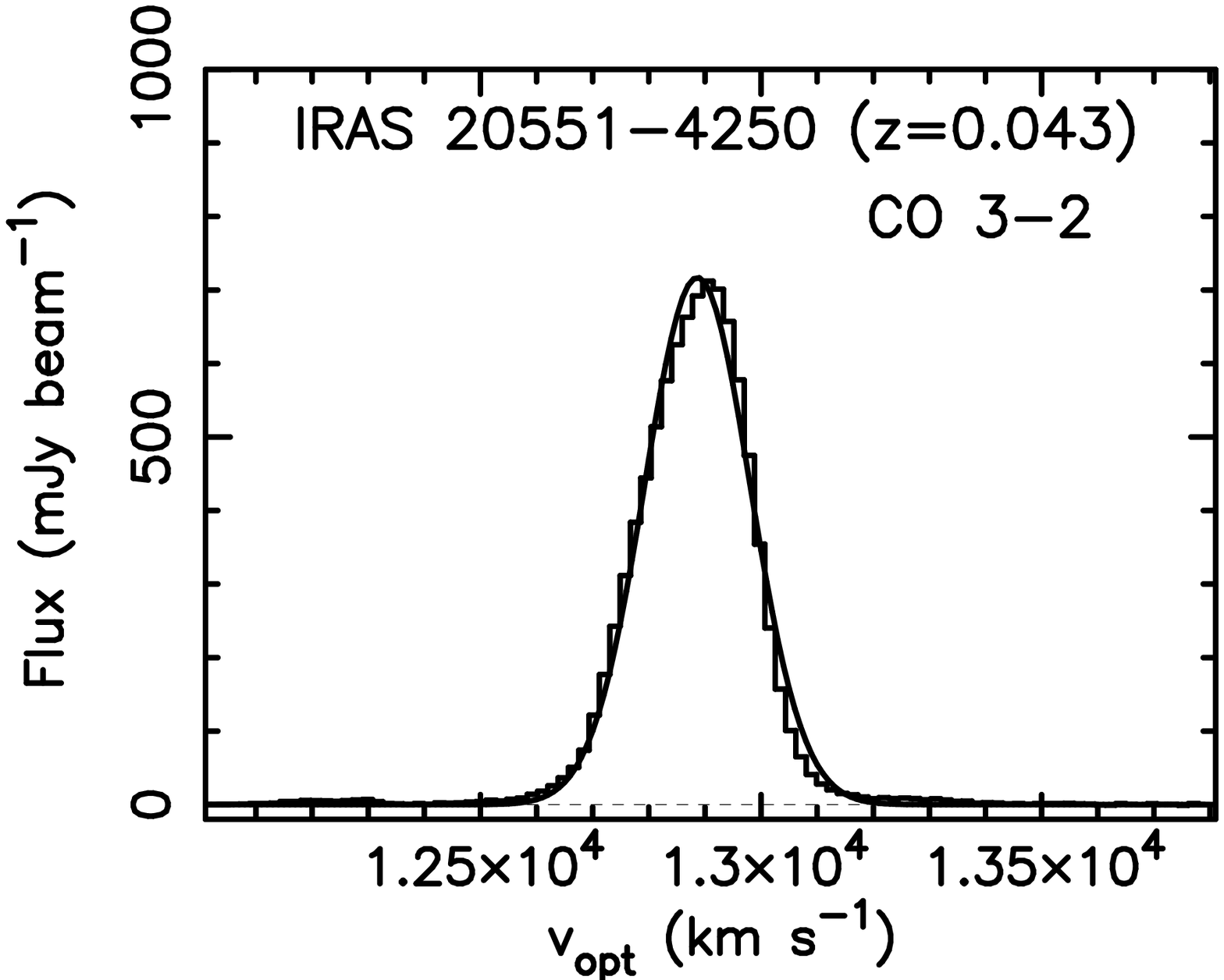} 
\includegraphics[angle=0,scale=.274]{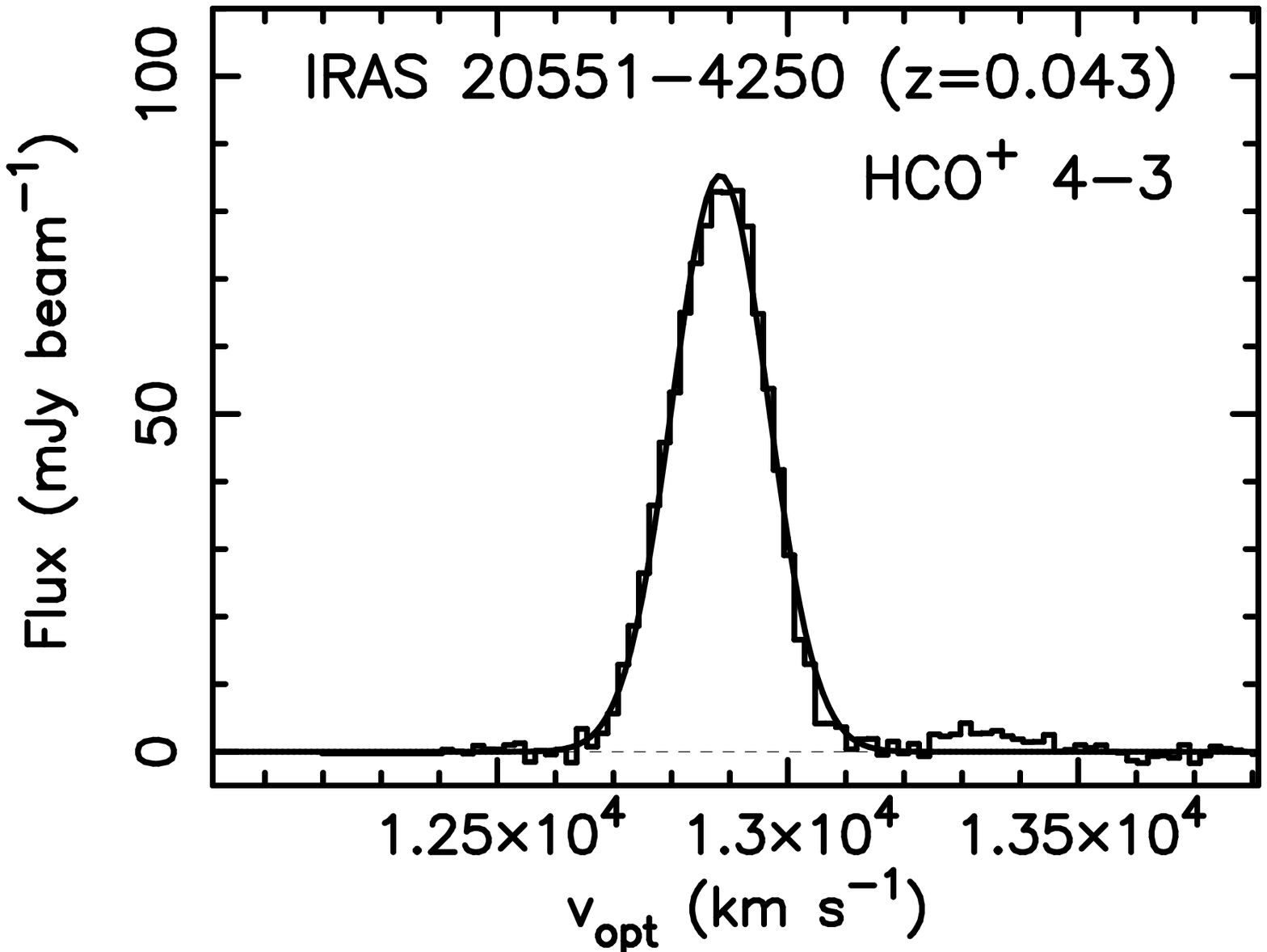} 
\includegraphics[angle=0,scale=.274]{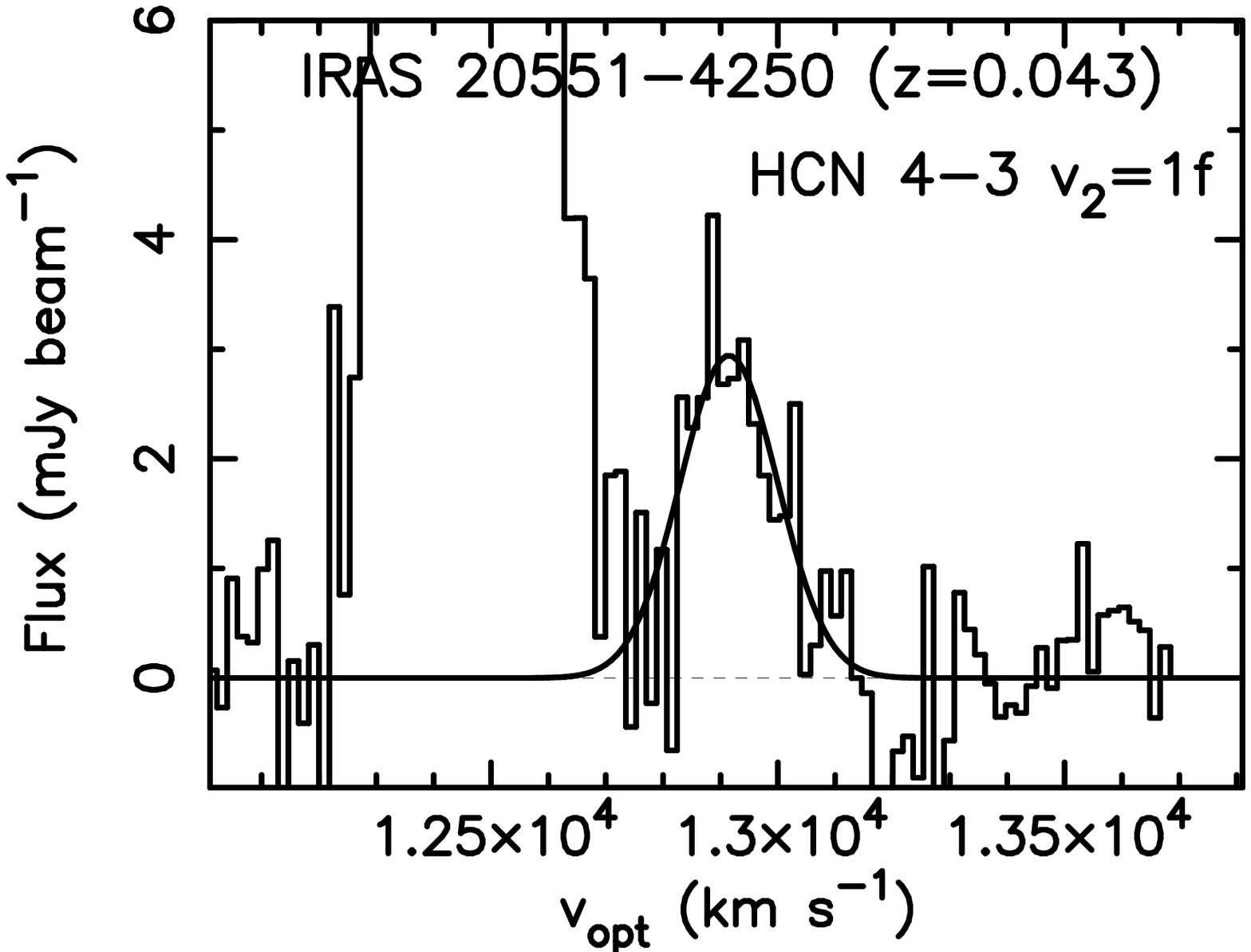} \\
\includegraphics[angle=0,scale=.274]{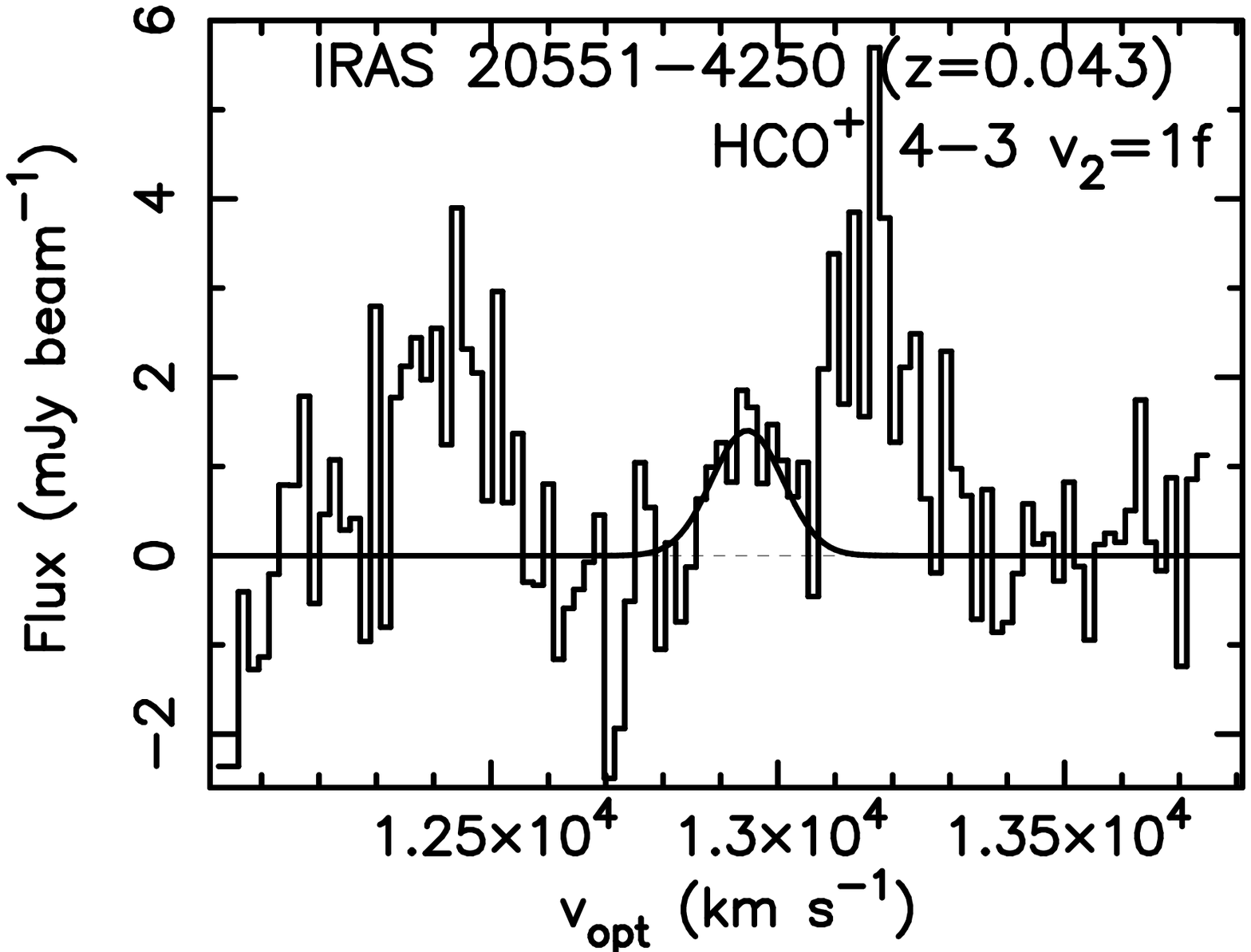} 
\includegraphics[angle=0,scale=.274]{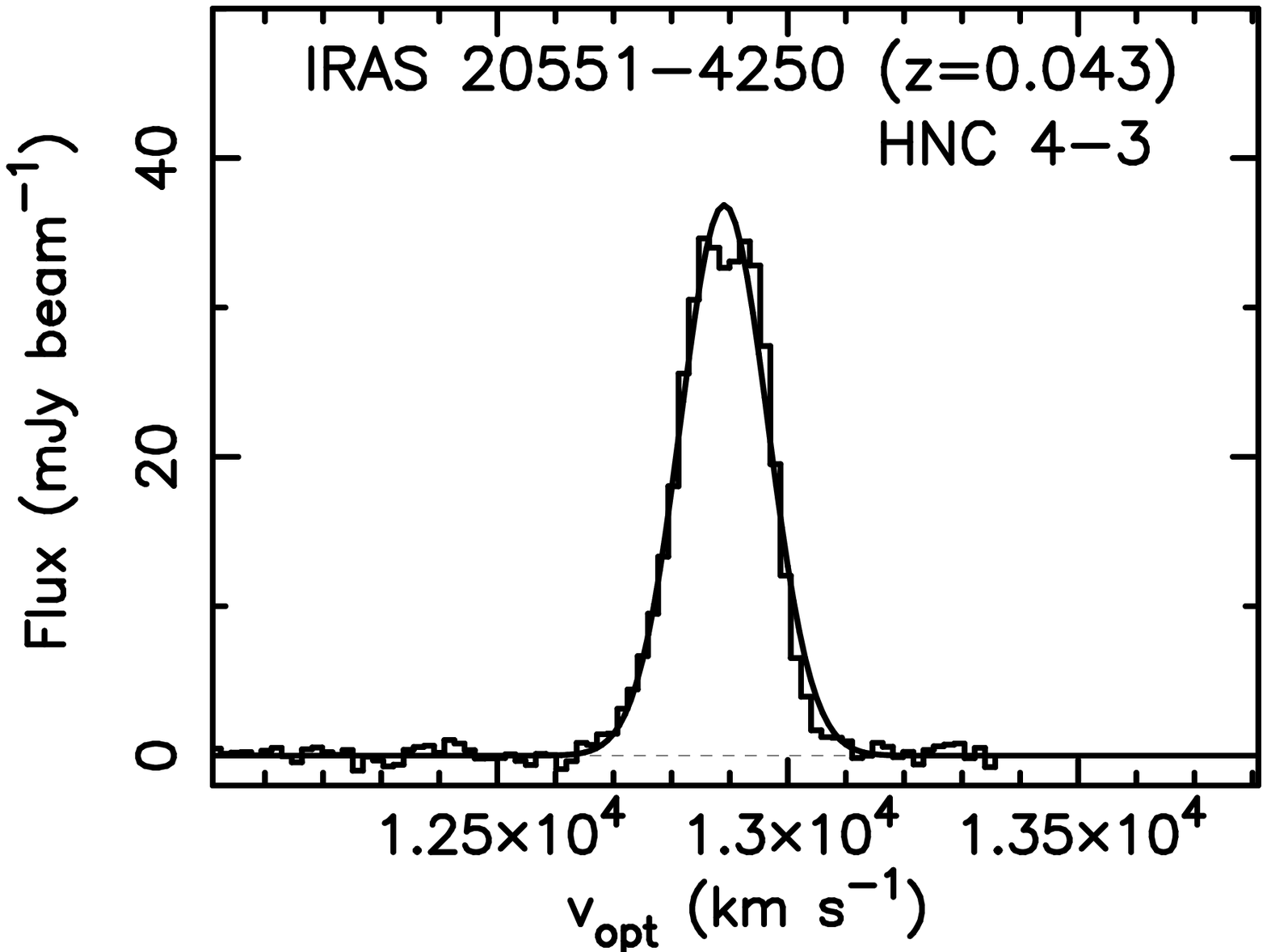} 
\includegraphics[angle=0,scale=.274]{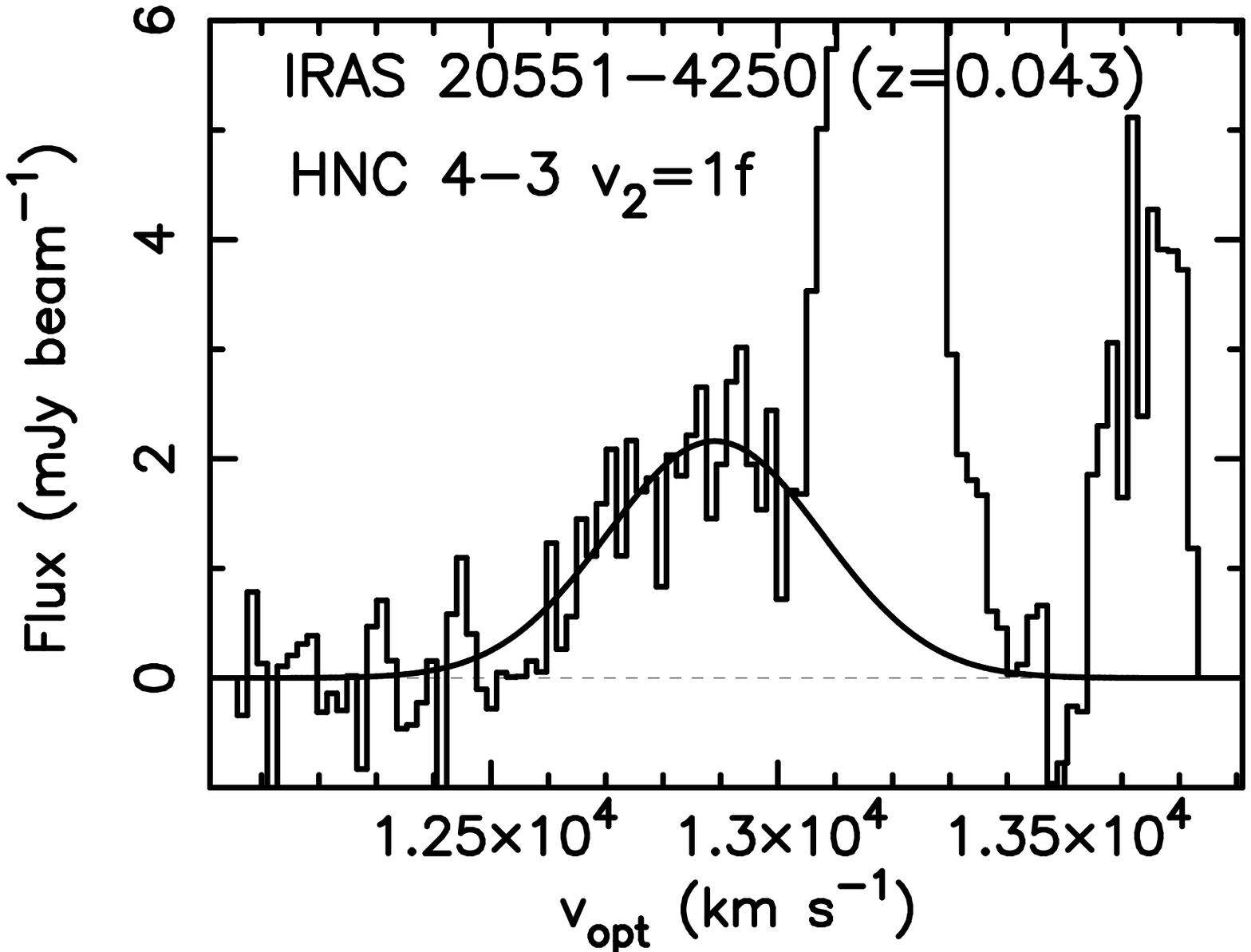} \\ 
\includegraphics[angle=0,scale=.274]{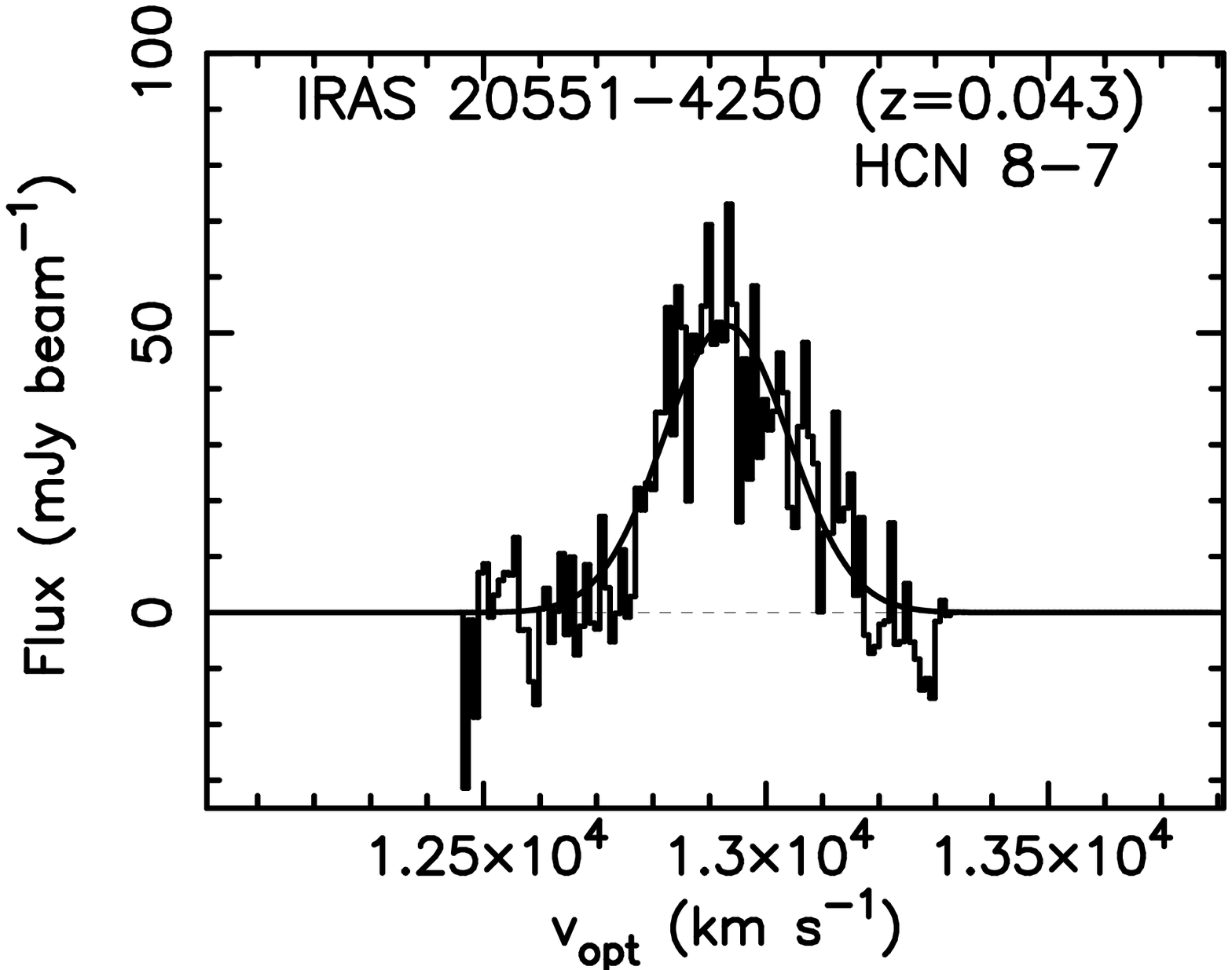}  
\includegraphics[angle=0,scale=.274]{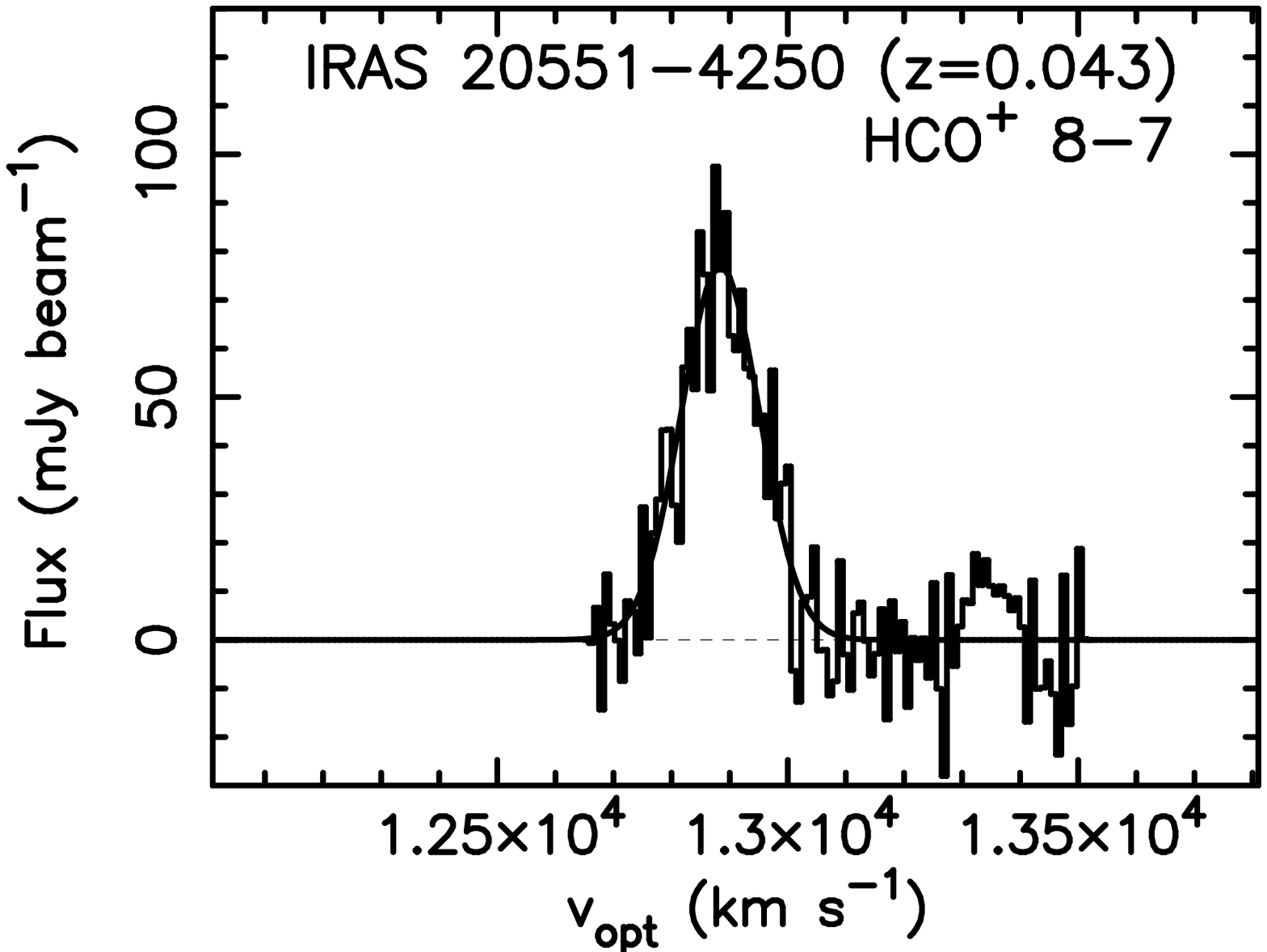} 
\includegraphics[angle=0,scale=.274]{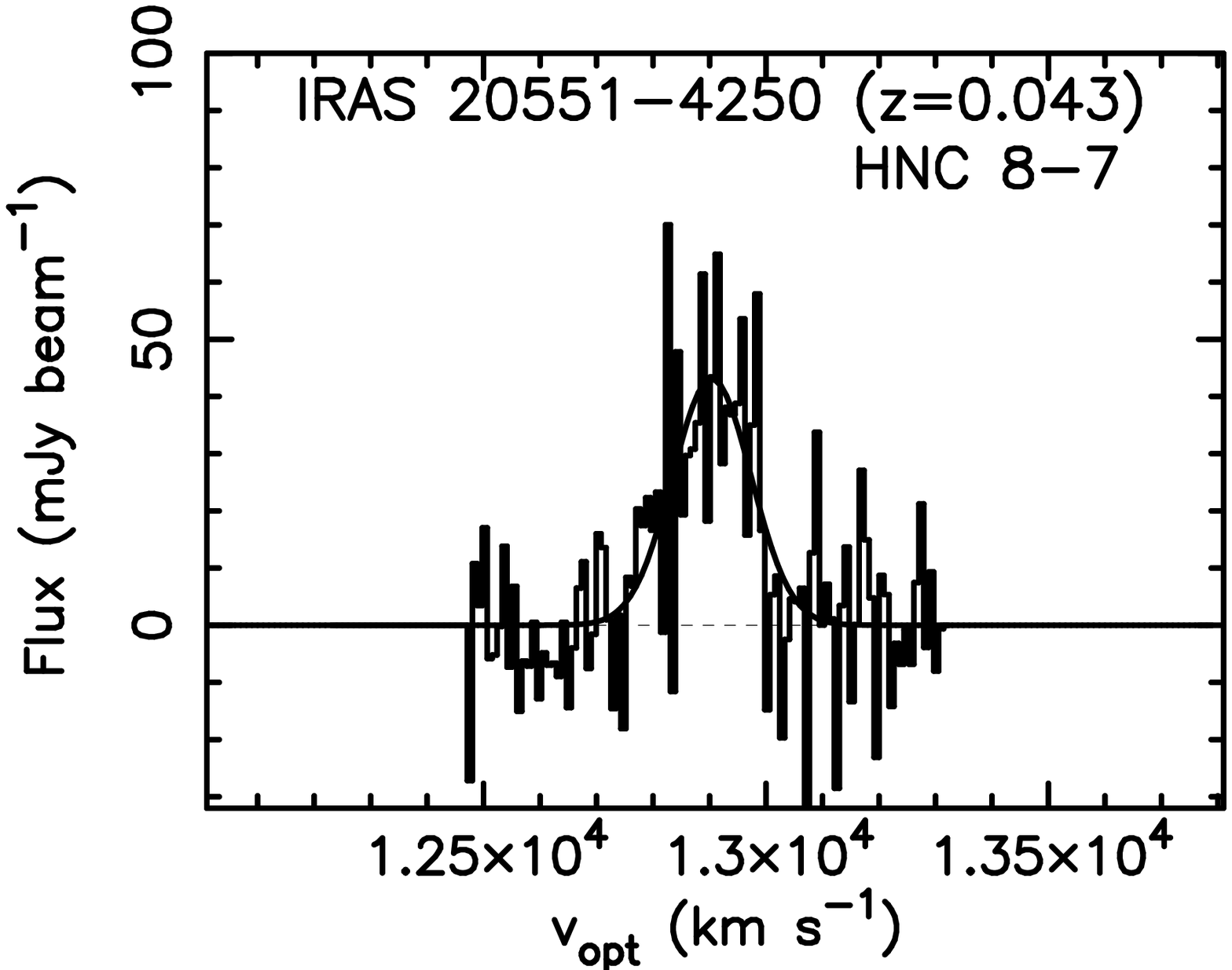} \\
\includegraphics[angle=0,scale=.274]{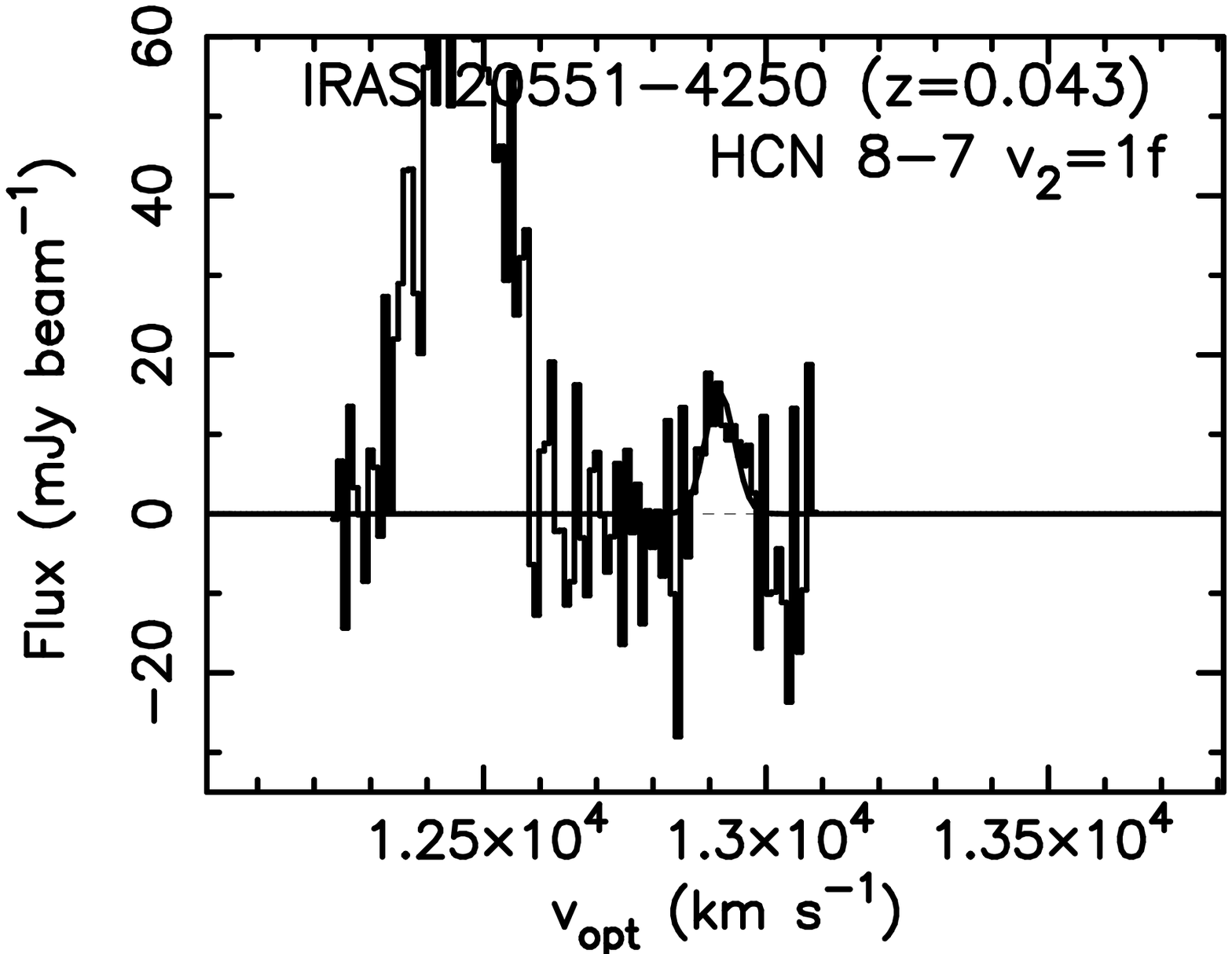} 
\includegraphics[angle=0,scale=.274]{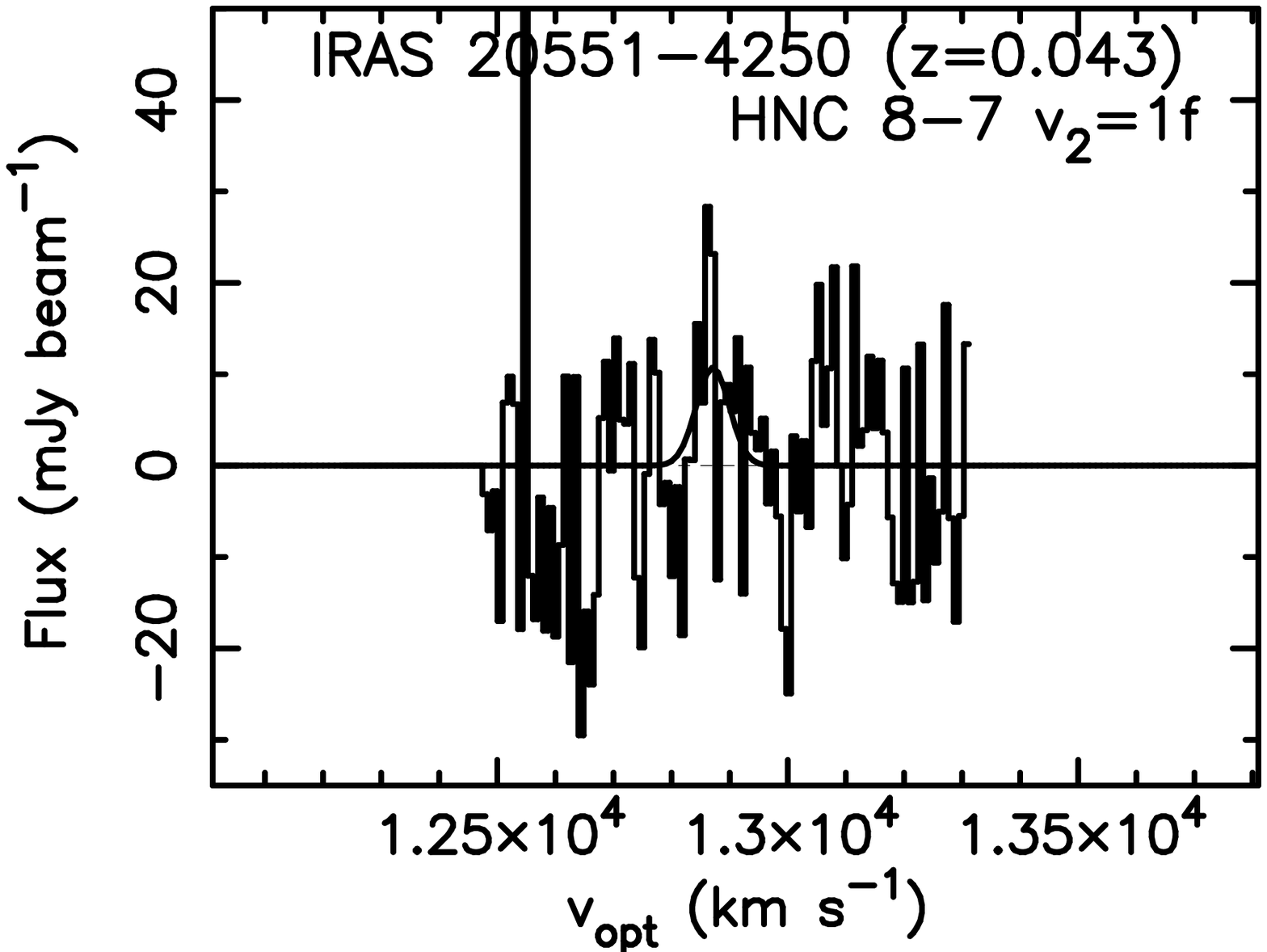} 
\includegraphics[angle=0,scale=.274]{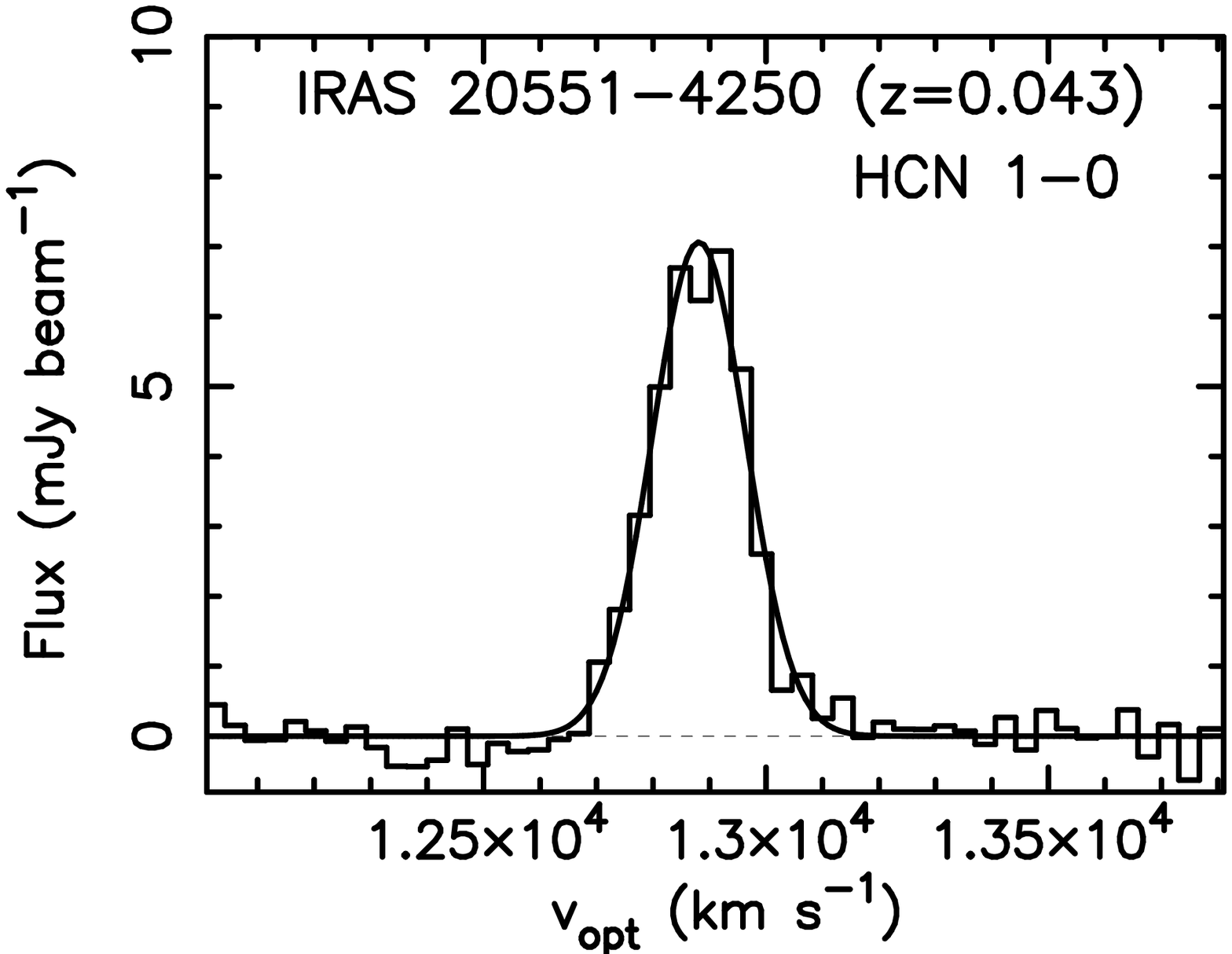} \\
\includegraphics[angle=0,scale=.274]{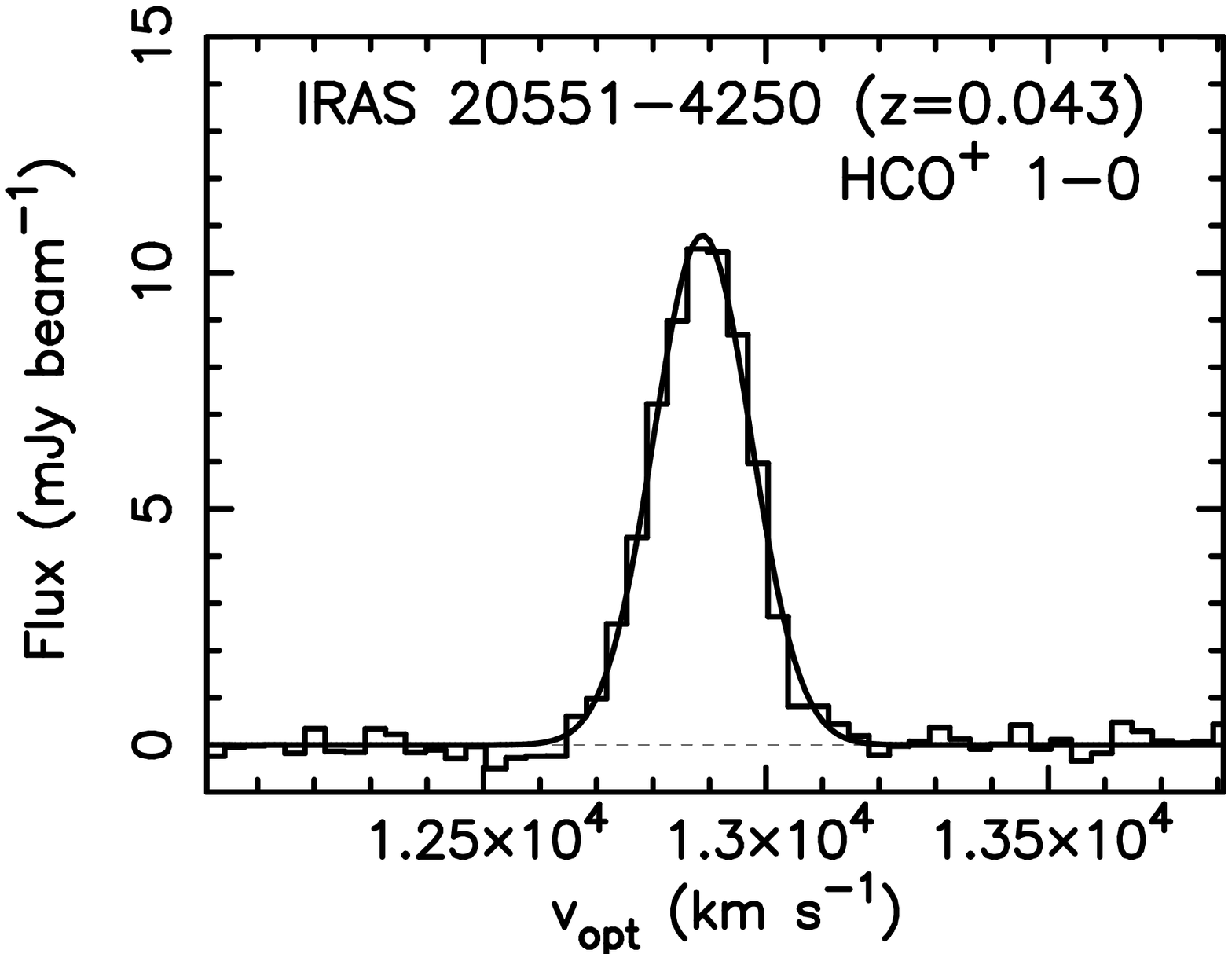} 
\includegraphics[angle=0,scale=.274]{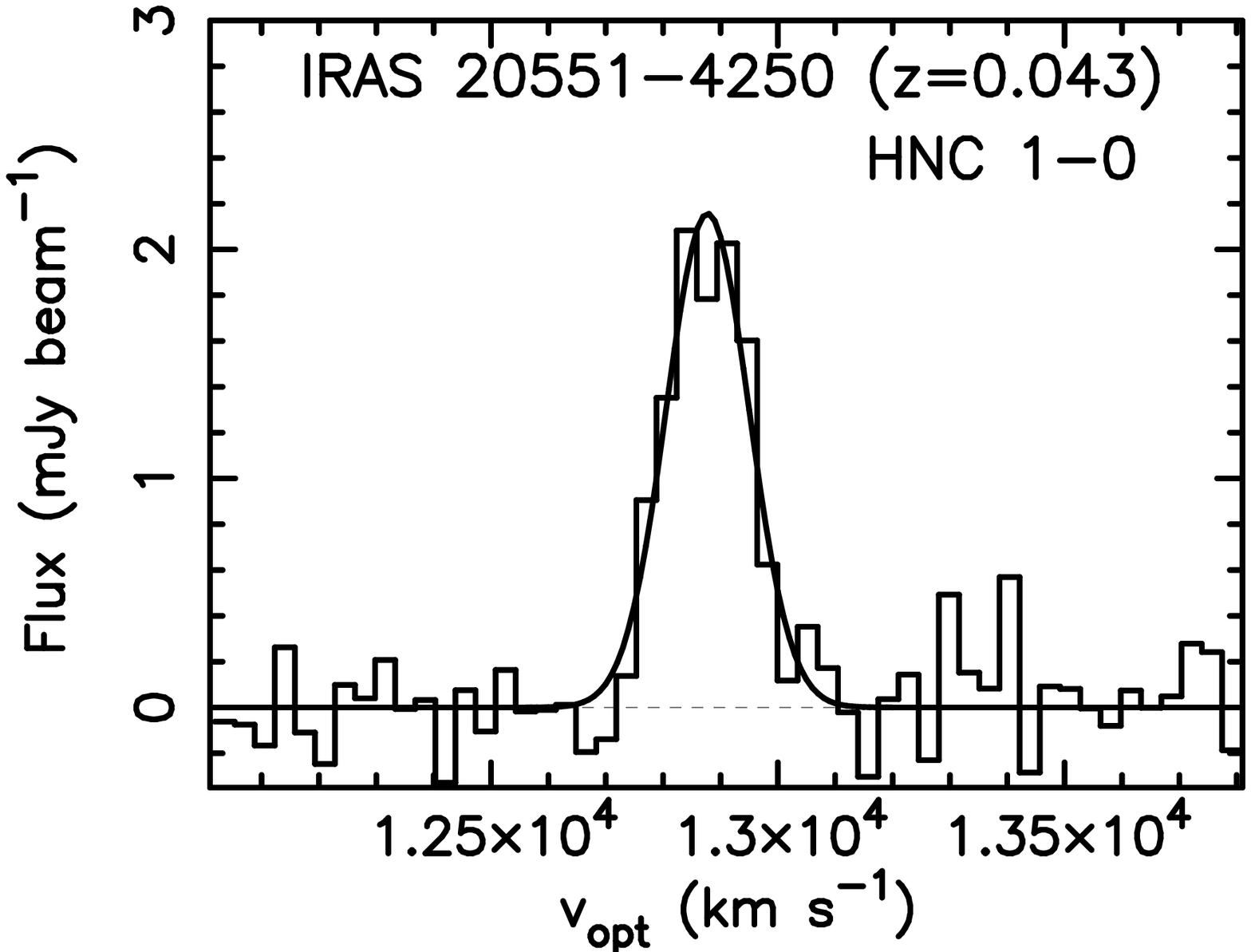} 
\includegraphics[angle=0,scale=.274]{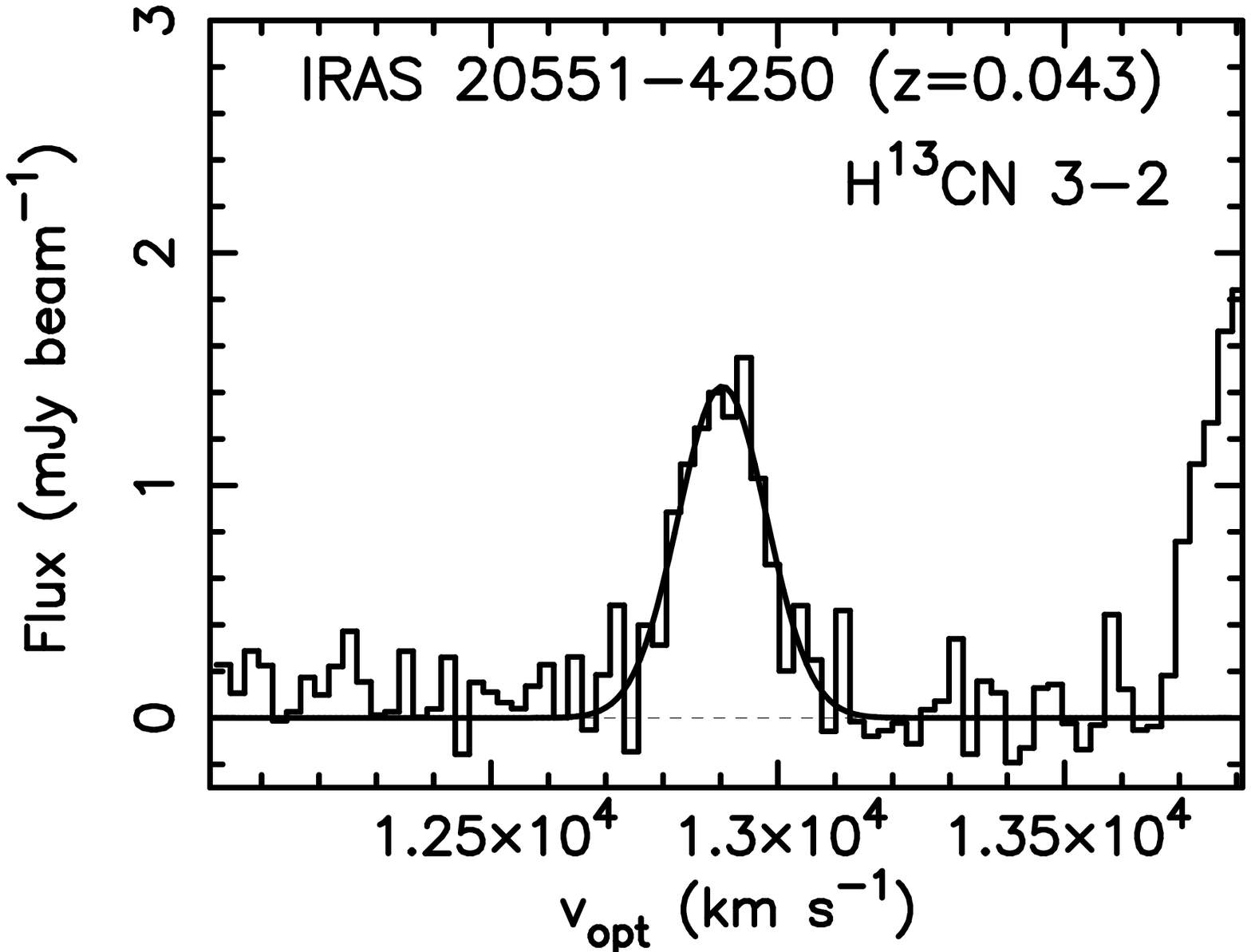} \\
\end{center}
\end{figure}

\clearpage

\begin{figure}
\begin{center}
\includegraphics[angle=0,scale=.274]{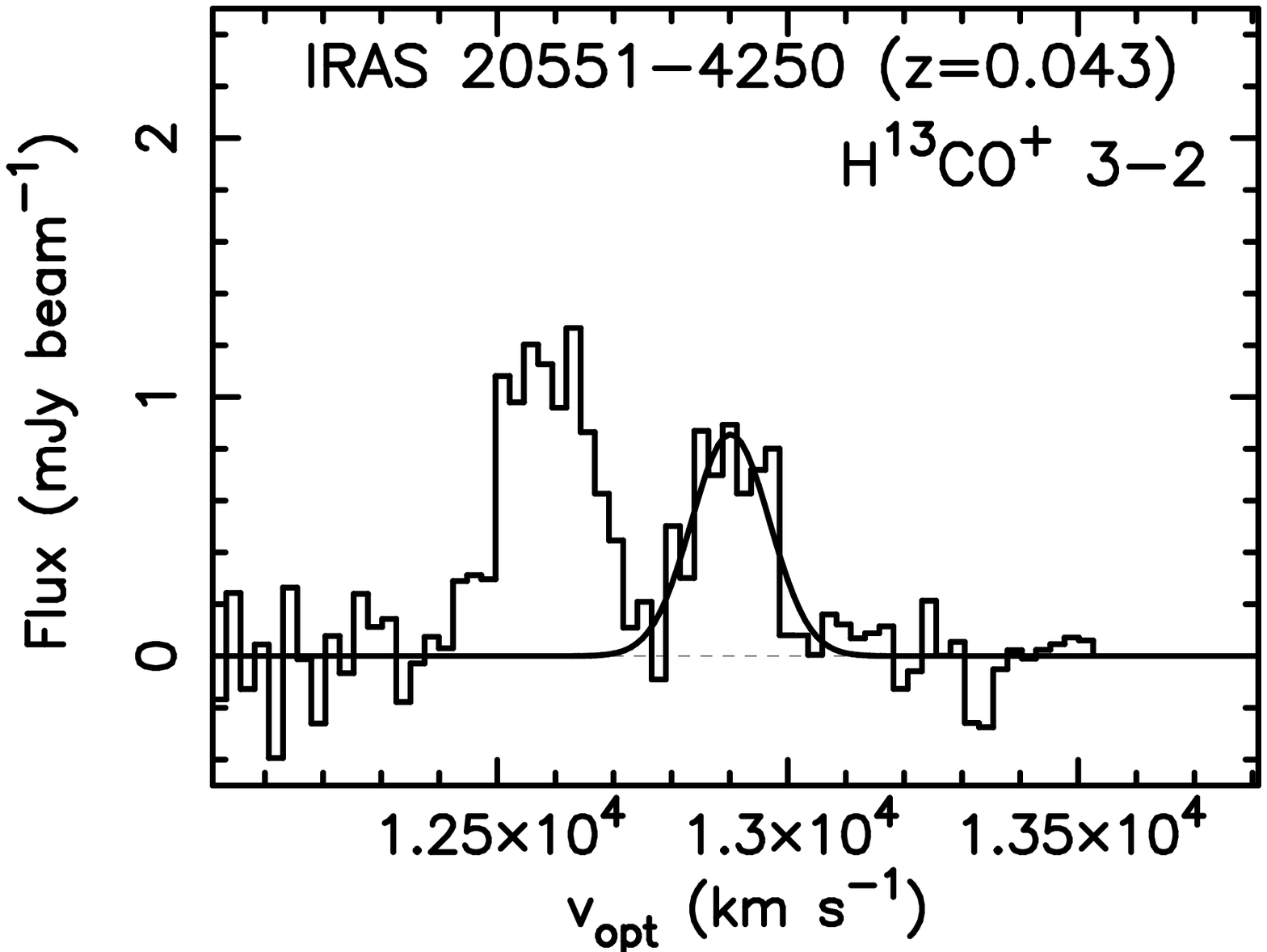} 
\includegraphics[angle=0,scale=.274]{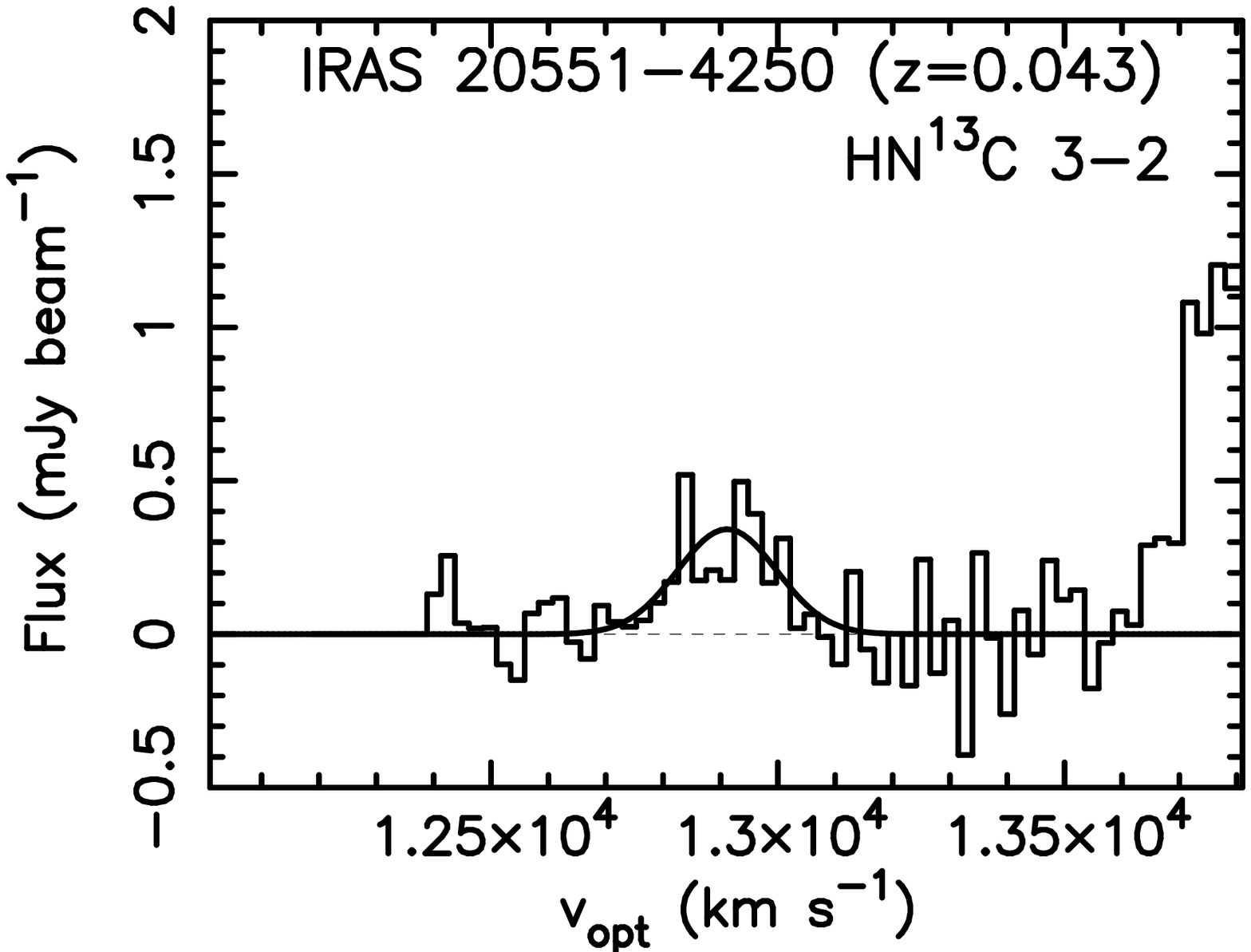} 
\includegraphics[angle=0,scale=.274]{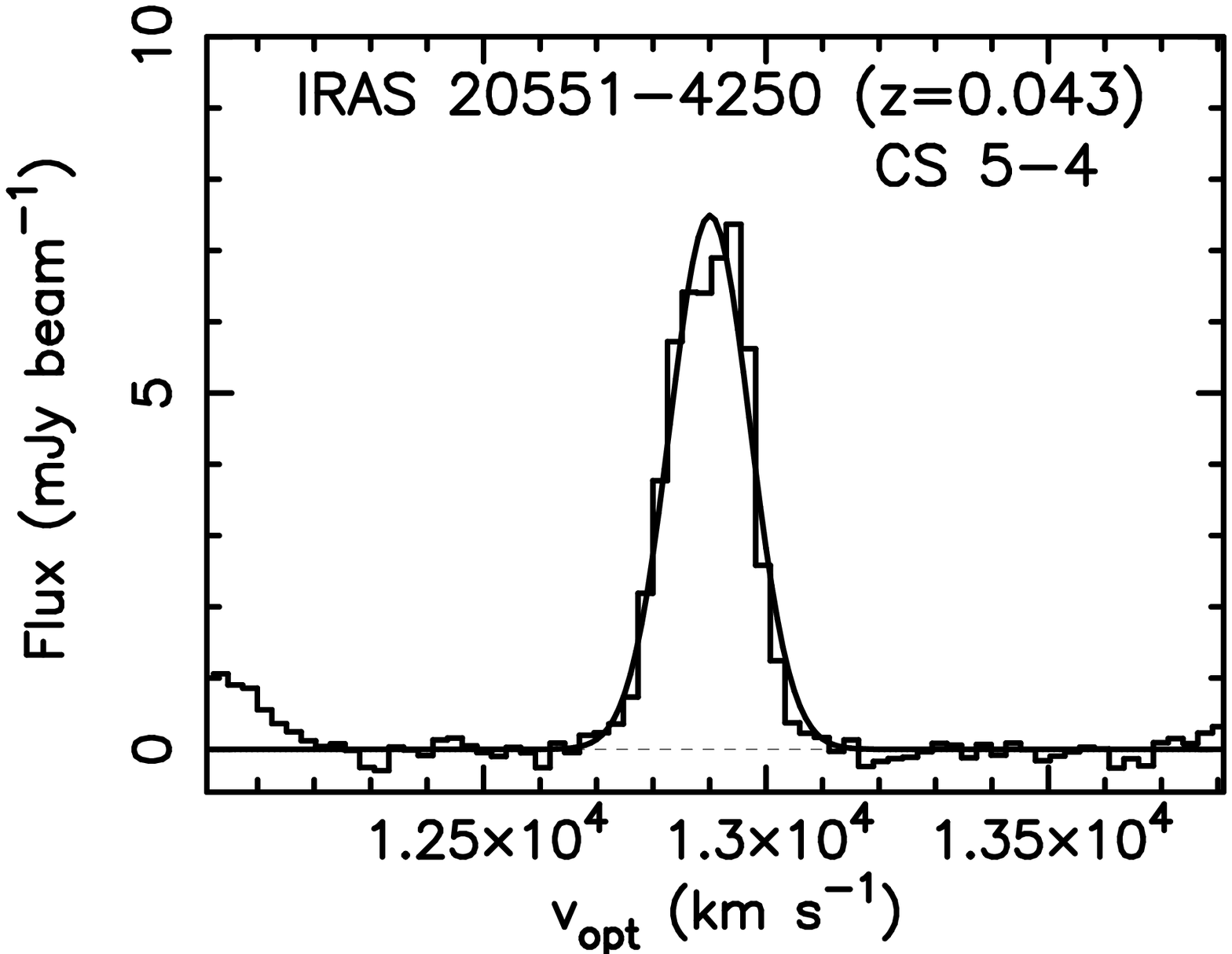} \\
\caption{
Spectra around individual molecular emission lines.
The abscissa is optical LSR velocity (v$_{\rm opt}$
$\equiv$ c ($\lambda$$-$$\lambda_{\rm 0}$)/$\lambda_{\rm 0}$), and the
ordinate is flux in (mJy beam$^{-1}$).  
Best Gaussian fits are overplotted with solid curved lines.
}
\end{center}
\end{figure}

\begin{figure}
\begin{center}
\includegraphics[angle=0,scale=.38]{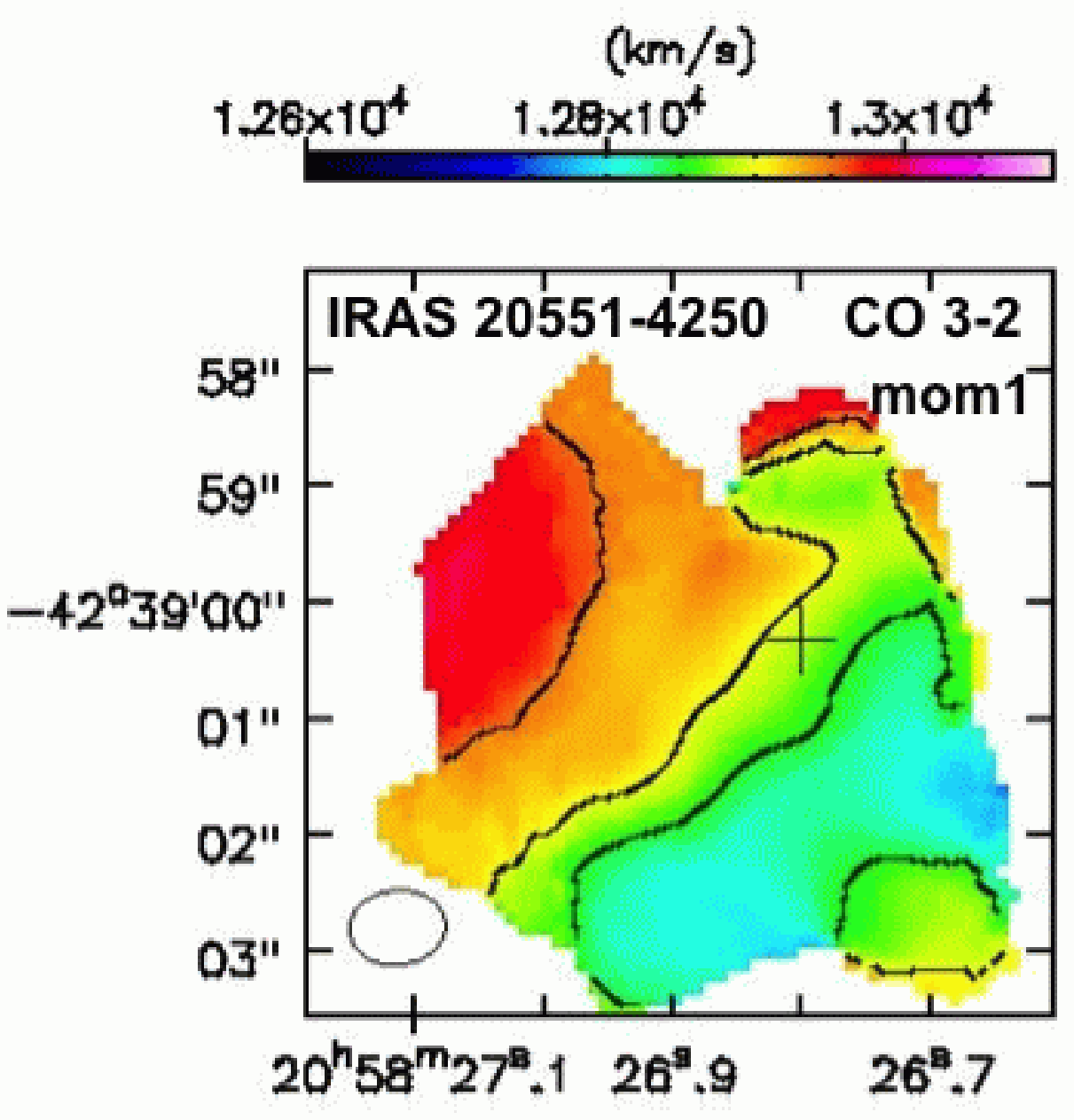}
\includegraphics[angle=0,scale=.38]{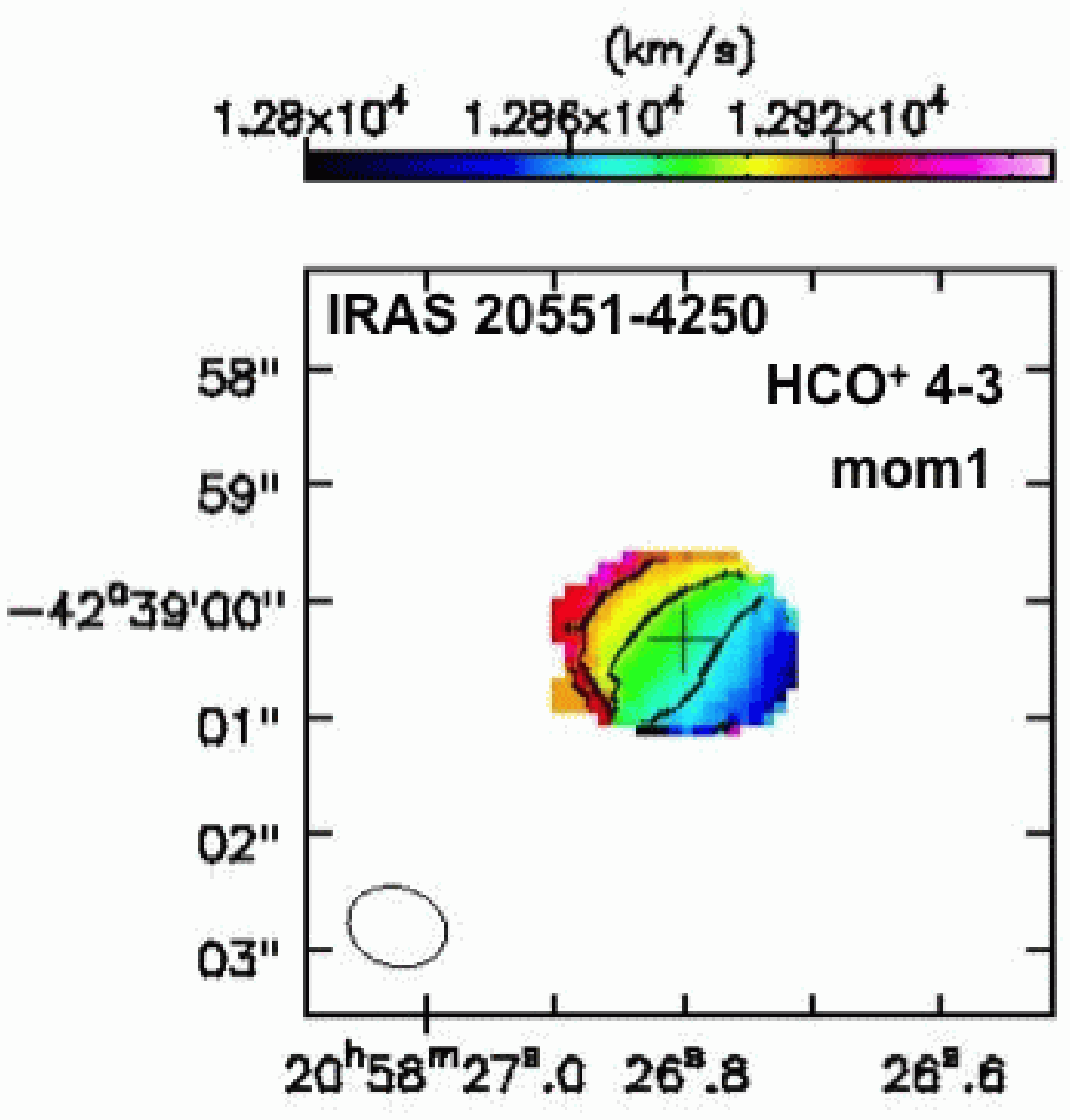}
\includegraphics[angle=0,scale=.4]{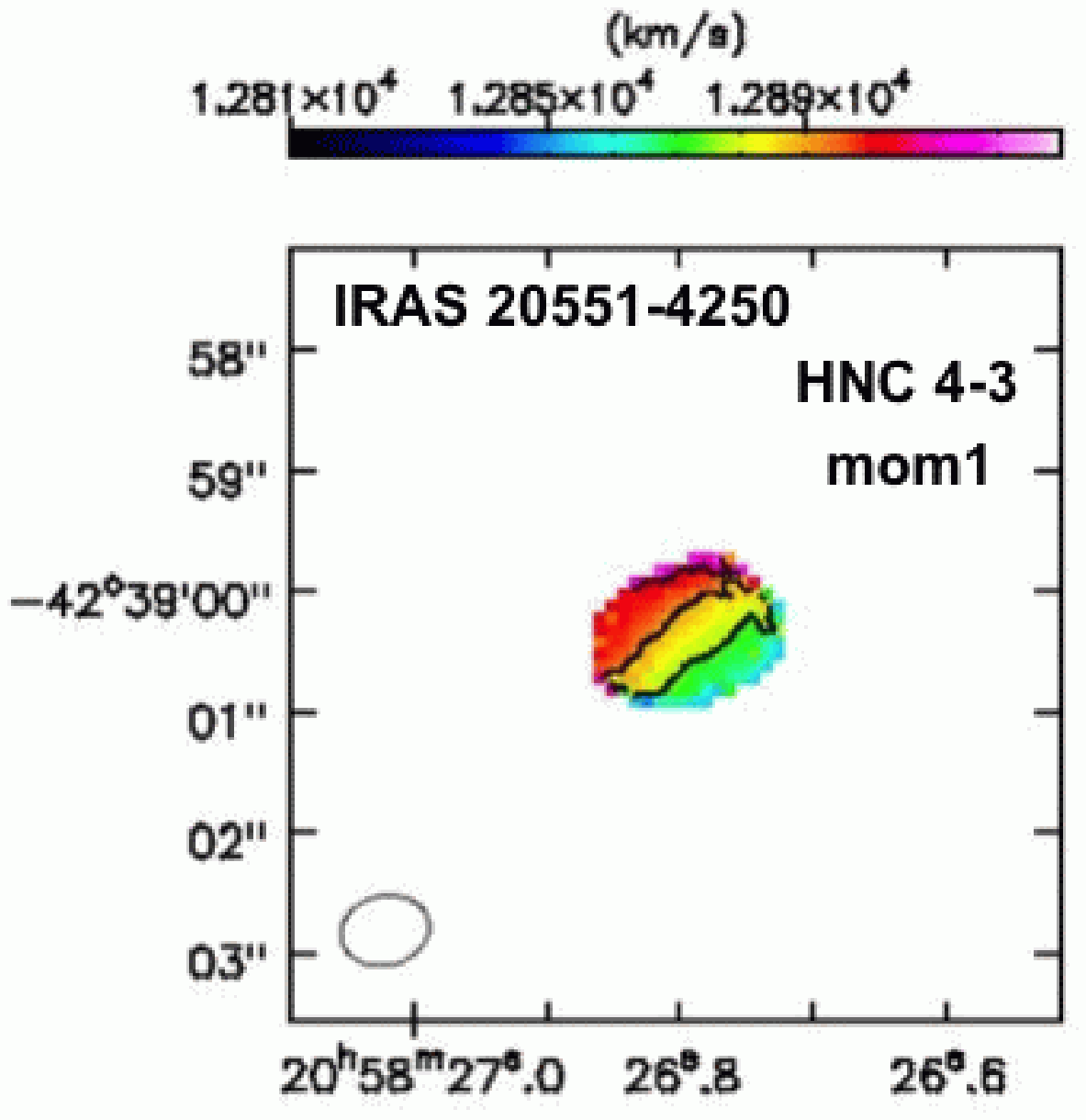} \\
\includegraphics[angle=0,scale=.38]{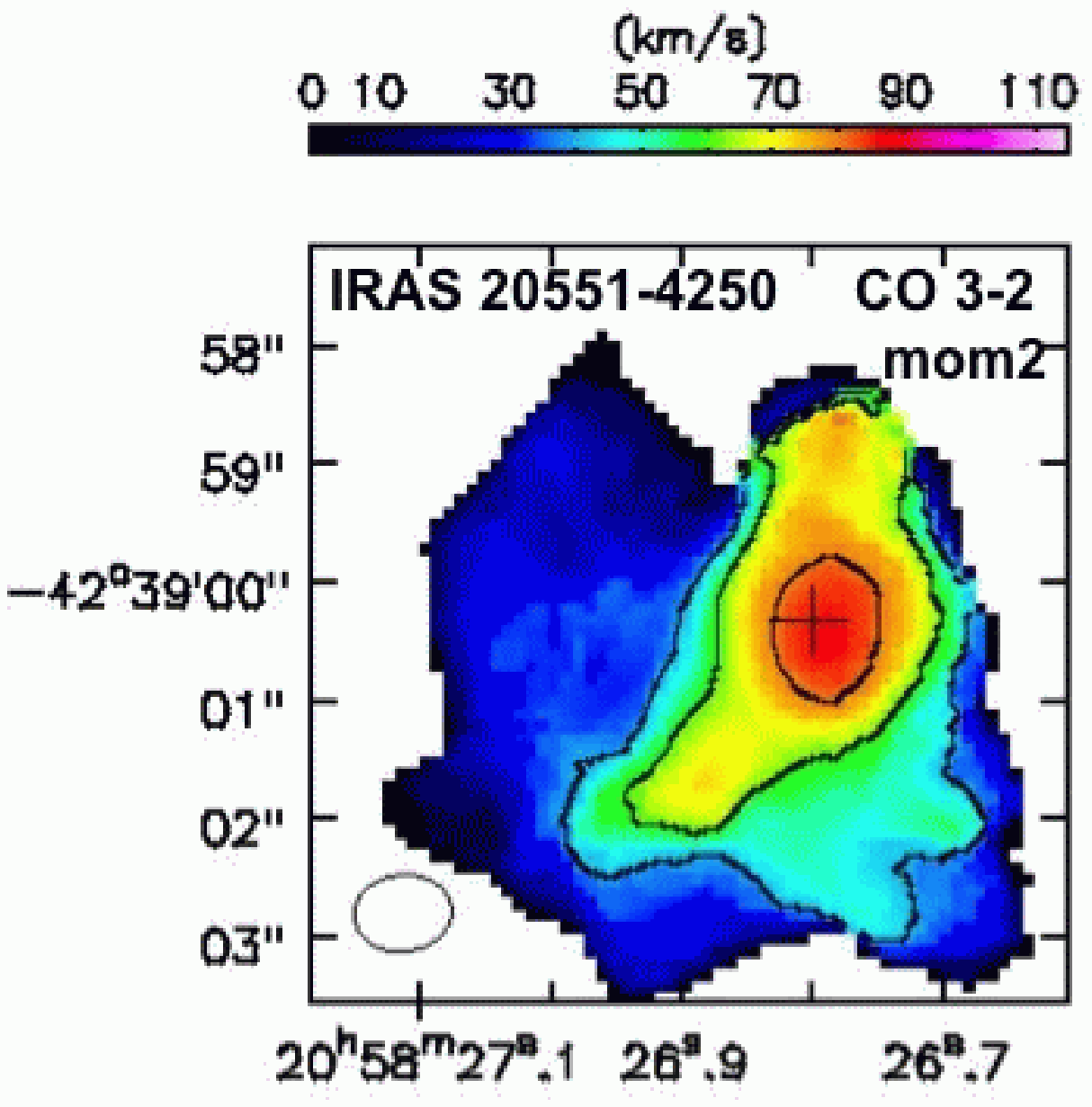} 
\includegraphics[angle=0,scale=.38]{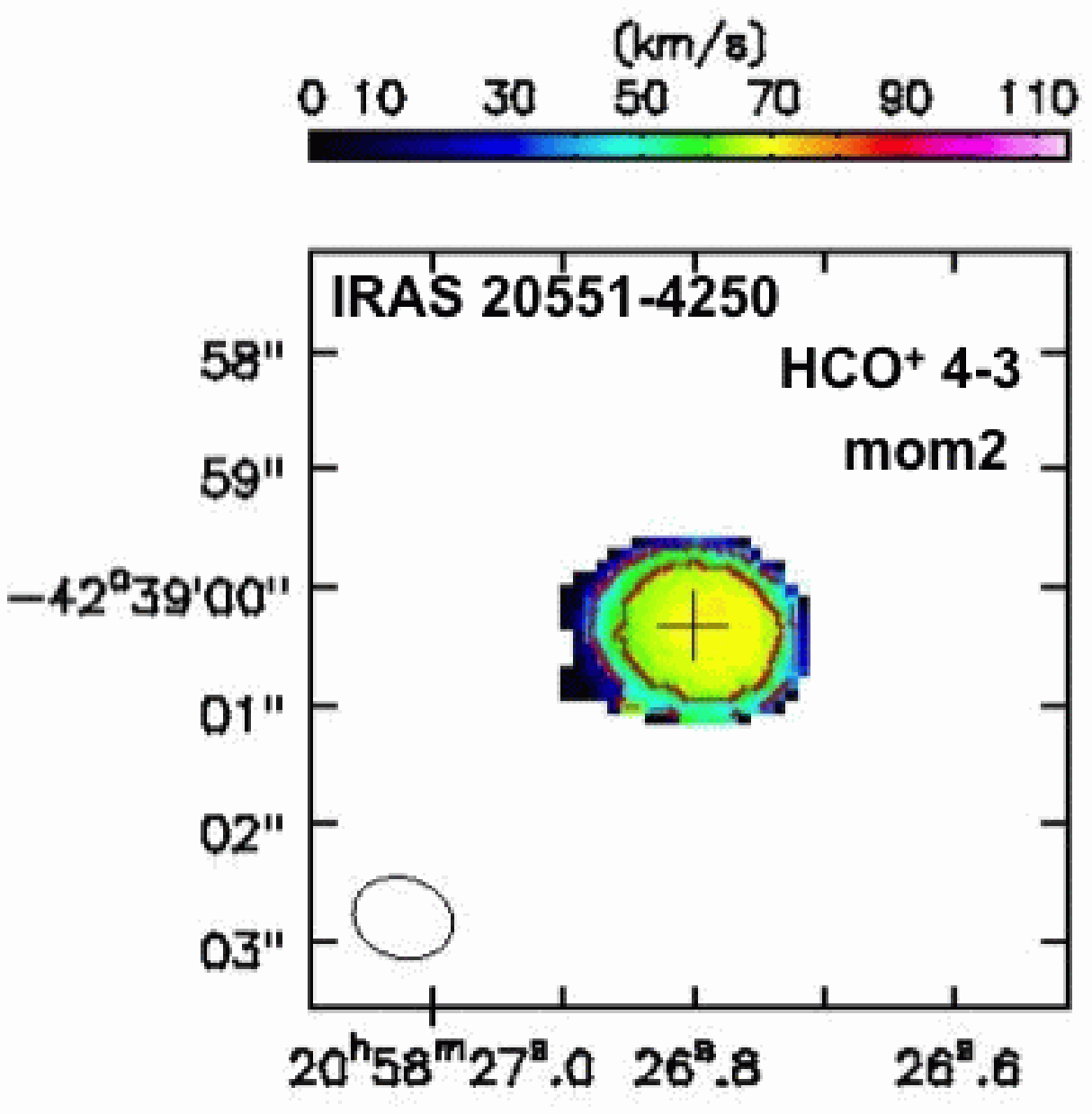}
\includegraphics[angle=0,scale=.4]{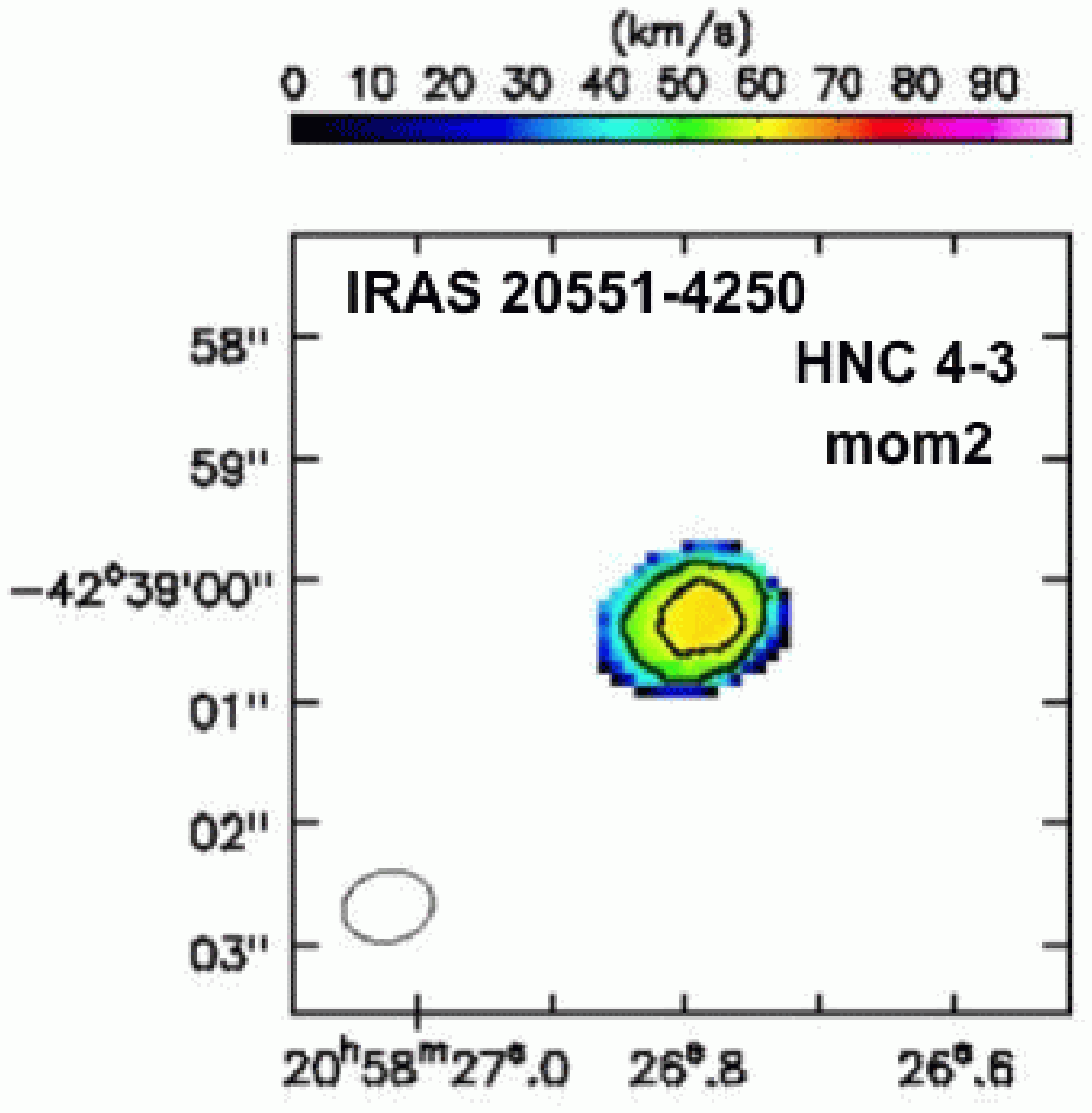} \\
\end{center}
\caption{
({\it {Top}}) : Intensity-weighted mean velocity (moment 1) maps for CO
J=3--2, HCO$^{+}$ J=4--3, and HNC J=4--3 emission lines in band 7. 
The abscissa and ordinate are right ascension (J2000) and declination 
(J2000), respectively. 
The contours represent 12850, 12900, 12950 km s$^{-1}$ for CO J=3--2, 
12870, 12895, 12920 km s$^{-1}$ for HCO$^{+}$ J=4--3, and 
12875, 12890, 12905 km s$^{-1}$ for HNC J=4--3.
({\it {Bottom}}): Intensity-weighted velocity dispersion (moment 2) maps. 
The contours represent 40, 60, 80 km s$^{-1}$ for CO J=3--2, 
40, 60 km s$^{-1}$ for HCO$^{+}$ J=4--3, 50, 60 km s$^{-1}$ for HNC J=4--3.
The centers of the CO J=3--2 images are slightly displaced from those of
HCO$^{+}$ J=4--3 and HNC J=4--3 images, to show whole extended
structures. 
Beam sizes are shown as open circles at the lower-left part.
We applied an appropriate cut-off to prevent the resulting maps
from being dominated by noise.
In the left two panels, the continuum peak positions are marked with crosses.
}
\end{figure}

\begin{figure}
\begin{center}
\includegraphics[angle=0,scale=.38]{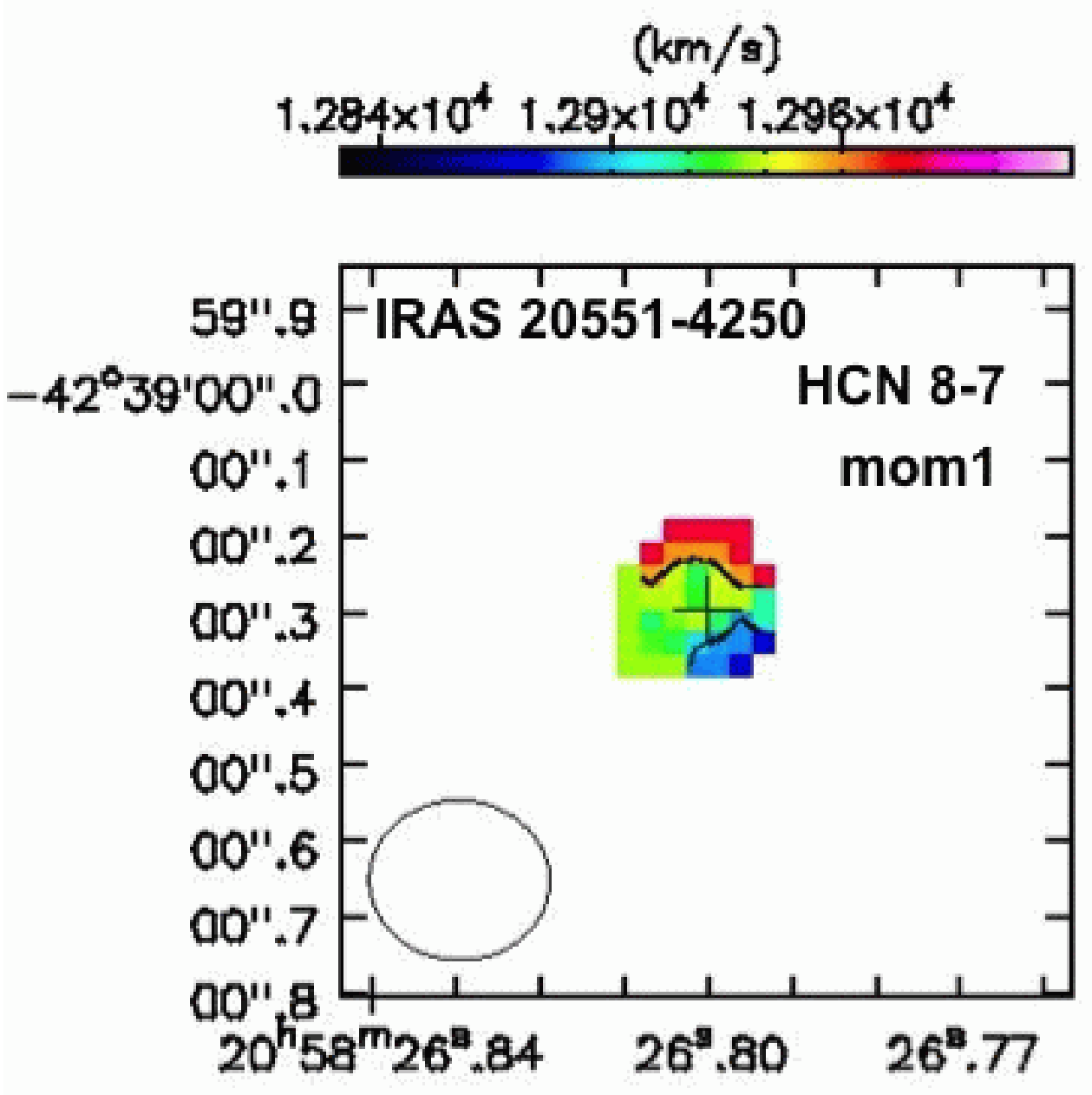}  
\includegraphics[angle=0,scale=.4]{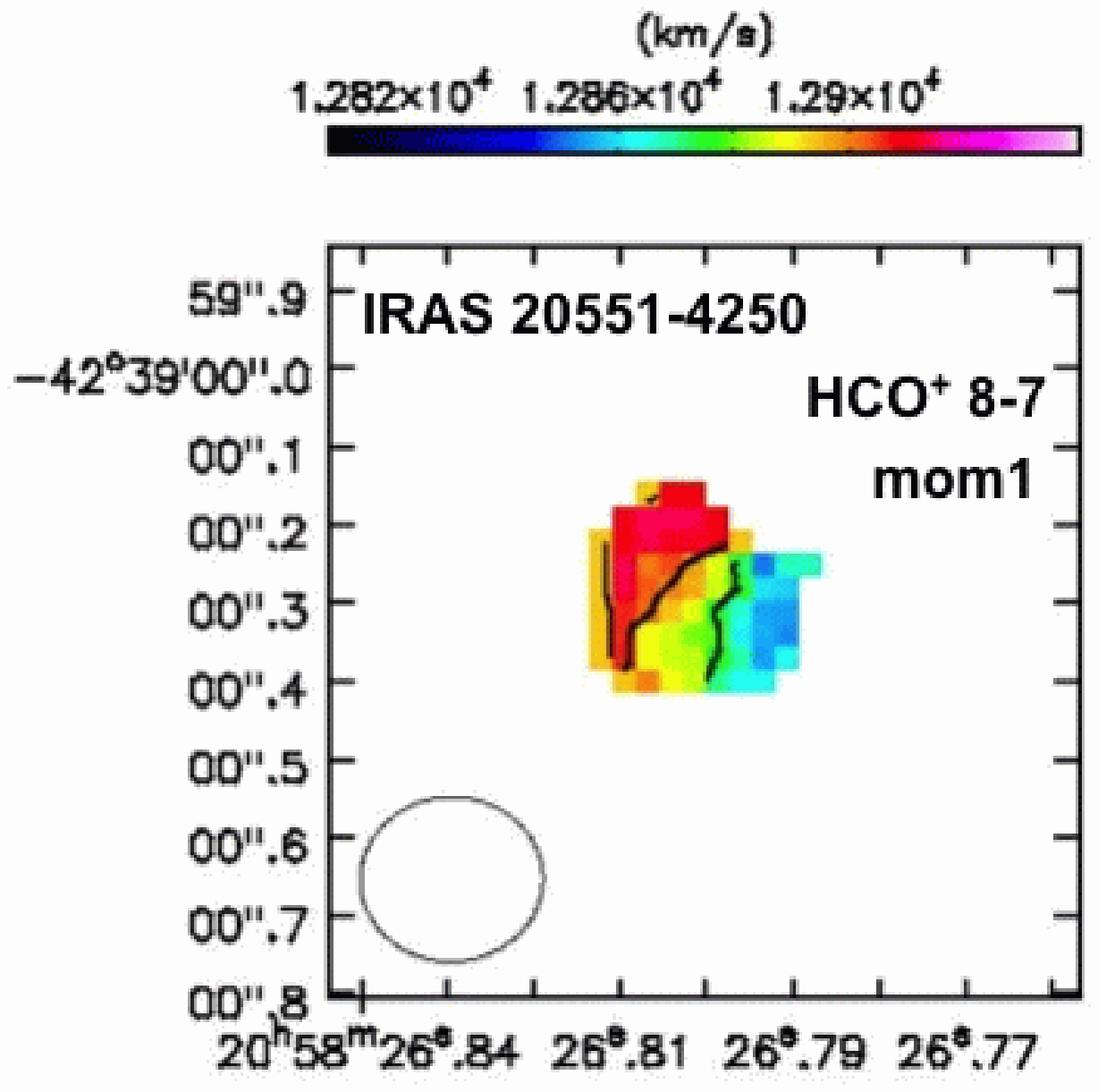}  
\includegraphics[angle=0,scale=.4]{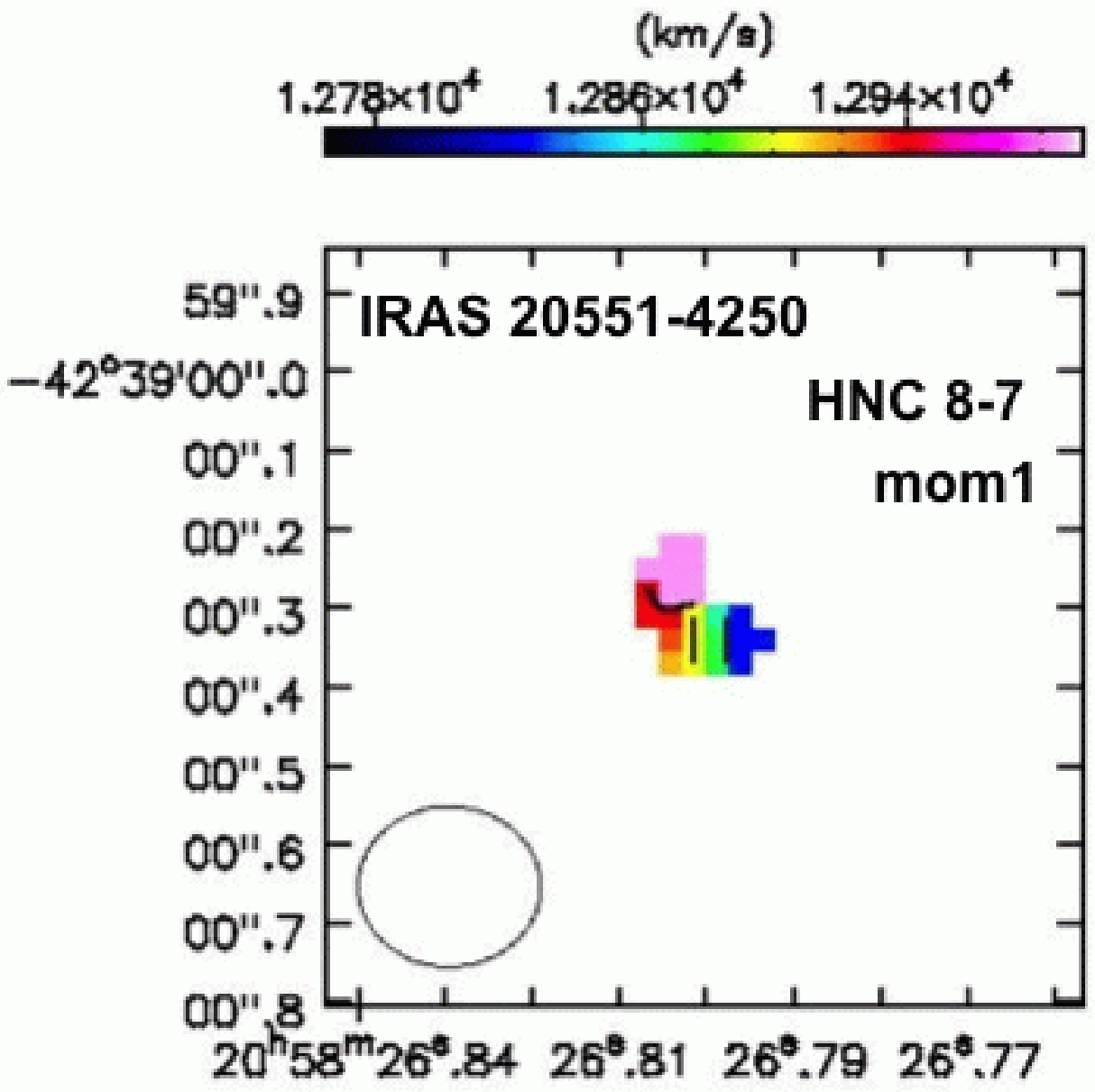} \\
\includegraphics[angle=0,scale=.38]{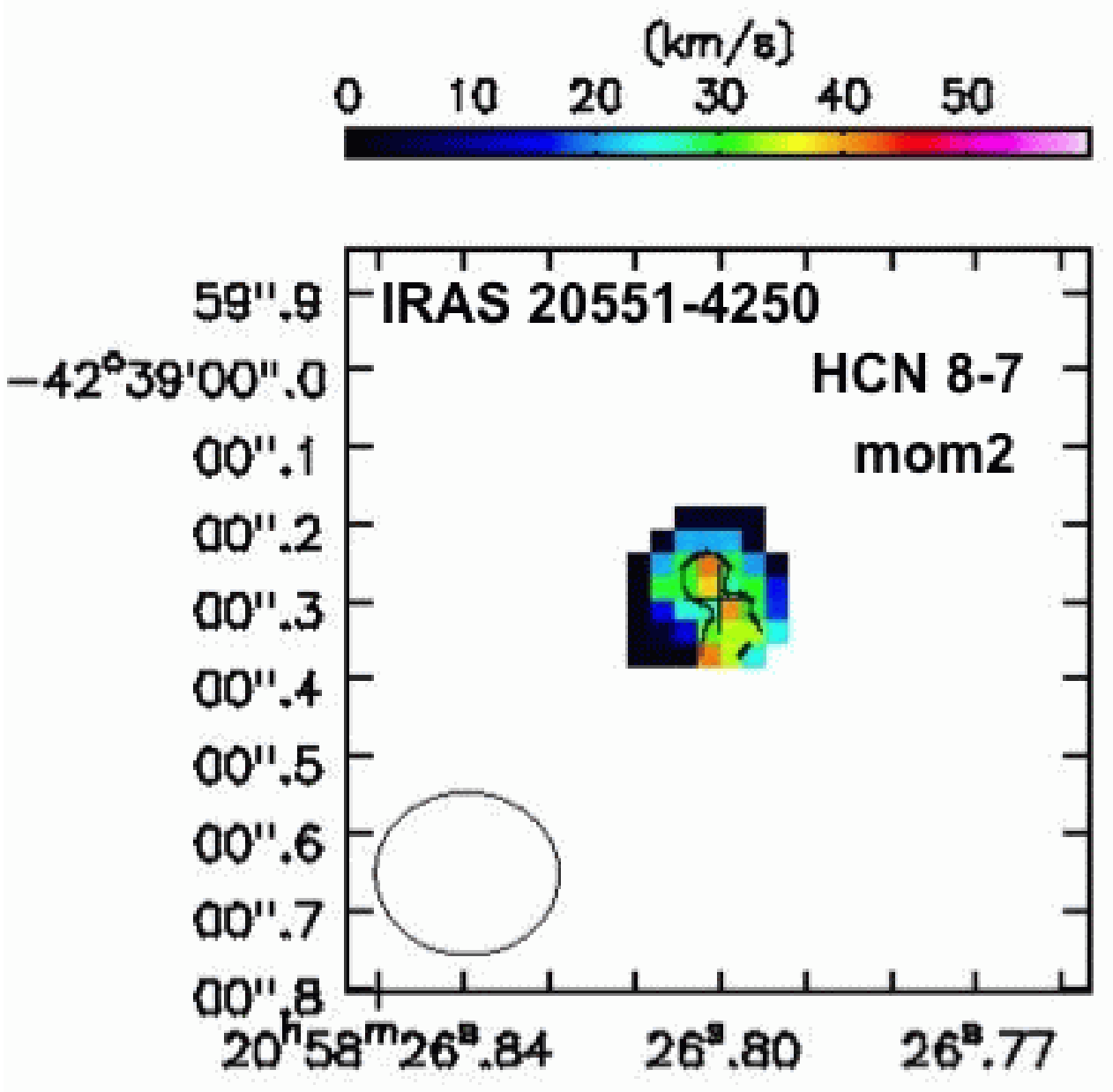}
\includegraphics[angle=0,scale=.4]{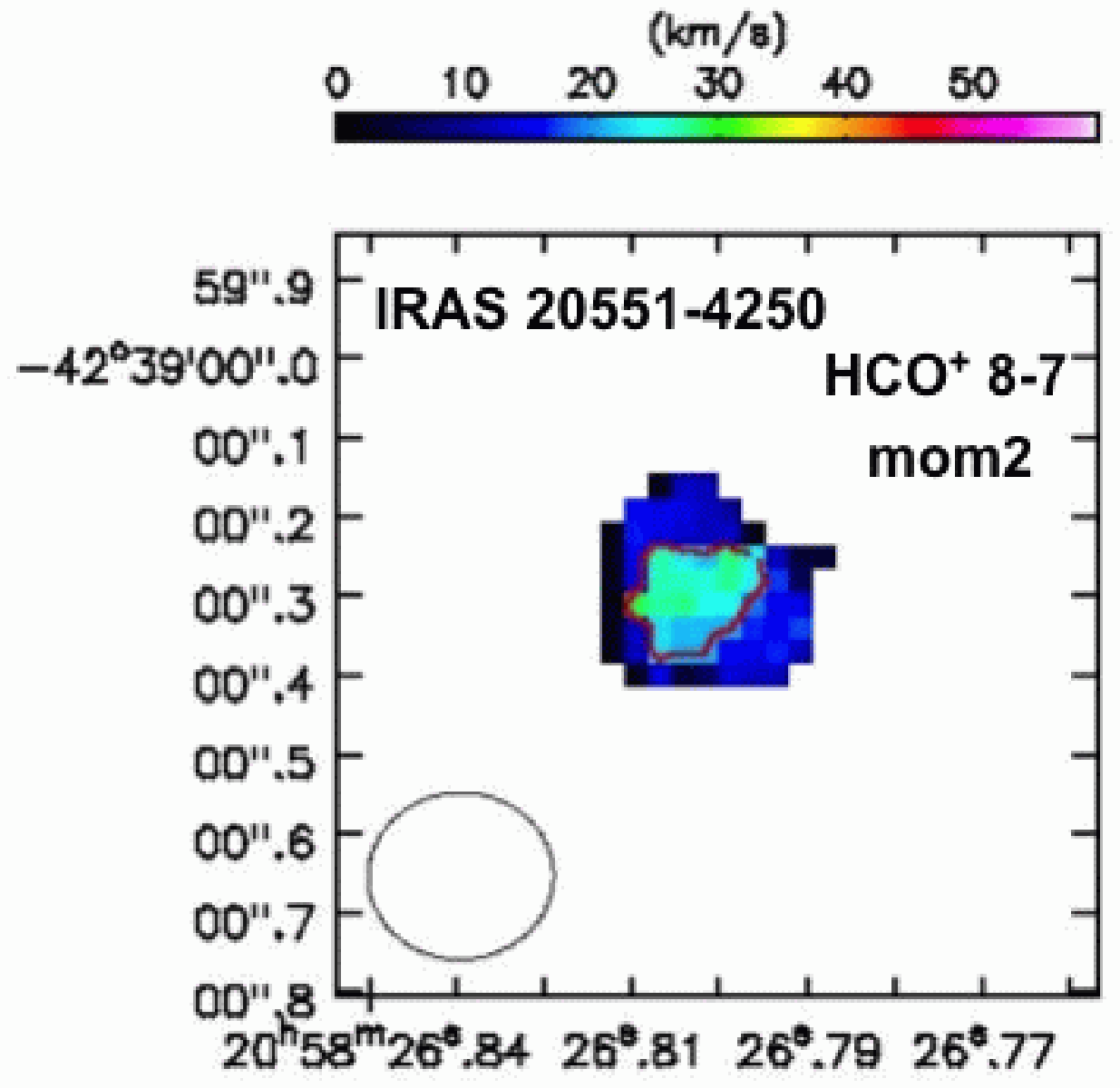}
\includegraphics[angle=0,scale=.4]{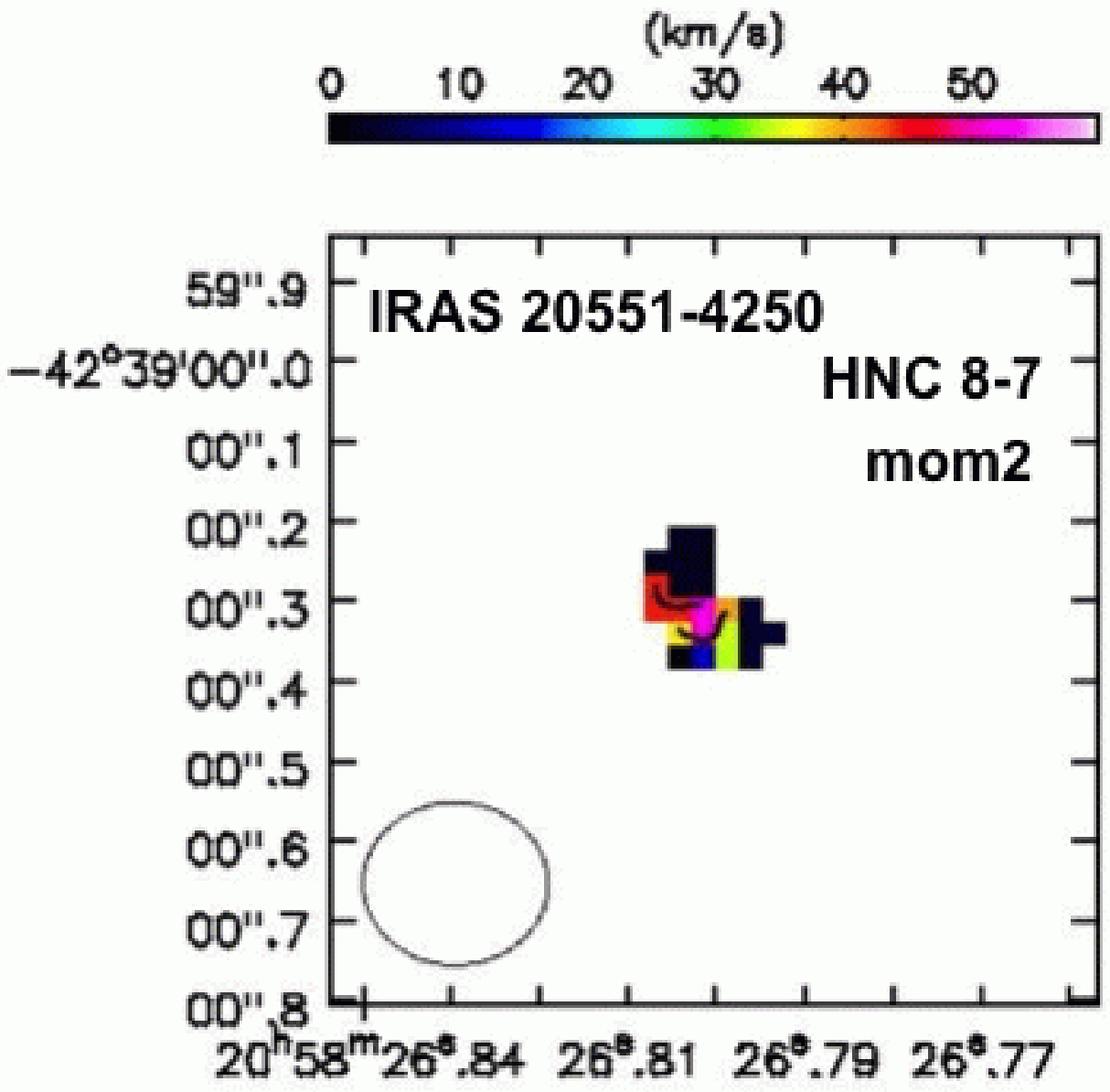} \\
\end{center}
\caption{
({\it {Top}}): Intensity-weighted mean velocity (moment 1) maps for 
HCN J=8--7, HCO$^{+}$ J=8--7, and HNC J=8--7 emission lines in band 9.
The abscissa and ordinate are right ascension (J2000) and declination 
(J2000), respectively. 
The contours represent 12900, 12950 km s$^{-1}$ for HCN J=8--7, 
12875, 12900 km s$^{-1}$ for HCO$^{+}$ J=8--7, 
and 12850, 12900 km s$^{-1}$ for HNC J=8--7.
({\it {Bottom}}): Intensity-weighted velocity dispersion (moment 2)
maps. 
The contours represent 30 km s$^{-1}$ for HCN J=8--7, 
20 km s$^{-1}$ for HCO$^{+}$ J=8--7, 
and 40 km s$^{-1}$ for HNC J=8--7.
Only smaller central areas are displayed because the achieved
beam size in band 9 was much smaller than those of bands 7 and 3.
Beam sizes are shown as open circles at the lower-left part.
An appropriate cut-off was chosen for these moment 1 and 2 maps.
In the left panels, the continuum peak positions are marked with crosses.
}
\end{figure}

\begin{figure}
\begin{center}
\includegraphics[angle=0,scale=.355]{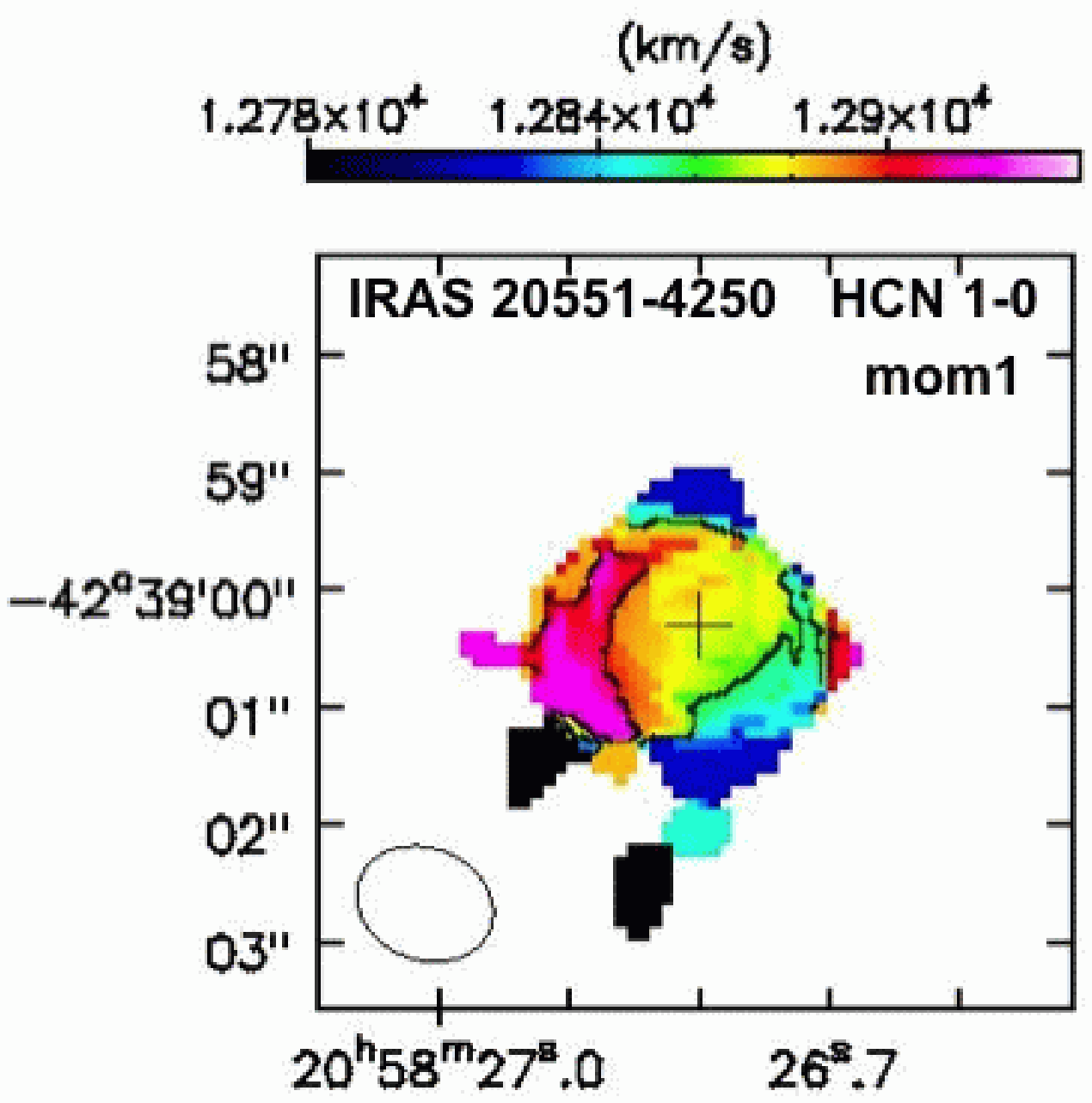}  
\includegraphics[angle=0,scale=.376]{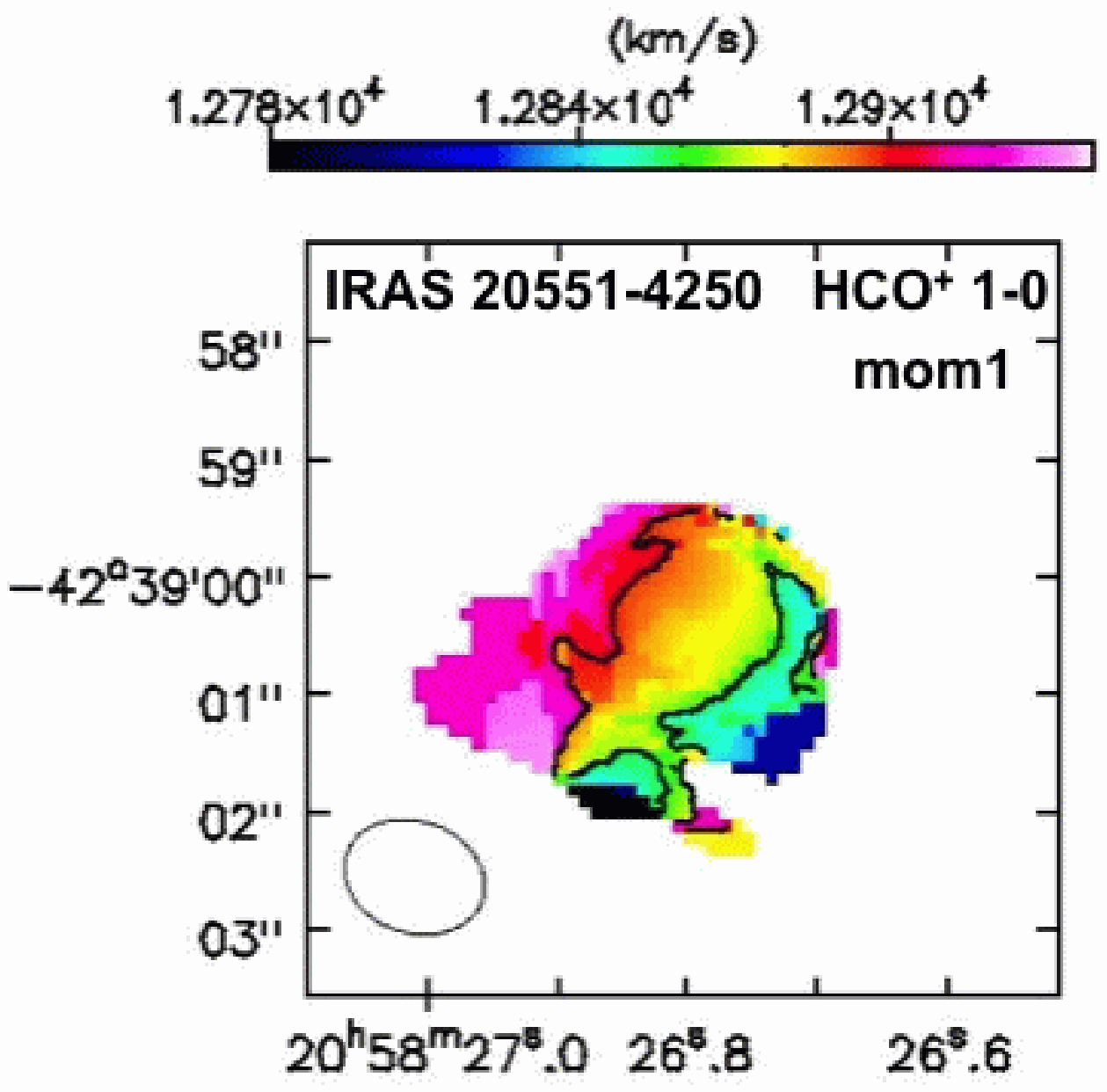}  
\includegraphics[angle=0,scale=.376]{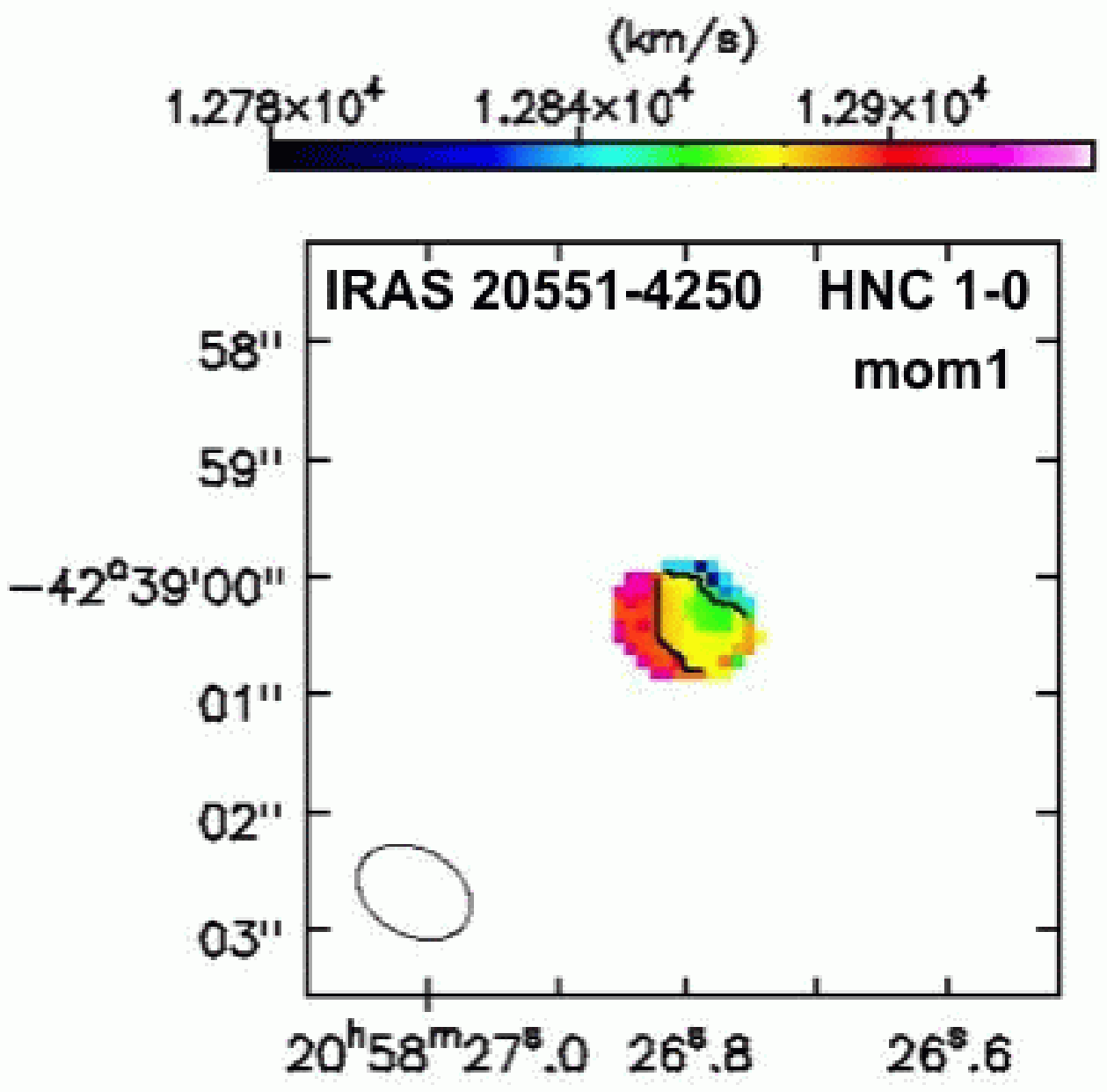} \\
\includegraphics[angle=0,scale=.355]{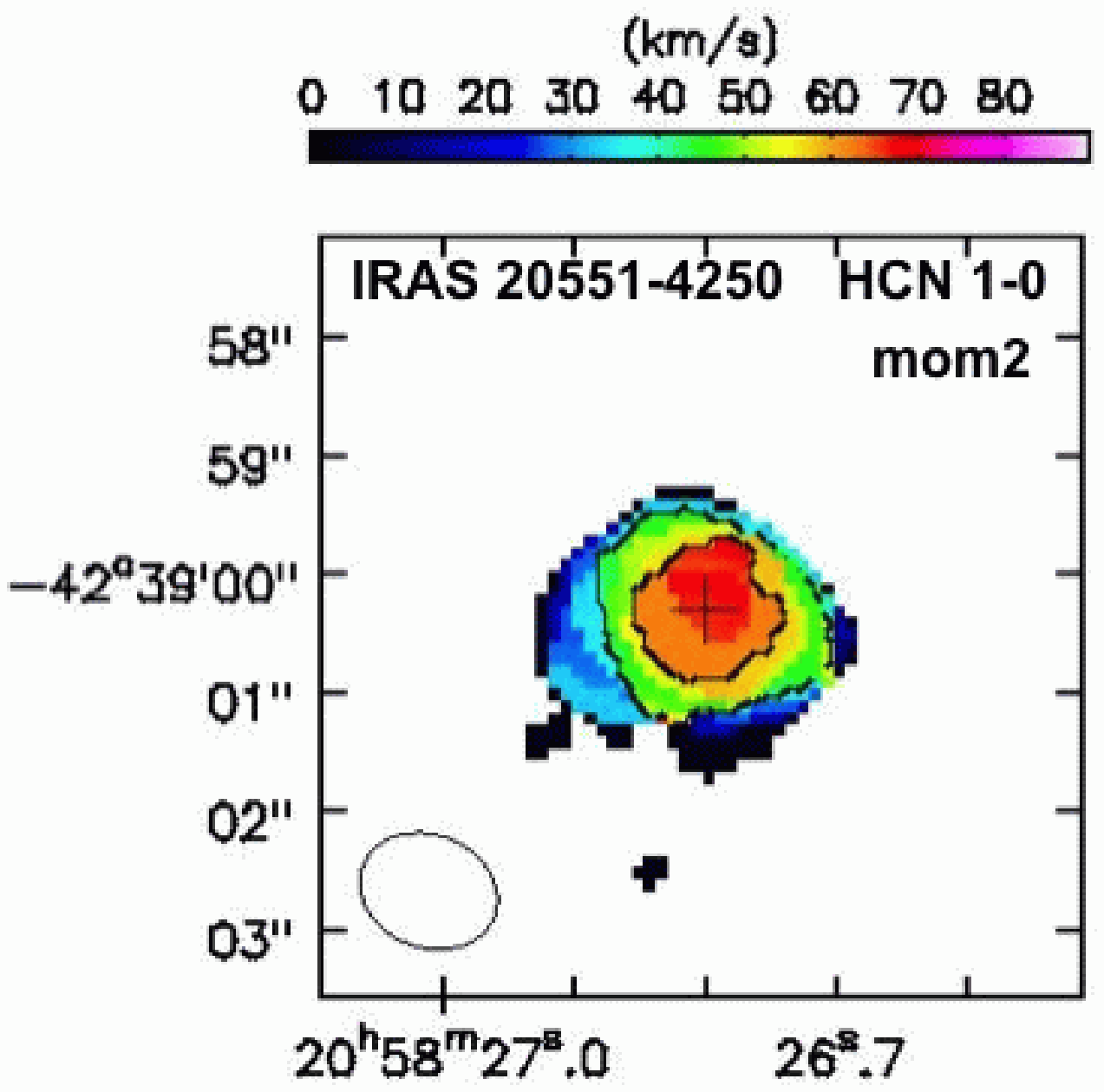}  
\includegraphics[angle=0,scale=.376]{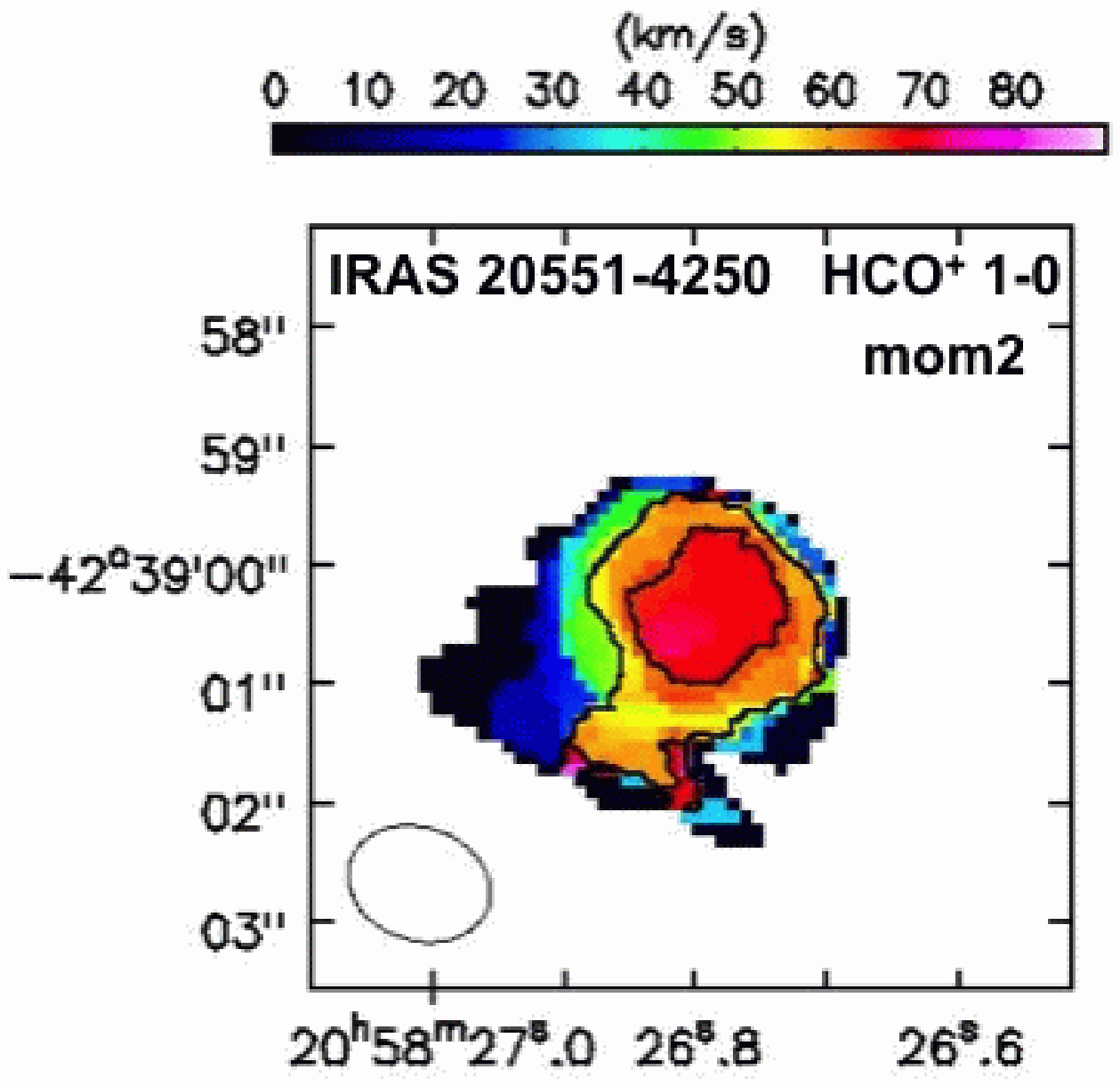}  
\includegraphics[angle=0,scale=.376]{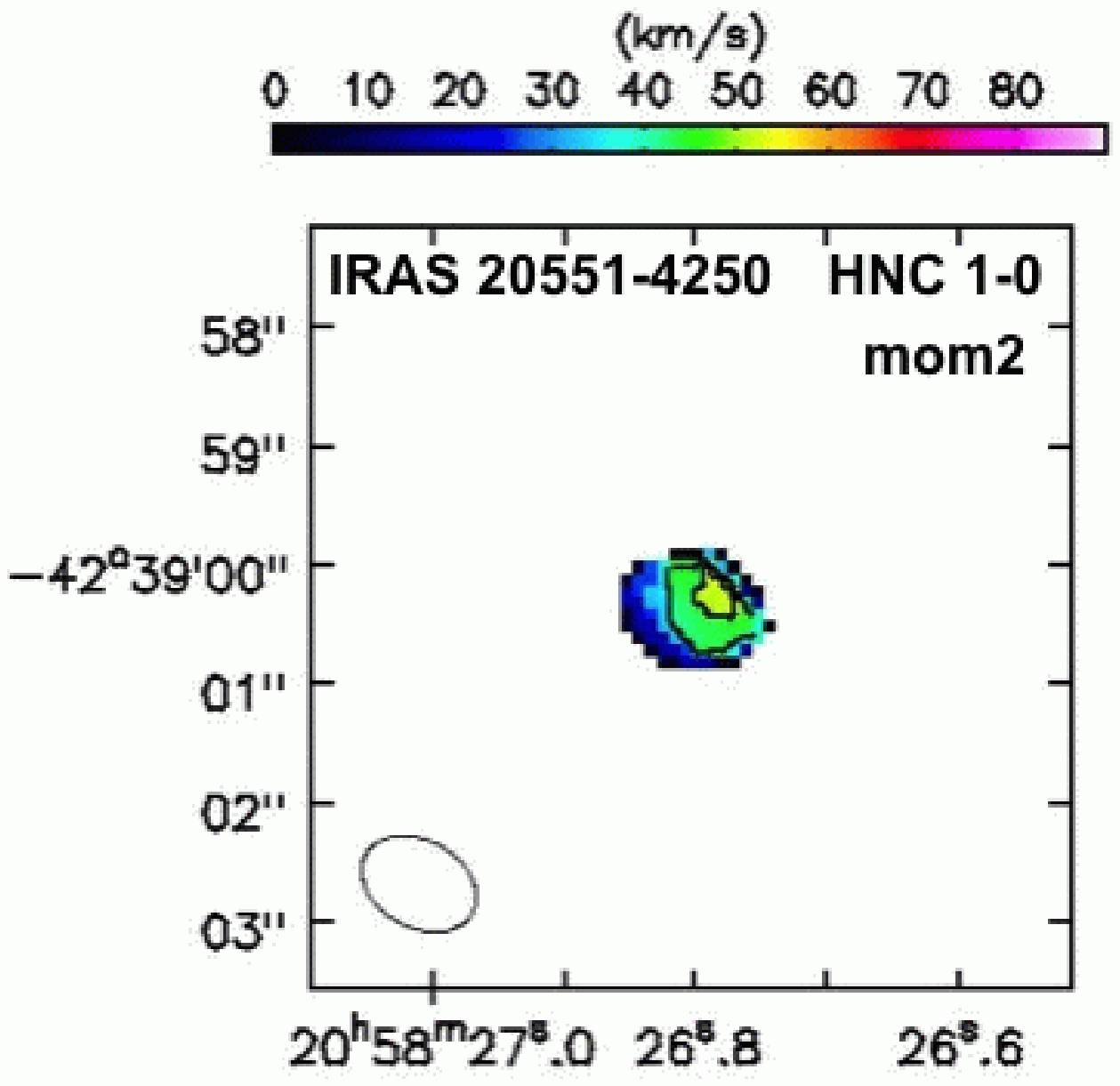} \\  
\end{center}
\caption{
({\it {Top}}) : Intensity-weighted mean velocity (moment 1) maps for HCN
J=1--0, HCO$^{+}$ J=1--0, and HNC J=1--0 emission lines in band 3. 
The abscissa and ordinate are right ascension (J2000) and declination 
(J2000), respectively. 
The contours represent 12860, 12900 km s$^{-1}$ for HCN J=1--0, 
12860, 12900 km s$^{-1}$ for HCO$^{+}$ J=1--0, and 
12860, 12890 km s$^{-1}$ for HNC J=1--0.
({\it {Bottom}}): Intensity-weighted velocity dispersion (moment 2) maps. 
The contours represent 40, 60 km s$^{-1}$ for HCN J=1--0, 
53, 66 km s$^{-1}$ for HCO$^{+}$ J=1--0, 
40, 50 km s$^{-1}$ for HNC J=1--0.
Beam sizes are shown as open circles in the lower-left region.
An appropriate cut-off was chosen for these moment 1 and 2 maps.
In the left panels, the continuum peak positions are marked with crosses.
}
\end{figure}

\begin{figure}
\begin{center}
\includegraphics[angle=0,scale=.38]{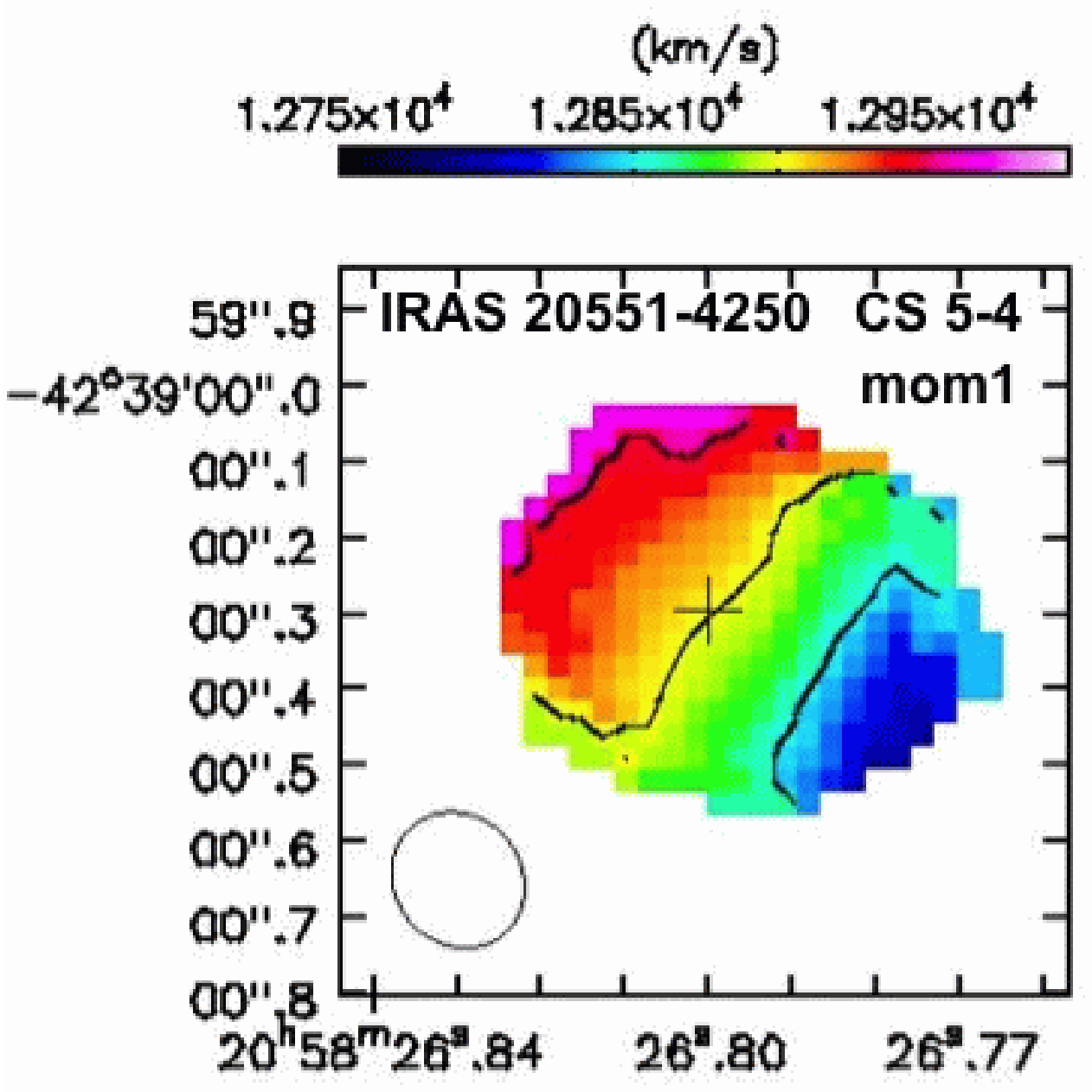}  
\includegraphics[angle=0,scale=.38]{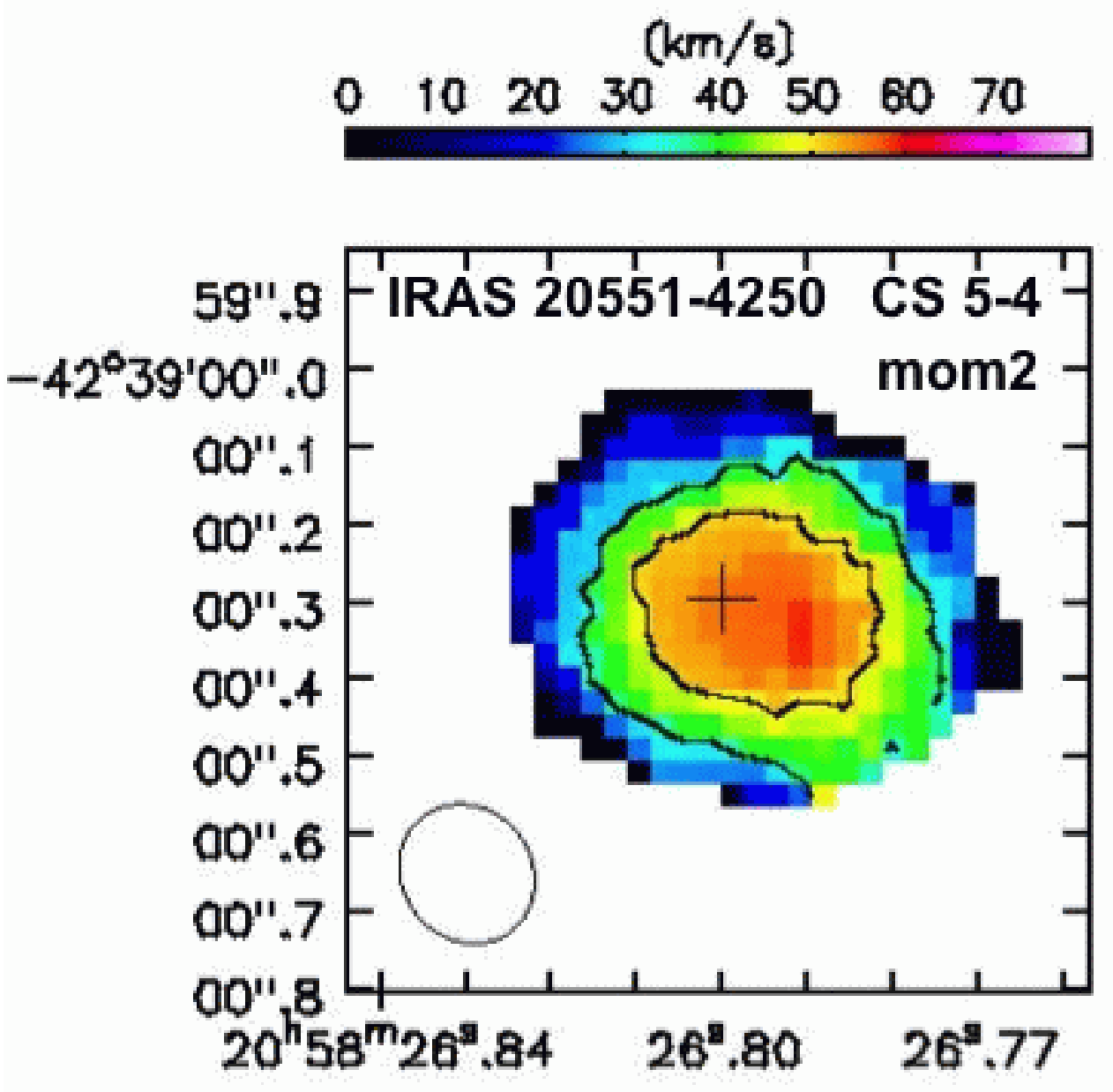}  
\end{center}
\vspace*{-0.3cm}
\caption{
({\it {Left}}): Intensity-weighted mean velocity (moment 1) map for 
CS J=5--4 in band 6.
The abscissa and ordinate are right ascension (J2000) and declination 
(J2000), respectively. 
The contours represent 12850, 12900, 12950 km s$^{-1}$.
({\it {Right}}): Intensity-weighted velocity dispersion (moment 2)
map. 
The contours represent 35, 50 km s$^{-1}$.
The image size is the same as that of band 9 data in Figure 6.
Beam sizes are shown as open circles in the lower-left region.
An appropriate cut-off was chosen for these moment 1 and 2 maps.
The continuum peak positions are marked with crosses.
}
\end{figure}

\begin{figure}
\begin{center}
\includegraphics[angle=-0,scale=.41]{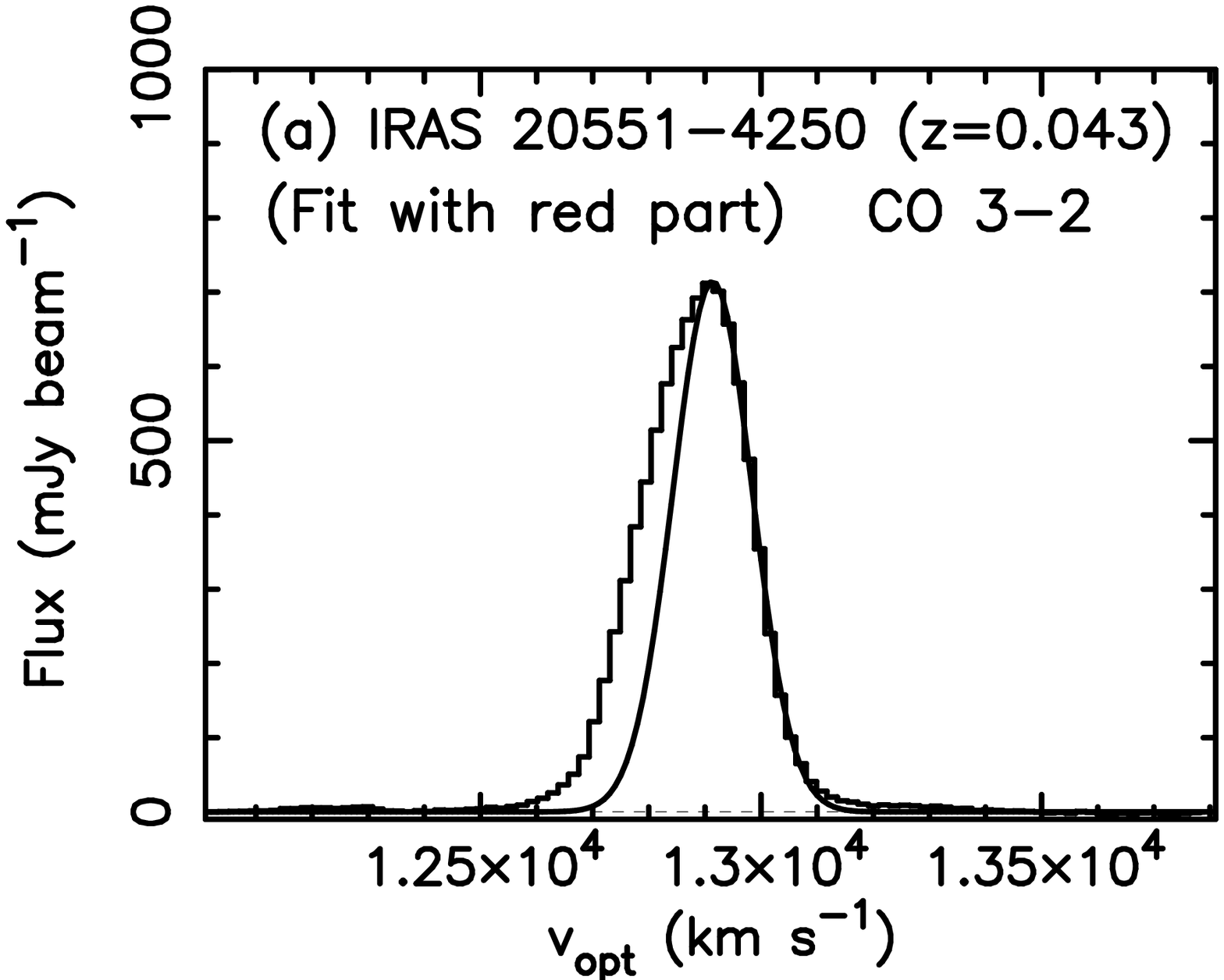} 
\includegraphics[angle=-0,scale=.41]{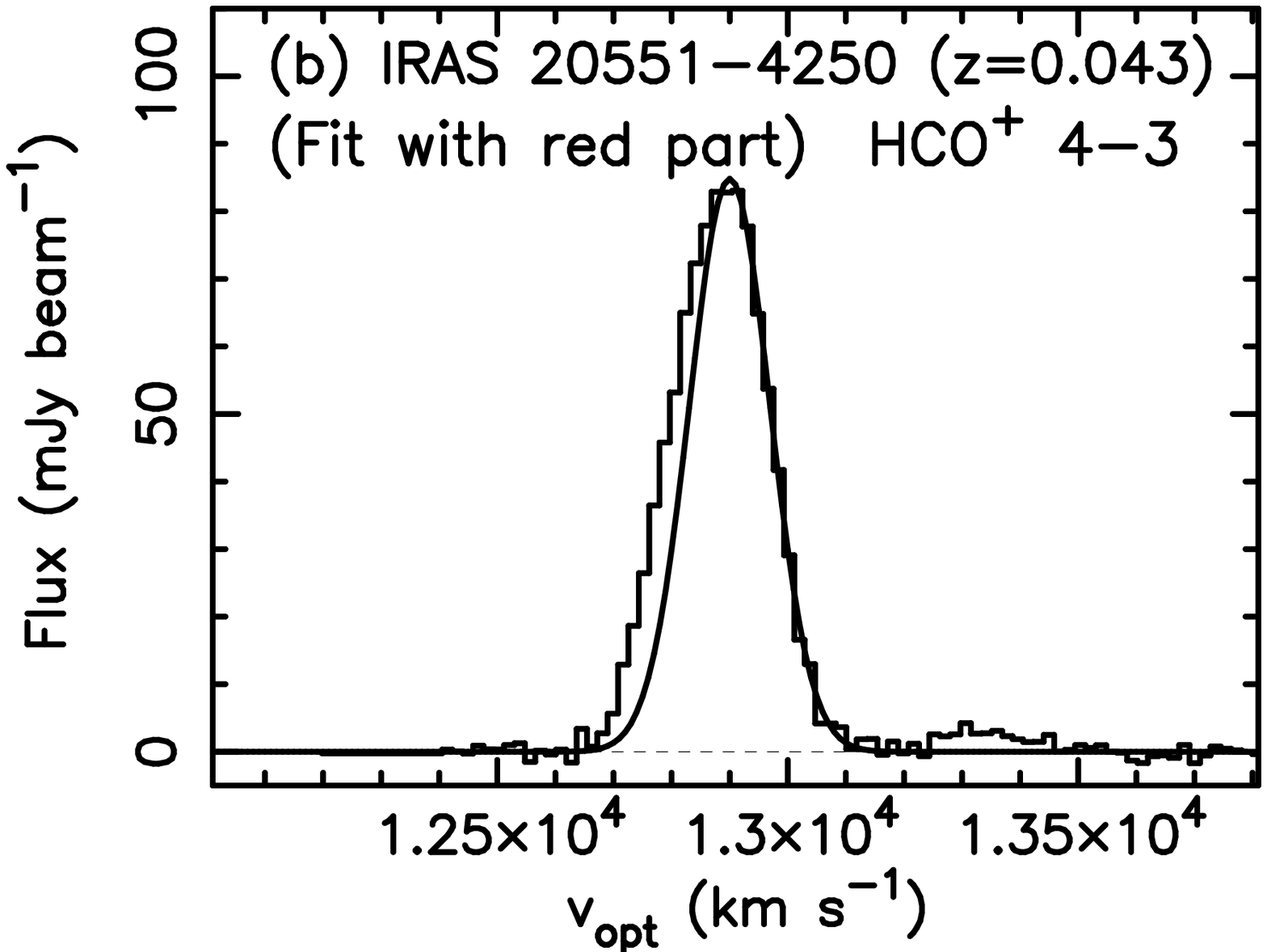} \\
\includegraphics[angle=-0,scale=.41]{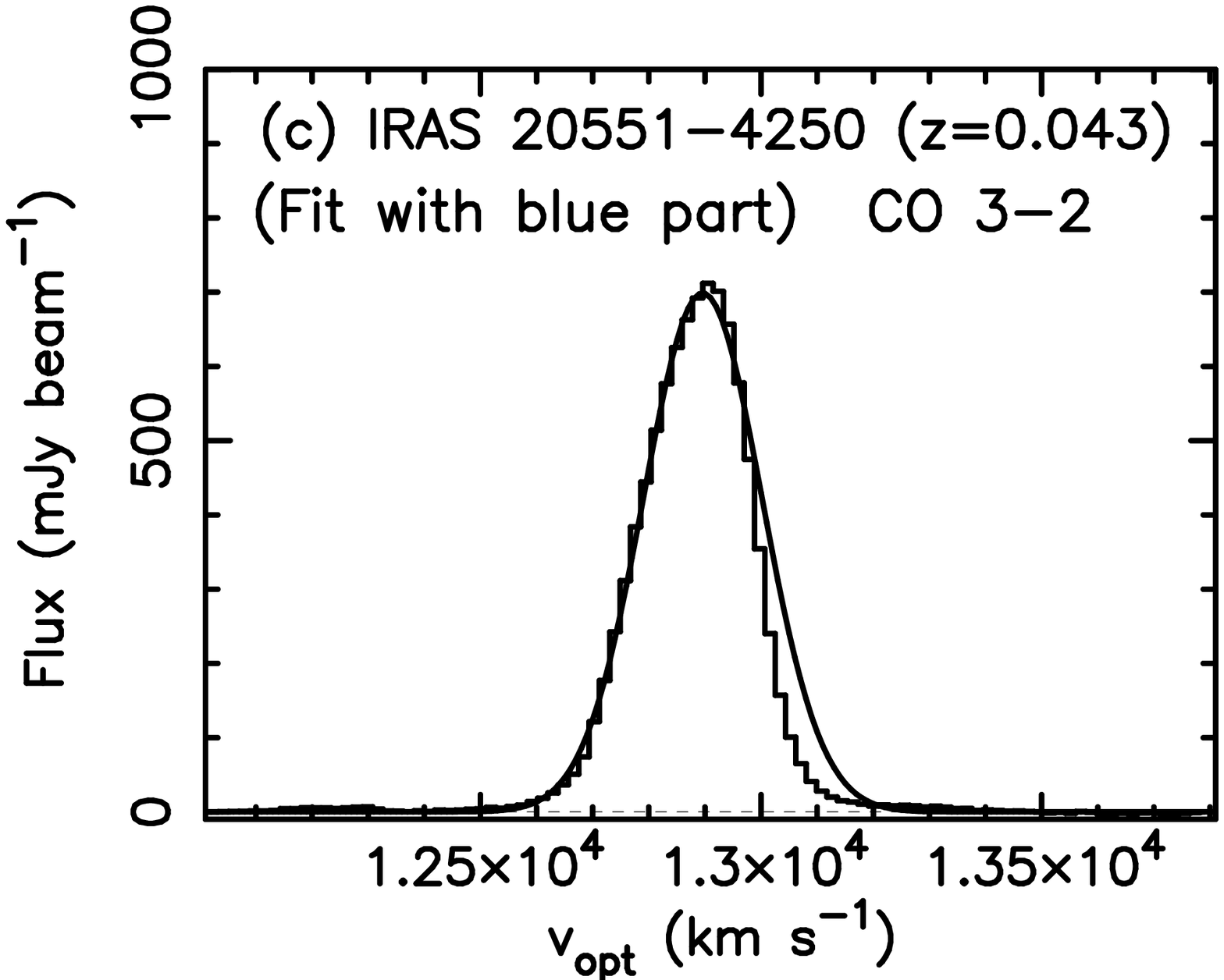} 
\includegraphics[angle=-0,scale=.41]{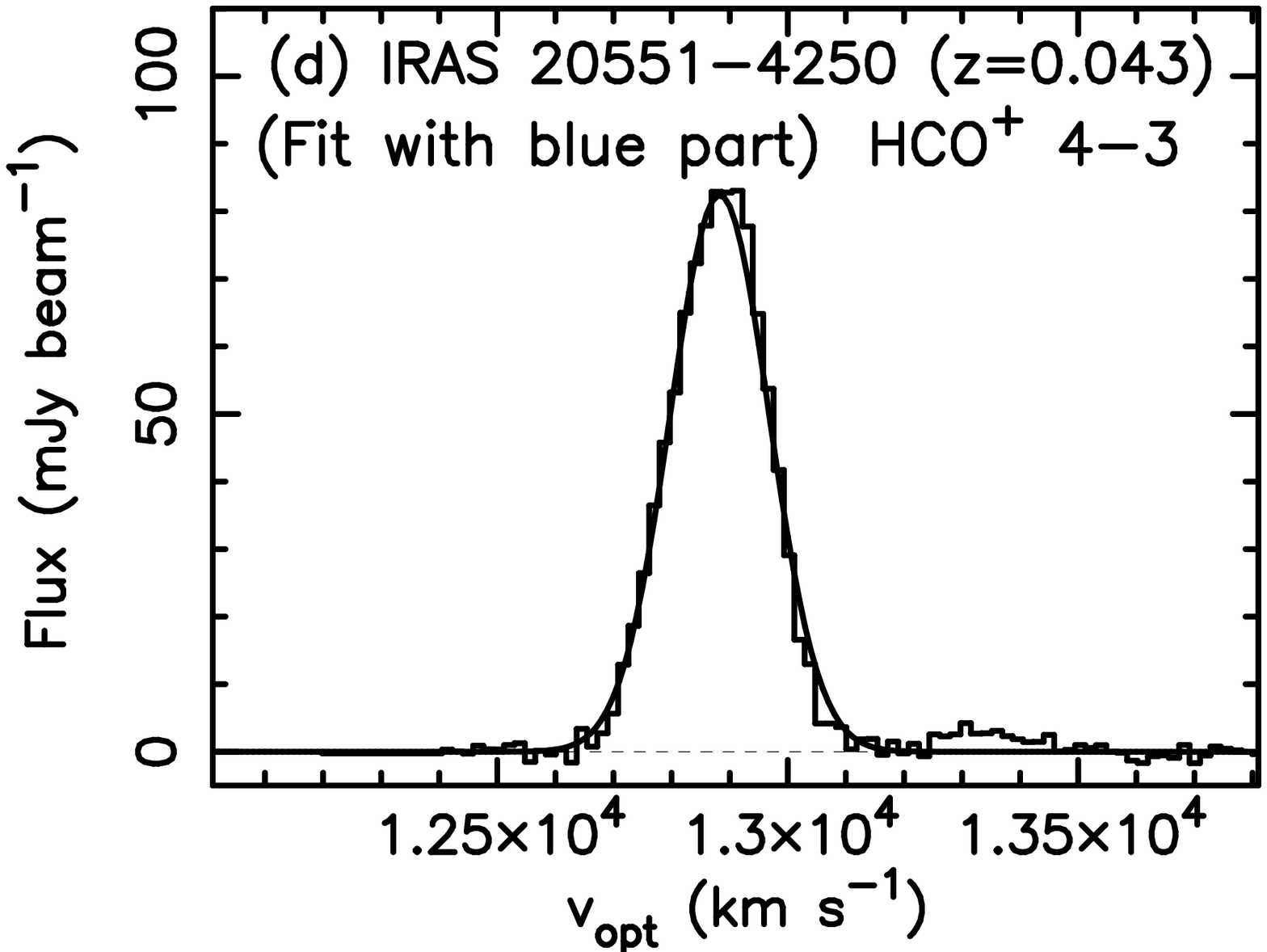} 
\vspace{-0.4cm}
\end{center}
\caption{
(a): Gaussian fit only for the redder part of the emission
peak for CO J=3--2 (v$_{\rm opt}$ $>$ 12905 km s$^{-1}$).
(b): The same fit as (a) for HCO$^{+}$ J=4--3 (v$_{\rm opt}$
$>$ 12905 km s$^{-1}$). 
(c): Gaussian fit only for the bluer part of the emission
peak for CO J=3--2 (v$_{\rm opt}$ $<$ 12905 km s$^{-1}$).
(d): The same fit as (c) for HCO$^{+}$ J=4--3 (v$_{\rm opt}$
$<$ 12905 km s$^{-1}$).
}
\end{figure}

\begin{figure}
\begin{center}
\includegraphics[angle=0,scale=.6]{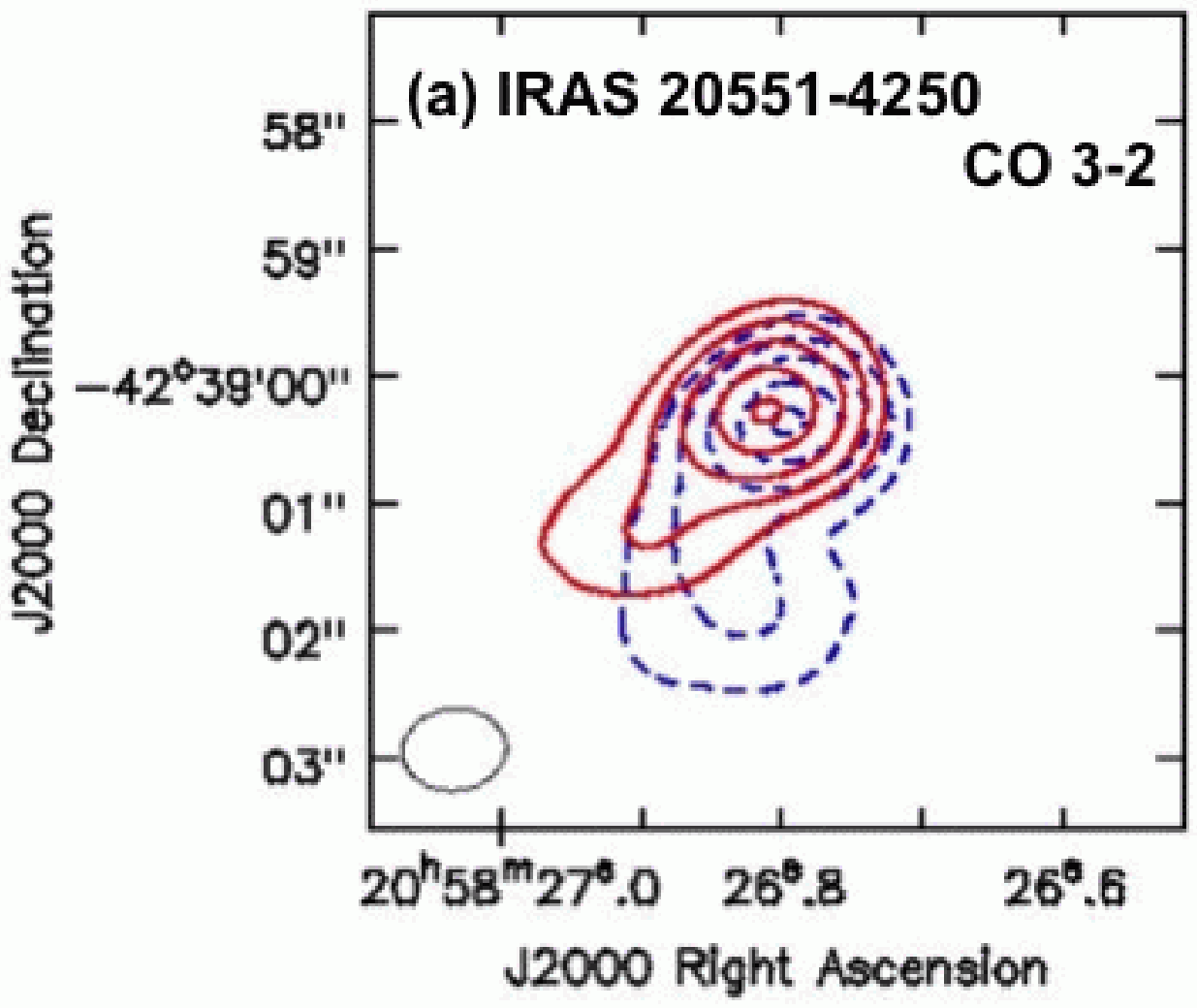} 
\includegraphics[angle=0,scale=.6]{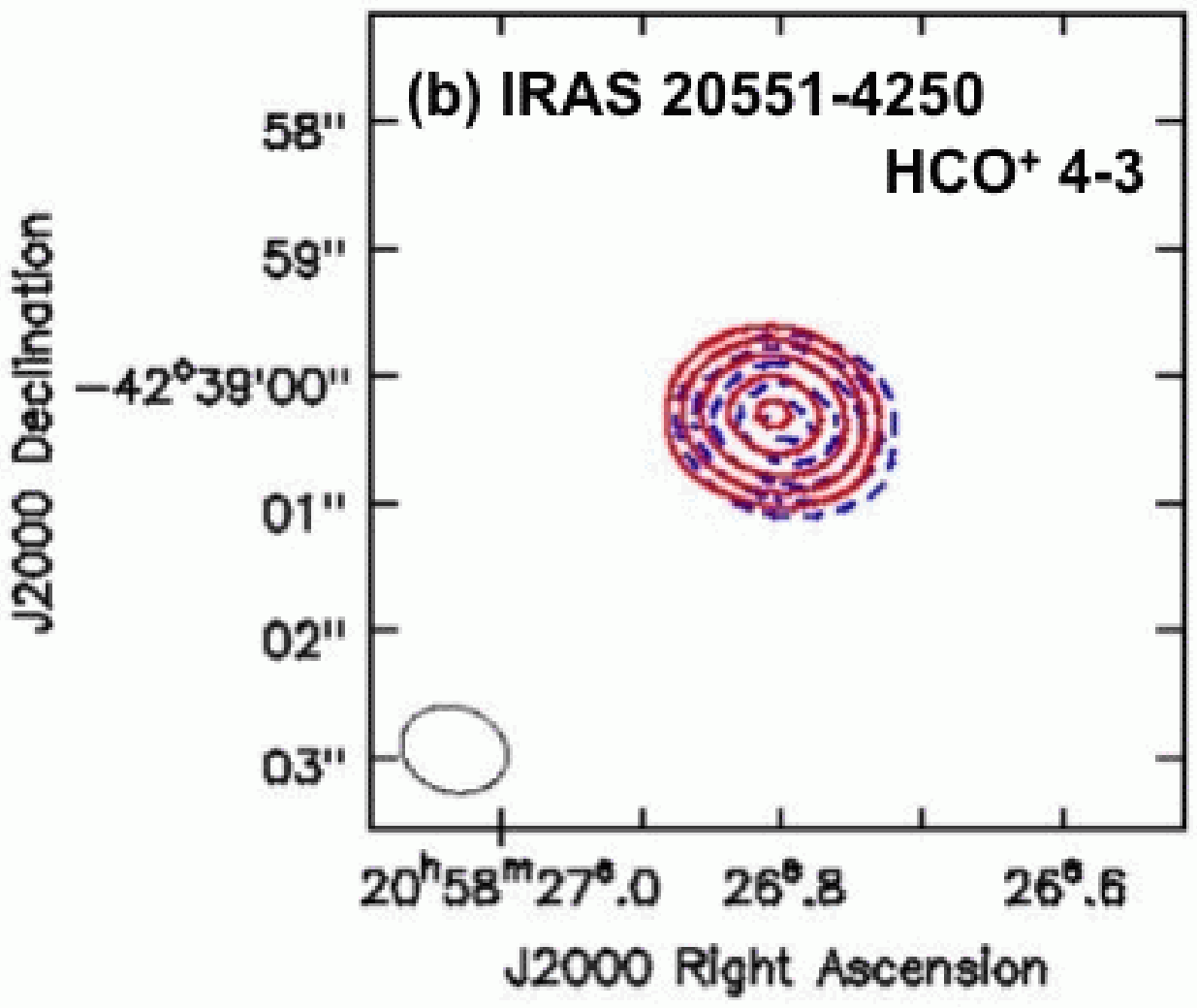} \\
\end{center}
\vspace*{-0.4cm}
\caption{
(a): Contours of the blue (v$_{\rm opt}$ $<$ 12905 km s$^{-1}$) and 
red (v$_{\rm opt}$ $>$ 12905 km s$^{-1}$) components of the CO J=3--2
emission line. 
The blue and red components are shown as blue dashed and red solid
contours, respectively. 
The contours represent 5$\sigma$, 10$\sigma$, 20$\sigma$, 40$\sigma$,
and 60$\sigma$ for both blue and red components.
The 1$\sigma$ level is 1.3 (Jy beam$^{-1}$ km s$^{-1}$) for the blue
component and 1.0 (Jy beam$^{-1}$ km s$^{-1}$) for the red component.
(b): Contours of the blue (v$_{\rm opt}$ $<$ 12905 km s$^{-1}$; blue
dashed lines) and red (v$_{\rm opt}$ $>$ 12905 km s$^{-1}$; red solid
lines) components of the HCO$^{+}$ J=4--3 emission line. The contours
represent 5$\sigma$, 10$\sigma$, 20$\sigma$, 40$\sigma$, and 60$\sigma$
(1$\sigma$ is 0.12 and 0.10 [Jy beam$^{-1}$ km s$^{-1}$] for the blue and
red components, respectively).
}
\end{figure}

\begin{figure}
\begin{center}
\vspace*{-1.5cm}
\includegraphics[angle=0,scale=.52]{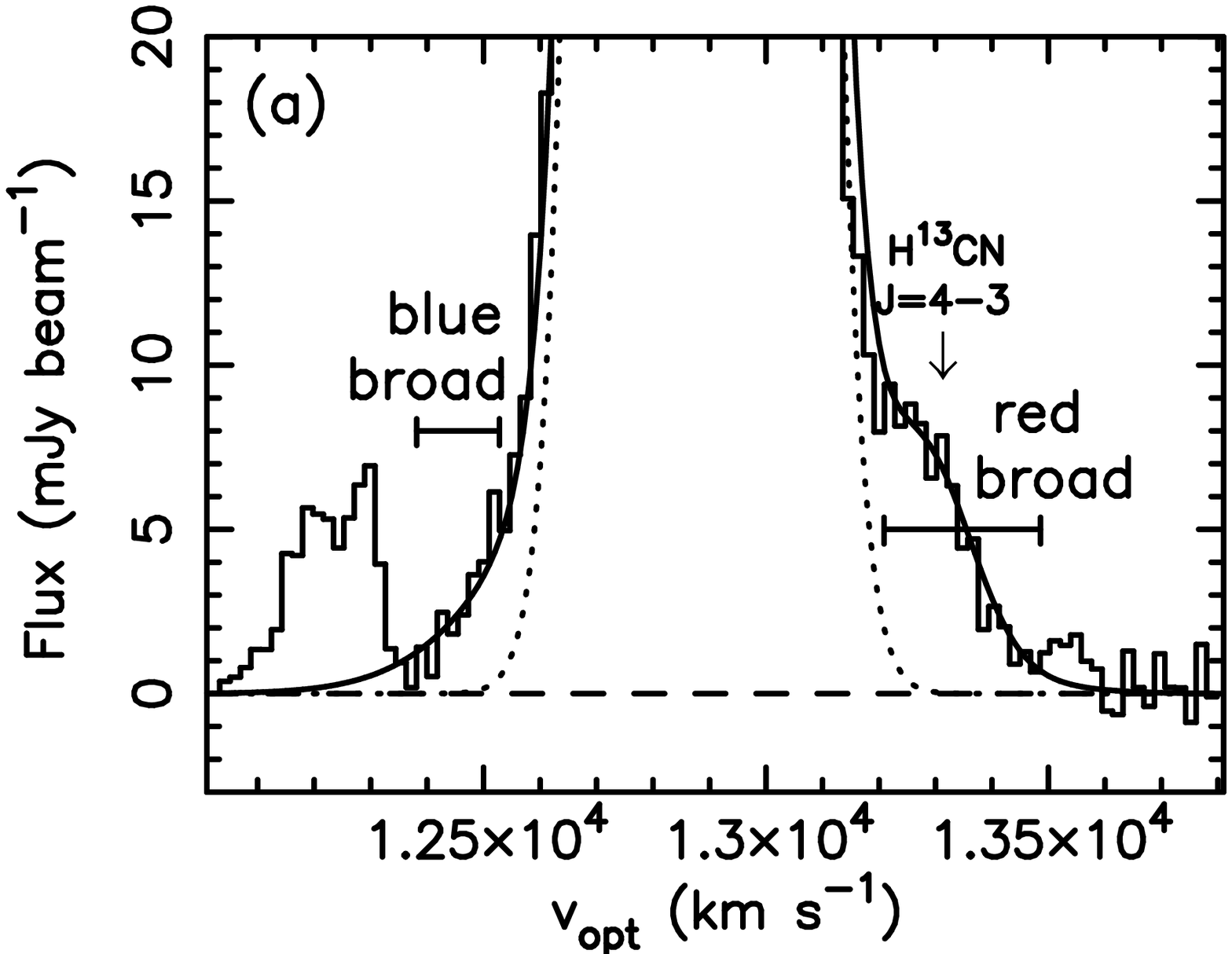} 
\includegraphics[angle=0,scale=.43]{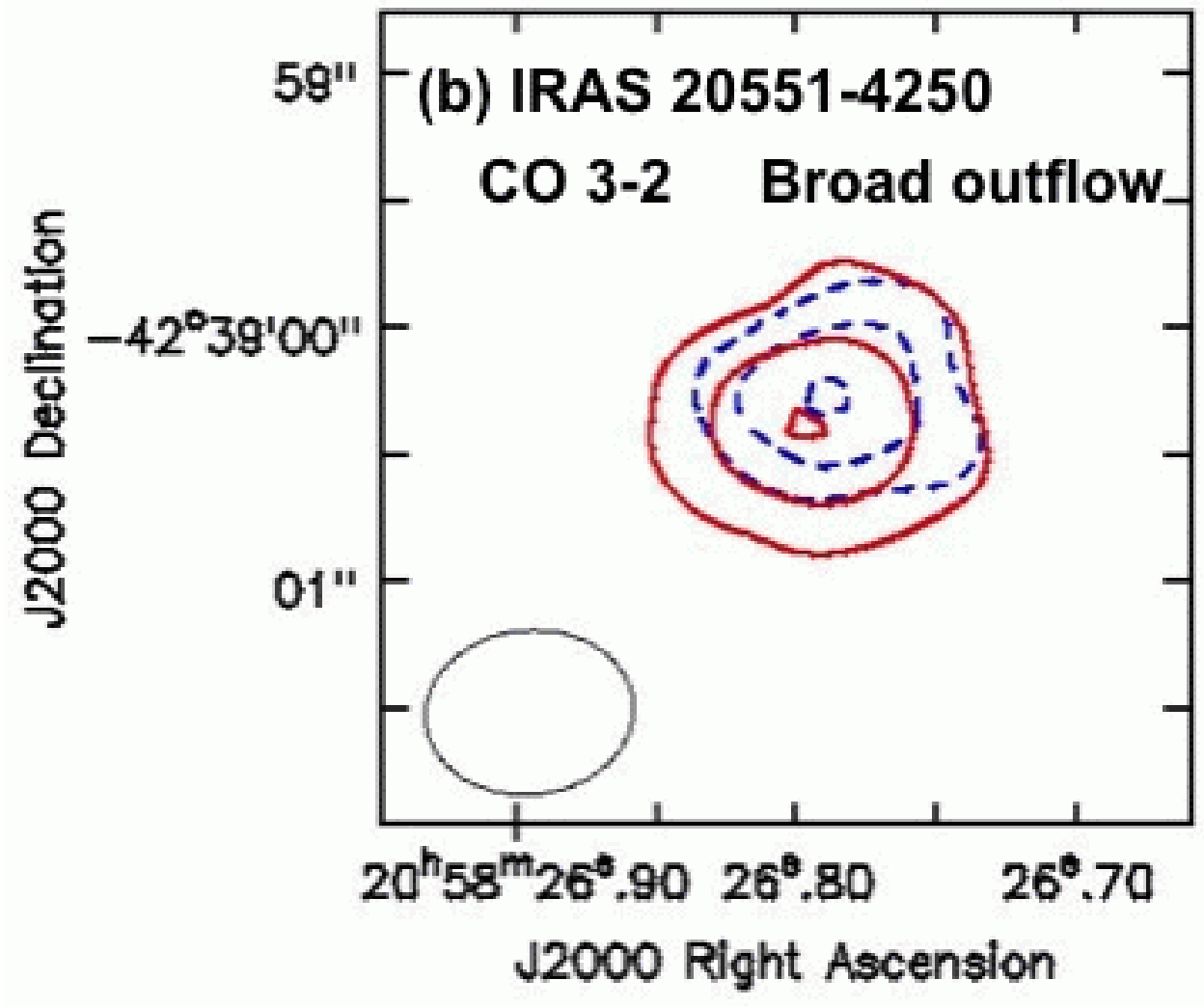} \\
\end{center}
\vspace*{-0.4cm}
\caption{
(a): Magnified spectrum at the bottom part of the CO J=3--2 emission line
and its best single Gaussian fit (using all CO J=3--2 emission line)
(the dotted curved line), to show the presence of a broad emission
line component. 
The expected frequency of the H$^{13}$CN J=4--3 line is indicated as a downward
arrow. 
The solid horizontal straight lines, inserted by two short vertical
lines, indicate the velocity range to create moment 0 maps of the 
blue and red {\it broad} emission components. 
The velocity ranges are v$_{\rm opt}$ = 12382--12528 km
s$^{-1}$ and 13209--13486 km s$^{-1}$ for the blue and
red {\it broad} components, respectively.
For the thick solid curved line,  
a Gaussian with FWHM = 540 km s$^{-1}$ and velocity peak at 12886
km s$^{-1}$ to represent the putative outflow-origin broad emission
component, and another Gaussian with FWHM = 190 km s$^{-1}$ to 
take into account the H$^{13}$CN J=4--3 emission line, are added to the
single Gaussian component shown as the dotted curved line.
(b): Contour maps of the blue (blue dashed line)
and red (red solid line) {\it broad} emission components, defined in
Figure 11(a). 
The contours represent 3$\sigma$, 5$\sigma$, 7$\sigma$ (1$\sigma$ is
0.055 Jy beam$^{-1}$ km s$^{-1}$) for the blue broad component,
and 4$\sigma$, 8$\sigma$, and 12$\sigma$ (1$\sigma$ is 0.10 Jy
beam$^{-1}$ km s$^{-1}$) for the red broad component.
For the red broad component, contamination from H$^{13}$CN J=4--3
emission line is likely to be present.
}
\end{figure}

\begin{figure}
\begin{center}
\includegraphics[angle=0,scale=.41]{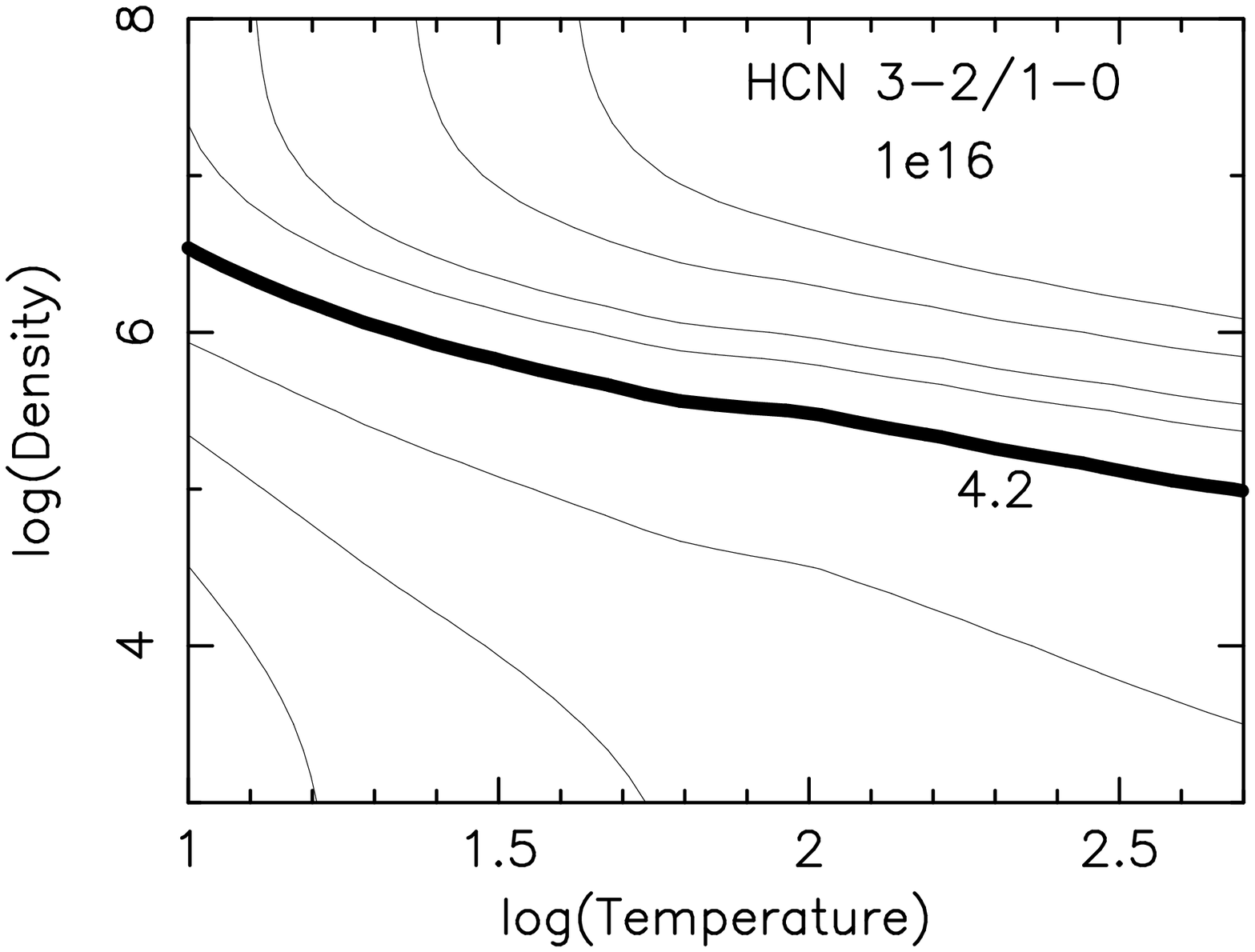}  
\includegraphics[angle=0,scale=.41]{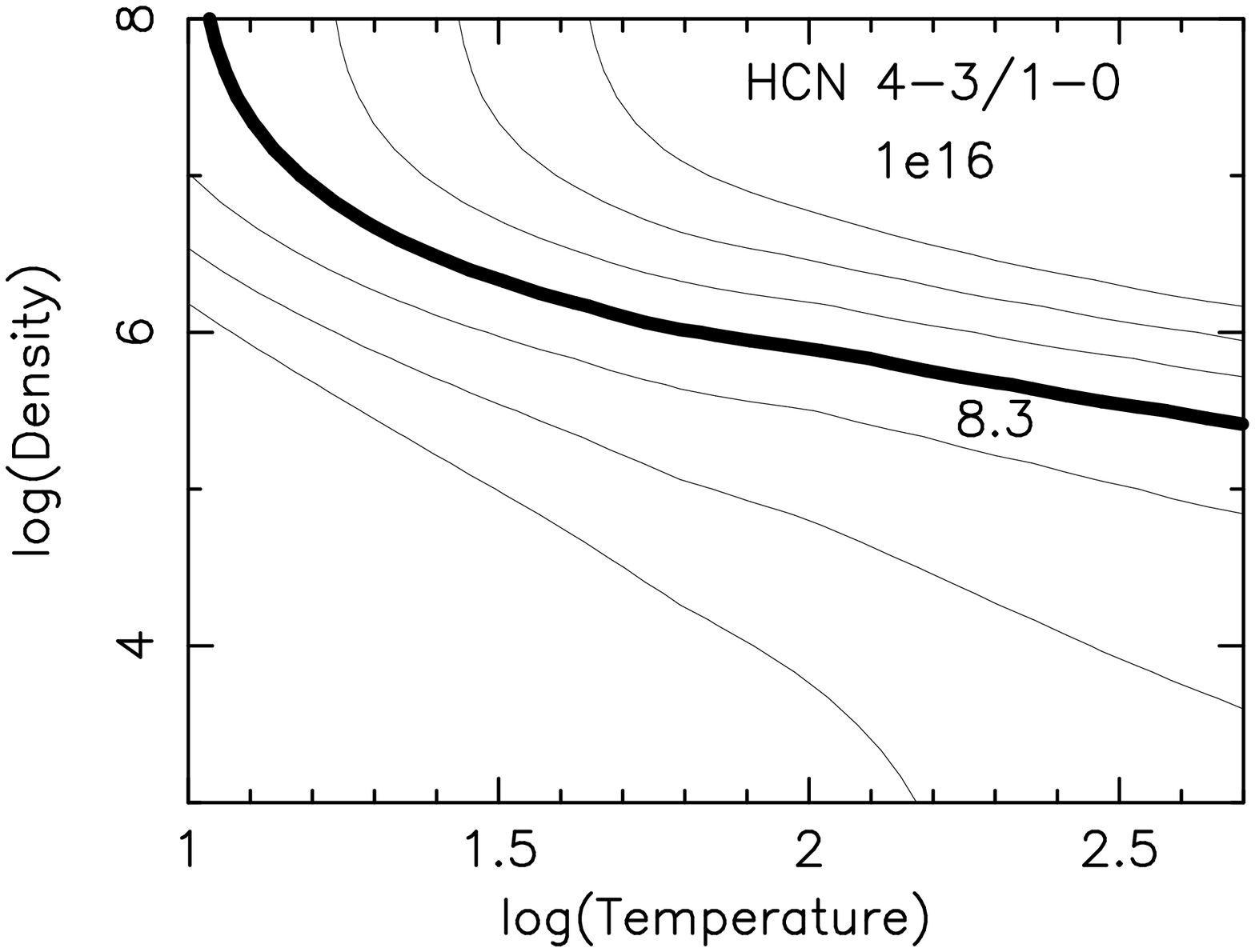} \\ 
\includegraphics[angle=0,scale=.41]{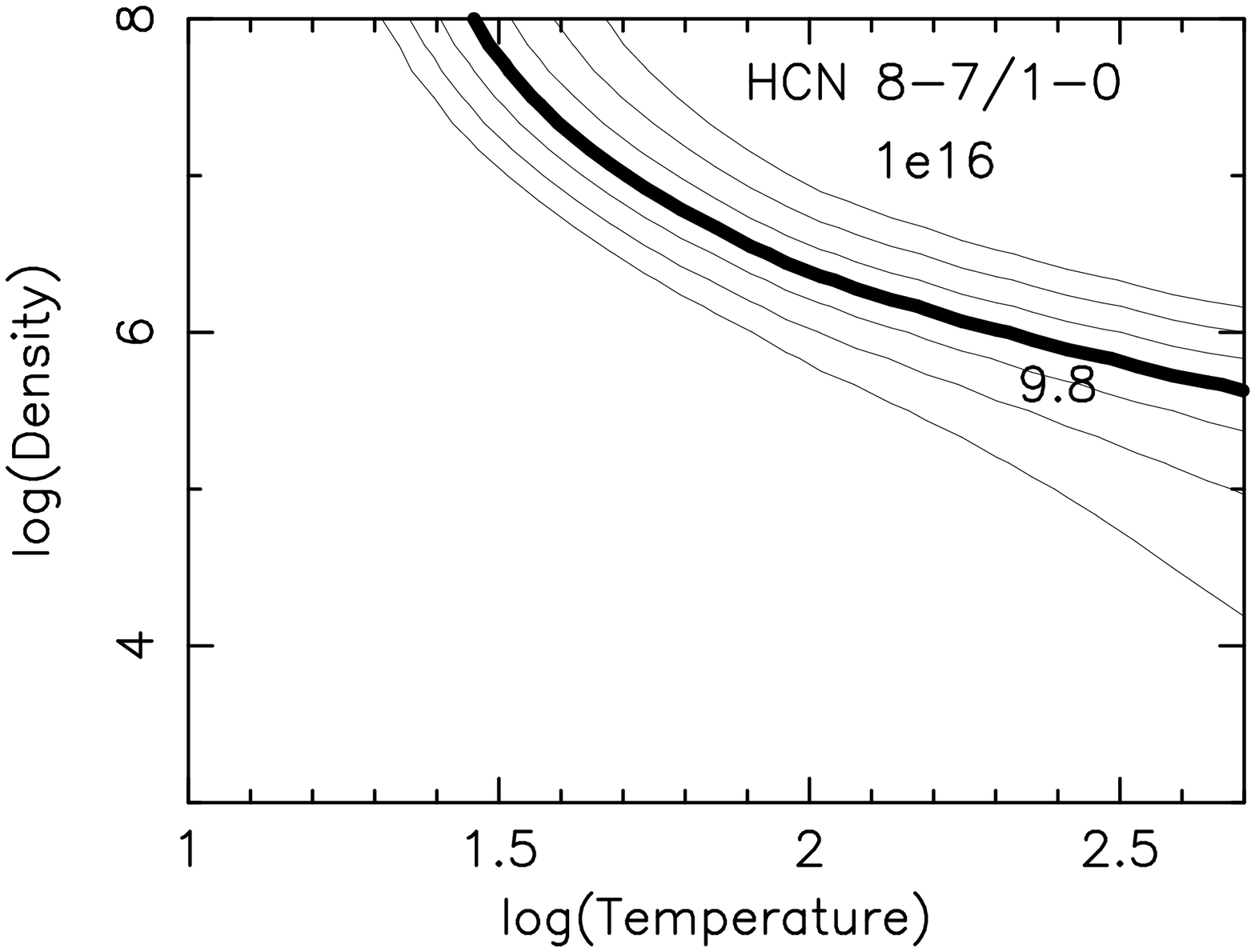}  
\includegraphics[angle=0,scale=.41]{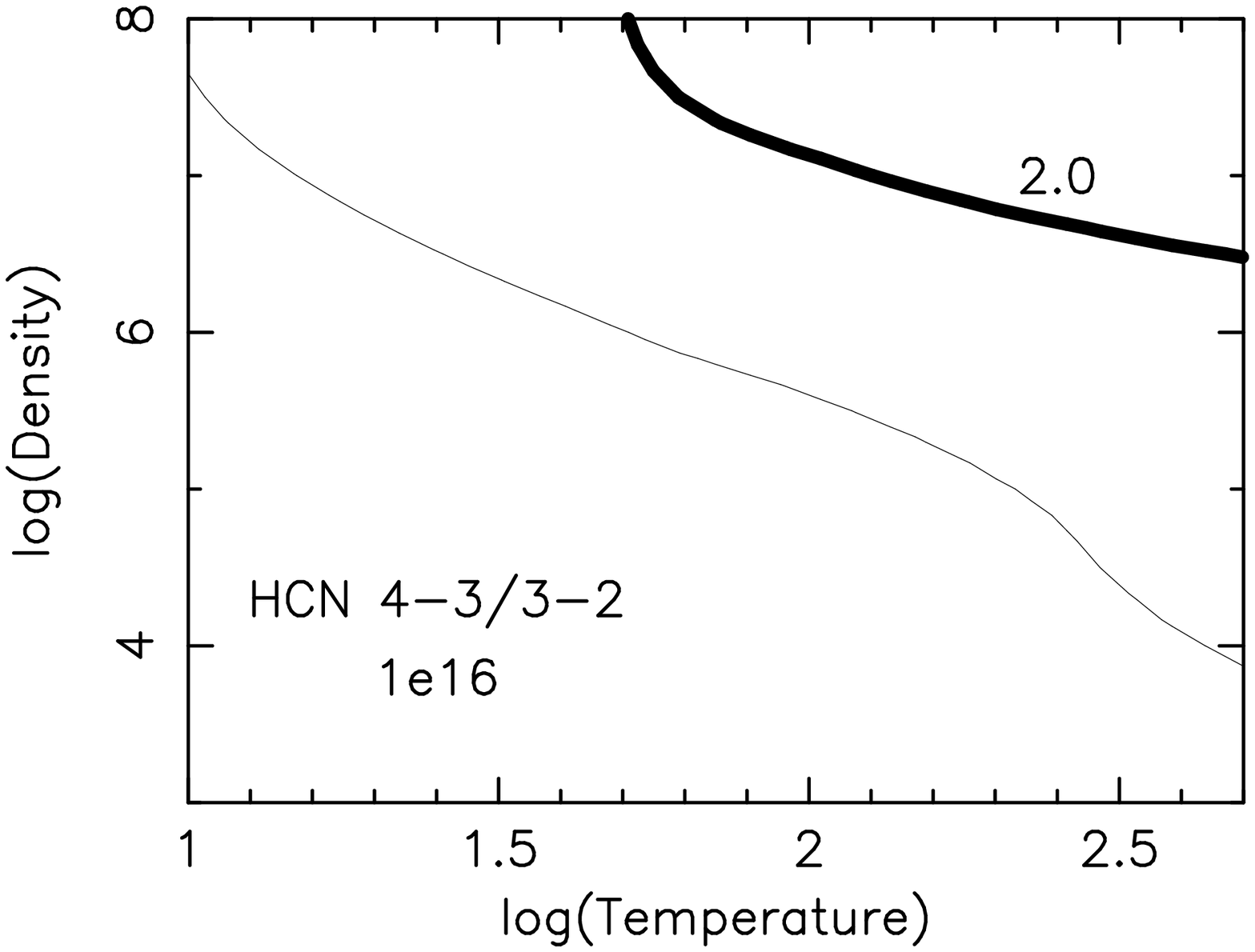} \\
\includegraphics[angle=0,scale=.41]{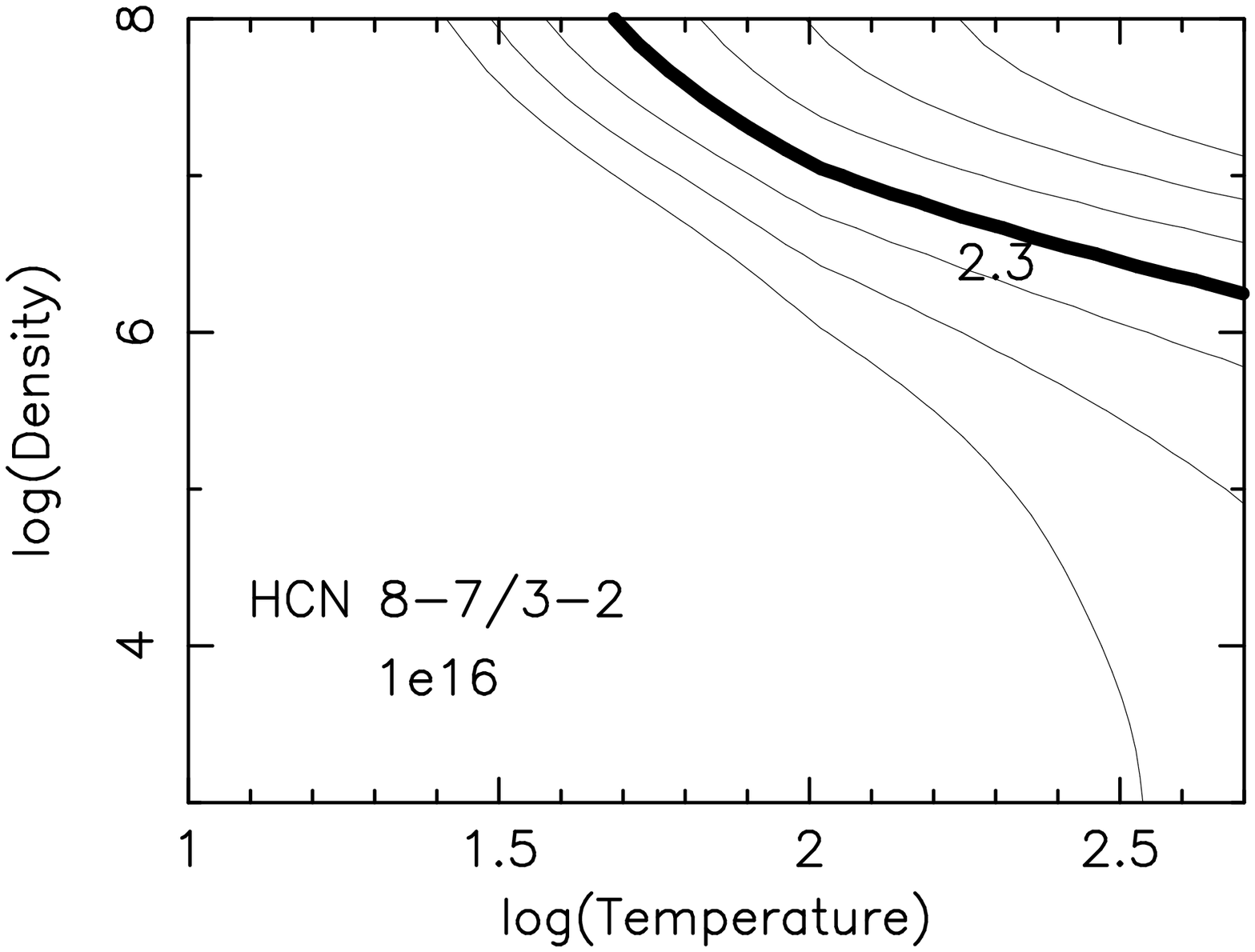}  
\includegraphics[angle=0,scale=.41]{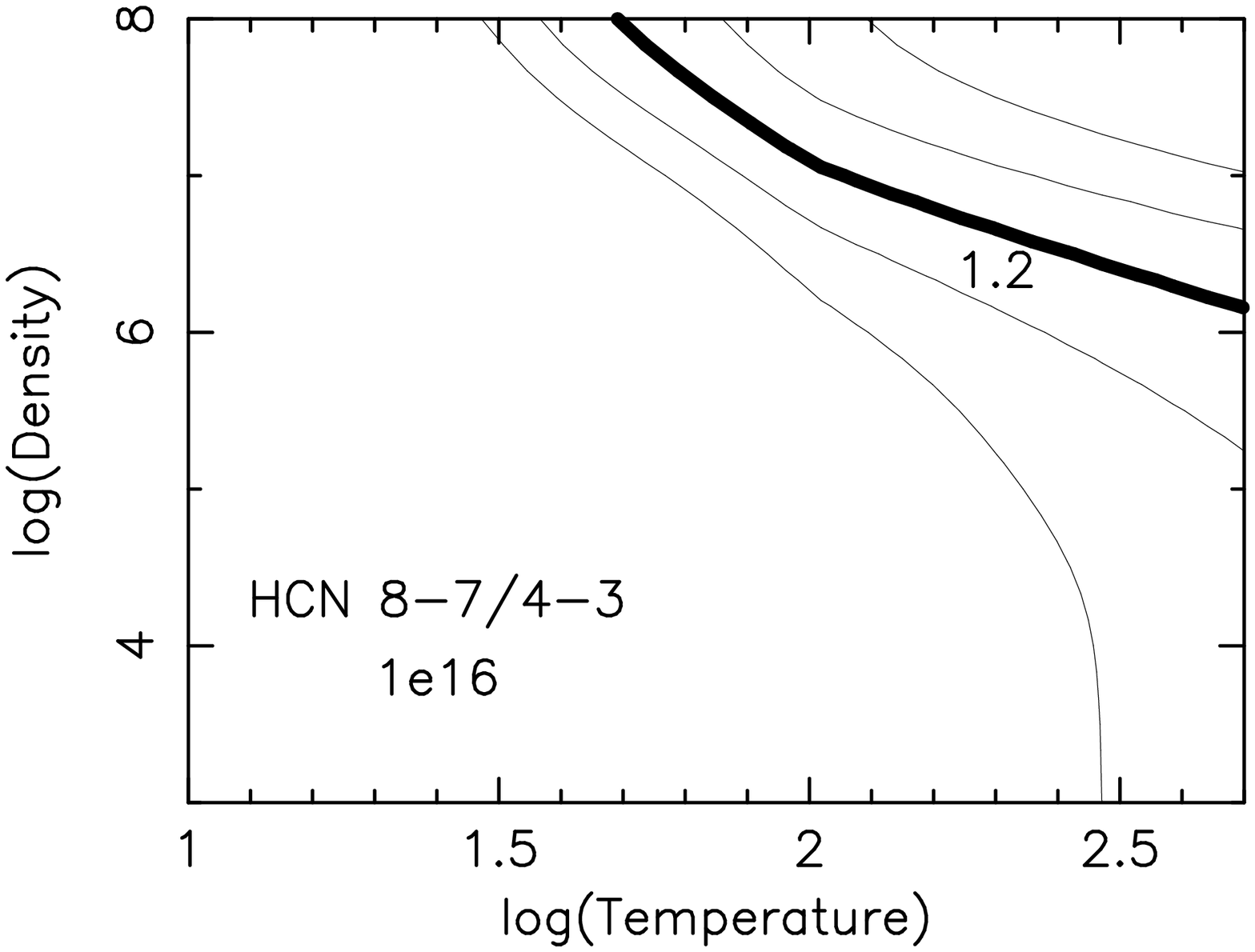}  
\end{center}
\end{figure}

\clearpage 

\begin{figure}
\begin{center}
\includegraphics[angle=0,scale=.41]{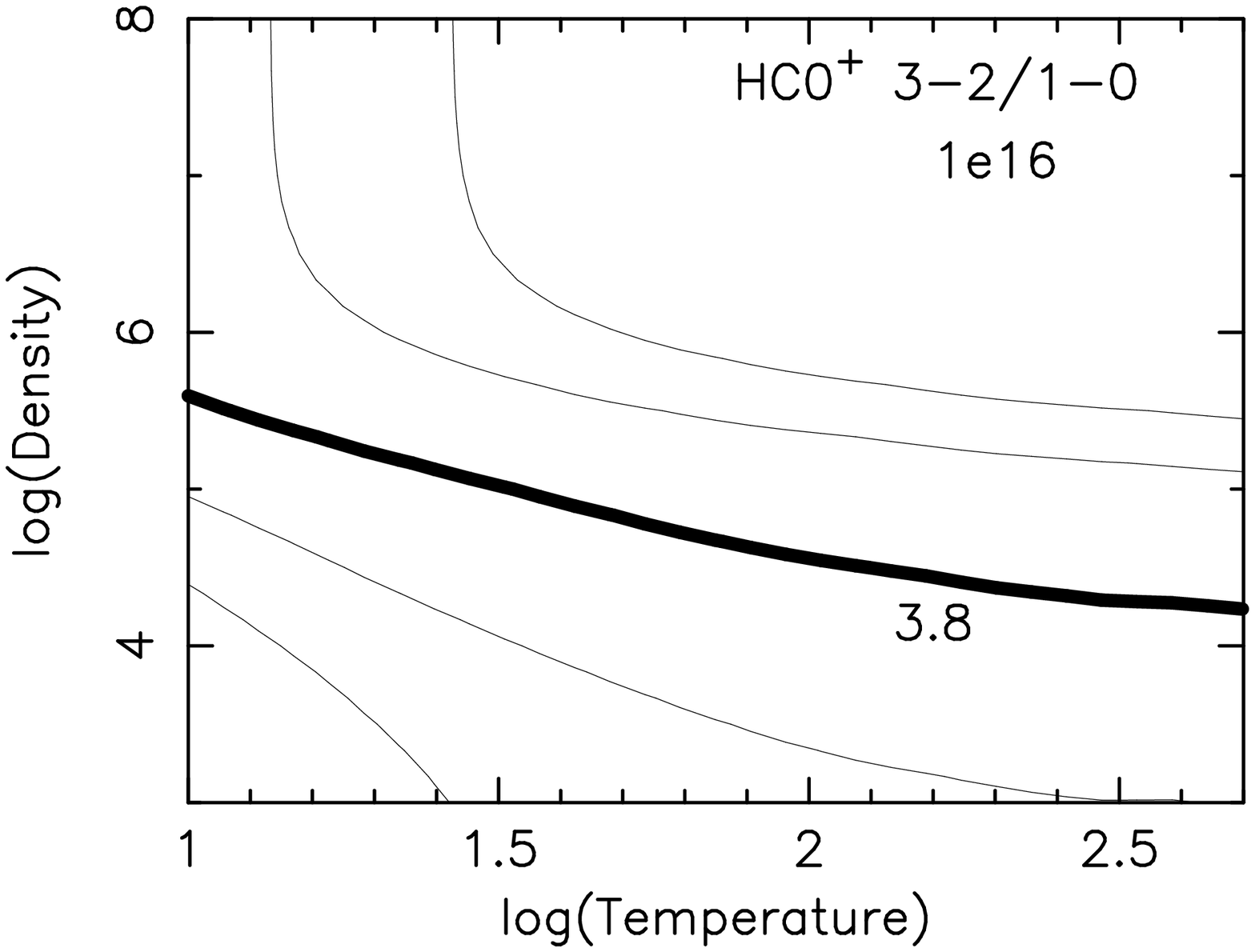}  
\includegraphics[angle=0,scale=.41]{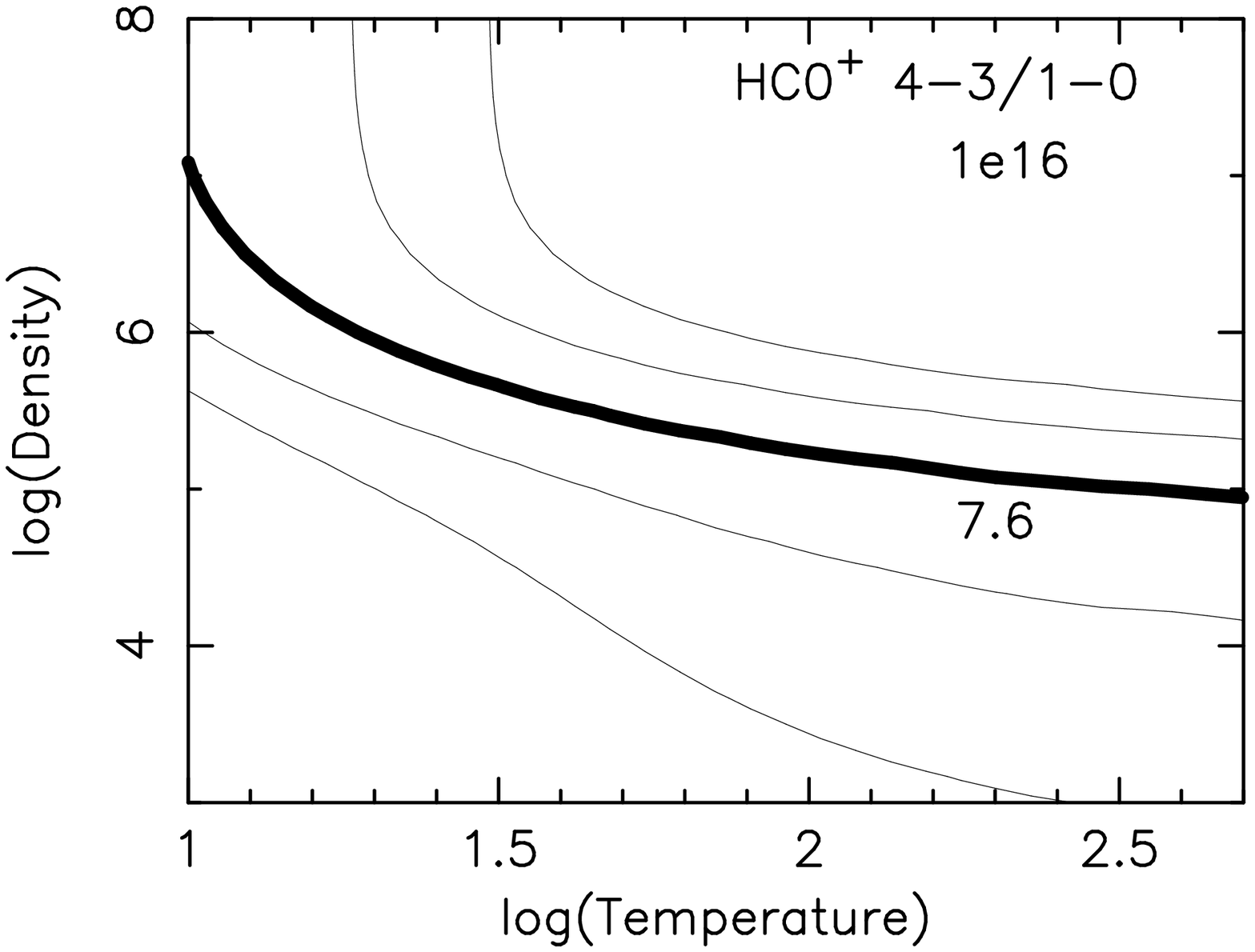}  \\
\includegraphics[angle=0,scale=.41]{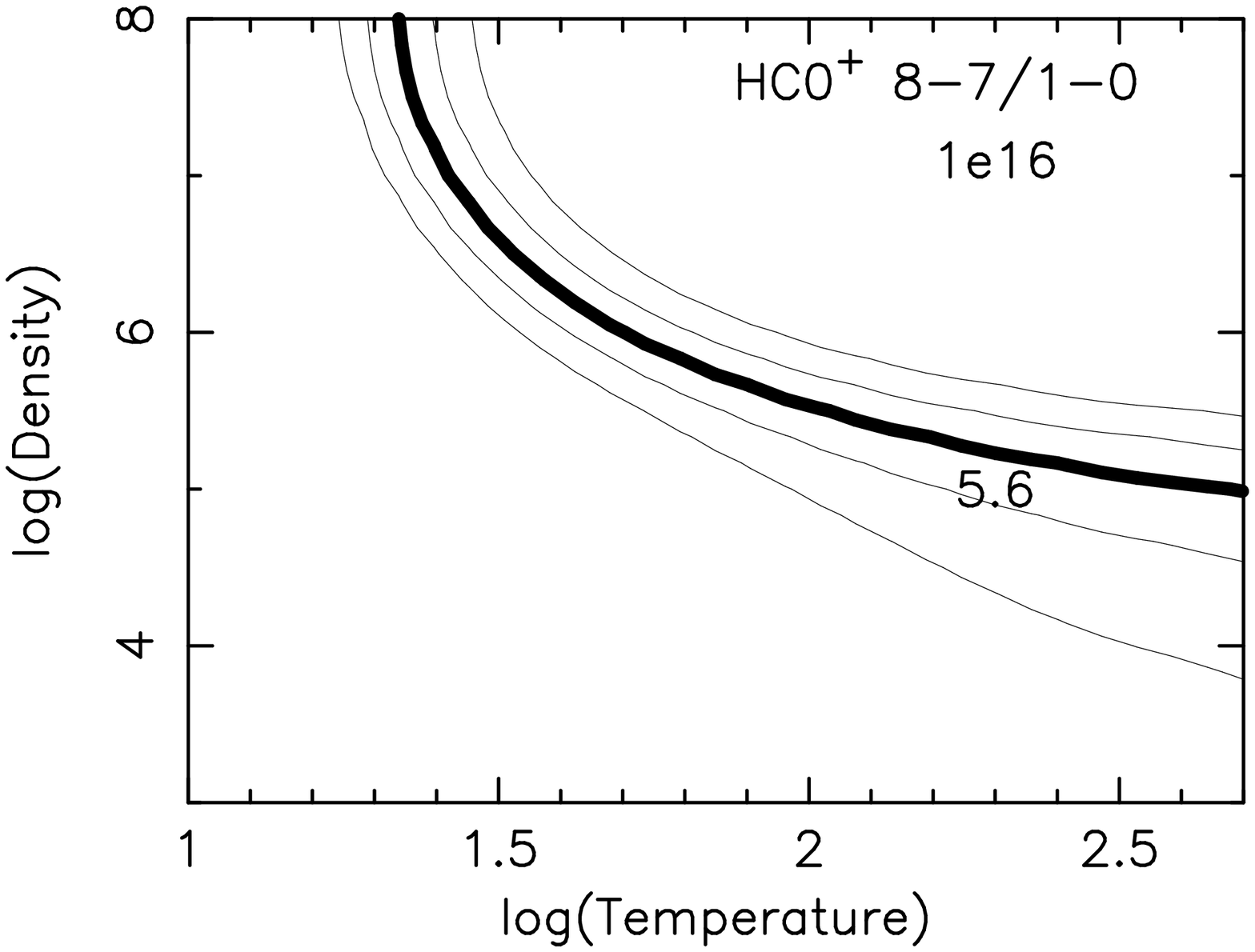}  
\includegraphics[angle=0,scale=.41]{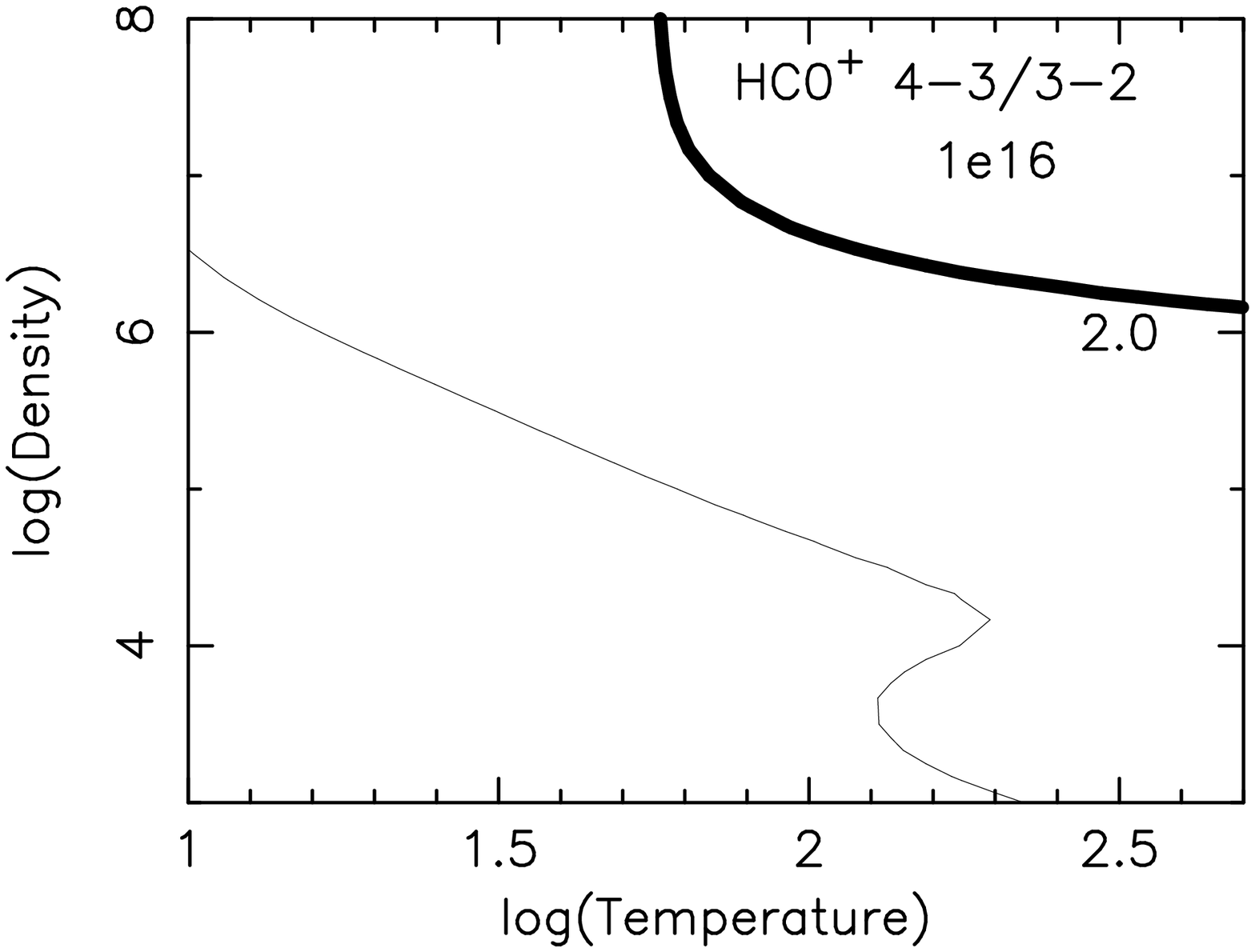}  \\
\includegraphics[angle=0,scale=.41]{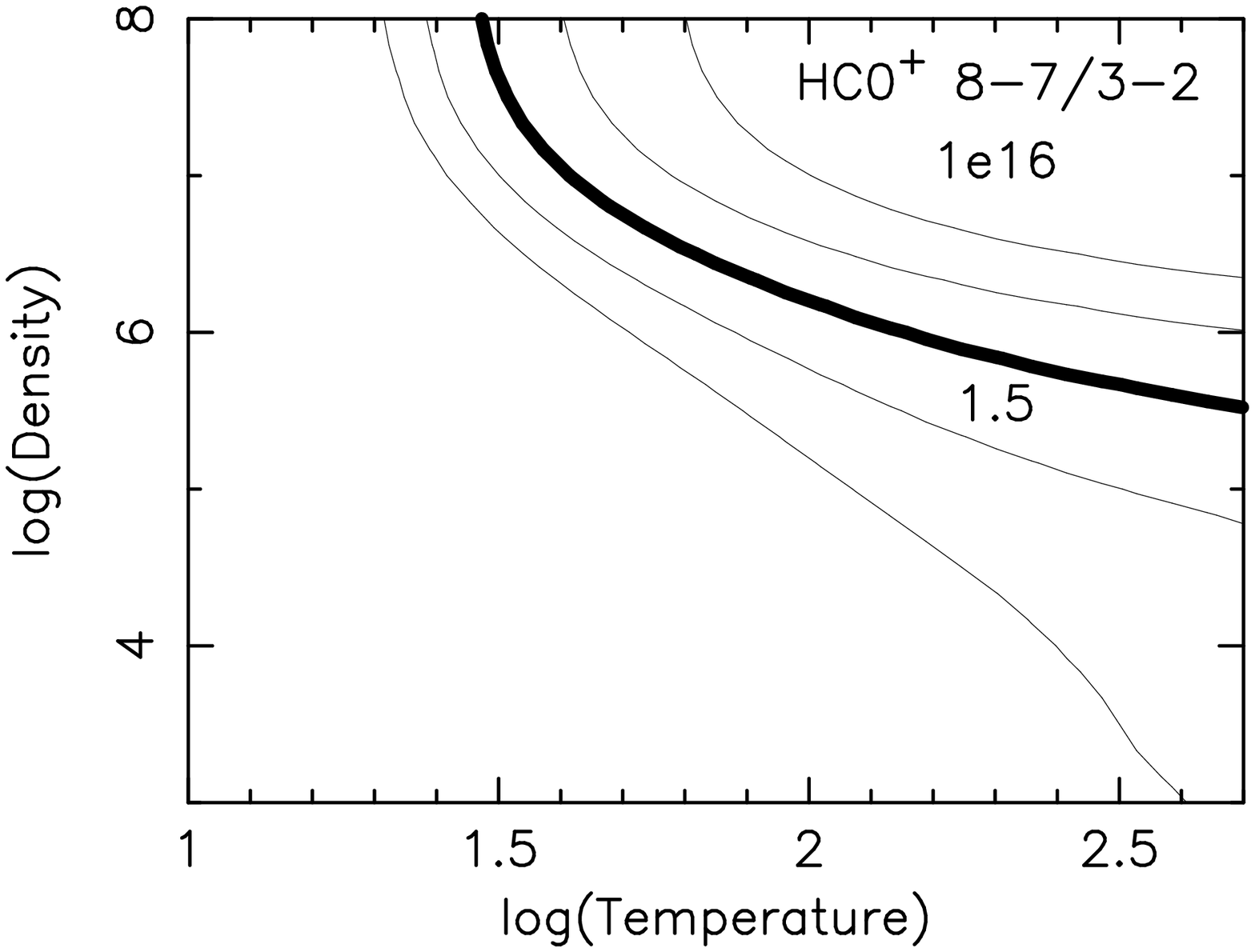}  
\includegraphics[angle=0,scale=.41]{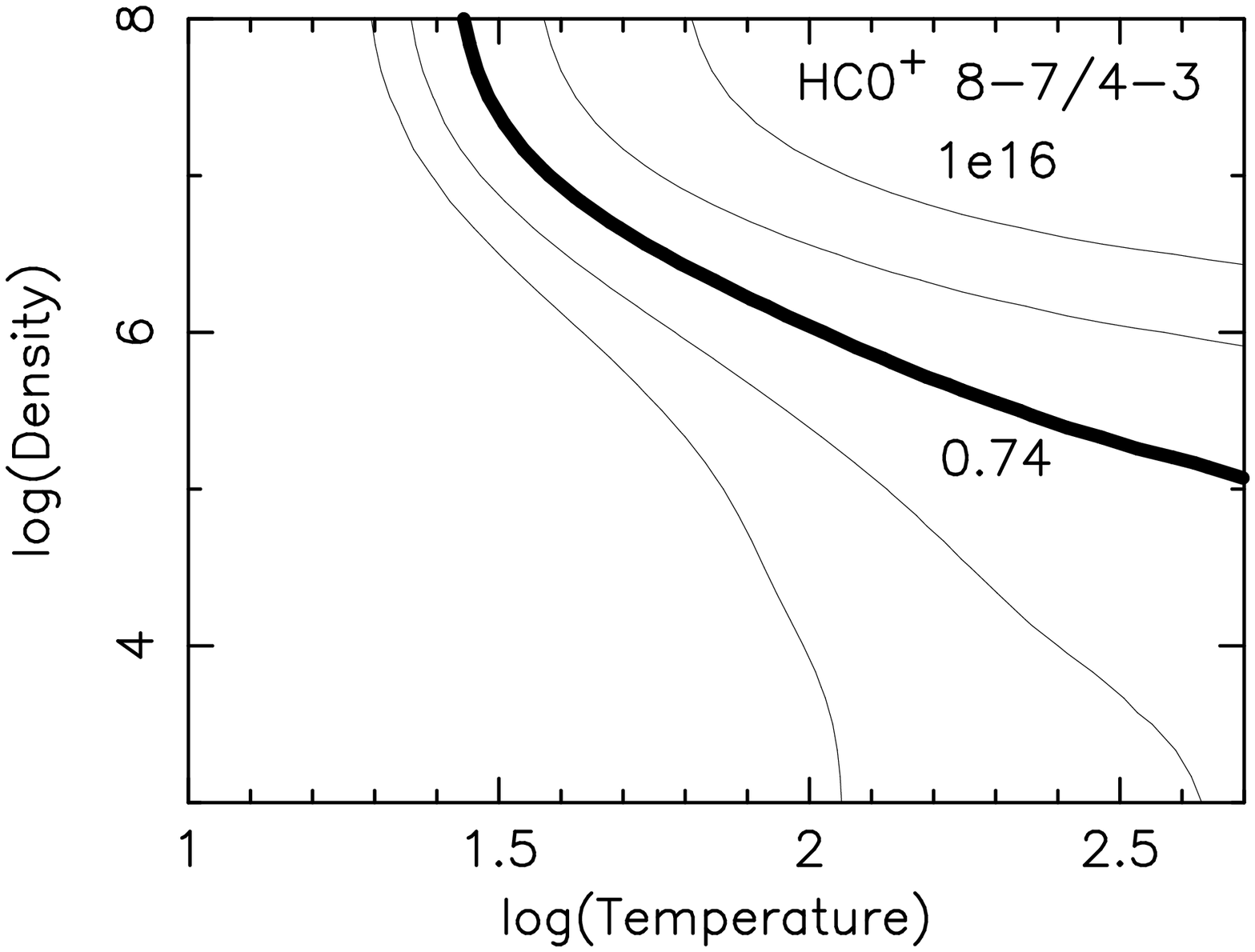}  
\end{center}
\end{figure}

\clearpage 

\begin{figure}
\begin{center}
\includegraphics[angle=0,scale=.41]{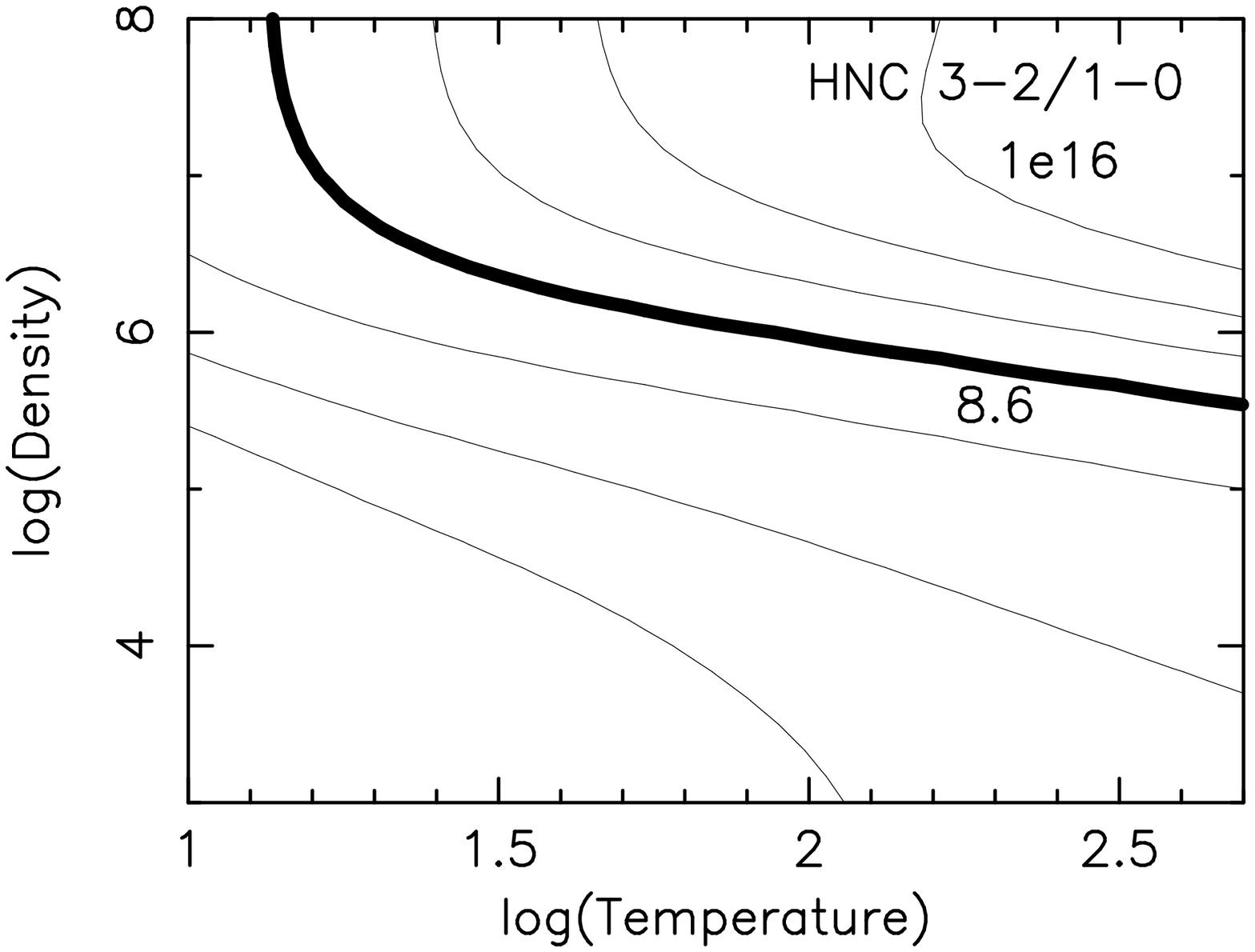}  
\includegraphics[angle=0,scale=.41]{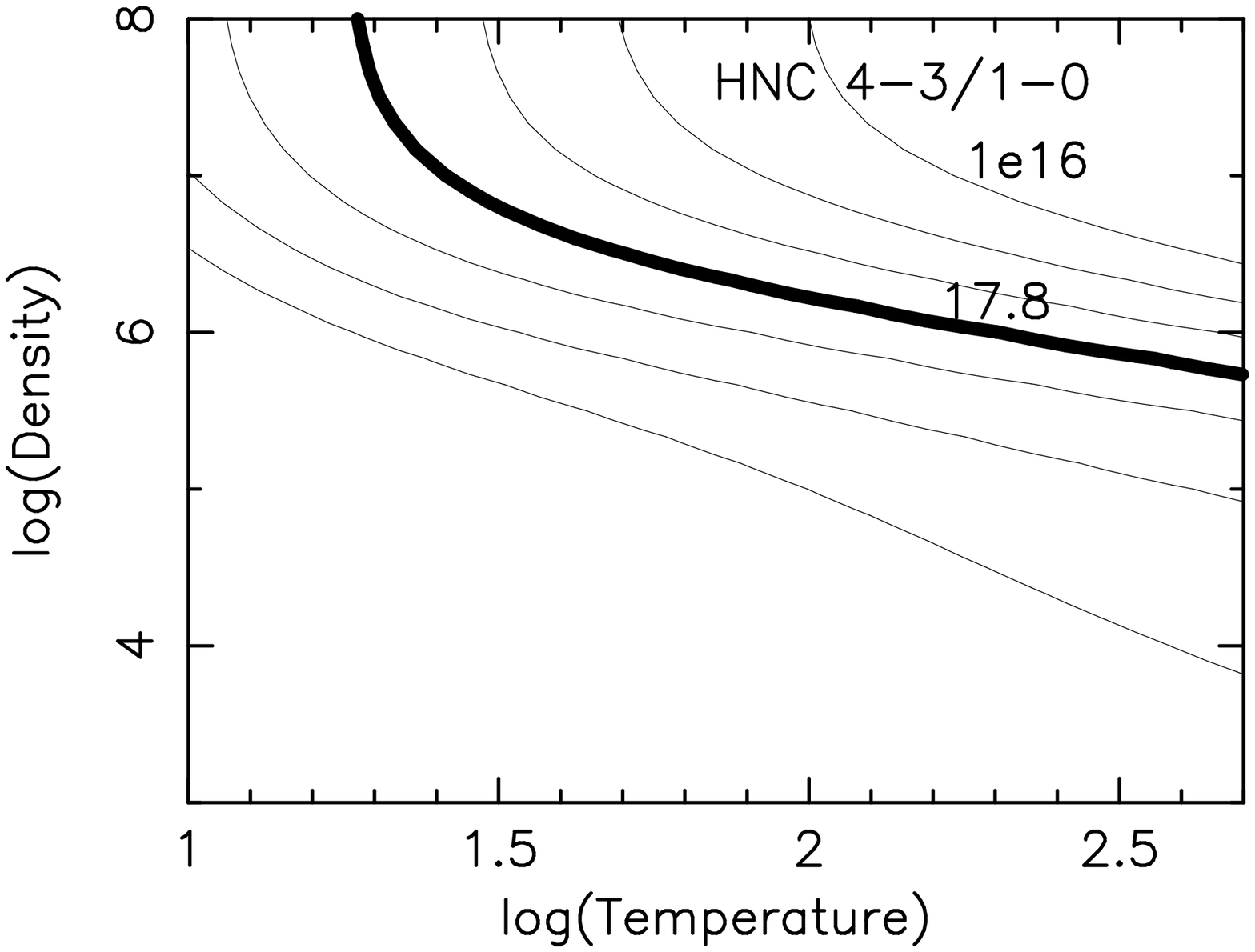} \\ 
\includegraphics[angle=0,scale=.41]{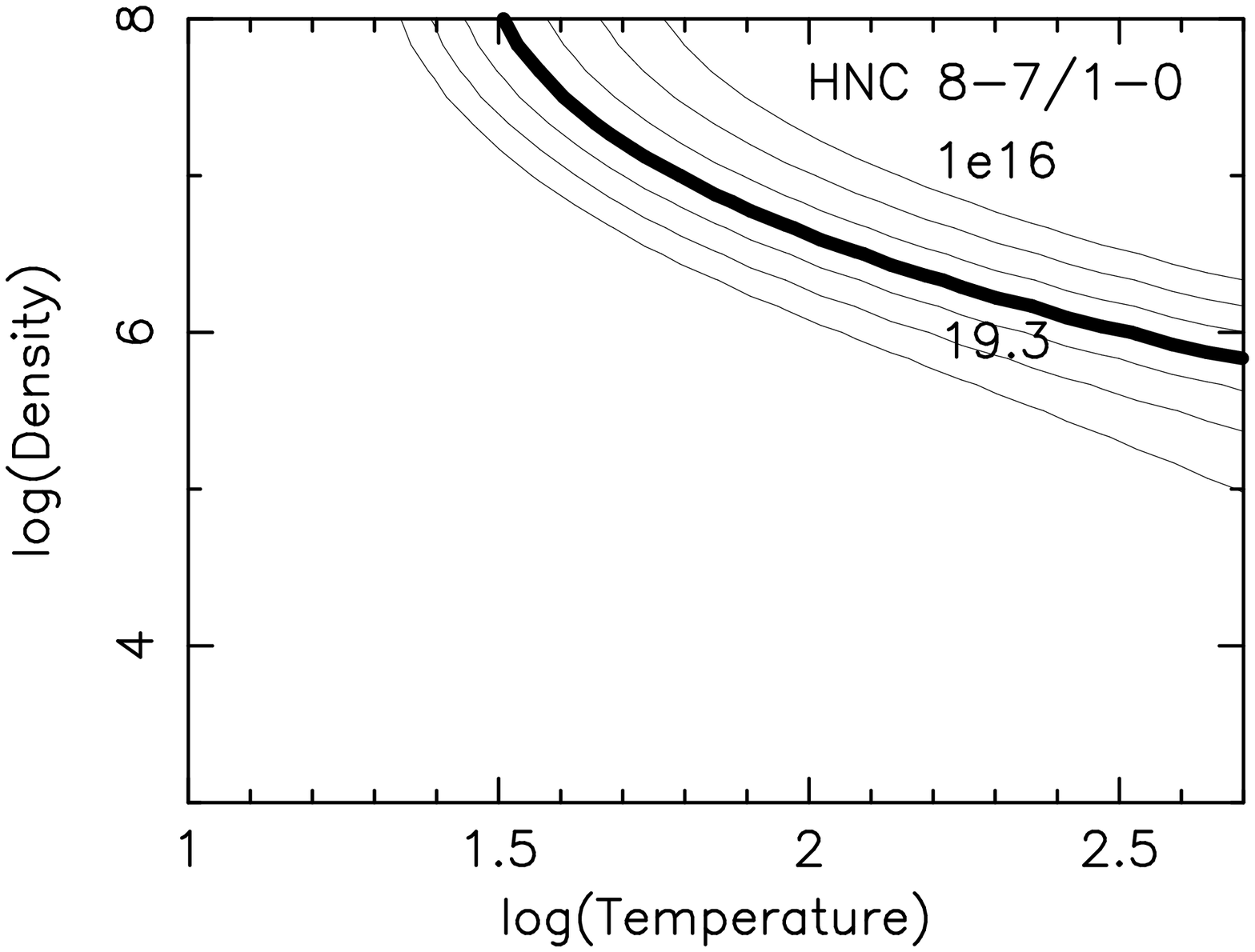}  
\includegraphics[angle=0,scale=.41]{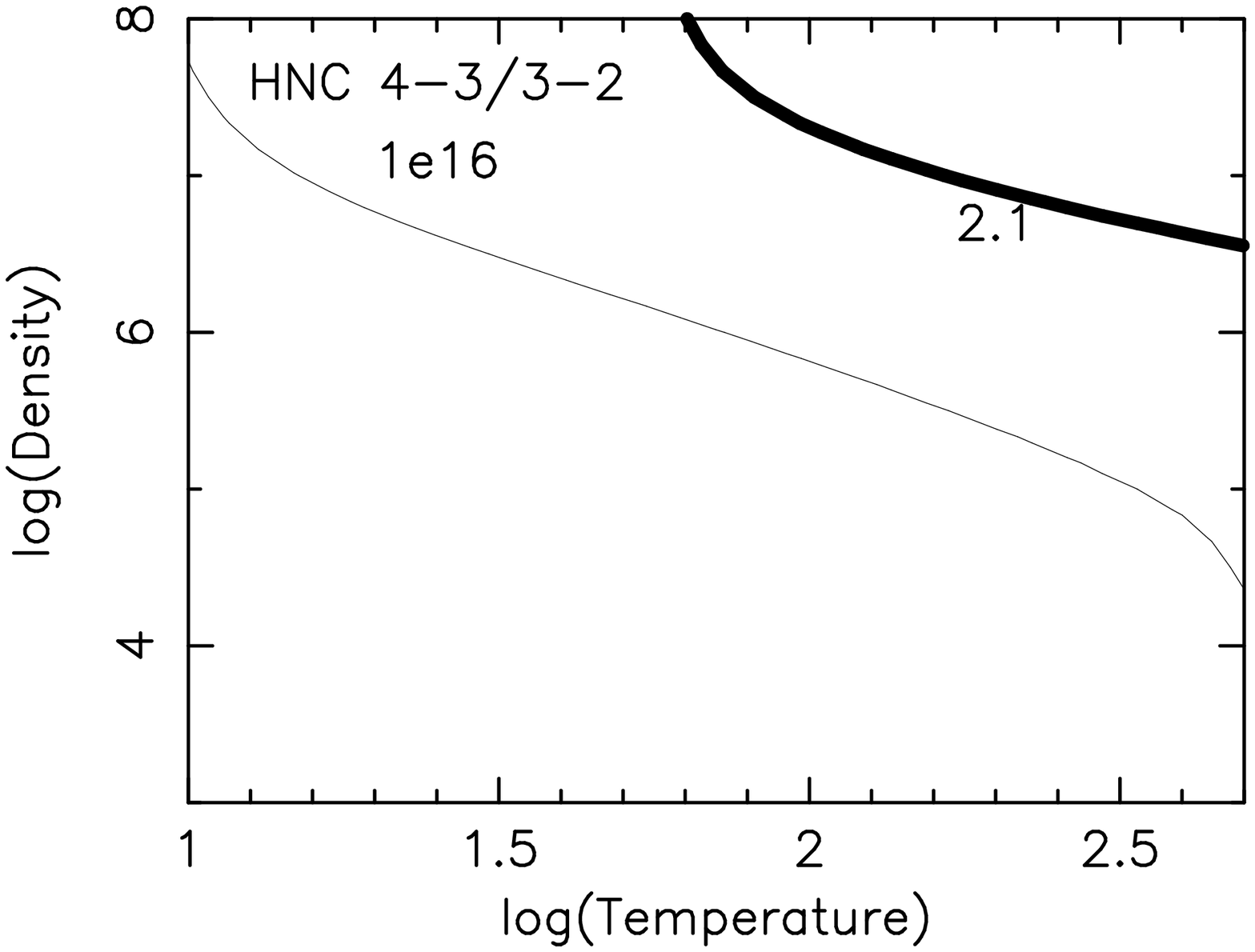} \\ 
\includegraphics[angle=0,scale=.41]{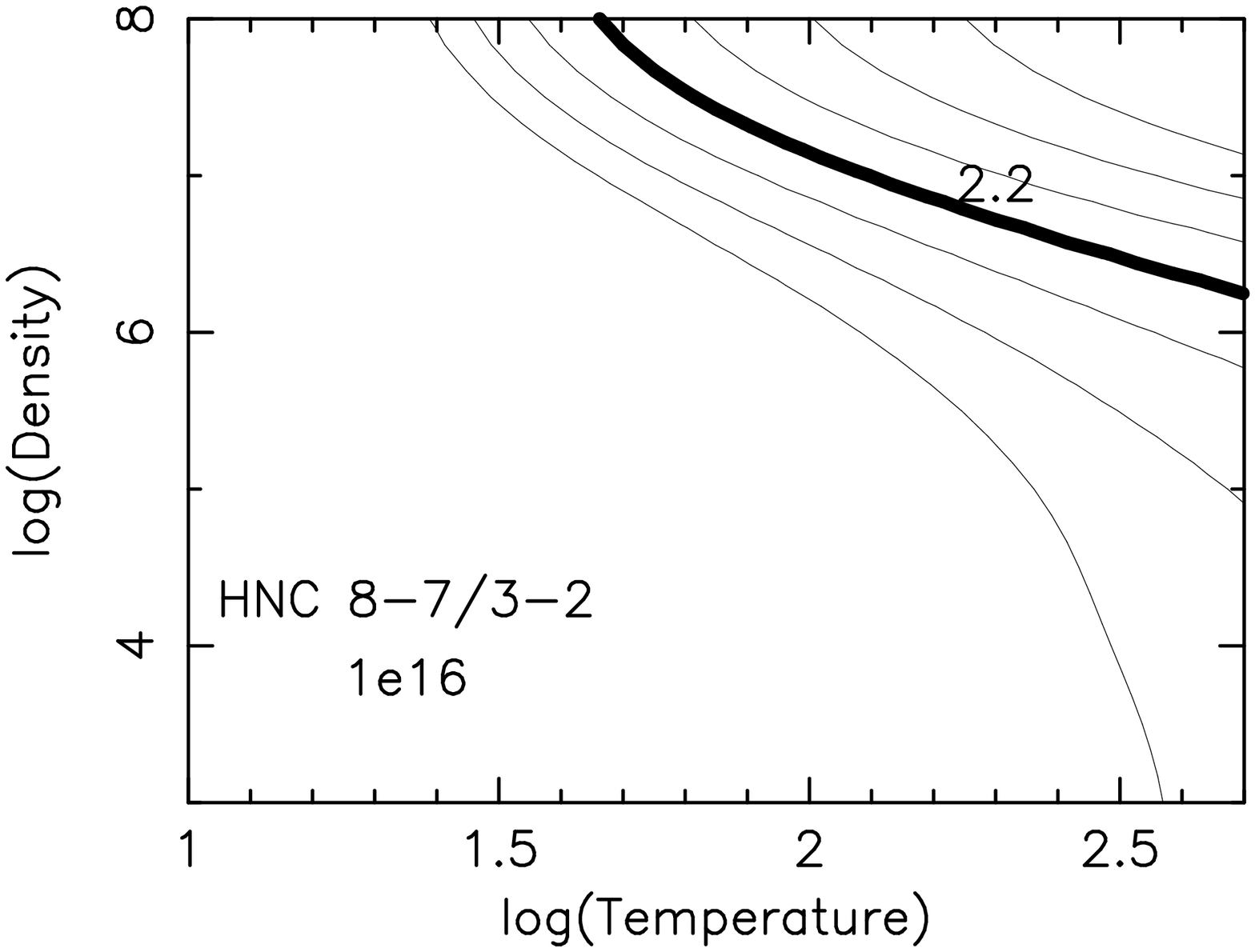}  
\includegraphics[angle=0,scale=.41]{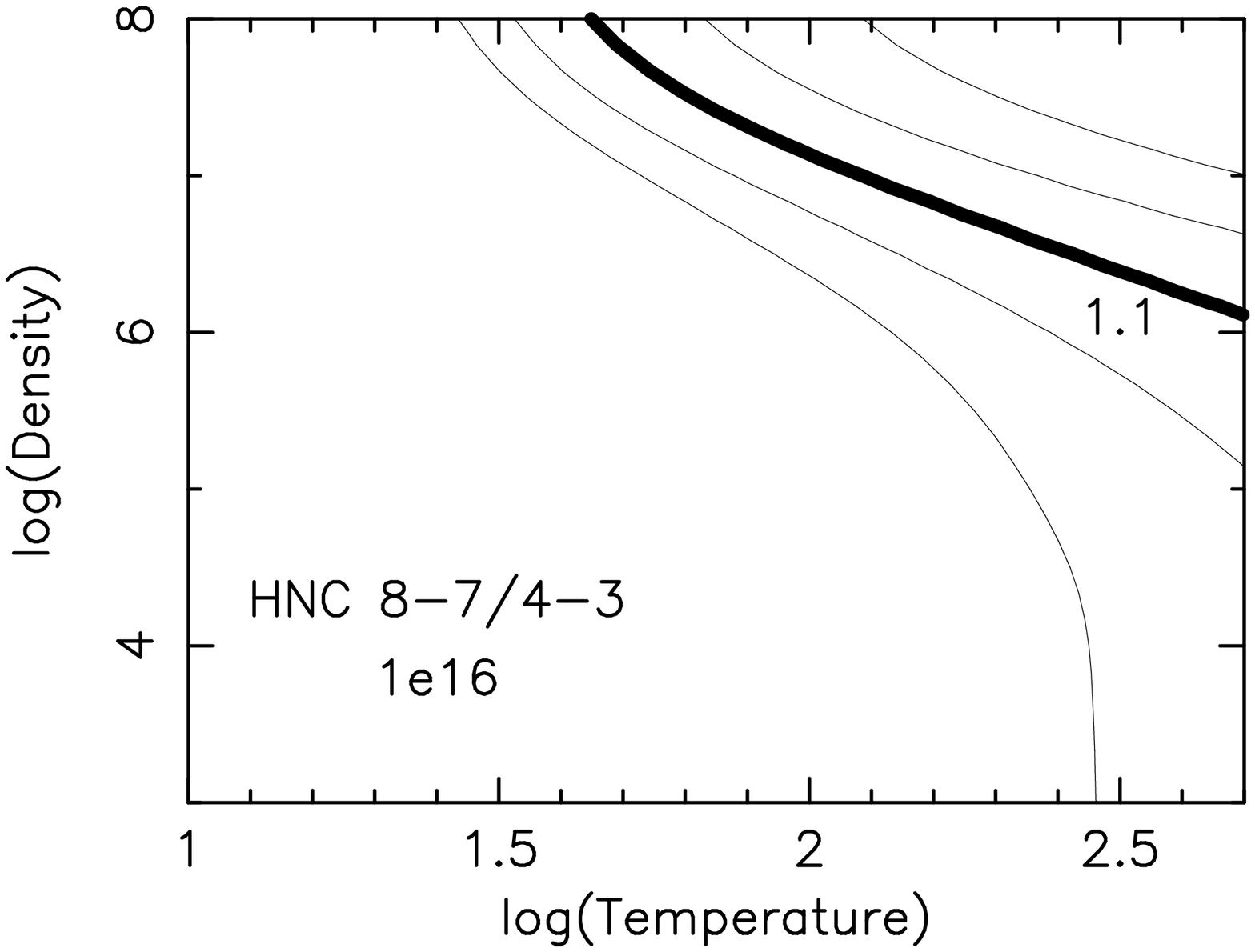}  
\caption{
RADEX calculation for the column density of 1 $\times$ 
10$^{16}$ cm$^{-2}$ for HCN, HCO$^{+}$, and HNC.
Contours are shown in step of a factor of 2, with higher (lower) values
toward the upper-right (lower-left) direction in all plots.
Parameters that reproduce the ratios of the observed fluxes in (Jy km
s$^{-1}$) are shown as thick curved lines with numbers.
}
\end{center}
\end{figure}

\begin{figure}
\begin{center}
\includegraphics[angle=0,scale=.7]{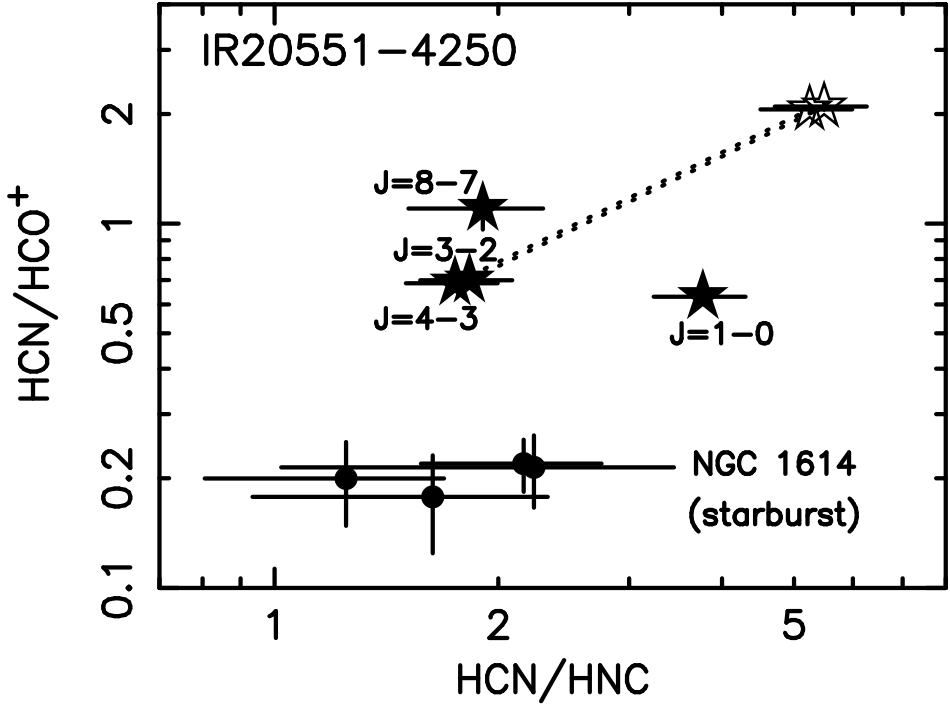} 
\end{center}
\vspace*{-0.5cm}
\caption{
Ratio of HCN-to-HNC flux in (Jy km s$^{-1}$) (abscissa) and 
HCN-to-HCO$^{+}$ flux in (Jy km s$^{-1}$) (ordinate) at 
J=1--0, J=3--2, J=4--3, and J=8--7.
While HCN and HCO$^{+}$ data were obtained simultaneously, 
HCN and HNC data were not. 
To calculate the HCN-to-HNC flux ratio, we added 10\% absolute flux
calibration uncertainty of individual ALMA observations.   
Observed ratios for IRAS 20551$-$4250 are shown as filled stars.
We assume the HCN flux attenuation by line opacity with a factor of
$\sim$3 at both J=3--2 and J=4--3 (minimum side of the estimated range
in $\S$4.2).
In this case, the flux attenuation of HCO$^{+}$ and HNC is insignificant 
(a factor of $\sim$1; $\S$4.2).
The line-opacity-corrected intrinsic flux ratios at J=3--2 and J=4--3
are shown as open stars and are connected to their observed ratios with
dotted lines.
In addition to IRAS 20551$-$4250, data at multiple locations of the
starburst-dominated galaxy, NGC 1614, are also shown as filled circles
for comparison.
}
\end{figure}

\begin{figure}
\begin{center}
\vspace*{-1.5cm}
\includegraphics[angle=0,scale=.7]{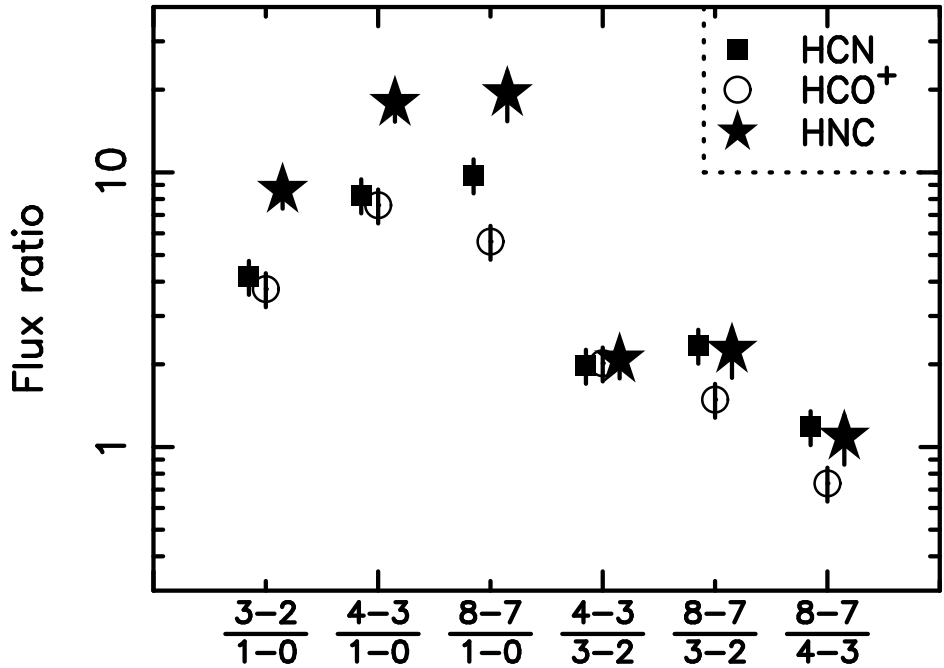} 
\end{center}
\vspace{-0.5cm}
\caption{
Flux ratios of high-J to low-J transition lines for HCN,
HCO$^{+}$, and HNC. 
Fluxes are in units of (Jy km s$^{-1}$).
}
\end{figure}

\begin{figure}
\begin{center}
\includegraphics[angle=0,scale=.45]{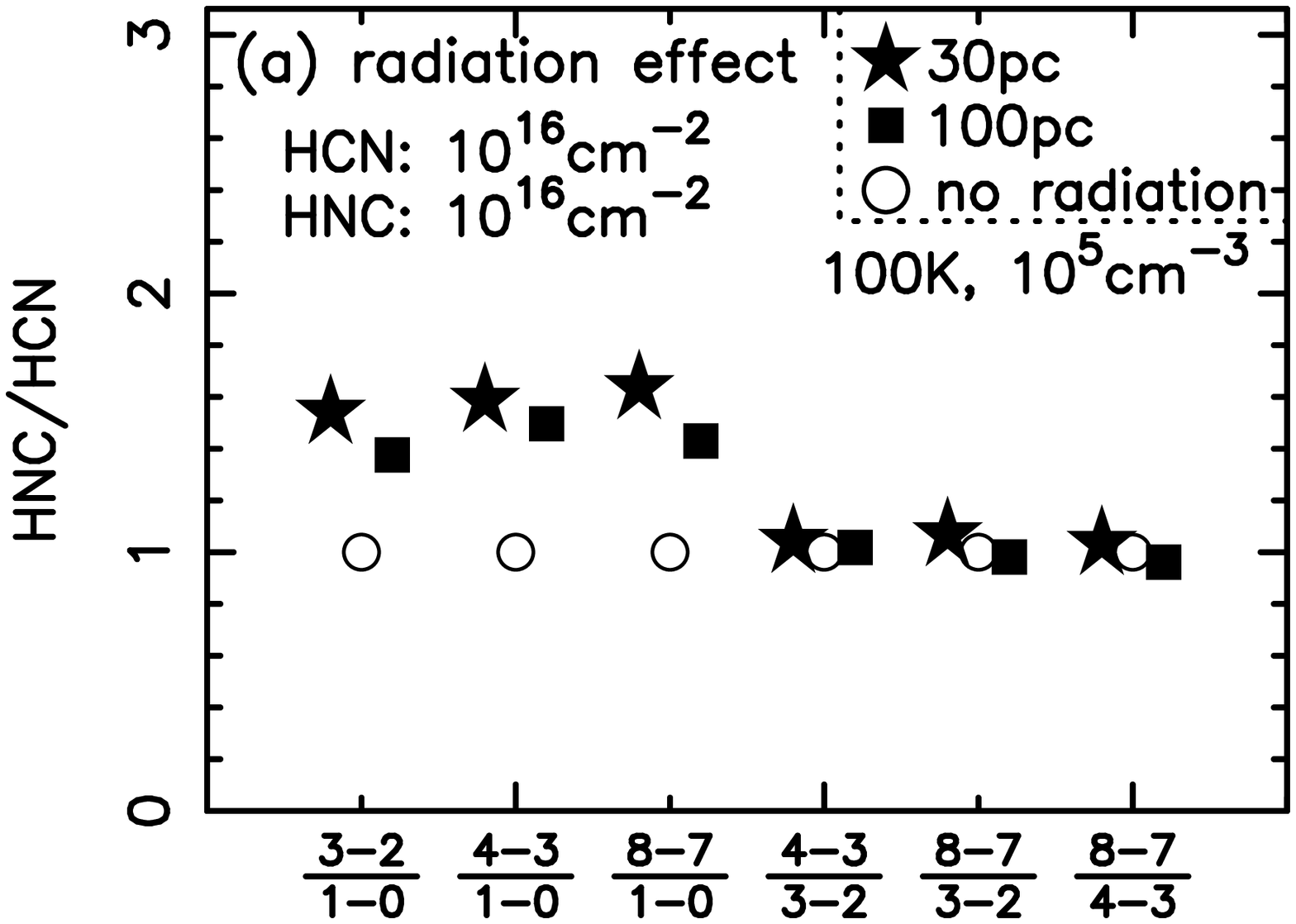} 
\includegraphics[angle=0,scale=.45]{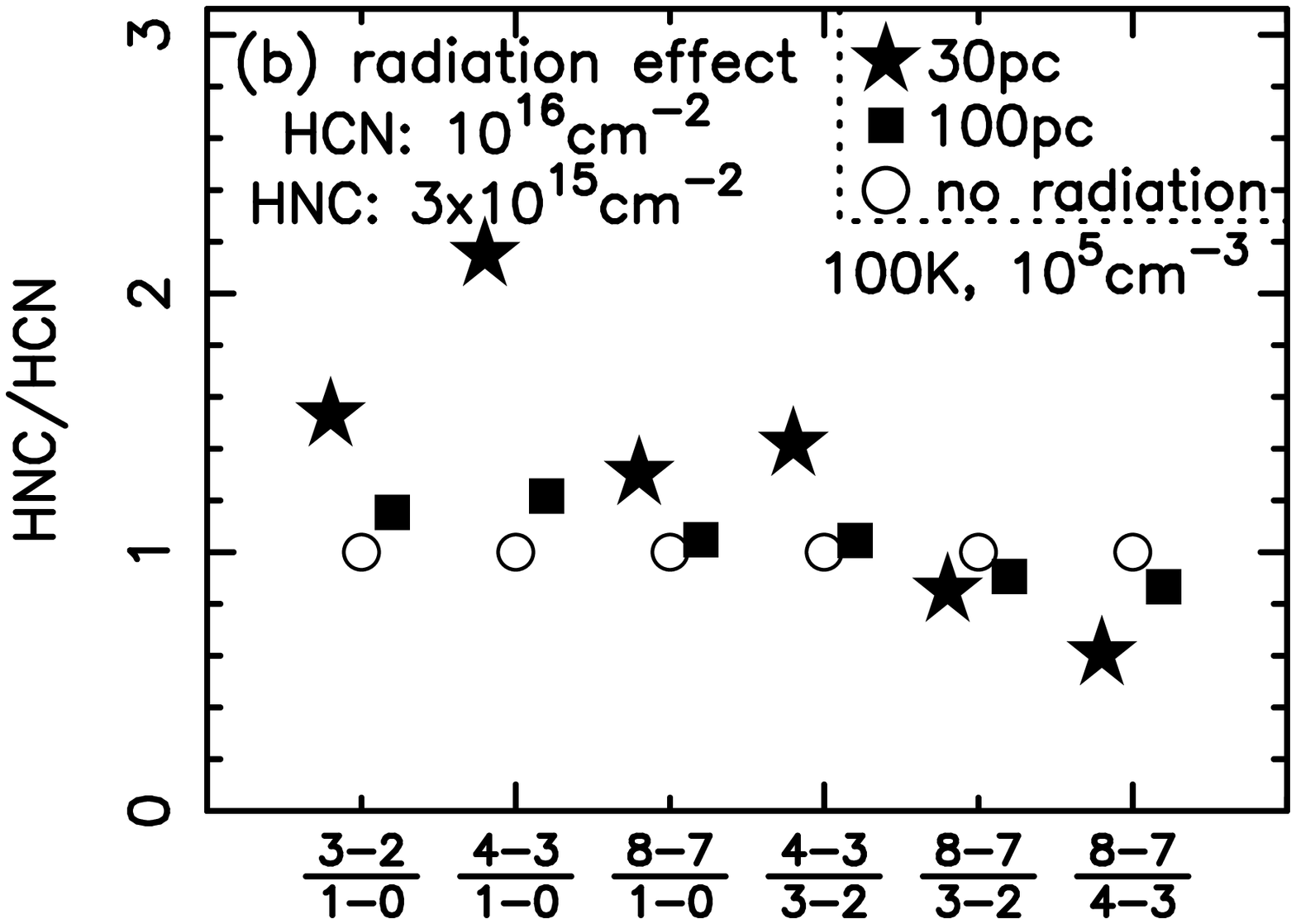} 
\end{center}
\vspace*{-0.5cm}
\caption{
Quantitative investigation of the effect of infrared radiation on high-J
to low-J flux ratios for HCN and HNC, using the MOLPOP software
\citep{eli06}.
The ordinate is the values of "high-J to low-J flux ratio of HNC,
divided by the same ratio of HCN".
The combination of ``high-J to low-J'' is ``J=3--2 to J=1--0'', 
``J=4--3 to J=1--0'', ``J=8--7 to J=1--0'', ``J=4--3 to J=3--2'',
``J=8--7 to J=3--2'', and ``J=8--7 to J=4--3''.
The values in the case of no external radiation (i.e., 
collisional excitation alone) are set to unity for all 
high-J to low-J combinations.
Thus, the ordinate corresponds to the increased factor of the values 
by the inclusion of infrared radiation. 
The molecular gas parameters are H$_{\rm 2}$ number density 
n$_{\rm H2}$ = 10$^{5}$ cm$^{-3}$, 
H$_{\rm 2}$ kinetic temperature T$_{\rm kin}$ = 100 K, and 
velocity width $\Delta$v = 300 km s$^{-1}$.
The $\sim$3 K cosmic microwave background emission is included.
Open circles: No external radiation.
Filled stars: Infrared radiating energy source with 
F$_{\nu}$ (Jy) = 0.0024 $\times$ $\lambda$($\mu$m)$^{2.0}$ is placed at
30 pc from the molecular gas. 
Filled squares: The same infrared radiating energy source is placed at 100 
pc from the molecular gas.
(a) The column density is 10$^{16}$ cm$^{-2}$ for both HCN and HNC.
(b) The HCN and HNC column densities are 10$^{16}$ cm$^{-2}$ and 
3 $\times$ 10$^{15}$ cm$^{-2}$, respectively.
}
\end{figure}

\begin{figure}
\begin{center}
\includegraphics[angle=0,scale=.7]{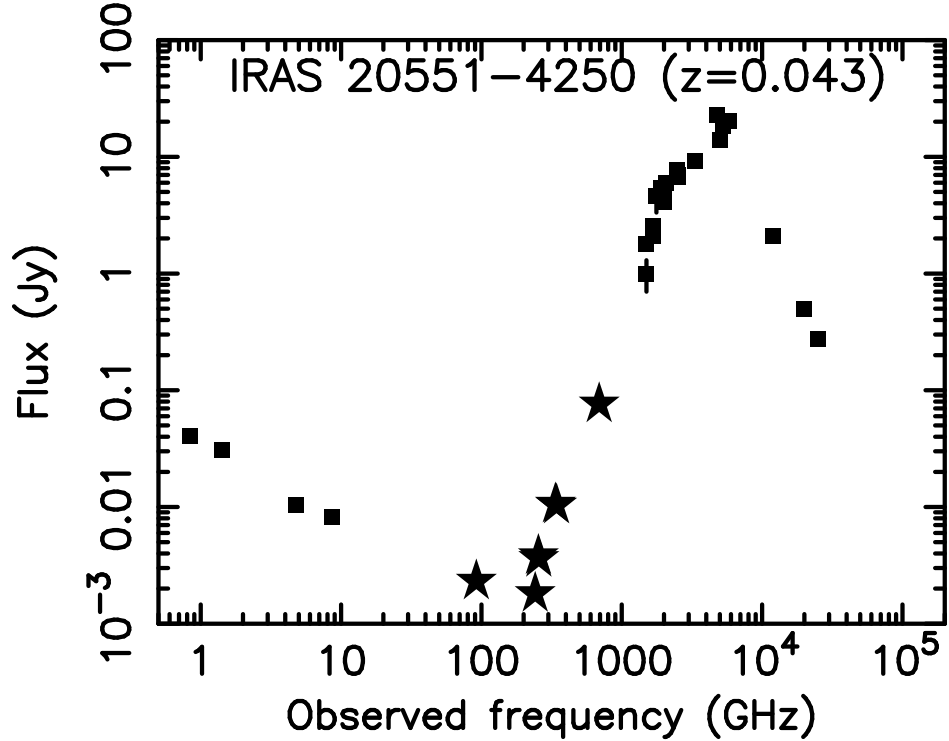} 
\end{center}
\caption{
Infrared to radio spectral energy distribution.
Filled stars: Our ALMA data, including \citet{ima16b}.
In band 7 (275--373 GHz), both our J43a and J43b continuum measurements
in ALMA Cycle 2 (Table 3) are plotted, but overlap in the
figure. 
Our ALMA Cycle 0 continuum measurements in band 7 \citep{ima13b} are not
plotted.  
Filled squares: From the literature.
Near-infrared 1--3 $\mu$m data are from 2MASS.
Infrared 10--200 $\mu$m data are from IRAS and ISO photometry 
\citep{spi02,bra08}.
Radio 1.4 GHz and 4--9 GHz data are from \citet{con96} and
\citet{geo00}, respectively. 
Radio 843 MHz data are from SUMSS (Sydney University Molonglo Sky
Survey, Version 2.1, 2008).
}
\end{figure}

\clearpage


\section{Appendix}

Integrated intensity (moment 0) maps of selected serendipitously
detected faint emission lines are displayed in Figure 17.
Figure 18 shows spectra around these lines and best Gaussian fits, 
except for ``SO$_{2}$ $+$ SO'' in Figure 17, because two lines
overlap at close frequency (Figure 2c). 
Figure 19 shows our RADEX calculations to reproduce the observed molecular
line flux ratios, in the case of HCO$^{+}$ and HNC column density of 3
$\times$ 10$^{15}$ cm$^{-2}$.

\begin{figure}[b]
\begin{center}
\includegraphics[angle=0,scale=.36]{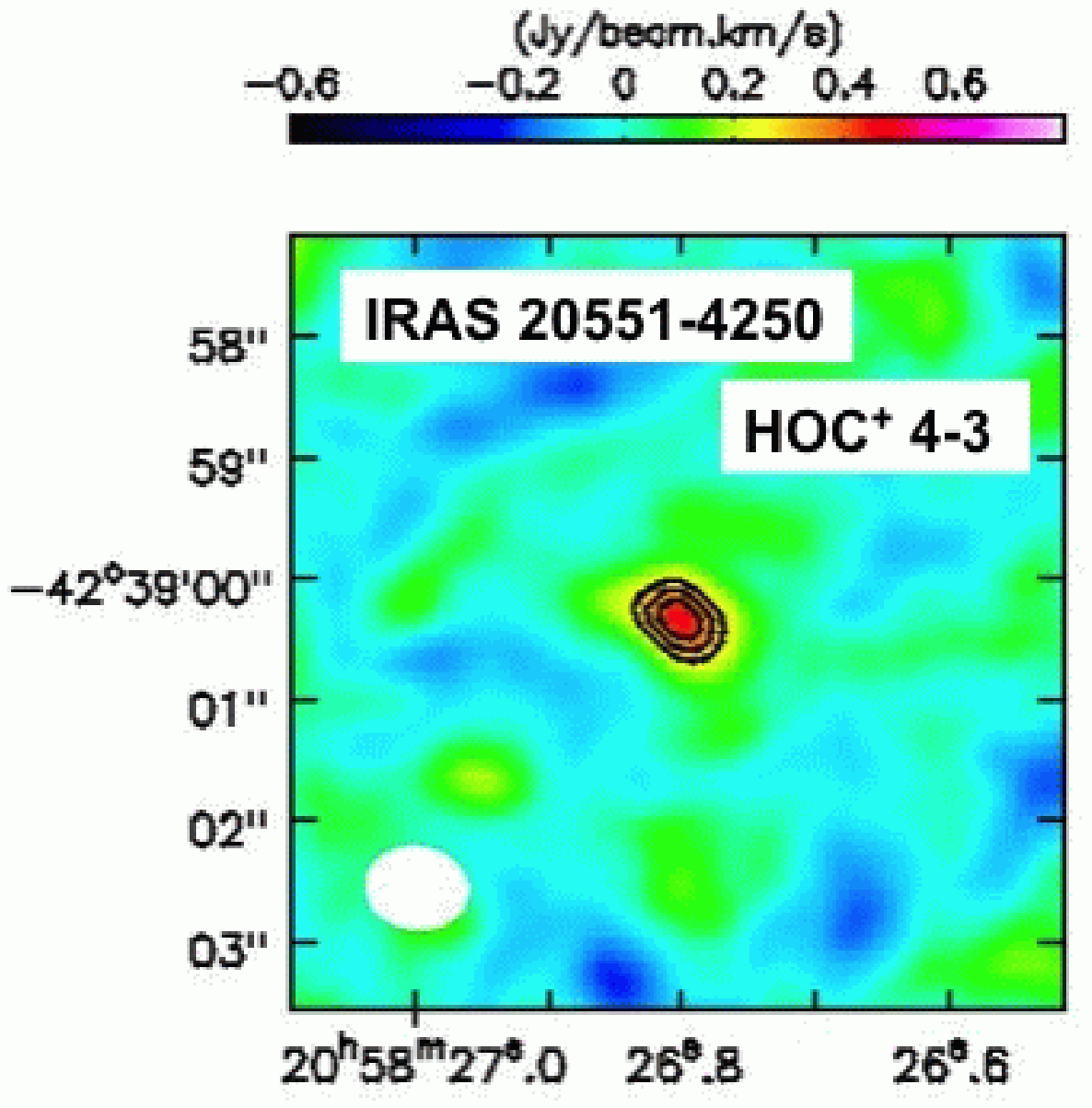} 
\includegraphics[angle=0,scale=.36]{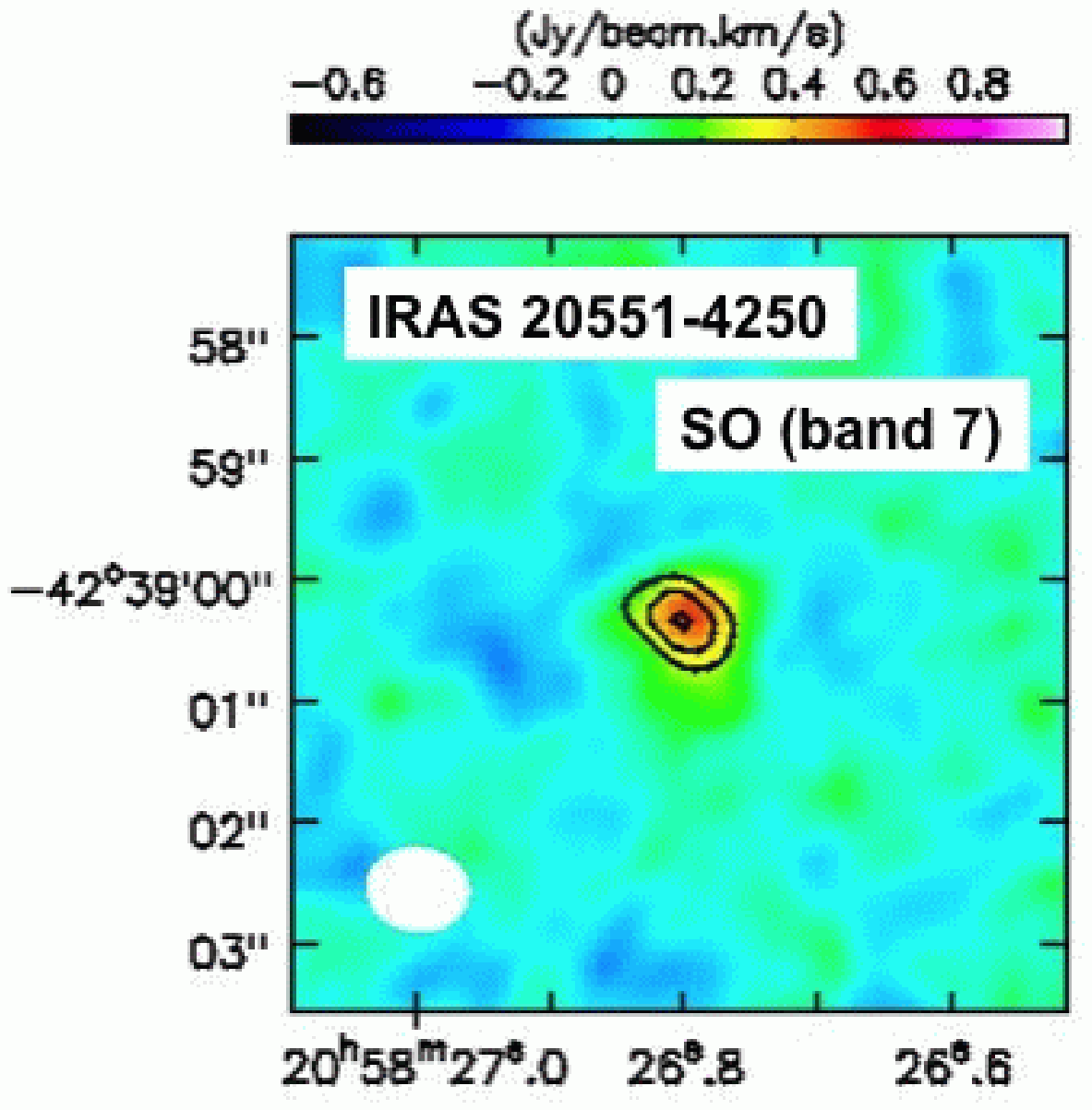} 
\includegraphics[angle=0,scale=.36]{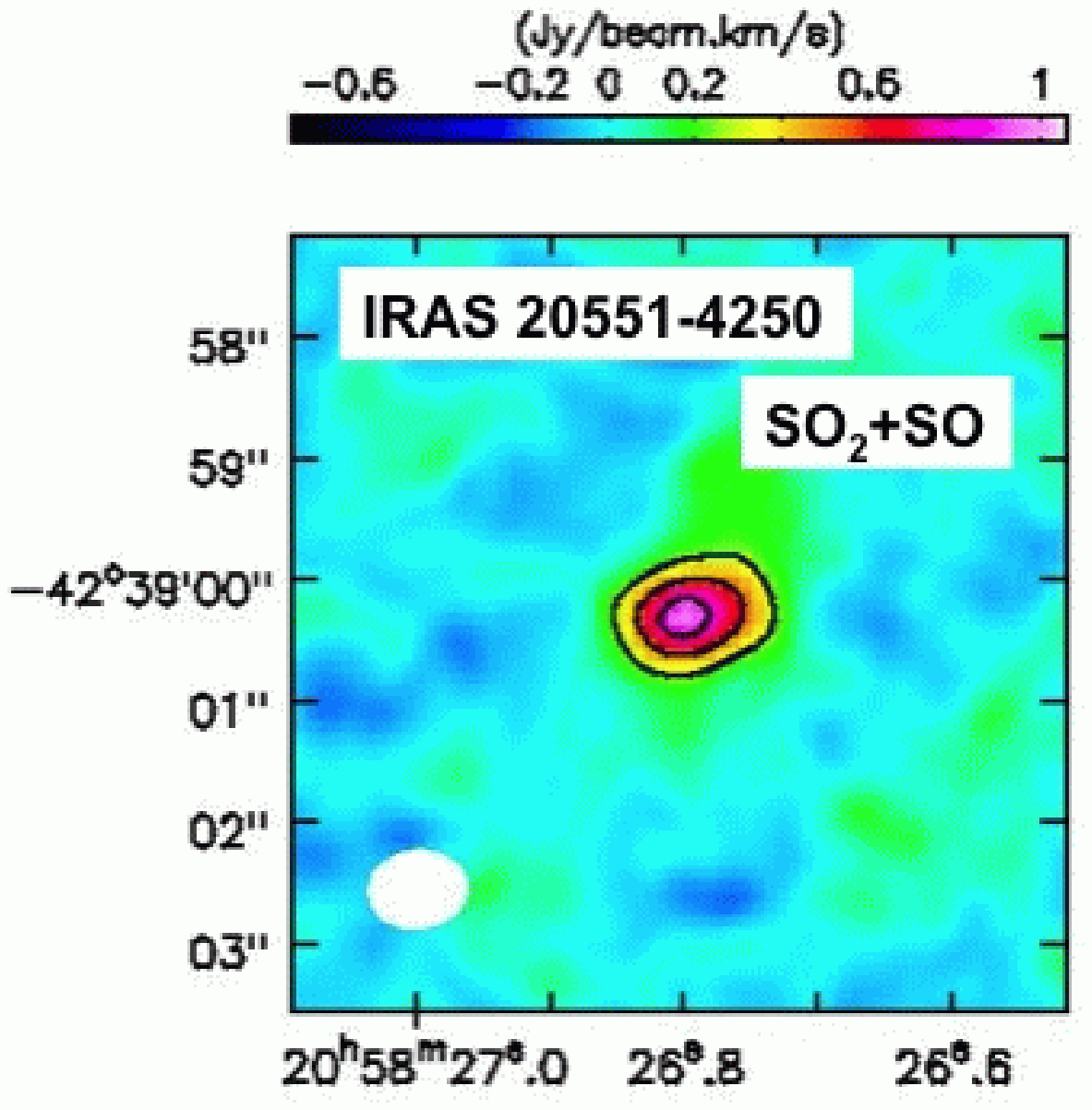} \\
\includegraphics[angle=0,scale=.36]{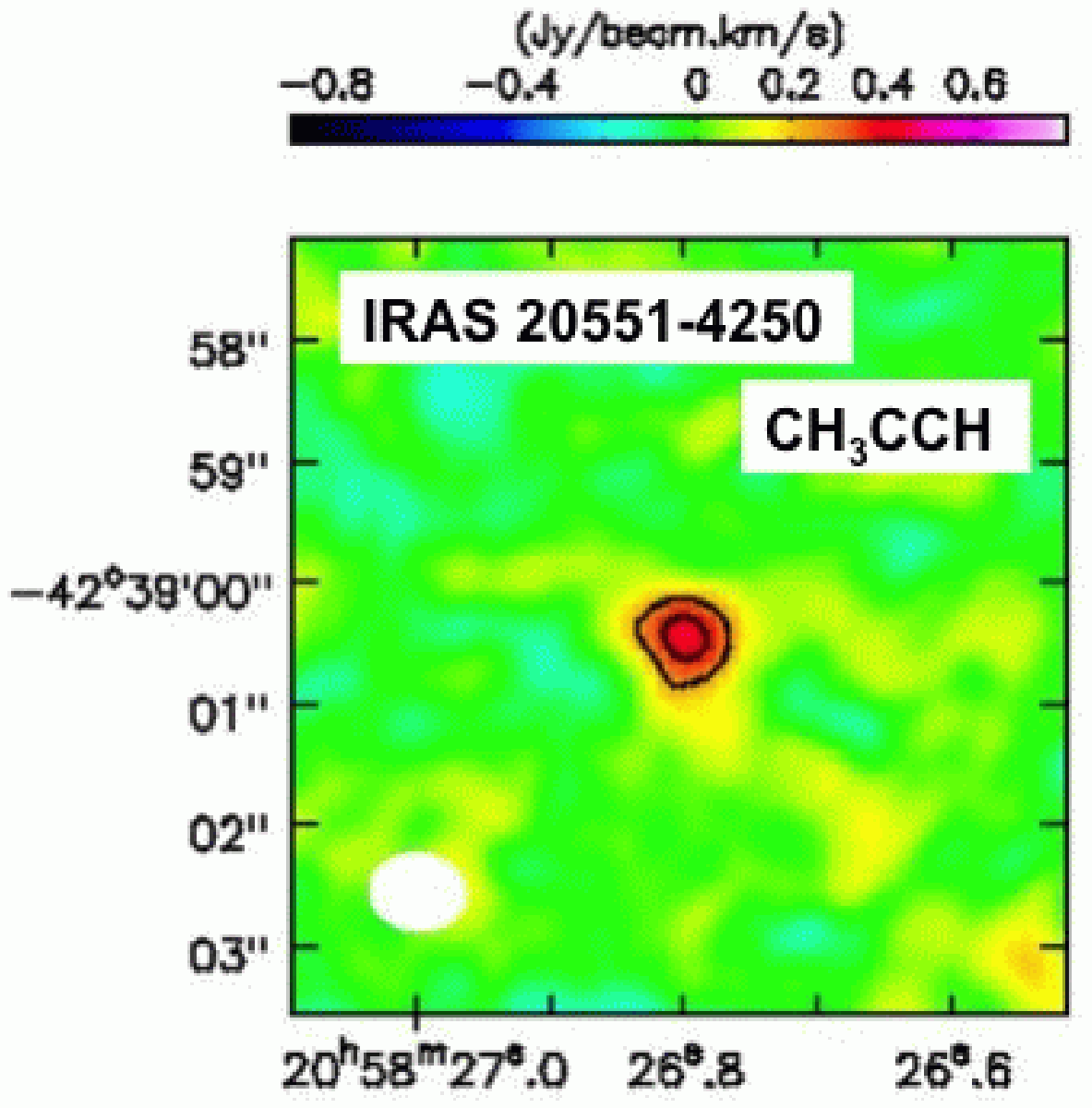} 
\includegraphics[angle=0,scale=.36]{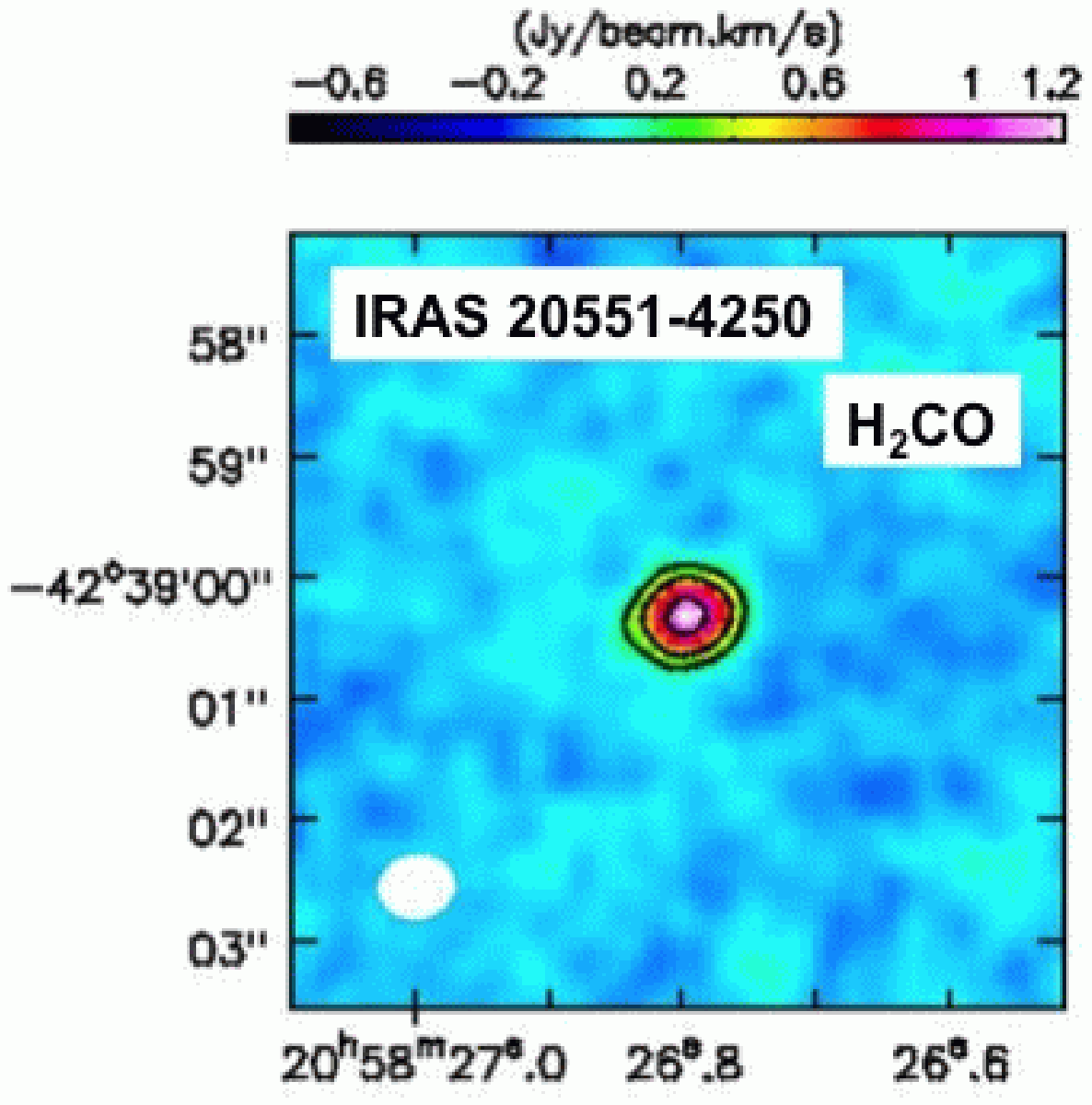} 
\includegraphics[angle=0,scale=.36]{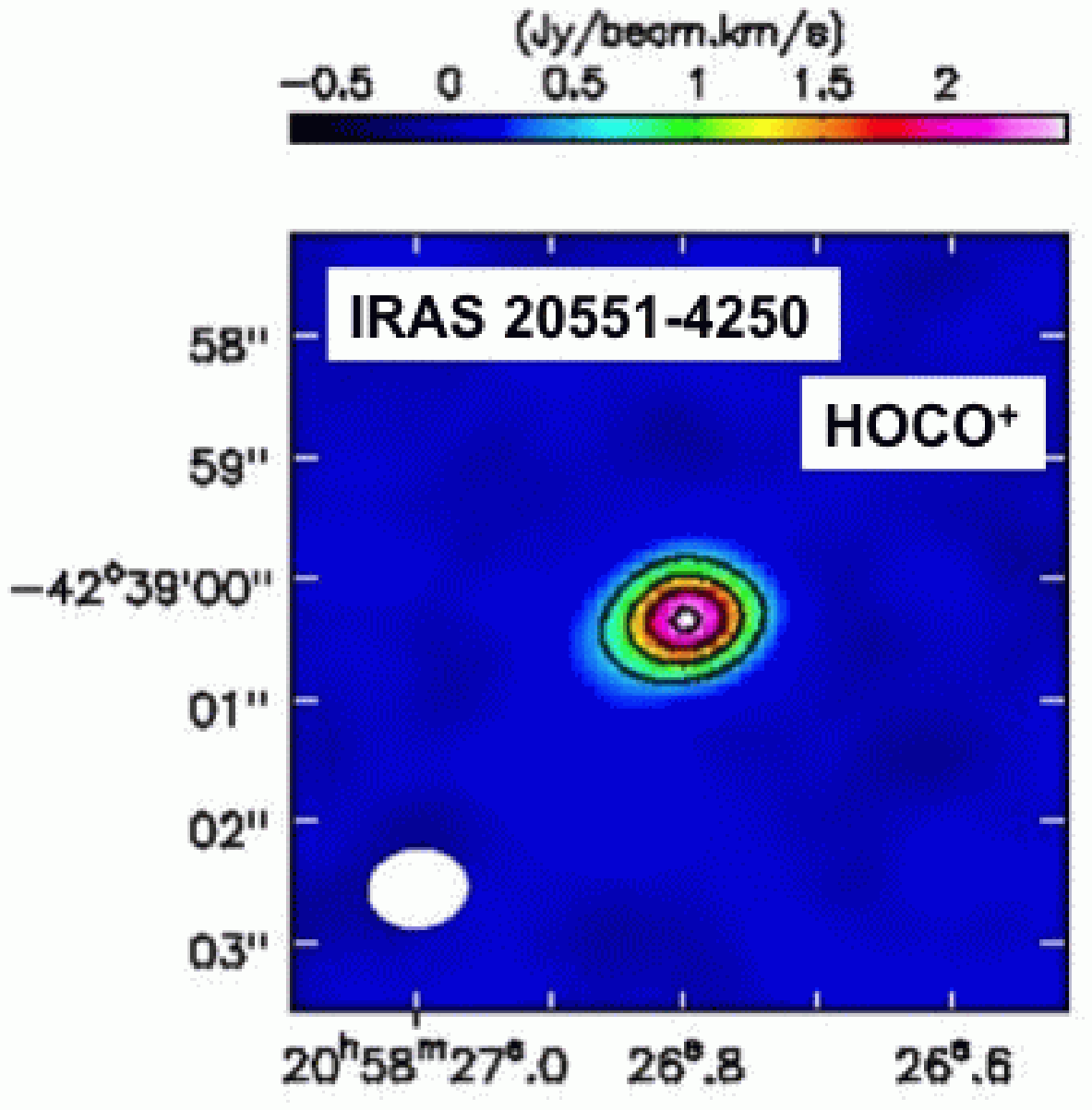} \\
\includegraphics[angle=0,scale=.36]{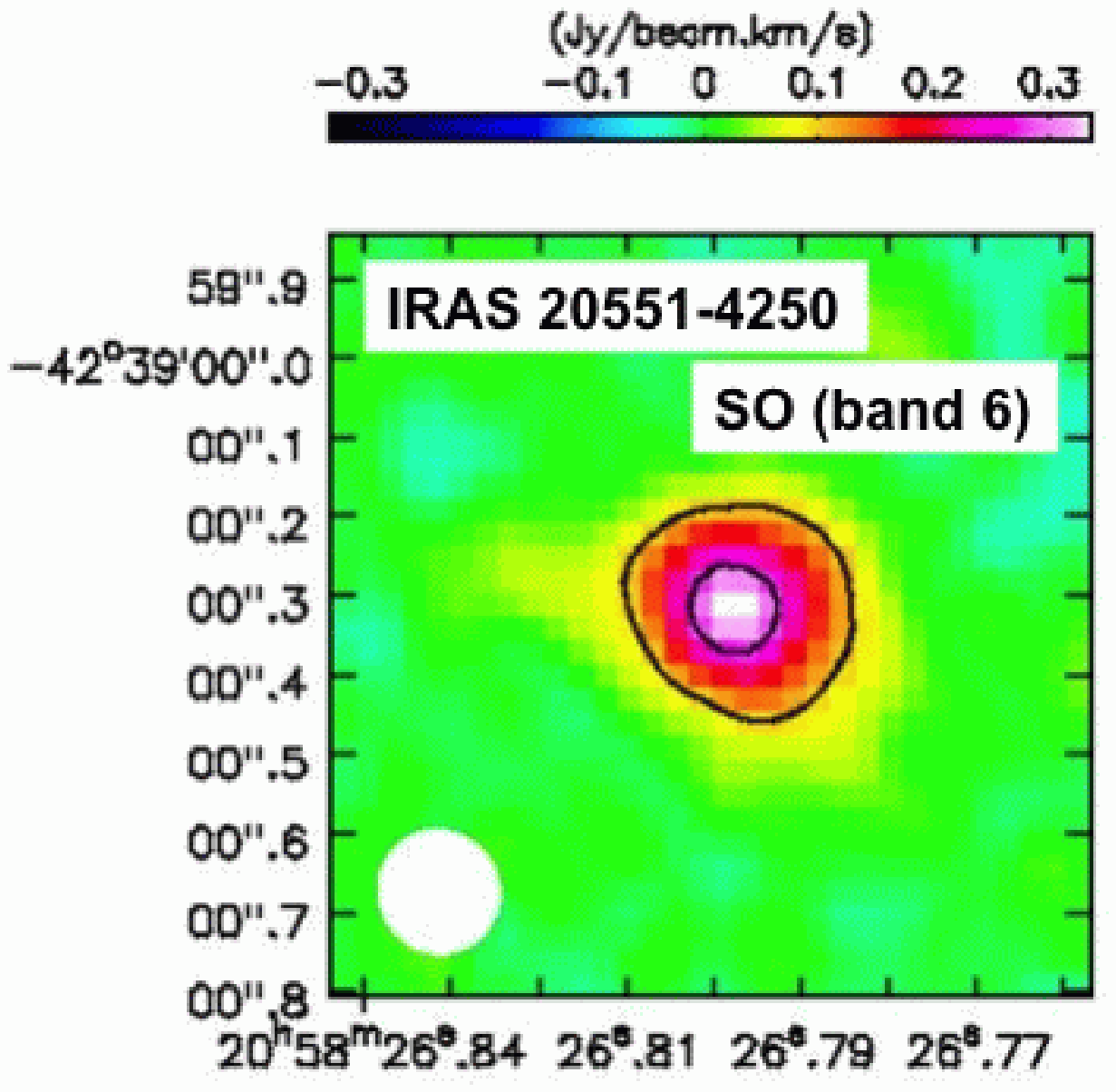} 
\includegraphics[angle=0,scale=.36]{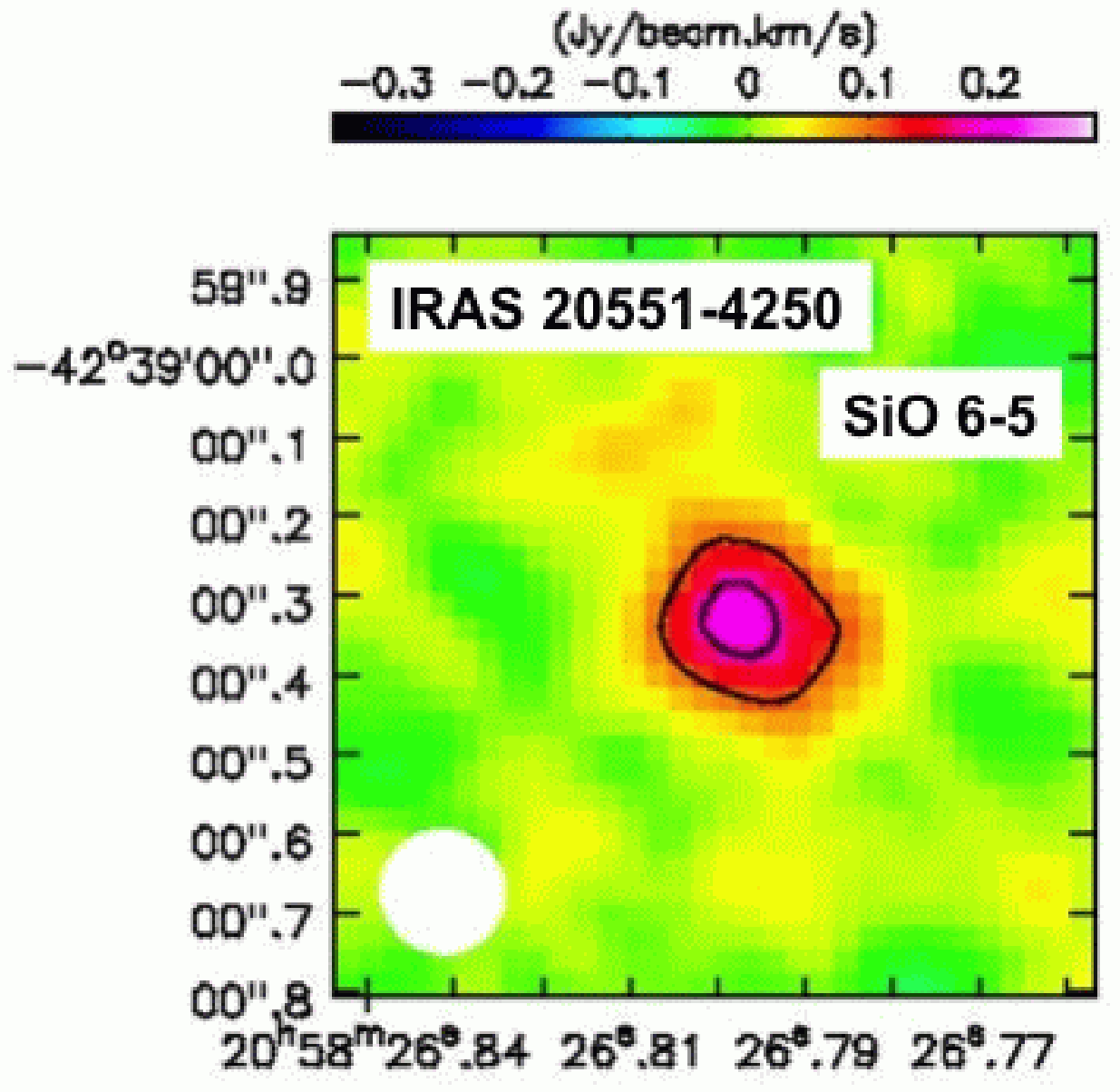} 
\includegraphics[angle=0,scale=.36]{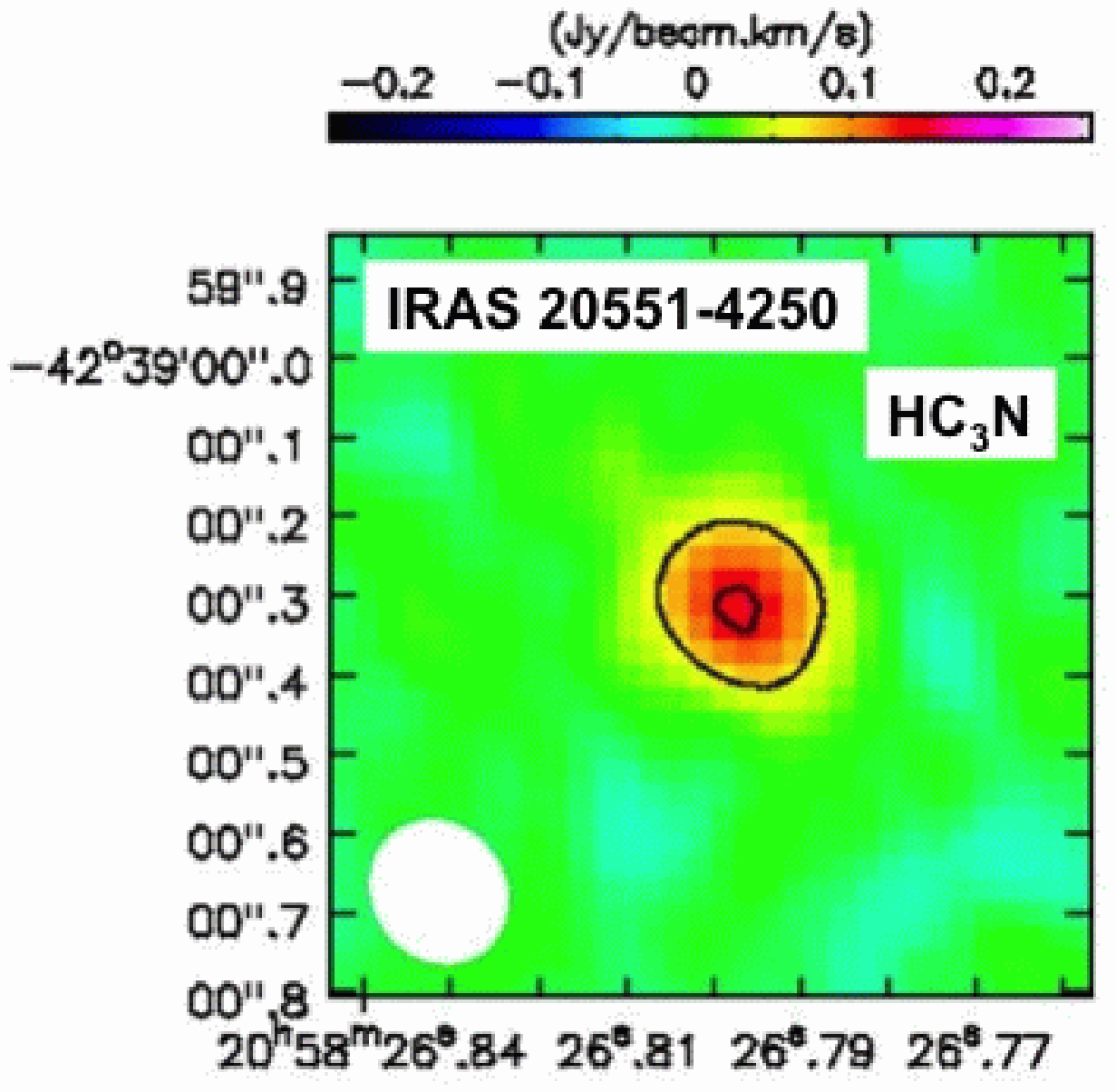} \\


\vspace{-0.1cm}
\caption{
Integrated intensity (moment 0) maps of the serendipitously detected
faint molecular lines in IRAS 20551$-$4250.  
Molecular lines detected in ALMA band 7 observations are displayed
first (first six images), followed by those in band 6 
(three images). 
The contours represent 
4$\sigma$, 5$\sigma$, 6$\sigma$ for HOC$^{+}$ J=4--3, 
4$\sigma$, 6$\sigma$, 8$\sigma$ for SO (band 7), 
4$\sigma$, 8$\sigma$, 12$\sigma$ for SO$_{2}$$+$SO, 
4$\sigma$, 6$\sigma$ for CH$_{3}$CCH, 
5$\sigma$, 10$\sigma$, 20$\sigma$ for H$_{2}$CO, 
10$\sigma$, 20$\sigma$, 30$\sigma$, 40$\sigma$ for HOCO$^{+}$, 
3$\sigma$, 8$\sigma$ for SO (band 6), 
3$\sigma$, 5$\sigma$ SiO J=6--5, and 
3$\sigma$, 6$\sigma$ for HC$_{3}$N J=27--26.
The 1$\sigma$ levels are different for different molecular lines, and
are summarized in Table 4. 
Beam sizes are shown as filled circles in the lower-left region.
The displayed areas differ depending on the beam size.
}
\end{center}
\end{figure}

\begin{figure}
\begin{center}
\includegraphics[angle=0,scale=.274]{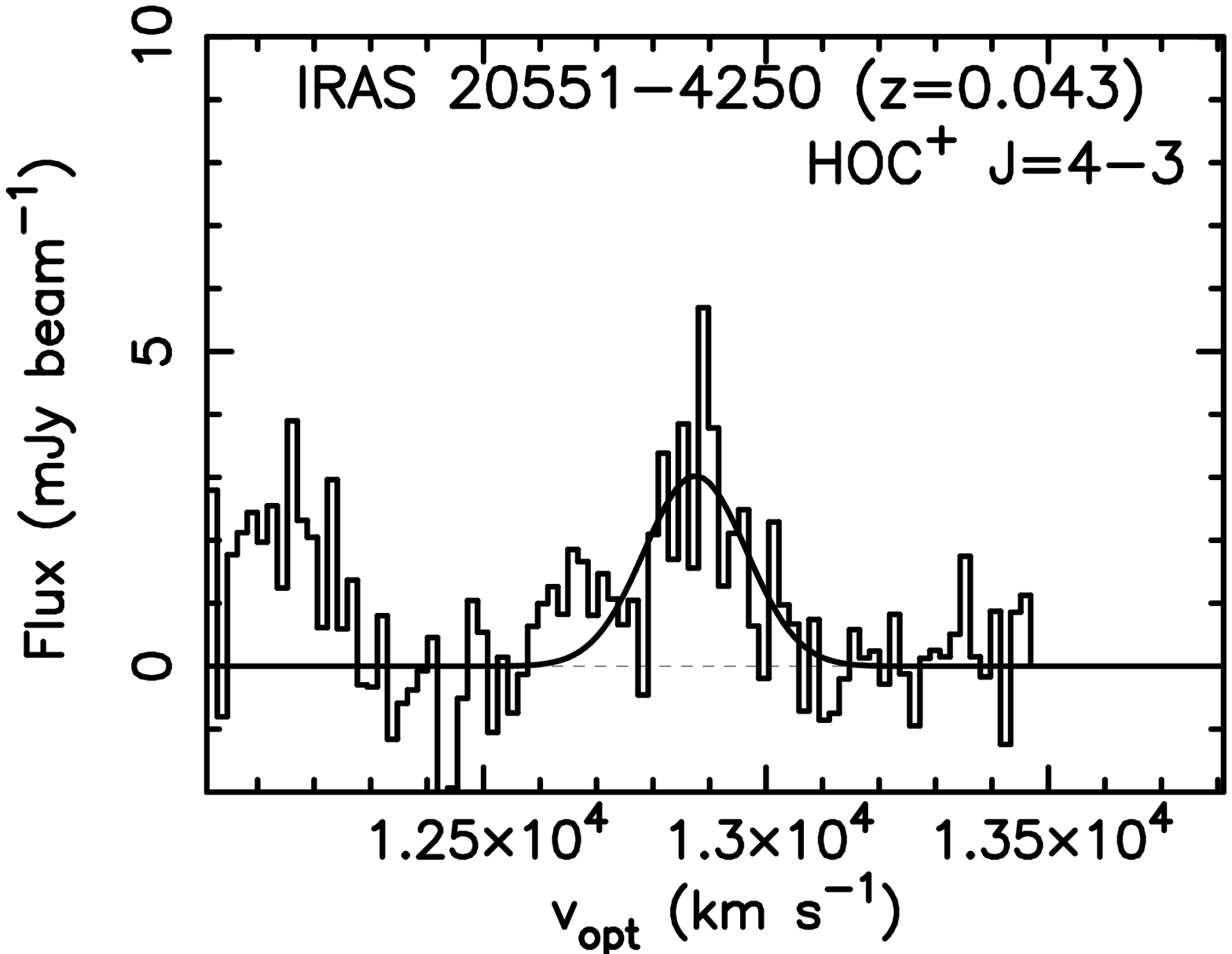} 
\includegraphics[angle=0,scale=.274]{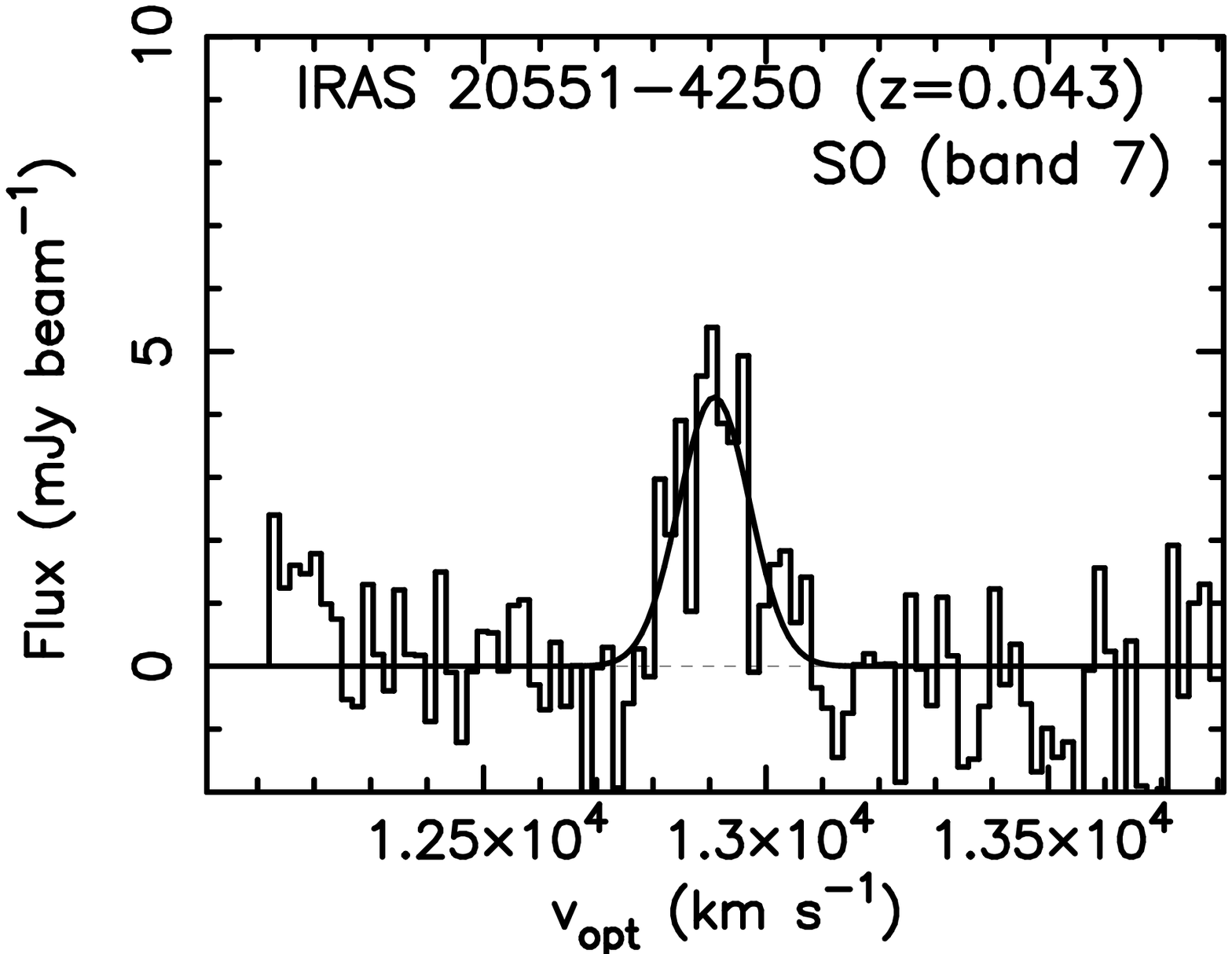} 
\includegraphics[angle=0,scale=.274]{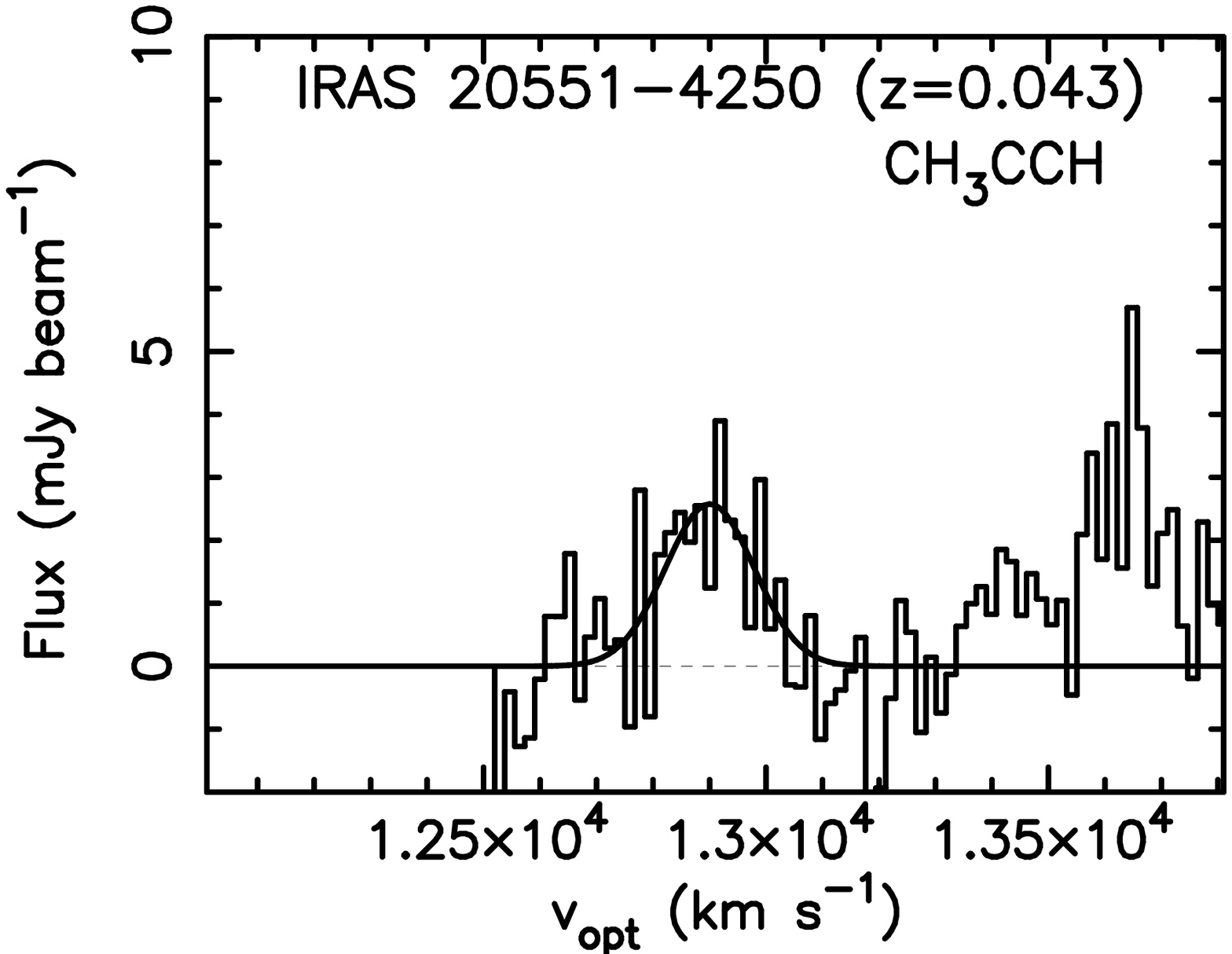} \\
\includegraphics[angle=0,scale=.274]{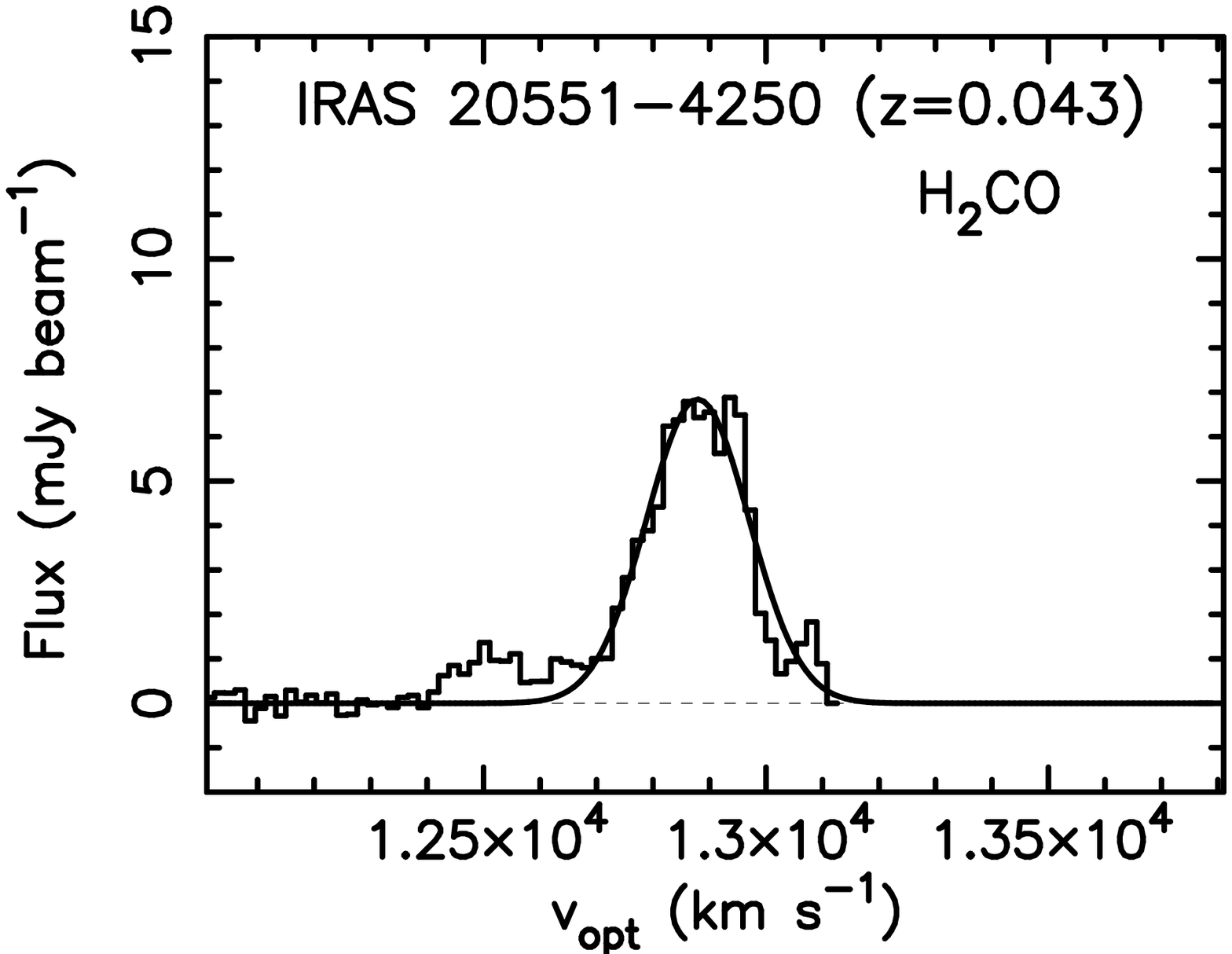} 
\includegraphics[angle=0,scale=.274]{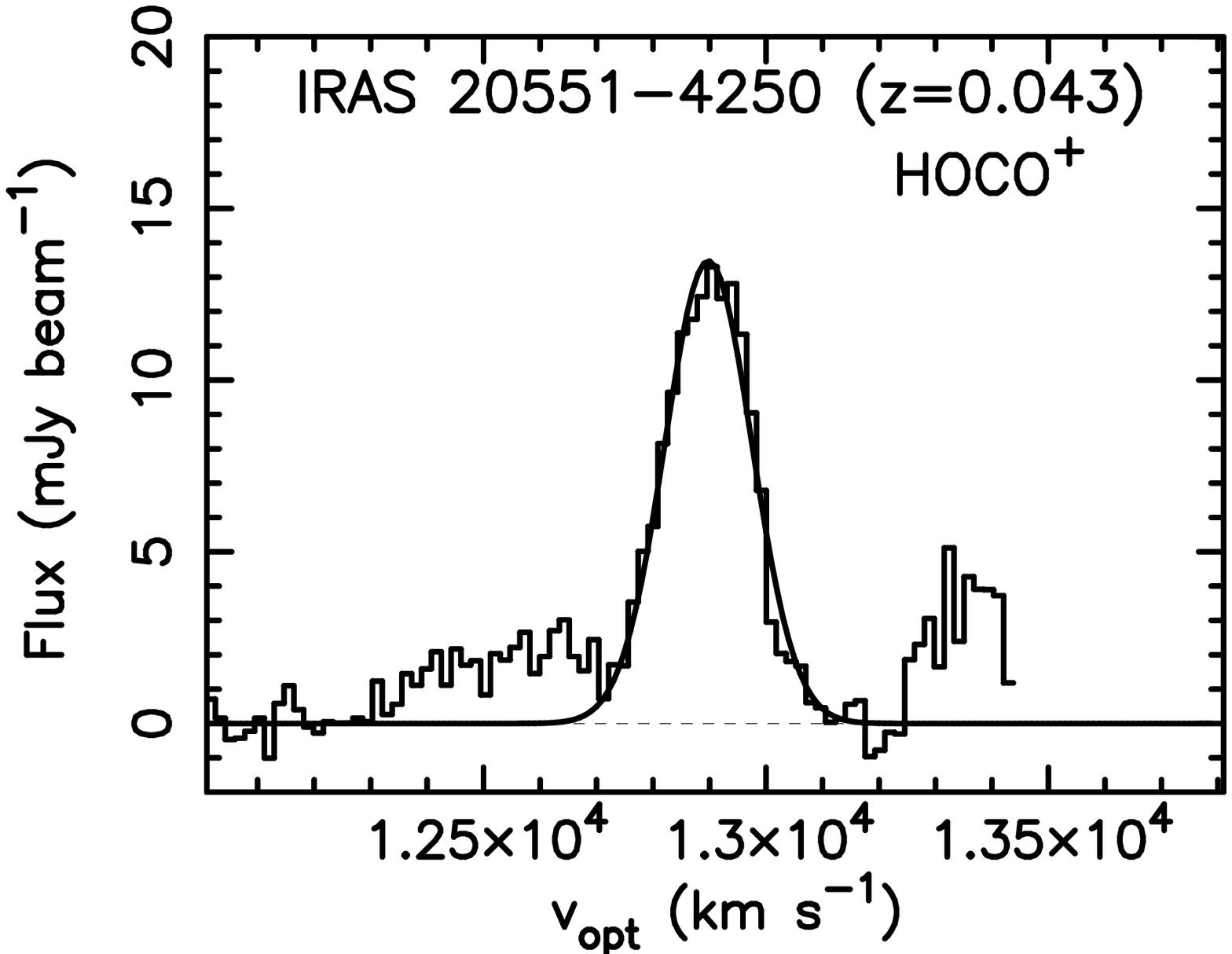} 
\includegraphics[angle=0,scale=.274]{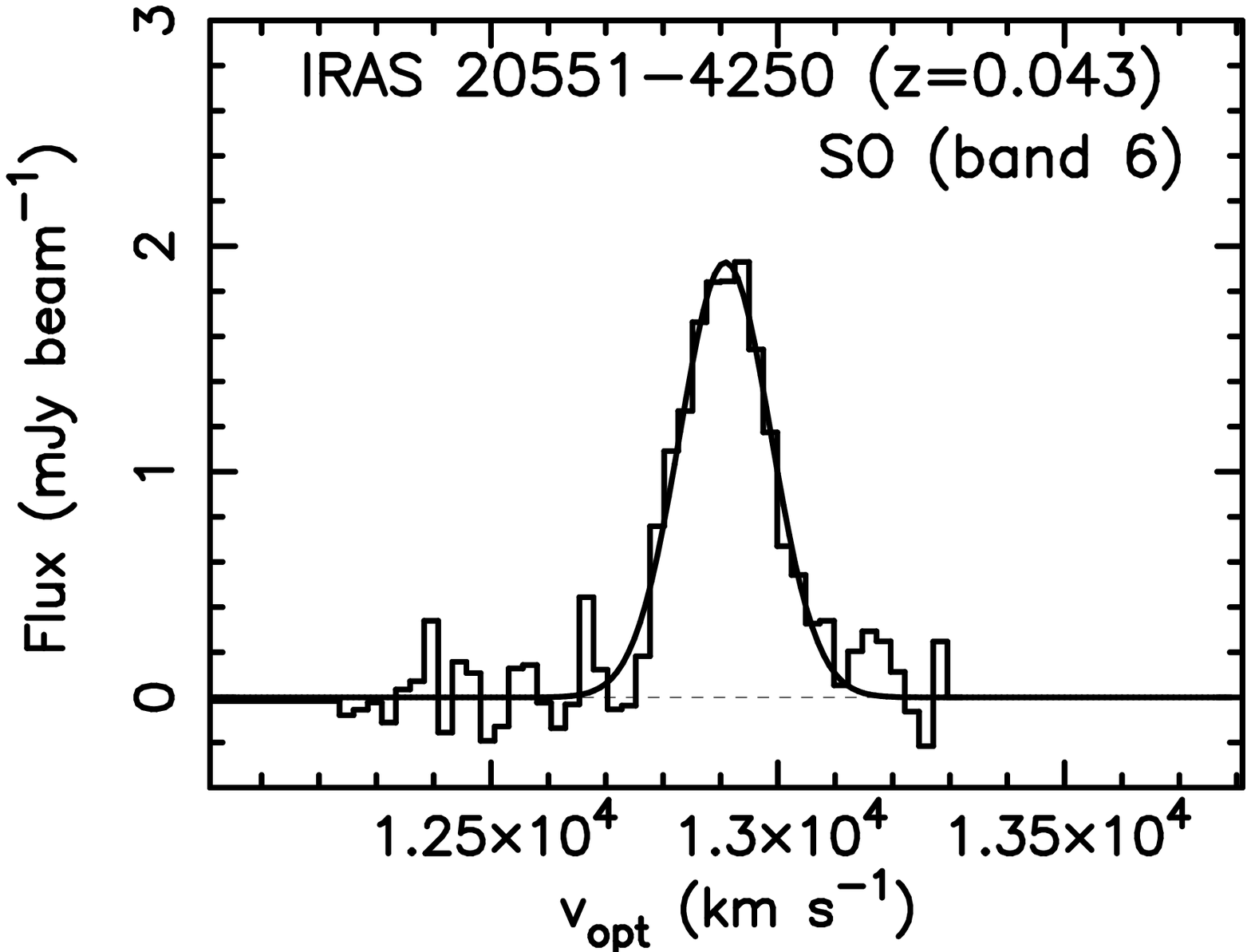} \\
\includegraphics[angle=0,scale=.274]{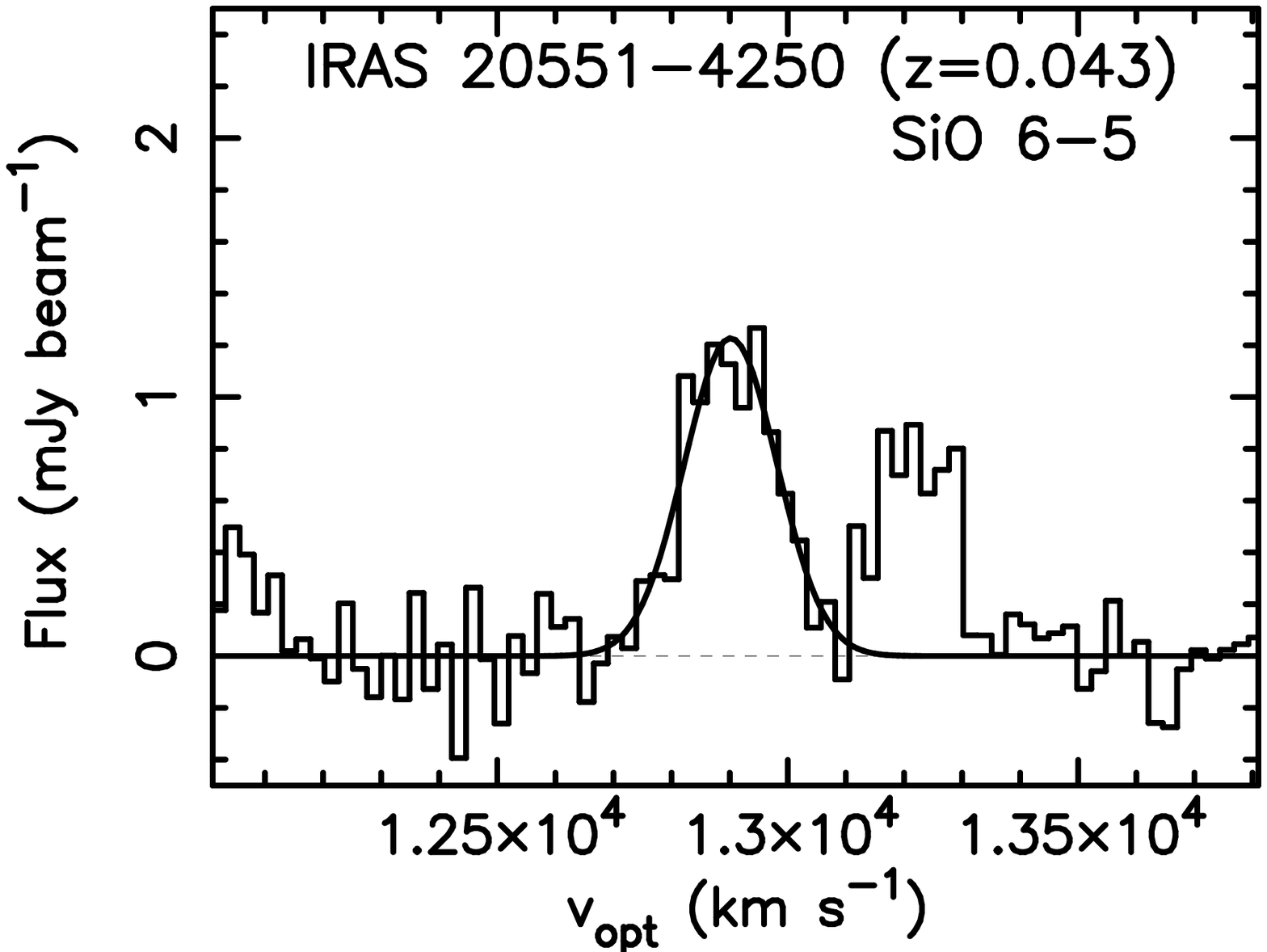} 
\includegraphics[angle=0,scale=.274]{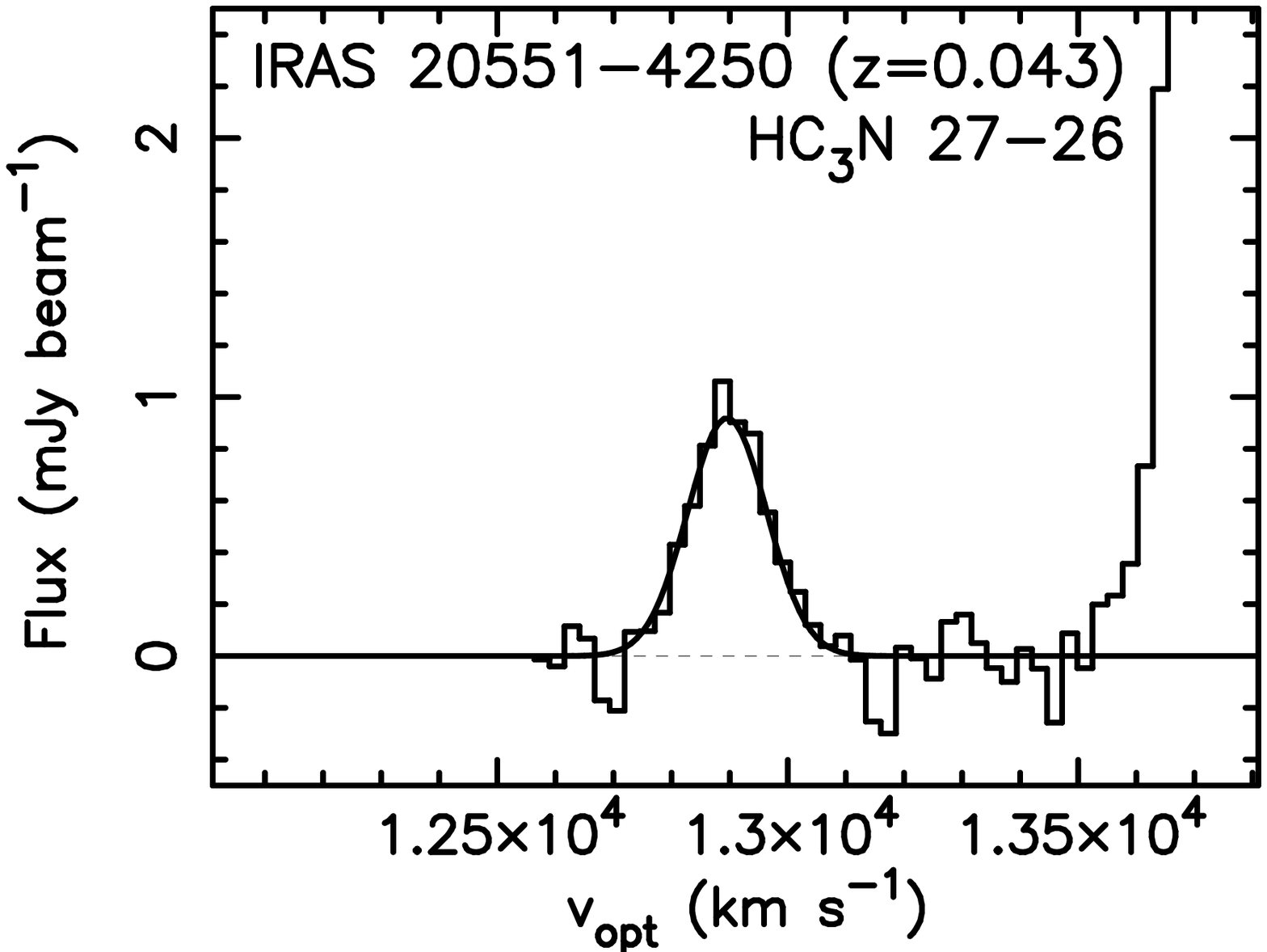} \\
\caption{
Spectra around selected serendipitously detected faint molecular
emission lines. 
The abscissa is optical LSR velocity (v$_{\rm opt}$
$\equiv$ c ($\lambda$$-$$\lambda_{\rm 0}$)/$\lambda_{\rm 0}$), and the
ordinate is flux in (mJy beam$^{-1}$).  
Best Gaussian fits are overplotted with solid curved lines.
}
\end{center}
\end{figure}

\begin{figure}
\begin{center}
\includegraphics[angle=0,scale=.41]{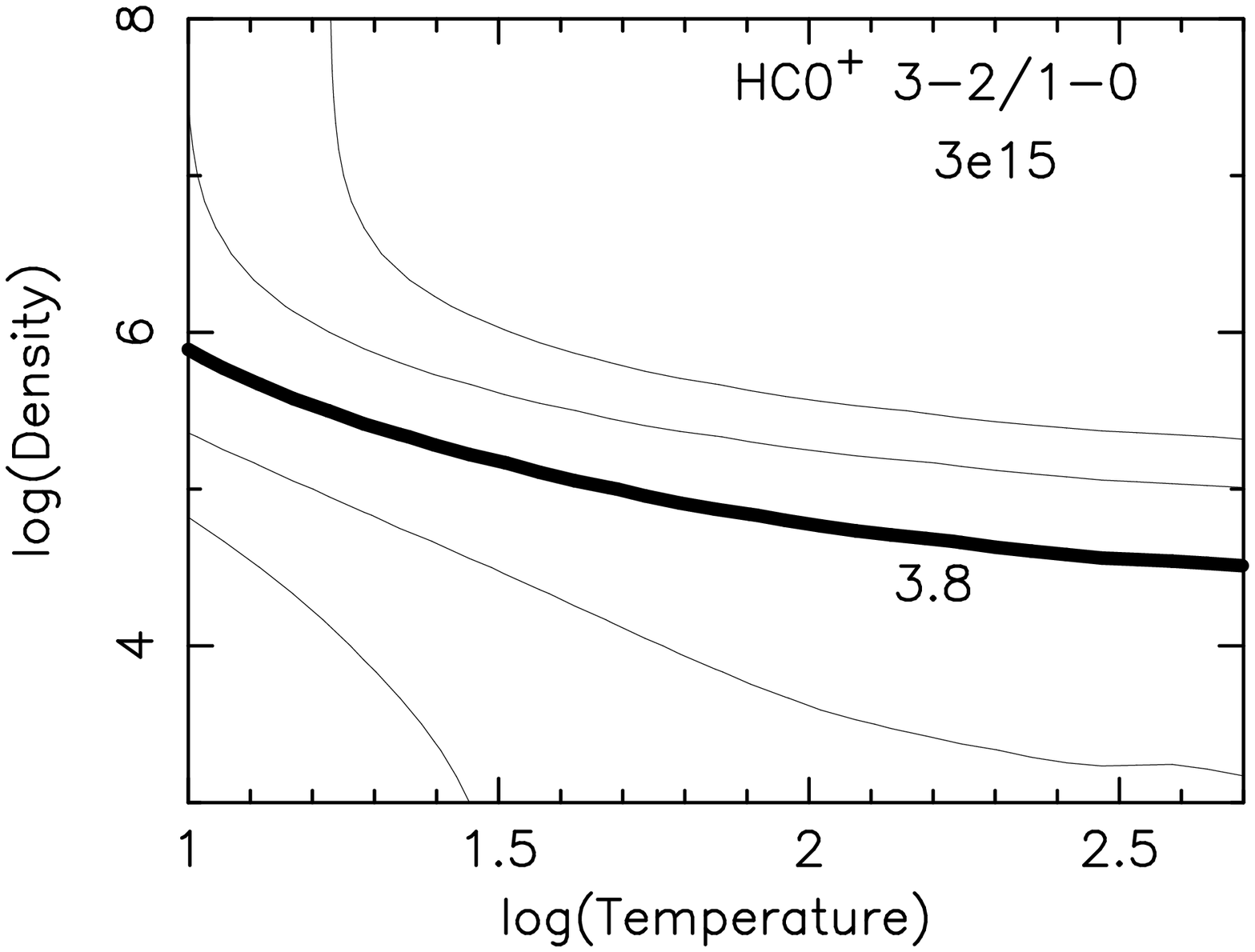}  
\includegraphics[angle=0,scale=.41]{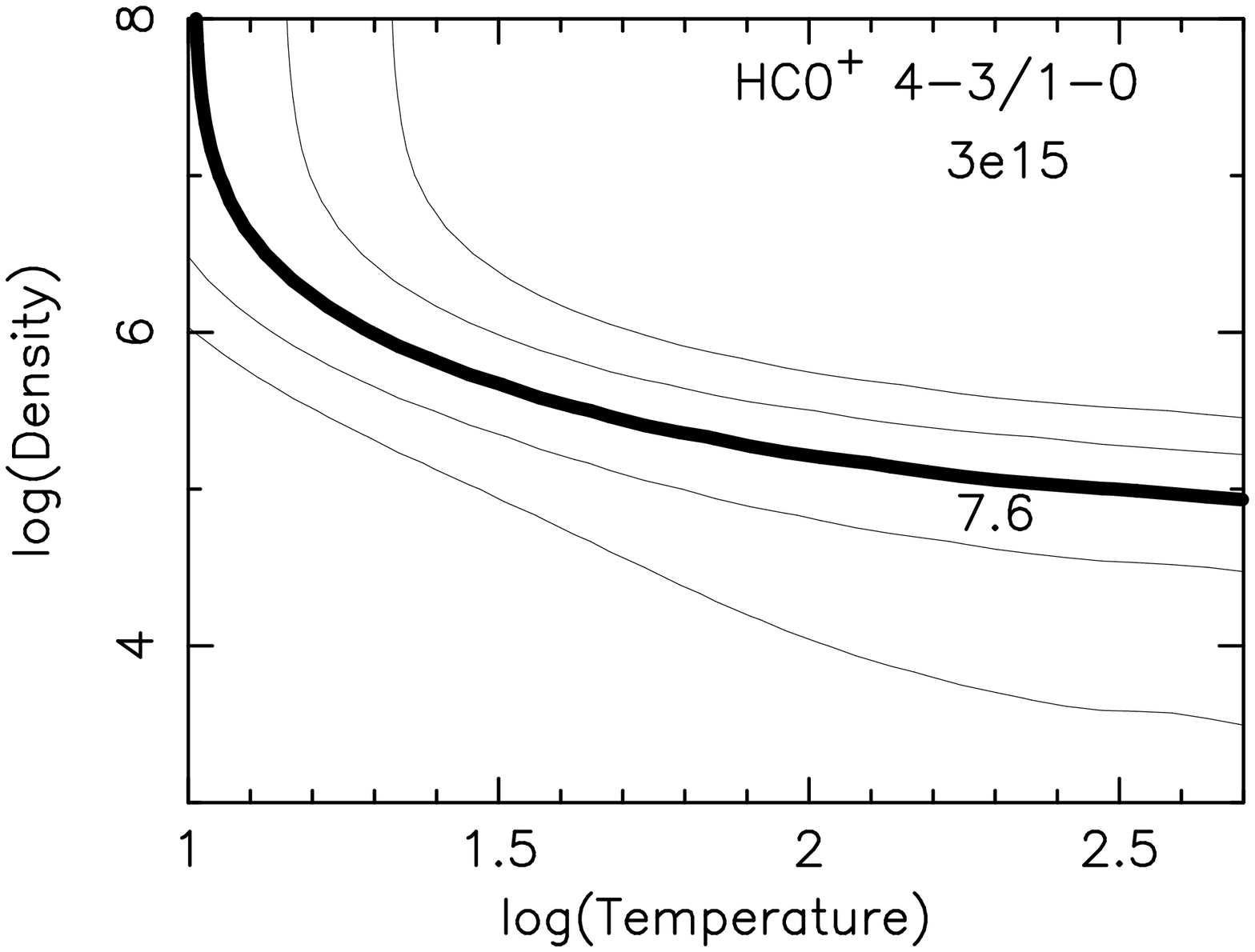}  \\
\includegraphics[angle=0,scale=.41]{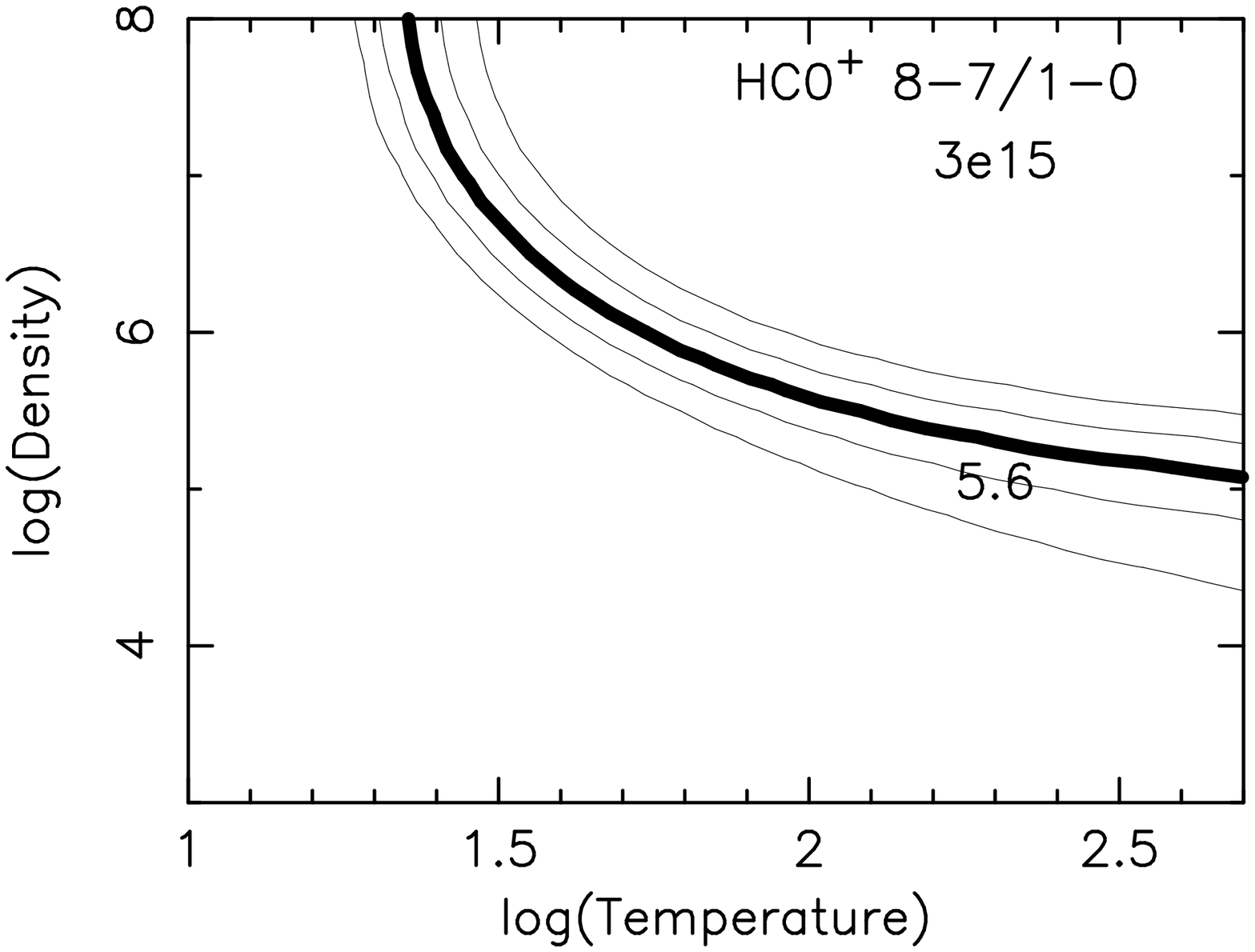}  
\includegraphics[angle=0,scale=.41]{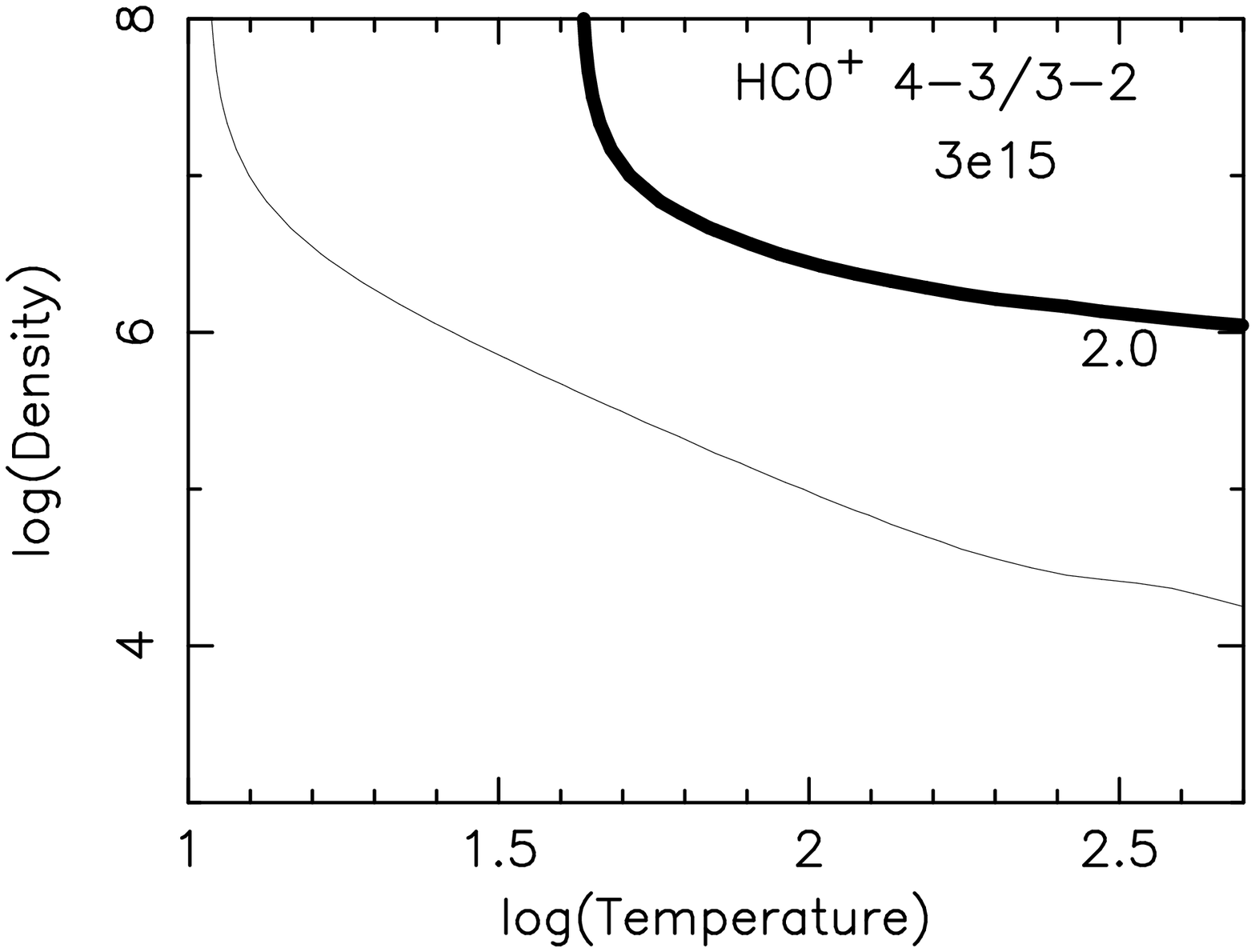}  \\
\includegraphics[angle=0,scale=.41]{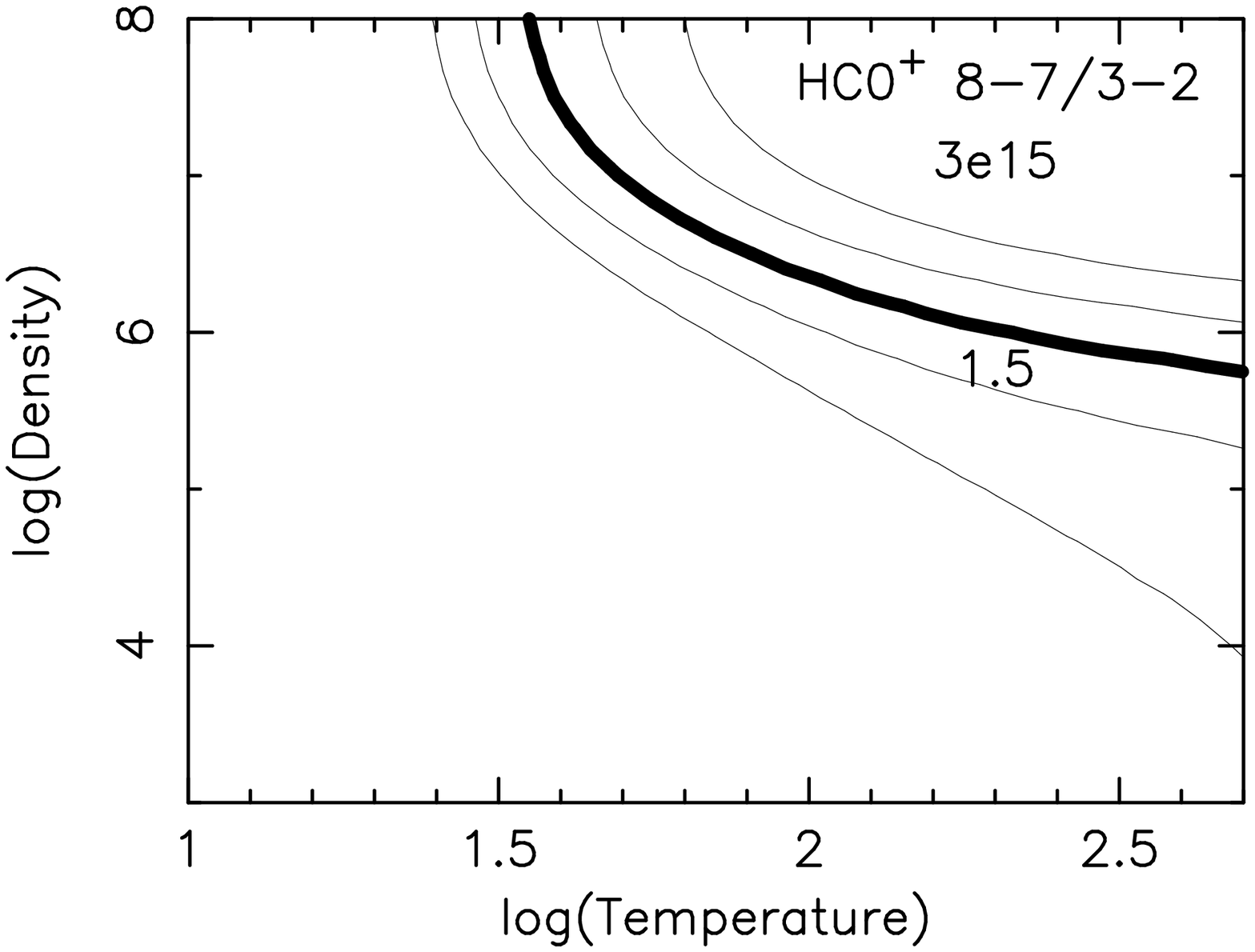}  
\includegraphics[angle=0,scale=.41]{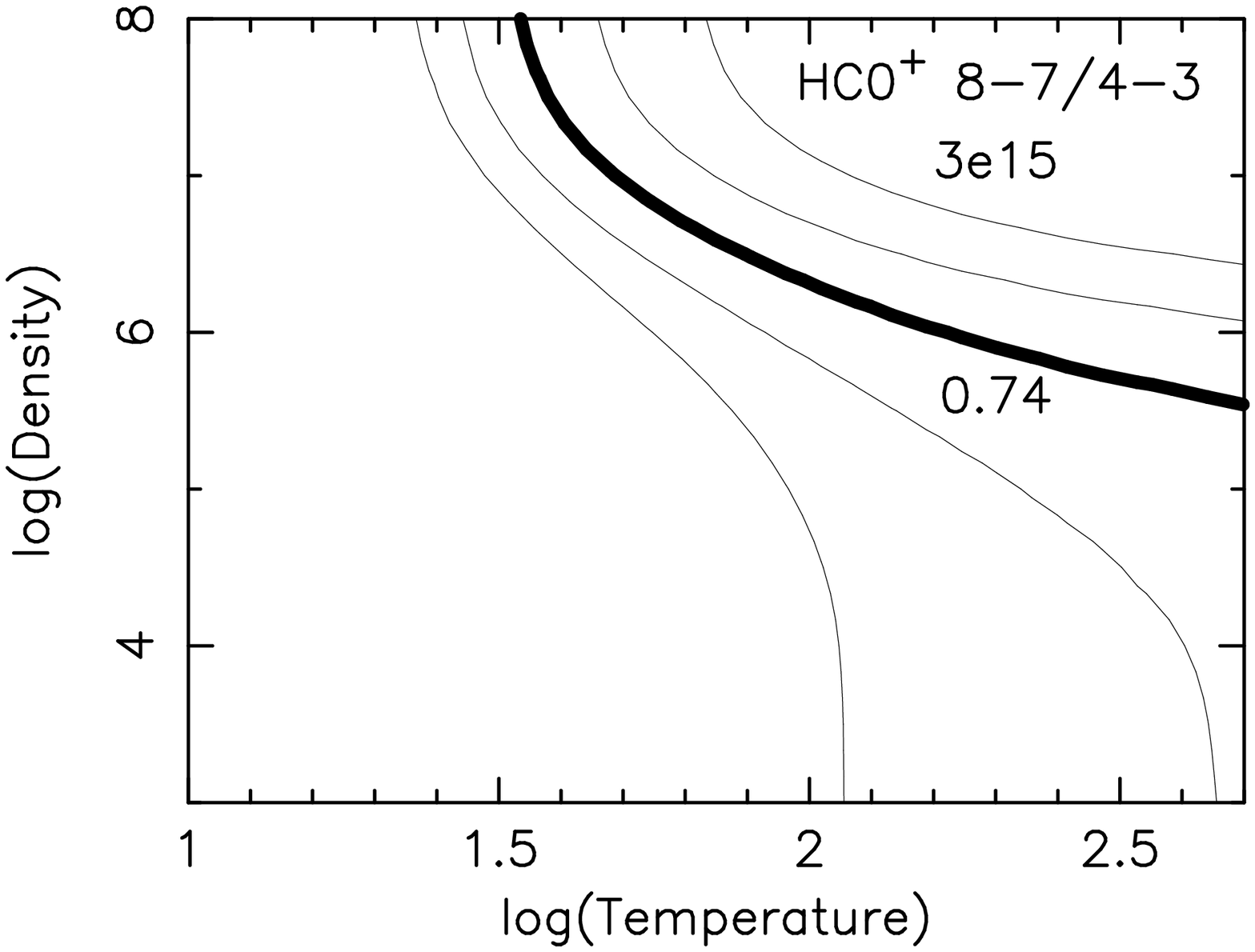}  
\end{center}
\end{figure}

\clearpage 

\begin{figure}
\begin{center}
\includegraphics[angle=0,scale=.41]{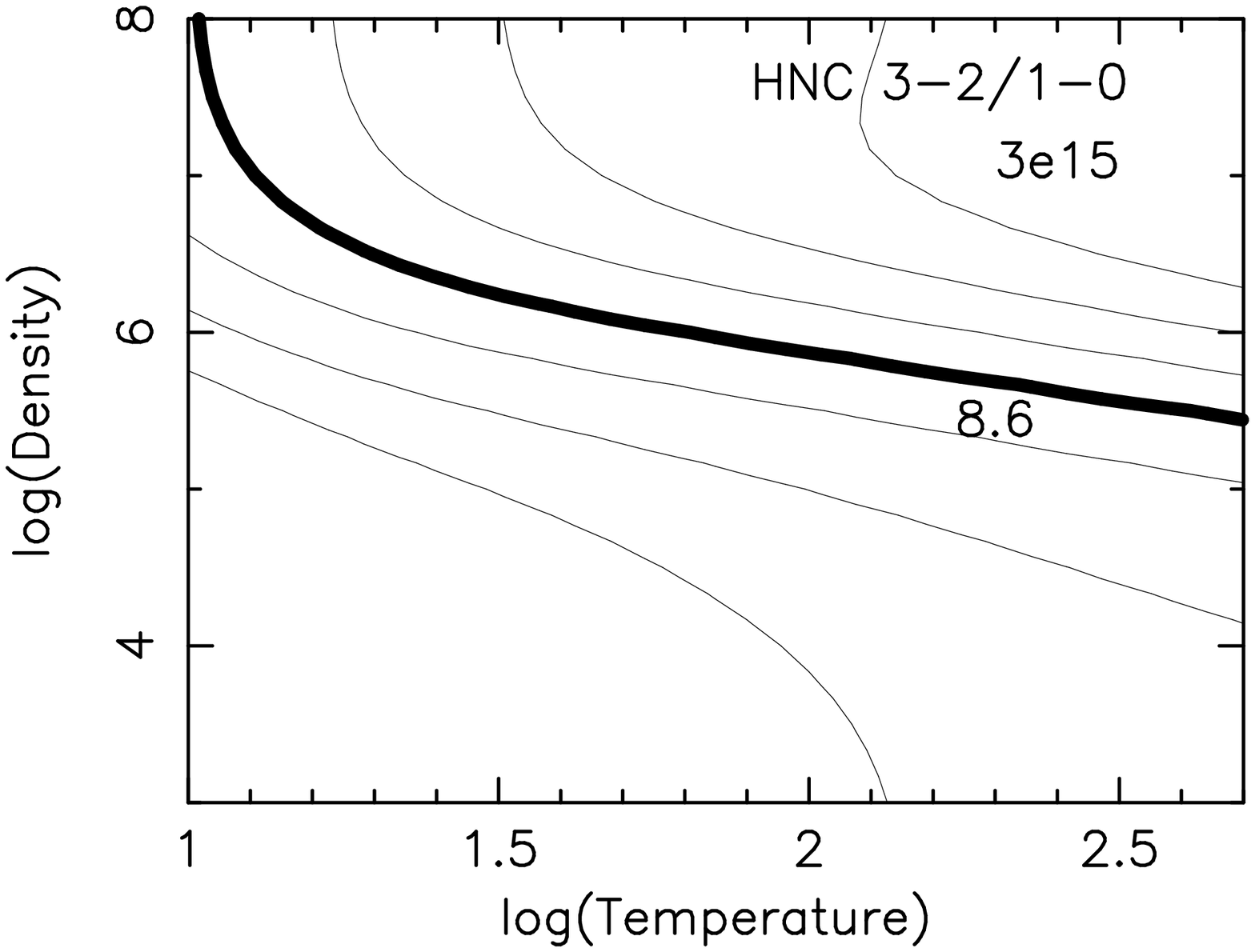}  
\includegraphics[angle=0,scale=.41]{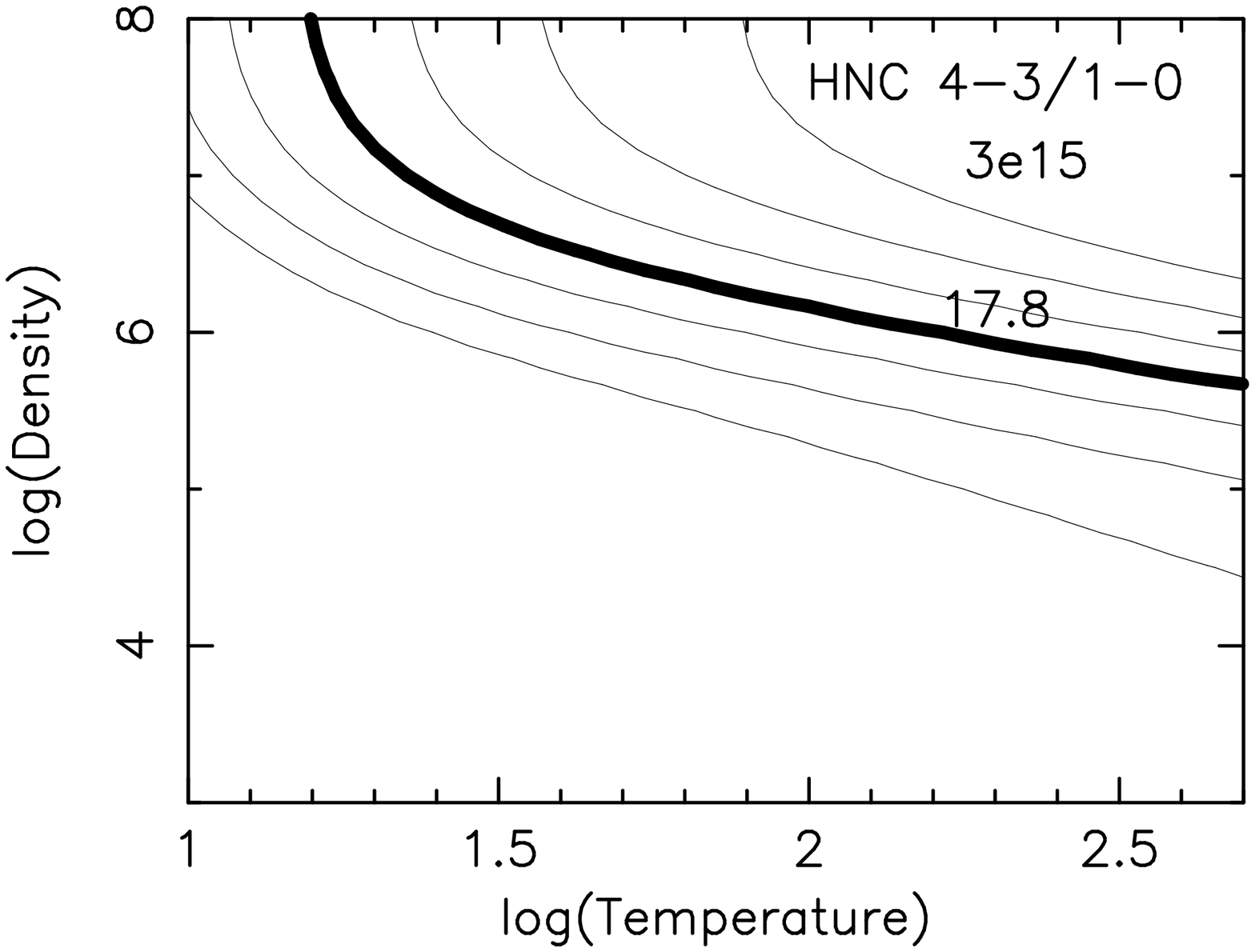} \\ 
\includegraphics[angle=0,scale=.41]{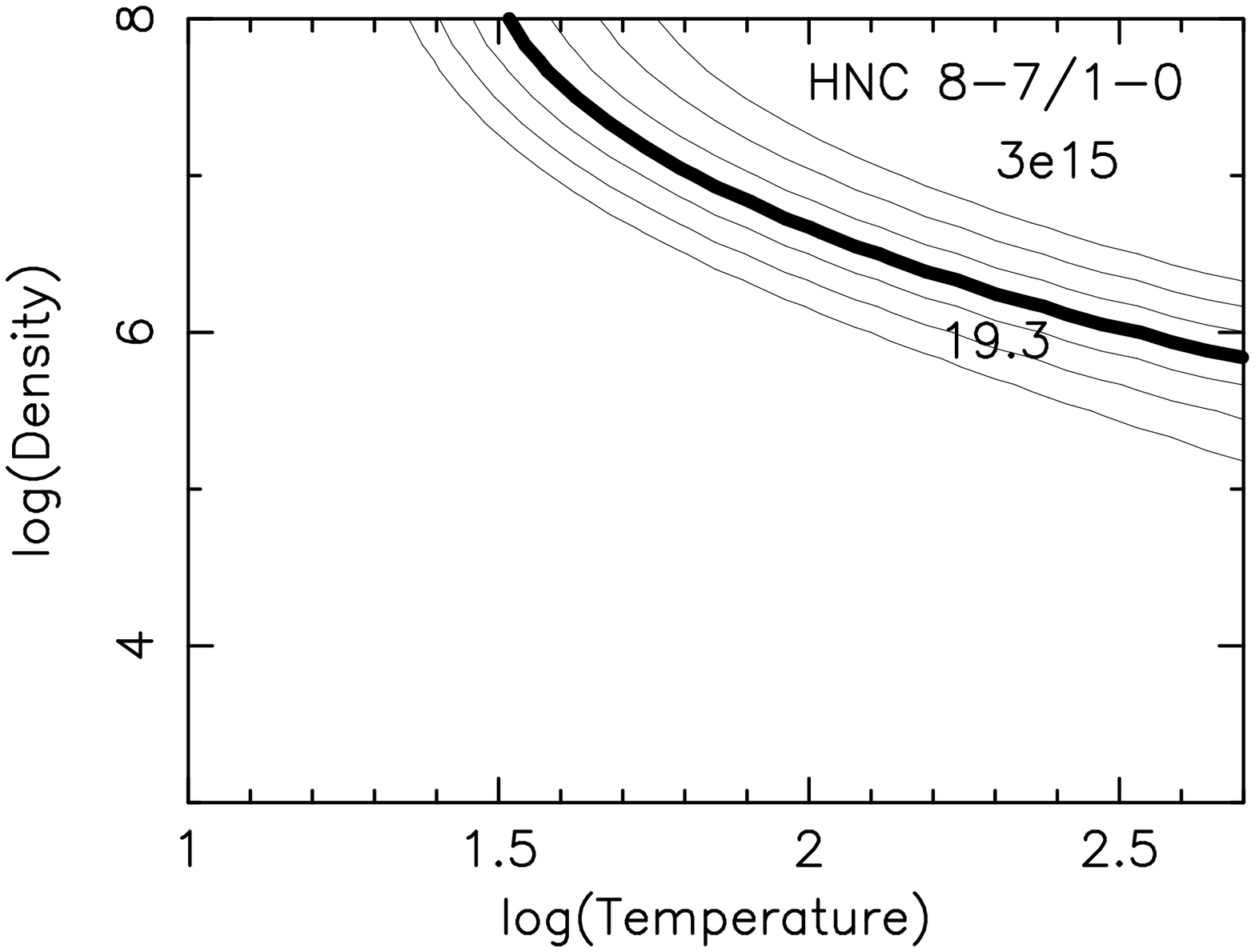}  
\includegraphics[angle=0,scale=.41]{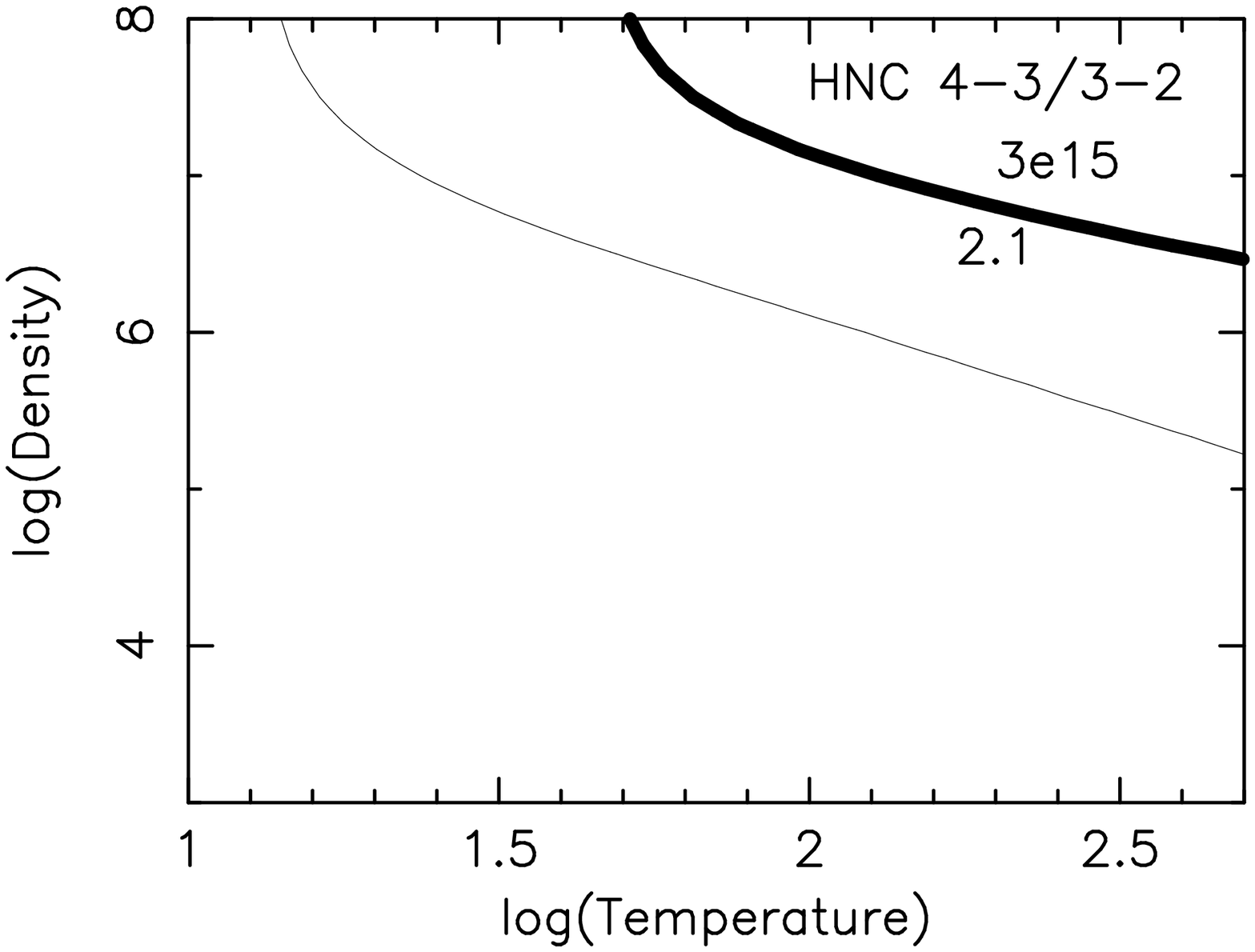} \\ 
\includegraphics[angle=0,scale=.41]{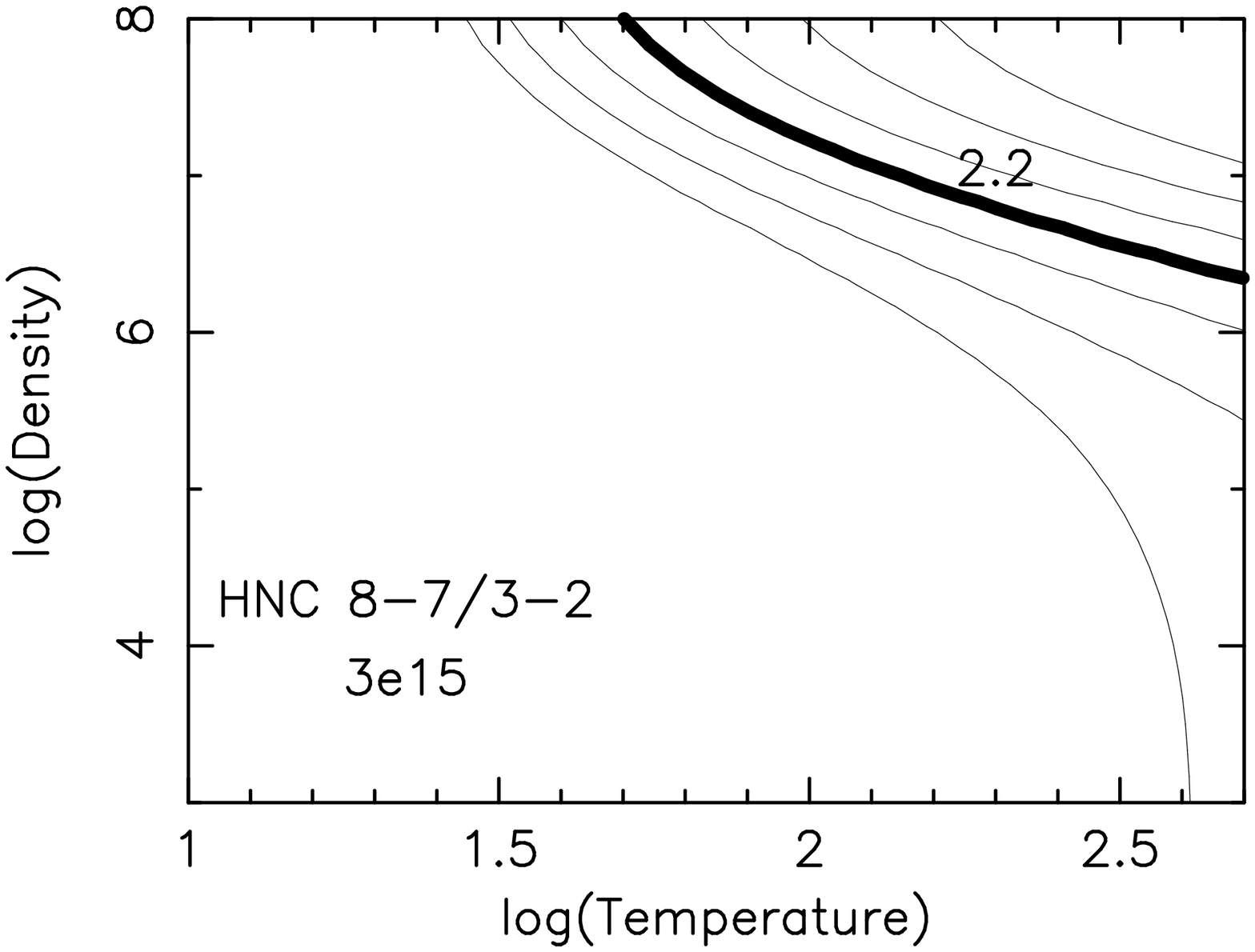}  
\includegraphics[angle=0,scale=.41]{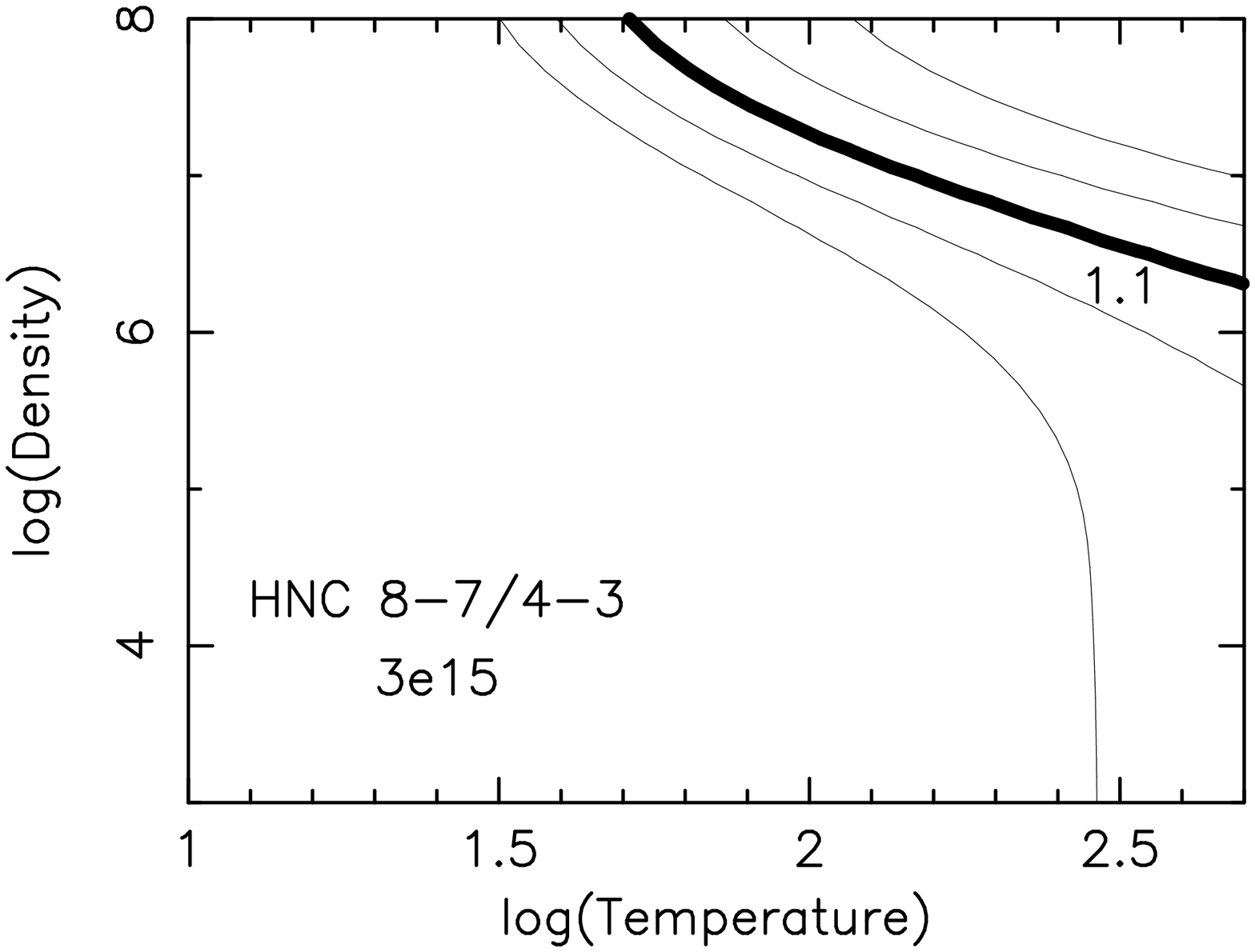}  
\caption{
RADEX calculation for the column density of 3 $\times$ 
10$^{15}$ cm$^{-2}$ for HCO$^{+}$ and HNC.
Contours are shown in steps of a factor of 2, with higher (lower) values
in the upper-right (lower-left) direction in all plots.
Parameters that reproduce the ratios of the observed fluxes in (Jy km
s$^{-1}$) are shown as thick curved lines with numbers.
}
\end{center}
\end{figure}

\end{document}